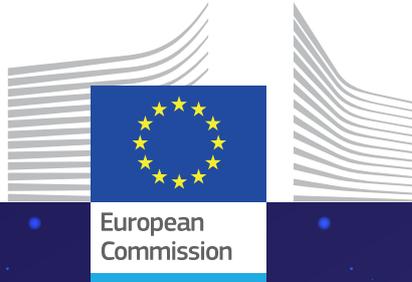

European Commission

# JRC SCIENCE FOR POLICY REPORT

## Data Innovation in Demography, Migration and Human Mobility

Opportunities and challenges of non-traditional data

Joint Research Centre

This publication is a Science for Policy report by the Joint Research Centre (J.R.C.), the European Commission's science and knowledge service. It aims to provide evidence-based scientific support to the European policymaking process. The scientific output expressed does not imply a policy position of the European Commission. Neither the European Commission nor any person acting on behalf of the Commission is responsible for the use that might be made of this publication. For information on the methodology and quality underlying the data used in this publication for which the source is neither Eurostat nor other Commission services, users should contact the referenced source. The designations employed and the presentation of material on the maps do not imply the expression of any opinion whatsoever on the part of the European Union concerning the legal status of any country, territory, city or area or of its authorities, or concerning the delimitation of its frontiers or boundaries.

Data Innovation in Demography, Migration and Human Mobility

Contact information
Knowledge Centre on Migration and Demography (KCMD)
Joint Research Centre, Via Enrico Fermi 2749, 20127 Ispra (VA), Italy
Email: EC-KCMD@ec.europa.eu

EU Science Hub
https://ec.europa.eu/jrc

JRC127369

EUR 30907 EN

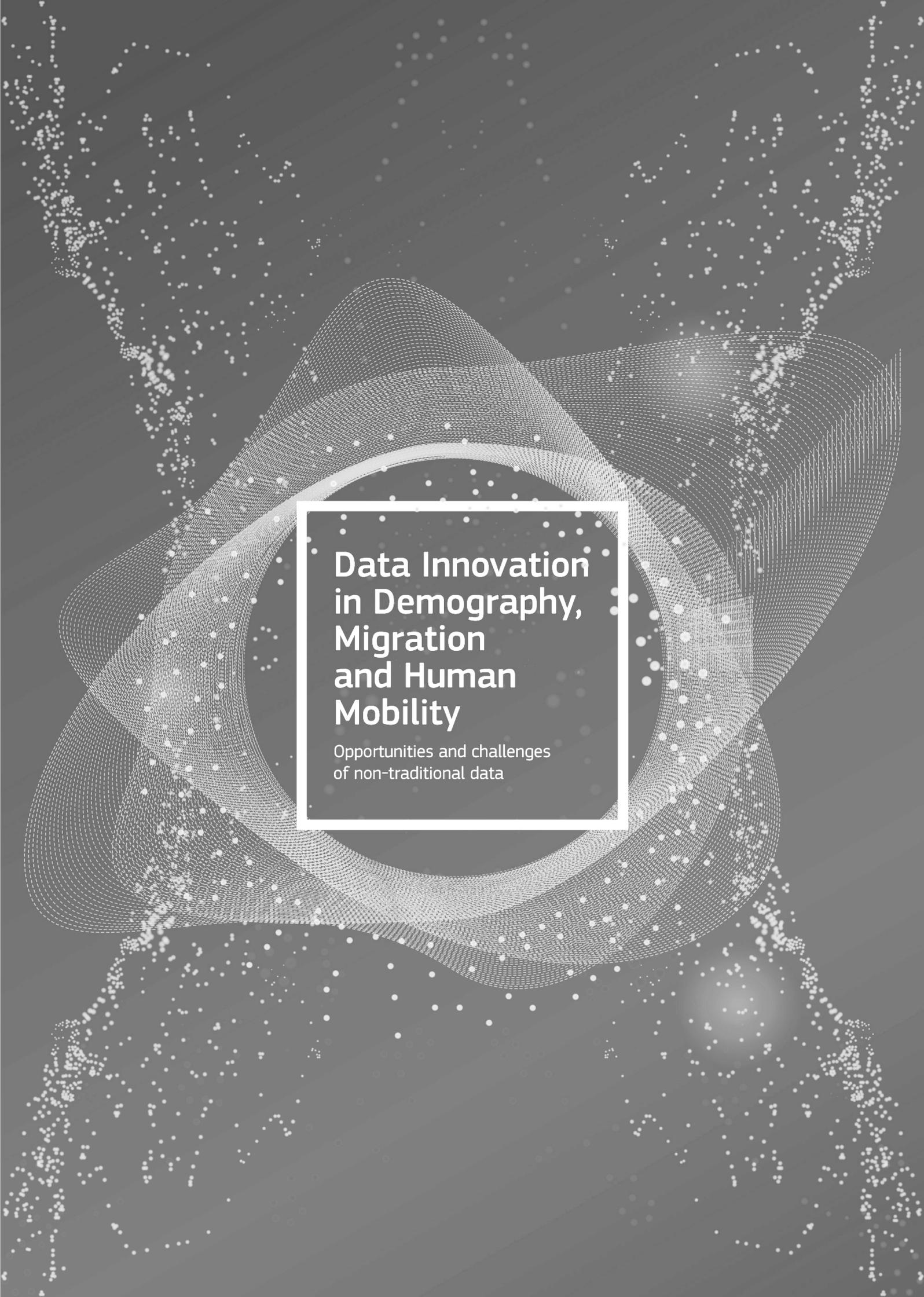

# Data Innovation in Demography, Migration and Human Mobility

Opportunities and challenges
of non-traditional data

# TABLE
# OF CONTENTS

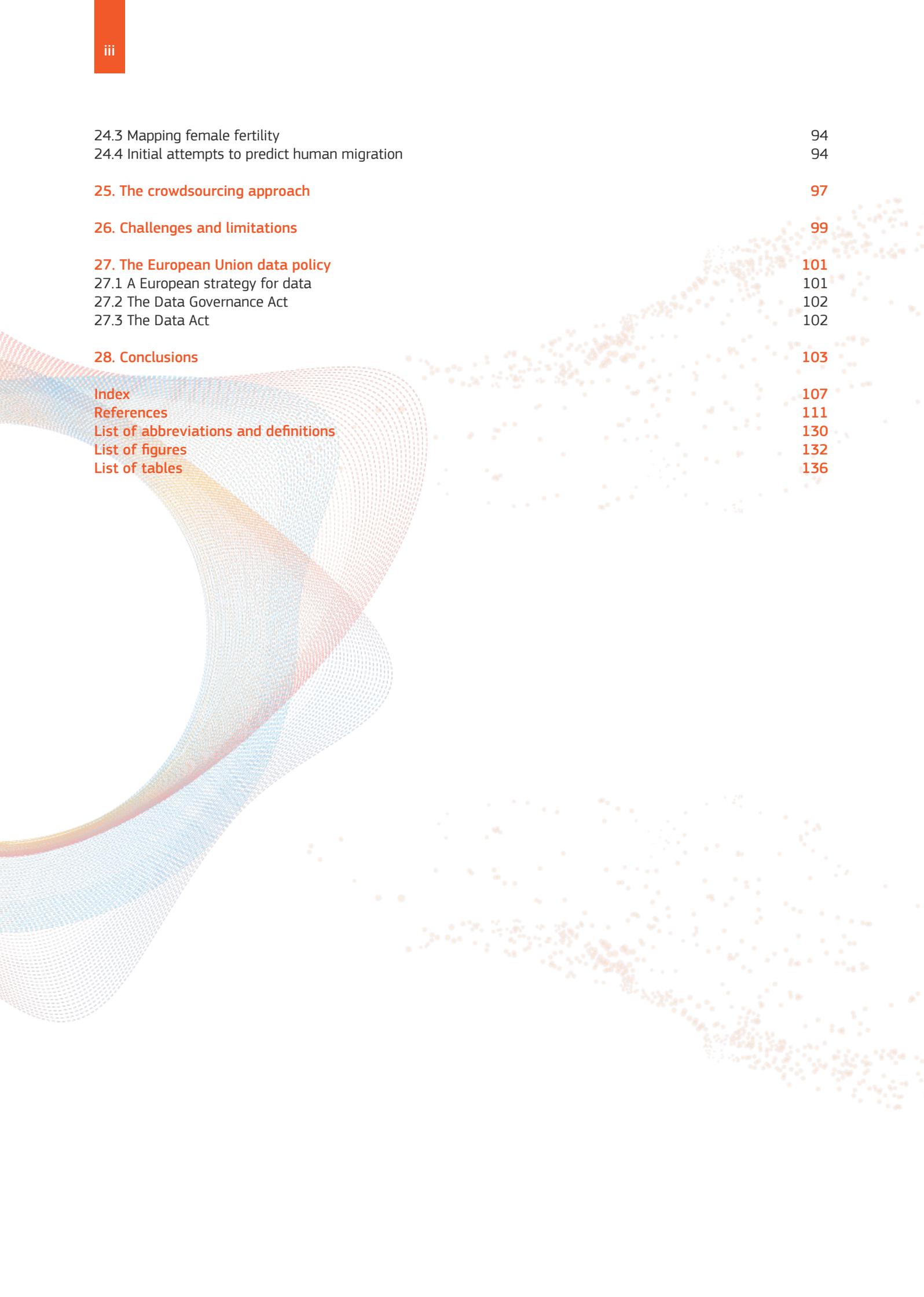

# ACKNOWLEDGEMENTS

The authors wish to thank the following colleagues of the European Commission's Knowledge Centre on Migration and Demography at JRC for an early reading of this report: Peter Bosch, Daniela Ghio, Anne Goujon, Nina Kajander, Sona Kalantaryan, Simon McMahon, Dario Tarchi and Michele Vespe.

The following experts are also kindly acknowledged for a critical reading of the manuscript: Iacopo Viciani (DG INTPA), Lewis Dijkstra and Magdalena Grzegorzewka (DG REGIO), Pawel Busiakiewicz, Laura Corrado, Lorenzo Franceschelli and Sabina Sirokovska (DG HOME), Yvo Wolman, Cristina Maier and Julie Sors (DG CNECT), Caoimhe Sheridan (DG JUST), Michele Amedeo (DG NEAR), Giampaolo Lanzieri (Eurostat), Alina Sîrbu (University of Pisa & Hum-MingBird), Ettore Recchi (Science Po, Paris & Migration Policy Centre-European University Institute), Ingmar Weber (Qatar Computing Research Institute QCRI, Hamad Bin Khalifa University), Jakub Bijak (University of Southampton & QuantMig), Francisco Rowe (Geographic Data Science Lab, University of Liverpool), Meghan Benton (Migration Policy Institute), Marcello Carammia (University of Catania), Laura McGorman and Alex Pompe (Data for Good at Meta).

The authors remain solely responsible for the content of this report and any errors or omission that could remain.

The authors wish also to thank Daniele Borio (JRC) for his help in dealing with the original LATEX template, and Davide Bongiardo (JRC) who designed the cover page and created the final layout design of this report.

# AUTHORS

This work was coordinated by:
Stefano M. Iacus

and written together with:
Claudio Bosco
Sara Grubanov-Boskovic
Umberto Minora
Francesco Sermi
Spyridon Spyratos.

# ABSTRACT

With the consolidation of the culture of evidence-based policymaking, the availability of data has become central to policymakers. Nowadays, innovative data sources offer an opportunity to describe demographic, mobility, and migratory phenomena more accurately by making available large volumes of real-time and spatially detailed data. At the same time, however, data innovation has led to new challenges (ethics, privacy, data governance models, data quality) for citizens, statistical offices, policymakers and the private sector. Focusing on the fields of demography, mobility, and migration studies, the aim of this report is to assess the current state of data innovation in the scientific literature as well as to identify areas in which data innovation has the most concrete potential for policymaking. Consequently, this study has reviewed more than 300 articles and scientific reports, as well as numerous tools, that employed non-traditional data sources to measure vital population events (mortality, fertility), migration and human mobility, and the population change and population distribution. The specific findings of our report form the basis of a discussion on a) how innovative data is used compared to traditional data sources; b) domains in which innovative data have the greatest potential to contribute to policymaking; c) the prospects of innovative data transition towards systematically contributing to official statistics and policymaking.

# EXECUTIVE SUMMARY

With the consolidation of the culture of evidence-based policymaking, the availability of data has become central to policymakers. However, in the area of population and migration policymaking the traditional data sources have not always been able to provide sufficiently detailed or updated information to meet the policy needs. Nowadays, the expansion of the Internet and digital technologies provides an opportunity to overcome some of these data gaps by providing large volumes of real-time and spatially detailed data on a range of demographic, mobility, and migration-related topics.

Innovative data (also called non-traditional data) is composed of data derived from an individual's digital footprint, from sensor-enabled objects, and/or can be inferred using algorithms. As such, innovative data may contain personal data, in which case their use must comply with the rules set down in the General Data Protection Regulation (GDPR). Innovative data, however, should not be taken as a definite category since the advent of new data sources or the loss of relevance of others is intrinsically subject to developments in digital technologies and digital services.

In focusing on the fields of demography, mobility, and migration studies, the aim of this report is to assess the current state of data innovation in the scientific literature as well as to identify areas in which data innovation has the most concrete potential for policymaking.

Furthermore, in the specific field of demography, this state-of-the-art review focuses on innovative data applications for measuring vital population events (mortality, fertility), migration, and human mobility as well as population change and population distribution. In addition, given the interdisciplinary nature of human mobility and migration studies, the scope of these two topics is expanded beyond the field of demography and into the fields of sociology and economics.

This study is based on a review of more than 300 articles and scientific reports, as well as numerous tools, that use one of the following non-traditional data sources for demographic, human mobility or migration research: social media (Flickr, Twitter, LinkedIn, Reddit and Meta family of apps such as Facebook and Instagram), Internet activity (Web search data, online news), mobile payment apps, mobile phone data, air passenger data, and satellite imagery.

## MAIN FINDINGS

The findings of our report can be summarised under three main points:

### Use of innovative data compared to traditional data

The current landscape of innovative data sources offers information that can be reused for a variety of demographic, mobility and migration-related topics. Specifically, this study found that innovative data is most extensively used to fill in the gaps in traditional statistics, with migration-related estimates at multiple temporal and spatial scales being the most frequent example. In some cases, non-traditional data has also shown potential in overcoming measurement errors in survey data, such as reconstruction errors and memory distortion. Overall, in comparison to traditional data, the *current* competitive advantage of innovative data sources lies in their greater geographic and temporal granularity, (near-) real time availability, and their extensive coverage which makes more immediate international comparisons possible.

This review also found a growing literature adopting mixed methodologies, based on the integration of traditional with innovative data, to study demographic and migration phenomena. This is an important methodological development which allows for new uses of traditional data sources when combined with non-traditional data sources. Among the literature surveyed, the examples of integration of mobile phone data and/or satellite imagery with (micro) census-derived population data was more frequently found. However, data linkage still remains an important challenge as the application of these methodologies brings into question issues such as licensing, privacy, and data protection.

Moreover, the report finds that the definitions of population, migration and human mobility in studies based on innovative data do not always comply with conventional definitions adopted by official statistics. One of the main reasons for this fluidity in definitions is that private sector data is based on algorithms that are often proprietary and do not necessarily reflect conventional statistical standards. Another possible explanation is that the scientific literature studying social phenomena using innovative data has been dominated by computer science and, therefore, has lacked conceptualisations common to social sciences.

### Domains in which innovative data can be used for policymaking

This review identifies the greatest potential of data innovation in the domains of "*situational awareness, nowcasting and response*" and that of "*prediction and forecasting*". Indeed, it was observed that all of the innovative data sources analysed have the potential to provide (almost) real-time, accurate, and detailed information of demographic trends and/or public opinion at different geographical scales. It was also found that innovative data, especially when combined with traditional data sources, can be used for prediction and forecasting. At the same time, a limited number of examples were observed using non-traditional data for cause and effect analysis.

Moreover, the report finds that data innovation has vast, although not yet fully exploited potential in the domains of impact assessment, evaluation and experimentation. The numerous examples of this potential range from satellite images supporting the monitoring and evaluation of family planning and health services by estimating fertility and mortality rates in developing countries to mobile phone data supporting impact assessment and evaluation of rural development programmes and urban planning by providing insights onto the population dynamics in specific areas.

## Data innovation transition: from exploratory phase to a regular use for official statistics and policymaking

The key to fully unleashing the potential of non-traditional data is creating the right conditions to enable the data innovation transition, that is to say, a transition from a phase of an exploratory use of innovative data to a phase of systematic use of innovative data for official statistics and policymaking.

This report shows that not all innovative data sources have been explored to the same extent in the demographic and migration-related literature. Some types of non-traditional data (e.g. mobile network operator data and satellite imagery) have been more extensively validated than others (e.g. social media or Google search data). Therefore, it can be argued that only some types of non-traditional data are potentially mature enough to exit the exploratory stage. However, and despite the scientific advances in the use of non-traditional data, important challenges (e.g. ethics, privacy, data governance, data quality, bias, data access, etc.) still hamper the advent of an actual data innovation transition. This report discusses a few of these challenges.

Legislation is undoubtedly a *sine qua non* condition of data innovation transition. Among other things, legislation has to regulate the access to data held by the private sector in a way that guarantees the individual's fundamental rights and the interest of the private sector. Legislation must also explore ways of allowing the national statistical offices to collect, analyse, and publish data from non-traditional data sources.

Yet, a favourable regulatory framework for data innovation transition alone is not sufficient. In order to enable safe and trusted data sharing and statistics, investments need to be made in developing operational models and secure technical systems too. Consequently, investments aimed at fostering collaborations between data owners and the private and public research sectors become equally important. Indeed, collaborations also trigger improvements in the ethics of data handling and preserving privacy that could in turn feed into the development of principles of trusted and smart official statistics for decision making. Likewise, it is also critical to overcome the current lack of data-literacy skills (in computer science, statistics, and social sciences) in order to empower the institutions with professional figures capable of assessing the quality of the data, develop methodologies and extract value from innovative data sources.

In order to enable the data innovation transition and harness its full potential for policymaking, the Commission's KCMD will continue to monitor and report on the state of data innovation applications in the fields of demography, human mobility, and migration as well as to foster collaborations between policymakers, statistical offices, experts and the private sector.

# 1. INTRODUCTION

We live in the age of ''*datafication*'', an age of exponential rise in data collection and storage driven by the growing use of digital technologies, e.g. mobile phones, the Internet of Things (IoT), and devices (Verhulst, 2021). Datafication has brought enormous economic opportunities for our societies. For example, the overall impact of the data market on the European Union (EU) economy – the so-called value of data economy – corresponded to 3% of the Gross Domestic Product (GDP) of the EU27 plus the UK area in 2020, and this is expected to grow.[1] At the same time, datafication has also brought tremendous opportunities for social development. It can enable more agile, efficient, and evidence-based policymaking by improving understanding of social phenomena and by increasing the efficiency of already limited resources in addressing social challenges (Hughes et al., 2016; Juech, 2021).

Large volumes of data are not new to demographers. Indeed, demographers have long collected and analysed large datasets (population and housing censuses, population registers, surveys), including population-generalisable survey data. These datasets traditionally represent key data sources for population and social studies. One of the strong points of these traditional data sources is that they offer representative and population-generalisable data derived from regulated and transparent processing methodologies and data governance models.

With the consolidation of the culture of evidence-based policymaking, the availability of data has become central to policymakers. Yet traditional data sources do not always contain sufficiently detailed or updated information to respond to the population and migration programming and policymaking needs. Some of the most frequently cited limits of traditional data sources for policymaking is the time-lag between the release of the data and policy needs which usually require timely evidence (UNESCAP, 2021a). For example, the population and housing census in most countries is generally conducted every 10 years. Due to COVID-19, the current 2020 census round is facing significant obstacles. Numerous countries are postponing their population and housing censuses with serious repercussions on the capacity to assess the impact of developmental policies at national and especially at sub-national levels (Mrkic, 2021). Even in advanced countries where the permanent census of population and housing is carried out (e.g. Italy) data is usually released with a 1 to 2 years' time-lag. Similarly, traditional data sources face difficulties collecting data on the so-called hard-to-reach populations such as populations affected by conflicts and natural disasters, with serious implications for emergency humanitarian assistance. In Chapter 2 there is a more in-depth discussion around data gaps relevant for policies addressing demographic change as well as for migration policy.

With the expansion of the Internet and digital technologies, large volumes of disparate (near) real-time data have become more available. Unlike traditional statistics, the innovative data is derived from an individual's digital footprint (Latour, 2007), from sensor-enabled objects, and/or inferred using algorithms. As such, this non-traditional data has offered a potential solution to some urgent policy needs such as the need for greater geographical and temporal granularity, extensive coverage and (near) real time data. At the same time, however, data innovation has brought about new challenges for social sciences, statistical offices, and policymakers, calling into questions analytical capacities, data governance models, etc.

This report considers *innovative data* as any digital representation of acts, facts, or information originated sourced by using digital technologies and whose primary purpose is not scientific research. As such, innovative data may contain personal data, in which case their use must comply with the rules set down in the General Data Protection Regulation (GDPR). The term ''*innovative data*'' is used interchangeably with ''*non-traditional data*'' in this report.

The proliferation of innovative data sources represents a turning point for social science, including demography, human mobility and migration studies requiring ''*new research training, new analytical techniques, and new ontological orientations in the research process*'' (Bohon, 2018; Metzler et al., 2016).

Likewise the use of data innovation has become a relevant and highly debatable aspects of the broader modernisation process for statistical offices.

At the EU level, the process of modernisation of official statistics has led to the adoption of the Regulation establishing a common framework for European statistics relating to persons and households (EU 2019/1700), as well as to launching of the European statistics on population European statistics on population (ESOP)[2] initiative which foresees a new regulatory framework for European statistics on population based on data from

---

1    https://datalandscape.eu/european-data-market-monitoring-tool-2018
2    https://ec.europa.eu/info/law/better-regulation/have-your-say/initiatives/12958-Da ta-collection-European-statistics-on-population-ESOP-_ en, last accessed 26 November 2021 administrative sources. In the domain of data innovation in particular, the European Statistical System has been developing the concept of "*Trusted Smart Statistics*". This TSS concept encompasses principles such as distribution of the computation to data sources, shared control (between the statistical offices and data holders) on the process of statistical production, full automatisation of the statistical production process, etc (Ricciato et al., 2020).

A series of initiatives has also been taken at the UN level in order to support the national statistics offices in developing frameworks for integration of official statistical production with innovative data. For example, the UN has established a Committee of Experts on Big Data and Data Science for Official Statistics (UN-CEBD)[3] with a mandate to develop handbooks and methodologies on the uses of various innovative data sources in official statistics. Moreover, the United Nations Economic Commission for Europe (UNECE) has set up a High-Level Group for the Modernisation of Official Statistics[4] to actively steer the modernisation of statistical organisations in the UNECE region, including guidance in the domain of data innovation.[5]

Until the advent of innovative data sources, statistical offices were largely the main actor governing the statistical production sector. With datafication, various new data players, especially from the private sector, have entered the data ecosystem. A positive by-product of this data ecosystem "*liberalisation*" has been the alignment of disparate actors around a common philanthropic goal (Juech, 2021). As a result, there has been a growing number of "*Data for Good*" initiatives that aim to foster the use of innovative data to have a positive social impact, commonly anchored to the achievement of 2030 Sustainable Development Goal (SDG).[6] In Chapter 3 some concrete examples are provided of different types of Data for Good initiatives also relevant in the domains of demography, human mobility, and migration.

In this context, this report aims to contribute to the discussion on the potential of non-traditional data for social development/social good, by providing the state-of-the-art of data innovation applications in the fields of demography, human mobility, and migration studies, although not an exhaustive list.

In the field of demography, this state-of-the-art review focuses on data innovation applications for measuring vital population events (mortality, fertility),[7] migration and human mobility as well as population change and population distribution. In addition, given the interdisciplinary nature of human mobility and migration studies the scope of these two topics was expanded beyond the field of demography and into the fields of sociology and economics.

Chapters 5-24 offer examples of data innovation applications from the following data sources: social media (Flickr, Twitter, LinkedIn, Reddit, and Meta family of apps such as Facebook and Instagram), Internet activity (Google trends, online news), mobile payment apps, mobile phone data (CDR, ODM, and XDR), air passenger data, and satellite imagery. This, however, is not an exhaustive list of data sources that have potential applications for data innovation as the advent of new data sources or the loss of relevance of others is intrinsically subject to developments in digital technologies and digital services.

It should be emphasised that this is a non-exhaustive state-of-the-art review whose aim is to use concrete examples to highlight the potential of data innovation for advancing the knowledge of the above-mentioned topics. The authors are fully aware of the wide scope and interdisciplinary nature of population, human mobility and migration studies and may expand this review in future editions, also to sub-topics not currently covered.

A word of caution about definitions is also necessary. The overview of studies based on innovative data highlights that the definitions of the above-mentioned phenomena, especially migration- and mobility-related ones, do not always comply with conventional definitions adopted by official statistics. One of the main reasons of this variation in definitions is that the data held by the private sector are based on algorithms that are often proprietary and do not necessarily reflect conventional statistical standards. Another possible explanation is that the scientific literature studying social phenomena with innovative data has been dominated by computer science and, therefore, lacks conceptualisations common to social sciences.

In fact, as the example of definitions suggests, there are still numerous challenges and limitations linked to the use of innovative data sources, including issues related to ethics, privacy, data governance, and so on. After an individual review of caveats in the usage of each type of innovative data, Chapter 26 summarises concisely the common challenges and shortfalls while in Chapter 27 there is a brief discussion of how these challenges have been managed at the EU level using specific data policies. Finally, Chapter 28 concludes by summarising the main findings on the state of data innovation applications in the fields of demography, human mobility, and migration.

---

3    https://unstats.un.org/bigdata/
4    https://statswiki.unece.org/display/hlgbas/High-Level+Group+for+the+Modernisation+of+Official+Statistics
5    For an overview on the use of innovative data for official migration statistics see UNECE Task Force on Migration Statistics (2021).
6    https://sdgs.un.org/goals
7    We exclude other vital events such as marriage and divorce from the scope of this study.

# 2. THE DEMAND SIDE OF DATA: POLICY NEEDS AND DATA GAPS

Official statistics have always been the primary source of data in the domains of both demography and migration, sometime complemented by surveys and *ad hoc* data collections. Thanks to their reliability, the primary role of official statistics as the main reference in these two domains will remain unchanged in the future, at least in countries where well established statistical offices already exist. Yet, by nature, statistics need methodologies and definitions to be harmonised (which is typically not guaranteed across different statistical offices - for instance, see the key work done by Eurostat in the EU on *population* (Lanzieri, 2019b, 2021) and *migration* (Lanzieri, 2019a), and a solid validation that generally does not allow the timely availability of certain data.

Globalisation, recession or economic growth, climate change, relatively-fast societal changes (e.g. urbanisation, digitalisation, ageing society), and analogous macro-scale phenomena such as conflicts, extreme weather events, and humanitarian crisis emphasise the need for more timely but still reliable ways to capture and predict demographic and migratory behaviours on different geographic scales. There is indeed room for improvement in the data landscape in both the demography and migration domains. In addition, there is a clear policy request for timely and reliable evidence supporting policy-making and crisis-management. In fact, although by their nature official statistics represent a reliable source of information, mobile phones, social media, and anything else generating *digital traces* (Latour, 2007), as well as air-passenger data and satellite imagery offer an unprecedented opportunity. Obviously, every opportunity comes with its challenges. The main challenge of using nontraditional sources of data in the demographic and migration domains are generally connected to aspects such as ethics, privacy, truthfulness, reliability, accuracy, and representativeness (see Chapter 26).

Although *migration* is one of the three *demographic determinants* (together with *fertility* and *mortality*), this section considers the domains of demography and migration separately, aiming to provide an overview of both the policy framework and the current data gaps for each of them. Nevertheless, it is worth noting that the two domains very often overlap with each other, with the same data being relevant in both demography and migration. One ongoing initiative aiming to improve the appropriateness of data for the EU policymaking has been coordinated by the European Commission (EC)'s Knowledge Centre on Migration and Demography (KCMD), which was set up in June 2016. The main mission of the KCMD[8] is to provide independent scientific evidence for strengthening the European Commission's response to the opportunities and challenges related to migration, demography, and related policies. In strategic partnership with a number of the Commission's Directorate Generals and services, the KCMD makes sense of existing knowledge, addresses knowledge gaps and facilitates the uptake of its findings among EU policy-makers and other stakeholders.

## 2.1 DATA NEEDS AND DATA GAPS IN POLICIES ADDRESSING DEMOGRAPHIC CHANGE

In the mission letter to EC's Vice President Dubravka Šuica of December 2019, President Ursula von der Leyen clearly states the "*need to address some of the deeper changes in our society that have led to a loss of faith in our democracy on the part of some people. Those who feel left behind by progress and transition are the ones most likely to become disaffected. The root cause of this for many is more about demographic change than democratic structures*".[9] The letter recognises the key role of the demographic change in the EU and its potential threat to "*our democracy*". President von der Leyen assigns the portfolio "*Democracy and Demography*" to VP Šuica in the same letter.
In fact, together with the *green transition* and *digitalisation*, the *demographic change* is the third transformation shaping the future of Europe. In order to understand and face this transformation, "*affecting every part of our society: from the economy to healthcare, from migration to the environment*", official statistics and data from surveys need to be complemented by data from non-traditional sources.

Data gaps in the measurement and forecasting of the three demographic determinants or components (namely fertility, mortality, and migration) are very different for high-income/developed-countries compared to low-income/developing-countries. While the first can count on official statistics from national and supra-national institutions, the latter often lack this information or have outdated and sometimes inconsistent data available. Analogously, the data timeliness varies greatly across the two groups of countries.

---

8    https://knowledge4policy.ec.europa.eu/migration-demography_en
9    https://ec.europa.eu/commission/commissioners/2019-2024/suica_en

In the following part of this section, some of the main data gaps are introduced for each of the demographic determinants, through policy initiatives highlighting the need for not-yet-available data, and recent scientific publications proposing the use of non-traditional approaches to feel these gaps.

## Fertility

Together with migration, this is the demographic determinant on which most of the scientific literature based on alternative data sources focuses on. In Billari et al. (2018), the authors explain how in the 21st century, characterised by relevant short-term fluctuation in birth rates, monitoring and forecasting monthly fertility variation is of primary importance not only to support demographic models, but also by virtue of its role as an indicator of the general well-being of societies. The paper also supports the increasing importance of monitoring the short-term (e.g., monthly) fluctuations of fertility, as already outlined in Sobotka et al. (2005). In fact, although this information is already available in countries with a consolidated birth registration system, where National Statistics Institutes gather and timely publish monthly fertility data,[10] it is scattered or even missing in the rest of the world.
On the other hand, Ojala et al. (2017) focus on three main objectives: *i)* to illustrate the socio-economic differences in the circumstances around pregnancy and birth by using Google Correlate;[11] *ii)* to explain regional variation in fertility by using predictive models; and *iii)* to model fertility trends using aggregated web search. This type of analyses based on non-traditional sources of data is becoming more and more necessary for regional, national and federal administrations to manage and predict demographic phenomena such as ageing and depopulation (Aurambout et al., 2021), but also to preserve rural areas as stated in the EC's *long-term vision for the EU's rural areas.*[12]

## Mortality

As long ago as the year 2000, a global initiative lead by the Department of Demography at the University of California (Berkeley - USA), the MPIDR, and the French Institute for Demographic Studies (INED), began the development of a database on mortality under the name of Human Mortality Database (HMD) project.[13] Already at that time, the importance of this type of data was clear not only in terms of demographic modelling and analysis, but also for health surveillance and prevention. The recent COVID-19 pandemic has boosted the need for timely weekly statistics on deaths and the HMD has promptly responded by introducing the Short-term Mortality Fluctuations (STMF) data collection. At the same time, in Europe, Eurostat responded just as quickly by publishing the `demomwk` special data collection[14] covering all of the EU Member States (MSs) and the European Free Trade Association (EFTA) countries[15] at the regional level (up to NUTS3).[16] This dataset has provided the opportunity to demonstrate the impact of the pandemic in terms of excess mortality. Similar statistics have been made available[17] in the US by the Centers for Disease Control and Prevention (CDC) and at the global level there are other initiatives such as the Global Burden of Disease (GBD)[18] and the Population Observatories described by Pison (2005) who collect and publish data on mortality to support both researchers and policymakers. Unfortunately, this data is generally available only on an annual basis and not below the national level; this makes it impossible to monitor mortality on a finer space/time scale, especially in less developed countries where mortality is higher, as it is the share of hard-to-reach population. This is why non-traditional approaches such as that proposed by Carioli et al. (2017) or by Wagner et al. (2018) are worth to be further explored.

## Migration

The importance of capturing migration trends in a more timely and granular manner is outlined in terms of policy-making and crisis management in Section 2.2. Migration - in particular the *migratory flow* - is a key component of the demographic picture. Yet, it is also the most difficult to capture and forecast because more prone than fertility and mortality to short-term fluctuations. The complexity of the migratory phenomena, which emerges already from the debated definition of *migrant*, makes it almost impossible for it to be captured in a timely manner. In fact, according to the European Migration Network (EMN) glossary, *"a migrant" is defined in the global context as "a person who is outside the territory of the State of which they are nationals or citizens and who has resided in a foreign country for more than one year irrespective of the causes, voluntary or involuntary, and the means, regular or irregular, used to migrate".*[19] This means that at least twelve months are needed before a person who has moved to another country can be counted as a migrant. Alternative definitions introduce the concept of *intention* to move to another country for more than one year; but intentions are even more difficult to be measured. Going back to the EMN definition, we see how a further level of complexity of the migratory phenomenon emerges; migration can be voluntary or involuntary and may follow legal or illegal pathways (sometime a combination of the two).

---

10   For instance, see "The Human Fertility Database" (HFD), a joint project of the Max Planck Institute for Demographic Research (MPIDR) and the Vienna Institute of Demography (VID) https://www.humanfertility.org/cgi-bin/main.php
11   Google Correlate was shut down on 15 December 2019, as a result of low usage. It has been in practice replaced by Google Trends, though with different functionalities.
12   https://ec.europa.eu/info/strategy/priorities-2019-2024/new-push-european-democracy/long-term-vision-rural-areas_en
13   https://www.mortality.org/
14   https://ec.europa.eu/eurostat/cache/metadata/en/demomwk_esms.htm
15   https://ec.europa.eu/eurostat/statistics-explained/index.php?title=Weekly_death_statistics
16   https://ec.europa.eu/eurostat/web/nuts/background
17   https://data.cdc.gov/NCHS/Weekly-Provisional-Counts-of-Deaths-by-State-and-S/muzy-jte6
18   http://www.healthdata.org/gbd/2019
19   https://ec.europa.eu/home-affairs/pages/glossary/migrant_en

Furthermore, when we reduce the geographical scale of the analysis below the national level and the temporal one below three months (the United Nations (UN) defines migration as mobility lasting beyond three months - *short-term* - and beyond a year *long-term*), *migration* actually becomes *mobility*. Also Intra-EU migration is generally referred to as mobility in the EU policies and by scholars dealing with European migration (see, e.g., Boswell and Geddes, 2010; Geddes et al., 2020).

Migratory crises such as that started in 2013 in South Sudan, the 2015-16 Syrian refugees crisis, that of Rohingya in 2017, or that recently triggered by both the conflicts in the Tigray region and the withdrawal of the US troops from Afghanistan, just to list a few, have highlighted the need of policymakers and crisis-management organisations all around the world for timely data on forced human mobility. The same reasoning applies to natural disasters and extreme weather events forcing people to leave their houses. Recently the world was shocked by the images of Australian bushfires, flash floods in Indonesia, earthquakes in Jamaica, Russia and Turkey, as well as by the devastating effects of Cyclone Amphan in Bangladesh and India. In all these cases, traditional data could not allow to promptly monitor people movements, provide support and manage the humanitarian crises.

Clear data-gaps emerge also from some papers introducing the opportunities that nontraditional data offer in terms of scientific research and policy support (see, e.g., Tjaden, 2021). For example, in Cesare et al. (2018), the authors provide a broad overview of the benefits and challenges of using data from *digital traces* for social and demographic research, and review the state-of-the-art in the use of digital trace data in demography (including all three demographic components). Mrkic (2021) takes stock of a potential data gap due to the COVID-19 pandemic and in particular the disruption to the implementation of the 2020 World Population and Housing Census Programme.

Another context where non-traditional sources could potentially fill a significant data gap is the measurement of temporary population. Yet, before talking about *temporary population* we need to introduce two key concepts which are *de jure population* and *de facto population*. Starting with the Organisation for Economic Co-operation and Development (OECD)'s definitions:

- *de jure population*: is a concept under which individuals (or vital events) are recorded (or are attributed)

to a geographical area on the basis of the place of residence;[20]

- *de facto population*: is a concept under which individuals (or vital events) are recorded (or are attributed) to the geographical area where they were present (or occurred) at a specified time.[21]

Census data can be based on *de jure* population as well as on *de facto* population. In fact, there are no universally recognised standards. In other words, statistics can refer to the official place of residence or to the place where the individual was present at the time of the enumeration. In absolute terms, neither of the two concepts is better or more correct than the other. Indeed, it always depends on the context where it is used. Section 22.3 discusses this issue in the context of innovative data.

Whenever spatial-demography is considered at a very high space/time granularity, the *actual population* needs to be captured within a given area during a relatively short time period. For instance, the demographic composition of a city (or a neighbourhood) during the working hours (or overnight) needs to be known. This is where the concept of *temporary population* needs to be introduced. Digital traces, satellite imagery (Stathakis and Baltas, 2018; Rigalll-Torrent, 2010), and other different types of non-traditional data can help in capturing this demographic data. Spatial-demography and temporary population data is needed more than ever in urban contexts for efficient infrastructure planning and management, but also outside cities, in areas with a lower population density, where access to services (and a reasonable maximum distance from it) should be guaranteed to all inhabitants. Moreover, timely data about temporary population can be life-saving during crisis management. For instance, a natural disaster such as a flood or a hurricane or about mass-gathering events such as demonstrations, sport events, etc.

Panczak et al. (2020) provides a review of the literature on empirical methods to estimate temporary populations, whereas Jo et al. (2020) shows how spatial patterns of temporary populations estimated using mobile phone data show gender gaps in the use of urban spaces in Seoul.

## 2.2 DATA NEEDS AND DATA GAPS IN MIGRATION POLICY

Migration is a global phenomenon, but data describing this phenomenon is not available with the same granularity, frequency, and reliability for all parts of the world. There are more data gaps in developing countries and in hard-to-reach populations (i.e. undocumented migrants, missing migrants, etc.).

The *Global Compact for Safe, Orderly and Regular*

---

20    https://stats.oecd.org/glossary/detail.asp?ID=580. In other words, this is about the legal, not factual place of residence.
21    https://stats.oecd.org/glossary/detail.asp?ID=571.

*Migration,*[22] adopted in December 2018 by UN Member States at the Global Conference in Marrakesh (Morocco), recognised the need to collect and use accurate data as a basis for evidence-informed policymaking as being one of its main objectives.

Moreover, the UN's SDGs, and in particular targets 17.18 and 17.19 call for the provision of high-quality, timely, and reliable data disaggregated by, among other variables, the migratory status.

In Europe, the 2015-16 *refugee crisis* highlighted the need to support official statistics on migration with more timely data supporting the EU and the national governments in their response to instability.

Migration studies do not only consider migrants but sometimes focus on specific migrants' categories such as *asylum seekers, refugees,* victims of trafficking of human beings, etc. as well as on people who have moved but do not come under the definition of a migrant (e.g the *Internally Displaced Persons (*IDPs)). Sometimes the differences between the various categories are quite blurred and there are no unique definitions shared by national authorities and international organisations. Very often people who are forced to leave their houses move first within their own country (becoming IDPs) and, if they cannot find there what they are looking for, may move to a different country (becoming international migrants). Once in the new country, they might ask for asylum (becoming asylum seekers, i.e. a particular sub-group of migrant) and possibly receive a form of international protection (perhaps becoming refugees when this particular form of international protection is granted by the hosting country). The 1998 *UN recommendations on Statistics for International Migration - Revision 1* indeed represent a fundamental practical guide on how to collect migration statistics, providing definitions and describing methodologies. Yet, as stated in the 2018 concept note on its revision by the United Nations Statistics Division (UNSD),[23] ''[…] *changes in migration patterns, border control, as well as methods in data collection in the last 20 years have clearly pointed to the need for undertaking a review and update of the 1998 recommendations.*'' In the EU, the common rules for the collection and compilation of Community statistics on migration are set by EU 2007/826 (2007).

Starting from the very basic concepts, when looking at the gaps in migration data, it is necessary to distinguish between the following two dimensions to be captured:

- *stocks*: i.e. the overall populations counted at a given point in time (generally at the end of the year);

- *flows*: i.e. for instance the number of people entering or leaving a country over a reference period (it can be days, weeks, months, or even years), or the number of asylum applications lodged in a country over a given period, etc.

The main data gaps in the recording of stocks are with hard-to-reach third-country nationals who are unlawfully present on the territory (undocumented migrants, people with an expired visa, etc.) or with people in huge reception facilities with uncontrolled access (refugee camps, informal settlements, IDP camps, etc.), or even with intra-EU migrants (often referred to as *EU mobile citizens*) who are very difficult to count as in large part European citizens move informally within the Schengen area and they might do not register in host countries.

Obviously, also the geographic scope adopted to measure stocks is relevant. In Europe, for example, Eurostat provides official statistics on migrant residents by country of citizenship only at the national level; the same information is not available at the regional and local level.[24]

The geographic scope is a determinant in measuring flows too, but in this case the reference period is also a key parameter. The management of crisis situations, as for instance that due to the COVID-19 outbreak in 2020 or to the recent departure of the United States (US) troops from Afghanistan, just to mention a couple, requires timely flow data in order to coordinate humanitarian aids, deploy available resources and ultimately avoid the escalation of the crisis.

Moreover, temporary border closures might limit the circulation of workers across different countries. On this matter, the key role of immigrant workers[25] and migrant seasonal workers in the EU labour market highlighted the need to forecast the potential impact due to their unavailability on the different economic sectors during the first half of 2020. Yet, the current latency (from one to two years) and time frequency (typically annual) of both International Labour Organisation (ILO) estimates and Eurostat statistics do not allow to perform this type of analysis. Also in this case, the use of non traditional sources of data seems to be the most reasonable approach.

Another flow which is very hard to capture is represented by the so called *secondary movements*. According to the European Asylum Support Office (EASO), secondary movements occur when refugees or asylum-seekers move from the country in which they first arrived to seek protection or for permanent resettlement elsewhere (see also Tank, 2017). Secondary movements are of primary importance for the EU policy-making, since they are at the centre of the ongoing negotiations on the *New Pact on Migration and Asylum.*[26]

22    https://www.iom.int/global-compact-migration
23    https://unece.org/fileadmin/DAM/stats/documents/ece/ces/ge.10/2018/mtg1/UNSD_Recommendations_ENG.pdf
24    As a matter of fact, Eurostat does provide subnational data on education and employment disaggregated at foreign and native born level at most. See https://ec.europa.eu/eurostat/web/migration-asylum/migrant-integration/database.
25    https://knowledge4policy.ec.europa.eu/publication/immigrant-key-workers-their-contribution-europes-covid-19-response_en
26    https://ec.europa.eu/info/strategy/priorities-2019-2024/promoting-our-european-way-life/new-pact-migration-and-asylum_en

Also the *return migration* and in particular the *voluntary return* is very relevant for the policy side but cannot be completely captured by official statistics. According to the International Organisation on Migration (IOM)[27] *''There are two main forms of return migration: voluntary return and forced return. Data on forced return are usually collected by national and international statistical offices, border protection and immigration law enforcement agencies. The International Organization for Migration (IOM) collects data on assisted voluntary return and reintegration programmes that it implements worldwide.''* In Europe, Brexit first and COVID-19 pandemic later triggered a wave of voluntary returns of both EU and non-EU citizens to their country of origin, but because of the lack of reliable data, the size of this phenomenon has never been measured in a reliable way.

In Vespe et al. (2018), the authors take stock on the migration data landscape, outlining the main data gaps for the EU and explaining which are the dissemination issues, why some data is not-collected, and why other data is not accessible.

At the global level, the recent report from the UNECE's task force on the use of new data sources for measuring international migration (UNECE Task Force on Migration Statistics, 2021) provides a competent and critical overview on the use of non-conventional data to complement migration statistics, presenting the results of two surveys involving the member countries and describing the example of the US Census Bureau estimating Puerto Rican migration. Another key document is the *Guidance on Data Integration for Measuring Migration* published also by UNECE (United Nations Economic Commission for Europe, 2019); it describes risks and opportunities when two or more different data sources are combined to produce statistical outputs.

Many frequent and persistent gaps in international migration data (e.g. inconsistencies in definitions and measures, neglecting the drivers or reason for migration, undercoverage of major geographies and detailed demographic information about migrants, and a delay in timely representation of the data) are highlighted in Bircan et al. (2021). The paper propose also alternative solutions, recommending the use of non traditional sources of data to fill these gaps. Bijak et al. (2017) looks into similar issues when dealing with asylum-related data.

In 2016, EASO organized a first workshop on *Big data and early warning for asylumrelated migration*;[28] a year later, in collaboration with the IOM's Global Migration Data Analysis Centre (GMDAC), the KCMD organised a second workshop on *Big Data and Alternative Data Sources on Migration*,[29] gathering together researchers, data holders/providers, and policymakers (Vespe and Rango, 2017). This kind of initiatives have been taken at national and international level (i.e. the UN's *Big Data for Official Statistics*,[30] and the IOM's *Migration Data Portal*[31])

---

27  https://www.migrationdataportal.org/themes/return-migration
28  https://www.easo.europa.eu/file/big-data-and-early-warning-asylum-related-migration-workshop-agenda-and-presentations
29  https://knowledge4policy.ec.europa.eu/migration-demography/big-data-alternative-data-sources-migration_en
30  https://unstats.un.org/bigdata/
31  https://www.migrationdataportal.org/themes/big-data-migration-and-human-mobility

# 3. DATA FOR GOOD INITIATIVES

As highlighted in the introduction, the datafication of society has brought new data players into the data ecosystem, aligning them around a common philanthropic goal (Juech, 2021). As a result, in recent years, the number of "*Data for Good*" initiatives has been growing in order to leverage the potential positive social impact of non-traditional data, frequently anchored in the achievement of 2030 SDGs.

However, not all of the initiatives are the same, and can be distinguished according to the specific aspect of the data for social good that they aim to support. Building on the Porway (2021) classification[32] of the Data for Good ecosystem, the initiatives operate as:

- the dataset providers

- the storage providers

- the data governors

- the data talent providers

- the solution designers

- the diy software providers

- the data strategy providers

- the data talent trainers

- the community builders

- the responsible ai advocates

- the social data thought leaders

It should be noted that these initiatives are often fluid, contributing to two or more dimensions of the data for social good process (e.g. the UN Global Pulse initiative).

Likewise, in the majority of cases the social impact of these initiatives is cross-sectoral. For example, initiatives aiming at reducing inequalities are also relevant for the dimension of migration, fertility and reproductive health, mortality and COVID-19.

The aim of this chapter is to provide various concrete examples of initiatives which have directly or indirectly brought data innovation into the domain of demography, human mobility and migration studies. Specifically, the following paragraphs focus on examples of initiatives by dataset providers, data governors, and data strategy providers.

## 3.1 DATA SHARING INITIATIVES

The category of Dataset Providers includes initiatives that aim to "*create more data so that social impact organizations can create more data solutions so that social benefit increases*" (Porway, 2021). Consequently, the dataset providers, share data they have collected or that others have collected for the purpose of creating a positive social impact. These data sharing initiatives do not involve personal data but aggregated summary statistics.

A non-exhaustive list of some of relevant initiatives for demography, human mobility and migration is now presented.

- *Amazon Web Services (AWS) Open Data Registry*[33] allows government organizations, researchers, businesses, and individuals to store and publicly share data sets of any size. For example, one can find Meta's High Resolution Settlement Layers (HRSL) in this repository.

- *Copernicus,*[34] the European Union's Earth observation programme, provides access to a vast amount of global data from satellites and ground-based, airborne, and seaborne measurement systems with the aim of improving European citizens' quality of life and other significant aspects.

- *CUEBIQ Data for Good Programme*[35] provides access to anonymous, privacycompliant location-based data for academic research and humanitarian initiatives related to human mobility.

- *Data for Good at Meta*[36] (previously Facebook Data for Good) is a data sharing initiative that aims to empower partners with privacy-preserving data[37] that strengthens communities and advances social issues. Data for Good at Meta publish anonymised and aggregated data sets derived from multiple sources such as aggregated Meta user data, and targets surveys to Meta users as well as satellite data, etc. (also see Chapter 7). All publicly available Data for Good at Meta initiative are hosted on the UN's Humanitarian Data Exchange platform, which is described below.

---

32    https://data.org/news/charting-the-data-for-good-landscape/
33    https://registry.opendata.aws/
34    https://www.copernicus.eu/en
35    https://www.cuebiq.com/about/data-for-good/
36    https://dataforgood.facebook.com/
37    A privacy-preserving algorithm is a mathematical method that transforms the original data into synthetic data by adding different types of noise in a way that the results of certain statistical analysis remain valid and such that the re-identification of the original data records is virtually impossible.

- *Development Data Partnership*[38] is an initiative led by the World Bank that brings together international organisations and 25 data partners including companies such as Google, Meta, Esri, LinkedIn, Twitter, and Waze that share data for development and humanitarian projects.

- *Esri COVID 19 GIS Hub*[39] initiative, shares maps and geographic information systems (GIS) to help organisations respond to the COVID-19 crisis, maintain continuity of operations, and support the process of reopening. Esri has also implemented the ArcGIS Living Atlas.[40]

- *Global Human Settlement Layer* is The European Commission's open free data and tools for assessing human presence on the planet, using new spatial data mining technologies.[41]

- *Humanitarian Data Exchange (HDX)*[42] is a United Nations Office for the Coordination of Humanitarian Affairs (UN-OCHA) open platform for the sharing of humanitarian datasets, including those from non-traditional data sources

- *Humanitarian OpenStreetMap – HOTOSM*[43] is a data sharing initiative providing geospatial data for humanitarian action and community development through open mapping.

- *LUCA Data for Good*[44] is a data sharing initiative by Telefonica that offers connectivity data along with other external data sources to return the value of data back to the world and contribute to the UN Sustainable Development Goals for 2030.

- *Mastercard Center for Inclusive Growth*[45] is a sharing initiative blends opensource and proprietary data with a layer of insights based on Mastercard's highly aggregated and anonymised transaction data.

- *Qatalog*[46] is an open software platform developed by the UN Global Pulse initiative on which data from Twitter, Facebook and public radio talk shows can be queried, assigned, tagged, and analysed data from in order to obtain insights into the SDGs.

## 3.2 DATA GOVERNANCE INITIATIVES

The Data governors advocate for better data collection and access standards so that social impact organizations use data more effectively so that harm is reduced and benefit increases. In other world, these initiatives provide guidance about best practices for providing, storing, sharing and using non-traditional data for social good.

A non-exhaustive list of some of relevant initiatives for the demography, human mobility and migration is now presented.

- *Big Data for Migration Alliance (BD4M)*[47] aims to ''*accelerate the responsible and ethical use of new data sources and innovative methodologies to inform migration policies and programmes*'' by providing guidance and capacity-building support, crosssectoral collaboration, research, and advocacy.

- *Brighthive Data sharing solution*[48] aims to ''help networks of organizations increase their impact and value by creating the business, legal and technical framework needed to securely and responsibly link their data''.

- *Centre for Data Innovation*[49] ''*formulates and promotes pragmatic public policies designed to maximize the benefits of data-driven innovation in the public and private sectors. It educates policymakers and the public about the opportunities and challenges associated with data, as well as technology trends such as open data, artificial intelligence, and the Internet of Things*''.

- *The Global Partnership for Sustainable Development Data (data4sdgs)*[50] aims to improve government decision-making by harnessing the use of new technologies and data sources at scale. The data4sdgs activities also aims to foster a global movement of government, business, and civil society leaders in order to promote responsible data sharing and use, build public trust, and showcase pathways to success.

---

38   https://datapartnership.org/
39   https://www.esri.com/en-us/covid-19/overview/
40   https://livingatlas.arcgis.com/en/home//
41   https://ghsl.jrc.ec.europa.eu/
42   https://data.humdata.org/dataset/
43   https://www.hotosm.org/
44   https://luca-d3.com/data-for-good
45   https://inclusivegrowthscore.com
46   https://www.unglobalpulse.org/microsite/qatalog/
47   https://data4migration.org/
48   https://brighthive.io/
49   https://datainnovation.org/about/
50   https://www.data4sdgs.org

- *Data Sharing Coalition*[51] aims to drive cross-sectoral data sharing under control of the entitled party by enabling interoperability between data sharing initiatives and by strengthening individual initiatives. The coalition explores generic agreements for data sharing in order to achieve harmonisation in a fragmented landscape, defines and develops cross-sectoral use cases, and shares knowledge to support the development of existing and new data sharing initiatives.

- *EU Support Centre for Data Sharing*[52] aims to ''*facilitate data sharing, i.e. transactions in which data held by public sector or private sector are made available to other organisations public or private for use and re-use.*'' EU SCDS activities are focused on researching, documenting, and reporting data sharing practices, EU legal frameworks, and access and distribution technologies that involve novel models, as well as legal and technological challenges.

- *The Governance Lab (GovLab)*[53] aims at strengthening the ability of institutions – including but not limited to governments – and people to work more openly, collaboratively, effectively and legitimately to make better decisions and solve public problems. It provides the Data Collaboratives hub,[54] a resource on creating public value by exchanging data, collecting examples of data collaboratives from around the world, and providing information on how to develop and implement a data collaborative.

- *Global System for Mobile Communications Association (GSMA) #BetterFuture initiative*[55] is dedicated to sharing best practices on how mobile operators' data sharing initiatives are addressing the UN Sustainable Development Goals around the World.

- *UN Global Pulse*[56] initiative on big data and artificial intelligence for development, humanitarian action, and peace aims to ''contributes to a future in which big data and artificial intelligence are harnessed safely and responsibly for the public good'' by creating the right policy frameworks and strengthening ties with relevant communities of practice.

## 3.3 DATA STRATEGY INITIATIVES

The Data Strategy Providers ''*create more of the entire pipeline so that social impact organizations are able to apply data more effectively overall so that social benefit increases*'' Porway (2021). These initiatives build capacity in civil society by designing data strategies for nonprofits and government actors.

A non-exhaustive list of some of relevant initiatives for demography, human mobility and migration is now presented.

- Data-Pop Alliance[57] provides ''*support to governments and organizations in the process of designing and consolidating their digital transformations. This includes processes to conceive socio-technological innovations, as well as to support in developing data and digital strategies, contributing to building evidence and methods to tackle societal issues such as inequality and crime*''.

- *Flowminder*[58] focuses on ''*finding the appropriate combination of products and services that works in combination with the strengths and capacities of local partners to ensure value is delivered via the whole data value chain*''.

- *Open Data Institute (ODI)*[59] aims to ''*enable the development of data infrastructure in ways that benefit people, companies, governments and civil society . . . [by] increasing data flows around the data ecosystem, improving skills and capabilities, and encouraging innovation*''.

- Patrick J. McGovern Foundation (ex Cloudera Foundation)[60] aims to help non-profits unlock their capacity to apply data and artificial intelligence to inform more sustainable decisions and create transparency around global social challenges.

## 3.4 EU-FUNDED R&I PROJECTS ON DATA INNOVATION

Horizon 2020 (H2020) is the largest EU Research and Innovation programme, with funding of nearly 80 billion Euros made available in the period 2014 to 2020. Under this programme, the European Commission has also supported relevant projects for applying data innovation in

---

51   https://datasharingcoalition.eu/
52   https://eudatasharing.eu/homepage
53   https://thegovlab.org/
54   https://DataCollaboratives.org
55   https://sdgexplorer.gsma.com/
56   https://www.unglobalpulse.org
57   https://datapopalliance.org/transform/
58   https://www.flowminder.org
59   https://theodi.org/
60   https://www.mcgovern.org/

the fields of demography, human mobility, and migration. Some examples of these projects include:

- *DeepCube*[61] project aims to unlock the potential of big Copernicus data using Artificial Intelligence and Semantic Web technologies, for the purpose of addressing problems having great environmental and societal impact. The project develops a specific use case on climate induced migration in Africa combining data from diverse EO and non-EO data sources.

- *HumMingBird*[62] project aims to improve the mapping and understanding of changing migration flows by analysing patterns, motivations and new geographies. It also tests new methods in forecasting emerging and future trends. In the domain of data innovation, the project also aims to *"validate big data technologies to provide dynamic and novel evidence on various aspects/factors that may help estimate stock migration and migration flows"*.

- *ITFLOWS*[63] project aims to provide predictions and appropriate management solutions for migration flows in the European Union in the phases of reception, relocation, settlement and integration of migration by using multiple sources of information, including innovative data sources.

- *MOSAICrOWN*[64] project aims to enable data sharing and collaborative analytics in multi-owner scenarios in a privacy-preserving way, ensuring correct protection of private/sensitive/confidential information. Moreover, the project offers: *i)* a data governance framework able to capture and combine the protection requirements that can possibly be specified by multiple parties; *ii)* effective and efficient protection techniques that can be integrated with current technologies and that enforce protection while enabling efficient and scalable data sharing and processing.

- *QuantMig*[65] aims to advance the methodology of scenario generation and to improve the understanding of conceptual foundations of European migration flows. The knowledge base for scenarios includes a comprehensive review of key migration drivers in origin, destination and transit countries. Additionally, the project derives a distinctive set of custom-made harmonised statistical estimates of migration flows and develops innovative methods for simulating migration flows, describing scenario uncertainty, and providing early warnings.

- SoBigData++[66] aims to deliver a distributed, Pan-European, multidisciplinary research infrastructure for big social data analytics, coupled with the consolidation of a cross-disciplinary European research community, aimed at using social mining and big data to understand the complexity of the contemporary, globally interconnected society. The explanatories cover topics on societal debates and online misinformation, sustainable cities for citizens, demography, economics and finance 2.0, migration studies, sports data science, and the social impact of artificial intelligence and explainable machine learning.

61   https://deepcube-h2020.eu/
62   https://hummingbird-h2020.eu/
63   https://www.itflows.eu/
64   https://mosaicrown.eu
65   http://www.quantmig.eu
66   https://plusplus.sobigdata.eu//

# 4. HOW TO NAVIGATE/READ THIS REPORT

The following considers each data source in a separate section, though clearly some of them are connected (e.g., mobile network data, CDR, and XDR). A review of actual usage of these innovative data in terms of different aspects of demography and migration studies is presented in each section. While some fields of application can only be analysed using specific data sources, other topics can be explored using multiple data sources. Figure 1 contains a different way to explore a report by crossing data sources and fields of applications.

**FIGURE 1.** Synthetic view of data sources and their use in demography, migration, and human mobility considered in this report.

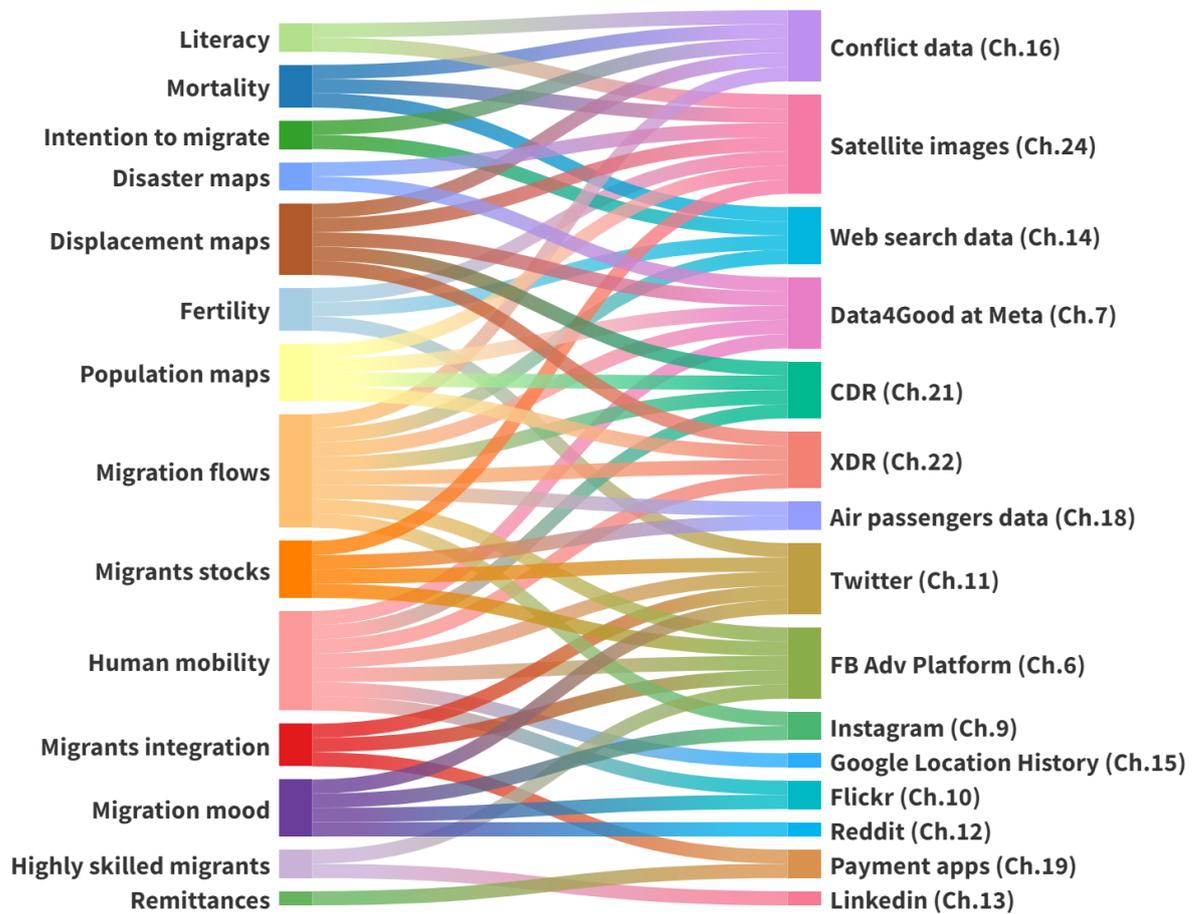

# 5. INNOVATIVE VS TRADITIONAL DATA FOR DEMOGRAPHY, MIGRATION AND HUMAN MOBILITY

The use of innovative data for demographic and migration research has evolved rapidly during the last decade. Innovative data has a lot of potential, especially in developing countries where official data is either outdated or not available. At the same time, it poses numerous scientific and ethical challenges. Zagheni and Weber (2015) proposed a general framework for extracting demographic information from non-representative Internet data. Cesare et al. (2018) discussed the benefits and challenges of using non-representative Internet data for social and demographic research. The use of innovative data for demographic research is not new. "*John Graunt, which is widely considered to be the father of demography, produced the very first life tables in the seventeenth century by leveraging data collected for marketing purposes to estimate the age distribution of potential customers in London (Cesare et al.,2018)*".

Innovative data may differ significantly from the traditional statistical data, see the comparison in Table 1. Firstly, innovative data is collected for various purposes not directly related to policy or administrative purposes as is the case for traditional data. Innovative data is generated actively or passively by the users of Internet applications, and by mobile devices and sensors. Traditional data is typical collected and/or produced by trained professionals. The methodology of data collection of traditional data is usually well established and transparent while the methodology used in collecting innovative data can be undisclosed due to potential commercial sensitivity. Consequently, innovative data sources are sometimes referred to as "*black boxes*" (Araujo et al., 2017; Mejova et al., 2018). Against this backdrop, the European Statistical System has been developing the concept of "*Trusted Smart Statistics*" calling, among other things, for full transparency and auditability of processing algorithms (Ricciato et al., 2020). The heterogeneity of innovative data is typically higher than that of traditional data and statistics, which are often harmonised in terms of definitions and data collection methodologies. The demographic representativeness of innovative data can be very low when this data is collected by the users of mobile devices and Internet applications. This is because people of different age, sex, educational background, and in different areas of the world are not using information and communications technologies to the same degree.

**TABLE 1.** Characteristics of innovative and traditional data.

|  | Innovative | Traditional |
|---|---|---|
| Purpose of data collection | commercial and other purposes | administrative/statistical purposes |
| Collected by | mobile devices, sensors, users of internet applications | trained professionals |
| Transparent methodology | rarely | always |
| Data heterogeneity | high | low |
| Demographic representativeness | can be low | high |
| Spatial and temporal completeness | varies | harmonized |
| Reference area | user/device location | administrative areas |
| Reference time period | undefined, continuous, can be real-time | defined |
| Timeliness | high | low |
| QC/QA mechanism | unknown | present |
| Data cost | varies but can be low | high |
| Stability/Sustainability of the data source | low | high |

**FIGURE 2.** Estimated penetration of social media by country and gender as of January 2021.
**Source:** WeAreSocial – https://datareportal.com/reports/digital-2021-global-overview-report

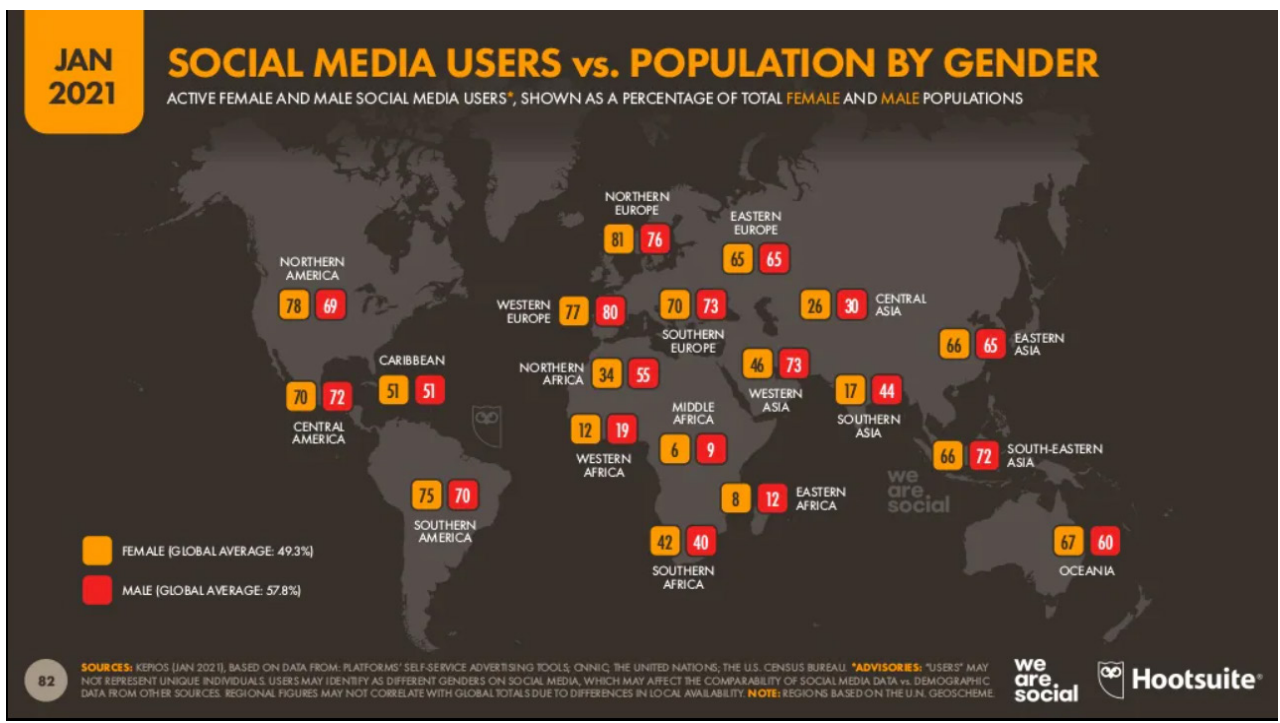

Moreover, the spatial and temporal completeness of innovative data might vary across different areas and time periods. For example, popular areas that are located in city centres are better represented in social media datasets compared to unpopular or sparsely populated areas (Antoniou et al., 2010; Spyratos et al., 2017; Spyratos and Stathakis, 2018). Contrary to social media data, the spatial and temporal completeness of other types of innovative data such as satellite images can be very high.

The reference area of innovative data is usually the location of the users or the devices that produce the data (except remotely sensed data from satellites) so it can range from local to global scale. The reference area of traditional data is usually administrative or statistical areas. The reference time period of innovative data can be unspecified and in the case of device generated data like mobile positioning data, it can be continuous and real-time. The timeliness of innovative data can be very high from real-time to few days depending on the processing required, for example, the time needed to aggregate and anonymise mobile positioning data. In the case of traditional statistics, several months are usually required from data collection to the publication of the statistics due to the procedures required such as the Quality Assurance and Quality Control mechanisms in place. In contrast to the traditional data, the quality of innovative data is typically not evaluated by formal quality assurance and control procedures (Spyratos et al., 2014). Since the innovative data was originally collected for other purposes and is being re-used for demographic and migration research purposes, the cost of data collection can be very low. However, the societal cost of generalising findings from innovative data sources that may represent a subset of the population of interest can be extremely high (Rowe, 2021a), especially when these generalisations are used as input in decision and policy making processes.

Finally, it is worth noting that given the absence of an agreement between researchers and innovative data providers, the data flow of innovative data can be interrupted or distorted at any time without prior notice due for example to change in the data provision methodology or policy by the data provider and disappearance of the source. In contrast, the provision of traditional data by statistical agencies is more sustainable and guaranteed due to the specific mandate and the legal commitments of the data providers.

Furthermore, several demographic and migration researchers have used Facebook data in recent years in aggregated and anonymized form. The advantages of Facebook data is that it is frequently updated, is rich in detail, has high spatial and temporal resolution, has almost global coverage, and importantly, is drawn from a sample of 2.9 billion monthly active users as of June 30, 2021.[67] It has therefore proved to be a very valuable source of information in multiple contexts. For example, Facebook data can be used to collect updated, cross-border, and detailed demographic and migration data in

---

emergency situations and in areas of the world where traditional demographic and migration data is not available or updated. However, even if almost one out of three global citizens has an active Facebook account, people of different age, genders, and origins are not equally represented (Spyratos et al., 2019), also see Figure 2. These inequalities need to be considered when using Facebook data for policy response, especially in countries where Facebook penetration is very low for women, elderly people, or minorities. Facebook data needs to be provided with the appropriate data protection, privacy, and ethical safeguards, to ensure that this data is truly and sufficiently anonymised and aggregated so that the risk of identifying individuals or vulnerable groups of individuals is almost zero. This means that if an individual can be identified, using all reasonable available means, then the aggregated and anonymised data will only have been pseudonymised and will thereafter fall within the scope of the General Data Protection Regulation (GDPR). Facebook has published anonymized, aggregated, rounded and with a confidentiality threshold of 1000 users statistical data derived from its users' activities directly though the Data for Good at Meta initiative as described in Chapter 7. In addition, a lot of demographic and migration research has been carried out by re-using data from Facebook's Advertising platform which is presented in Chapter 6 below. Similar limitations and corrections exist for all of the other Social Networking Site (SNS)s including Twitter, LinkedIn, etc., and will be discussed in the corresponding sections.

# 6. FACEBOOK ADVERTISING PLATFORM

The Facebook Advertising platform allows users to design targeted advertisements on the Meta family of applications by selecting the characteristics of the targeted audience. To this end, the Facebook Advertising platform provides statistics about the demographic characteristics and interests of its almost three billion monthly active user base. These characteristics of the Facebook users include age, gender, location and country of previous residence among others. Once users have selected the characteristics of the Facebook population that they wish to target in an advertising campaign, the advertising platform provides an estimate of the number of users that fulfil these characteristics. These estimates describe the Facebook users and not the general population. The advantages of these estimates are that they are timely, openly available, and cover most countries in the world. Several studies have tried to correct the selection bias in order to address key policy and research questions (Zagheni et al., 2017; Spyratos et al., 2019; Alexander et al., 2020). A recent study by Grow et al. (2021) using large-scale, cross-national online survey concluded that the Facebook Advertising platform classifies its users' sex, age, and region of residence correctly to a great extent (86% to 93%).

**FIGURE 3.** Distribution of Facebook users in Uganda by age and gender. The cyan coloured bar shows the % of population according to the 2019 UNDESA Projections. The blue bar shows the % of Facebook users in 2019.

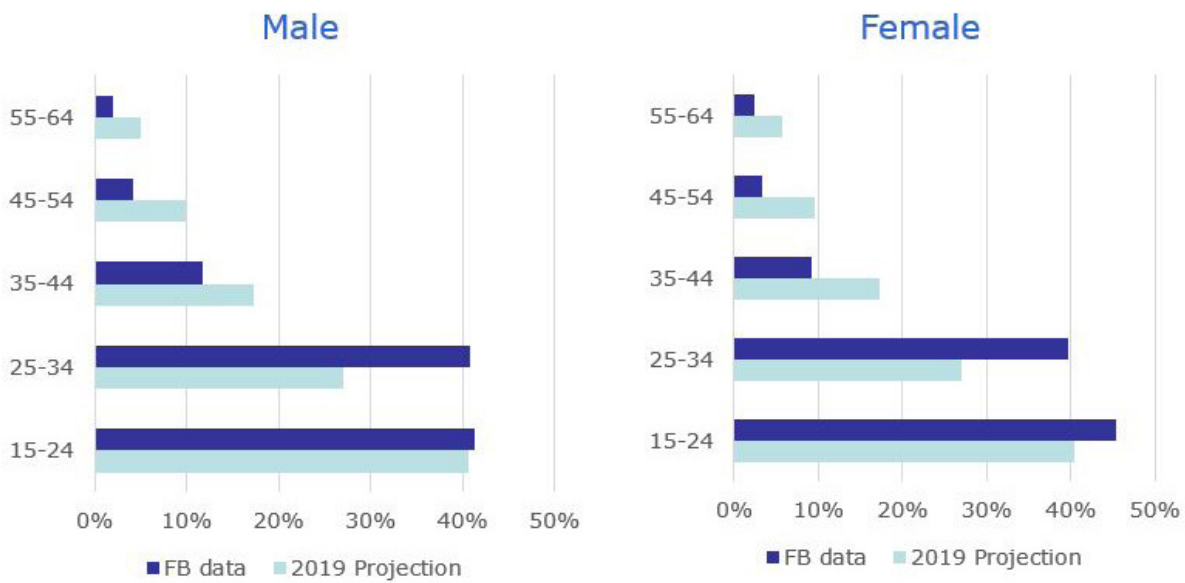

**FIGURE 4.** Distribution of Facebook users in Uganda by gender and education level in 2019.

| | Facebook enrolled or attained, Age 15-64 | | Distribution by education, age, gender IIASA Projection (2019) | | Distribution by education, age, gender IIASA (2015) | | Distribution of education attainment for persons aged 15 years and above. Uganda National Household Survey, 2009 | |
|---|---|---|---|---|---|---|---|---|
| | **Male** | **Female** | **Male** | **Female** | **Male** | **Female** | **Male** | **Female** |
| *Primary* | 2,0% | 2,6% | 50,9% | 54,3% | 55,3% | 58,5% | 60,7% | 65,9% |
| *Secondary* | 80,4% | 81,8% | 42,3% | 40,7% | 38,4% | 36,8% | 28,4% | 25,4% |
| *Tertiary* | 17,6% | 15,6% | 6,8% | 5,1% | 6,3% | 4,7% | 10,9% | 8,7% |

**FIGURE 5.** Model used to estimate the ''True'' Stock Number of European Migrants in the UK. Source: (Rampazzo et al., 2021)

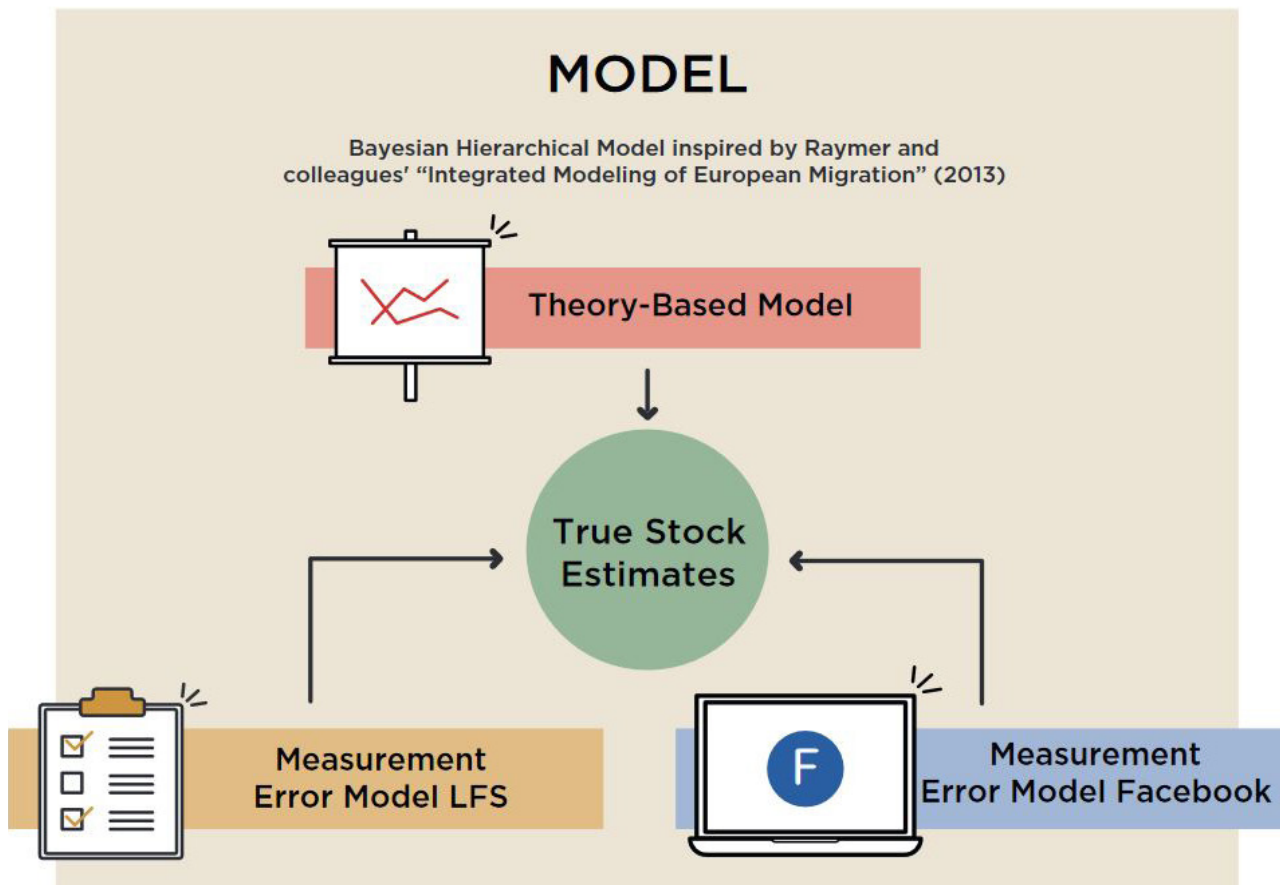

Statistics from the Facebook Advertising platform cannot replace official statistics because they are not representative, and the exact methods Facebook uses to classify its users are not known and might change over time. As described by Facebook in its Help Centre,[68] *''estimates from the Facebook Advertising platform aren't designed to match population, census estimates or other sources and may differ depending on factors such as: a) How many Meta/Facebook apps and services accounts a person has; b) How many temporary visitors are in a particular geographic location at a given time; c) Facebook user-reported demographics.''* In addition, Facebook often modifies its classification algorithms and metrics to improve them and make them more useful for advertising purposes. These changes need to be considered when estimates from the Facebook Advertising platform are reused for scientific purposes, and in particular for trend analysis, in order to avoid any data misinterpretation.

People of different ages, genders, origins, and in different countries are not using Facebook to the same degree. For example, as Figure 3 shows, Facebook users in the age-group 25-34 are overrepresented in Uganda. In addition, as shown in Figure 4, people with Secondary and Tertiary education in Uganda are overrepresented compared to people with Primary education. However, after carefully understanding and addressing the underlying selection bias and assessing the validity of the Facebook-derived estimates by comparing them with data from reliable sources, these estimates can be used for trend analysis and early-warning purposes. To provide more solid evidence on whether findings based on Facebook data are also valid for the general population, collaboration with Facebook is required to understand how data from the Facebook Advertising platform is produced.

Zagheni et al. (2017), and Spyratos et al. (2019) proposed two methodologies for correcting the selection bias of Facebook users and estimating the stock of Facebook ''migrants/expats'' in the real population. Alexander et al. (2020), Rampazzo et al. (2021), and Culora et al. (2021) combined Facebook data with traditional survey data to nowcast migrant stocks in the US, the United Kingdom (UK) (see Figures 5 and 6) and both the EU and US respectively. Dubois et al. (2018) used the Facebook advertising platform data to compare the interest profiles of different migrant groups to that of the German host population. Based on the interest similarities, the study derived an assimilation score and observed that European migrants have a higher assimilation score than Arabic-speaking migrants and Turkish speakers. Dubois et al. (2018) also concluded that the assimilation score varies between demographic subgroups with younger and more educated men scoring the highest. Palotti et al. (2020) used Facebook's advertising platform to estimate the number and the socio-economic profile of refugees and migrants from Venezuela to their neighbouring countries at national and sub-national levels.

**FIGURE 6.** Comparison of Facebook, Labour Force Survey, and model estimations (absolute numbers in thousands) of European migrants aged 15 or older for the years 2018 and 2019. LFS data is shown with 95% confidence intervals, and model estimates are shown with the interquartile. Source: (Rampazzo et al., 2021)

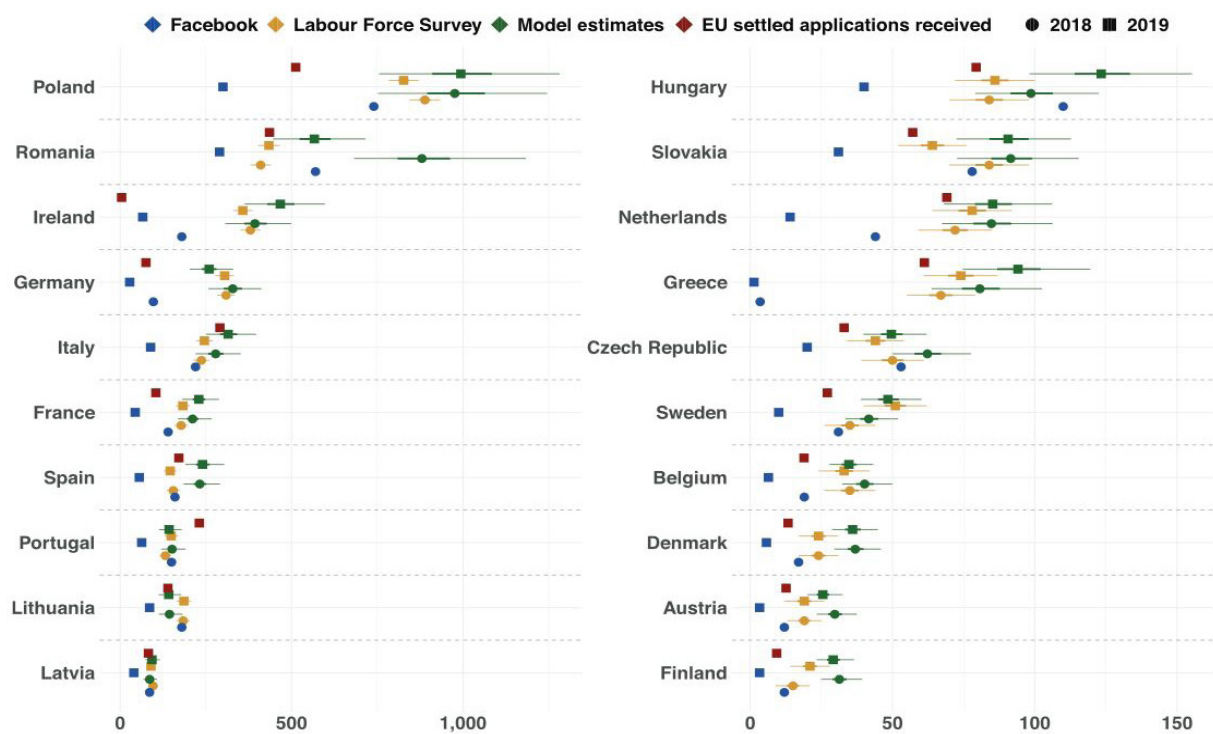

Garcia et al. (2018) and Fatehkia et al. (2018) demonstrated that data from the Facebook advertising platform can be used to estimate the digital gender gap at global scale. The Gender Divide in the usage of Facebook captures standard indicators of Internet penetration and gender equality indices in education, health, and economic opportunity (Garcia et al., 2018). Fatehkia et al. (2020) used Facebook's advertising platform data to capture geographic variations in wealth and poverty by gender in two low and middle income countries, the Philippines and India. Furthermore, Spyratos et al. (2018) demonstrated that gender inequalities in the use of Facebook across most of the countries in the world are highly correlated ($R^2 = 0.67$, p<0.001) with the UN Gender Development Index.

Rama et al. (2020) used the Facebook advertising platform data as a Demographic Tool to measure potential rural-urban inequalities in Italy. The authors found a good correlation between Facebook-derived and official statistics describing the rural-urban divide, and in addition they were able to provide insights into phenomena that have not been captured by official statistics yet. Finally, Rampazzo et al. (2018) demonstrated the feasibility of using Facebook advertising platform data to obtain an estimate of the mean age of childbearing.

This information exists in the official registers in developed countries, but it does not always exist in developing countries and so it can be used to fill data gaps for countries where official statistics are missing or not up-to-date.

## 6.1 NOWCASTING MIGRANT STOCKS AND FLOWS

Nowcasting[69] migrant stocks and flows can improve migration governance. Despite decades of calls by the international community to improve international migration statistics, the availability of timely and disaggregated data about short and long-term migration at the global level is still very limited. The Spyratos et al. (2019) study, drawing on statistics about the Facebook family of apps users who have lived abroad and applying a sample bias correction method, estimated the number of Facebook ''migrants''[70] in 119 countries of residence by age, gender, and country of previous residence. The proposed correction method estimates the probability of a person being a Facebook user based on his/her age, sex, and country of current and previous residence. By comparing Facebook-derived migration estimates the study successfully captured the increase in Venezuelan migrants in Colombia and Spain in 2018 as show in Figure 7.

---

69   The term nowcasting is used to indicate "the prediction of the present, the very near future, and the very recent past state" of an indicator. The term is a contraction of "now" and "forecasting" and originates in meteorology. Nowcasting estimates are used to monitor the state of a system (e.g., the economy) in real-time as a proxy of the value of an official statistics before it is released.

**FIGURE 7.** Stocks of international migrants and Facebook migrants from Venezuela in Spain. Source: Spyratos et al. (2019).

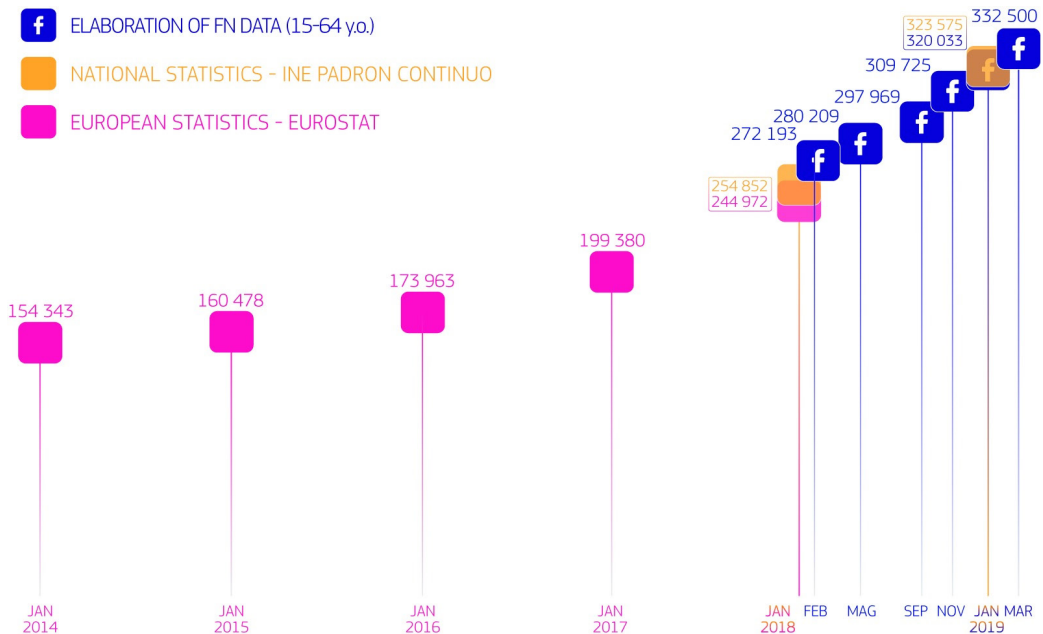

## 6.2 STUDYING MIGRANTS TRAVEL BEHAVIOUR

The KCMD study Spyratos et al. (2020) explored differences in the travel behaviour of migrant groups using anonymised, aggregated, rounded and with a confidentiality threshold of 1000 users statistical data from the Facebook advertising platform. Several studies have associated reduced geographical mobility with increased risk of social exclusion as well as socio-economic and psychological well-being. Therefore, geographical mobility can be used as a proxy for understanding the socio-economic statues of different migrant groups. For example, Figure 8 shows the correlation between the per capita annual incomes of individuals in the US by country of birth with the percentage of frequent travellers in the US by country of previous residence. The authors modelled the travelling behaviour of Facebook users grouped by countries of previous and current residence, gender, and age. The study found strong indications that the frequency of travel is lower for Facebook users migrating from low-income countries and for women migrating from or living in countries with high gender inequality. Such mobility inequalities impede the smooth integration of migrants from low-income countries into new destinations and their well-being. Moreover, the reduced mobility of women who have lived or currently live in countries with conservative gender norms capture another aspect of the integration which refers to socio-cultural norms and gender inequality.

---

70    By the term ''migrant'' we refer to the Facebook users who have been classified by the Facebook advertising platform as ''people who used to live in country X and now live abroad'', source: https://www.facebook.com/adsmanager/

**FIGURE 8.** Correlation between income and travel behaviour in the US. Source: Spyratos et al. (2020).

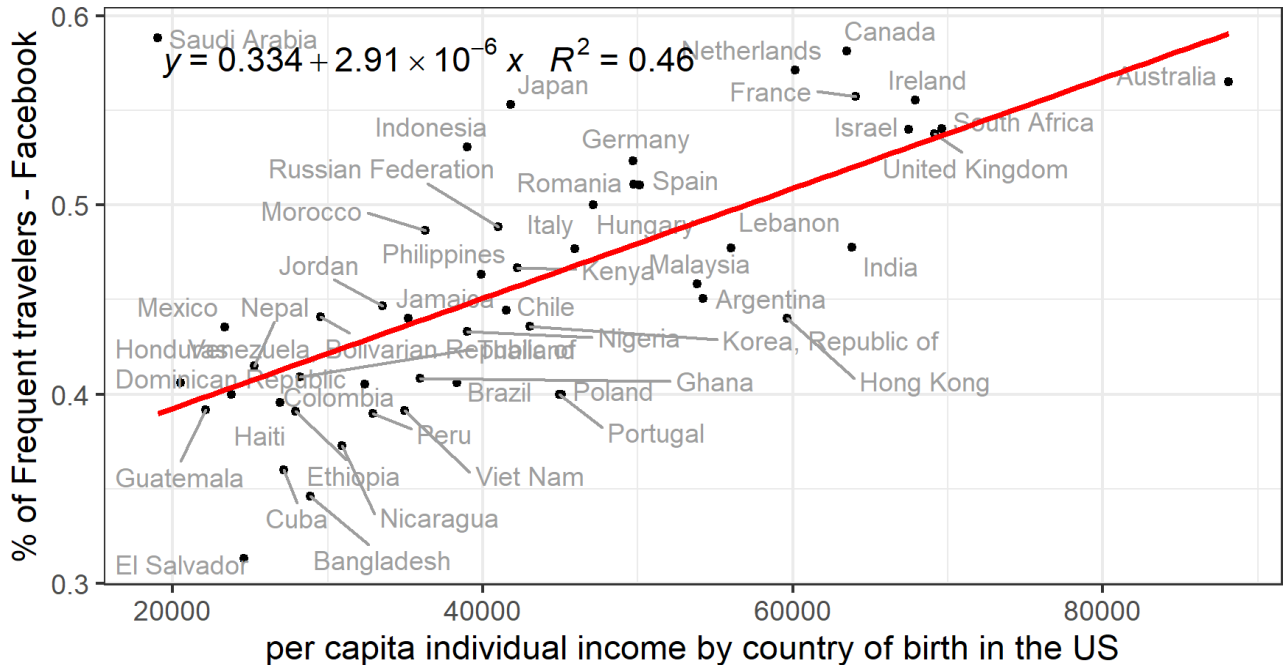

# 7. DATA FOR GOOD AT META

Data for Good at Meta (previously Facebook Data for Good)[71] is a data sharing initiative that aims to empower partners with privacy-preserving data algorithms that strengthens communities and advances social issues. Data for Good at Meta publish anonymised and aggregated data sets derived from multiple sources such as aggregated Facebook and Instagram user data and target surveys to Facebook users as well as satellite data, etc. These datasets aim to address humanitarian challenges such as disease outbreaks, natural disasters, and food insecurity. Some data sets are publicly available while others require data-sharing agreements. Various datasets which are relevant for demographic and migration research are presented below.

## 7.1 DISPLACEMENT MAPS

Displacement Maps[72] estimate how many Facebook users were displaced by a crisis, and where those people are in the period after the onset of the crisis event. In particular, Facebook classifies a user as displaced if his/her most common night location have changed after the crisis-trigger event. The night location is taken as a proxy of the home address of a user. Jia et al. (2020) study the use of Facebook displacement maps during the Mendocino Complex and Woolsey fires in California. The study concluded that even if the Facebook user base is biased in age and other demographic factor, these maps could reveal the trends, magnitude, and spatial clustering of population displacement in a timely manner with adequate representativeness out of the total population an intensively used area such as California. Acosta et al. (2020) used Facebook's Displacement Maps to quantify the dynamics of human mobility after Hurricane Maria in Puerto Rico in September 2017. These maps are created ad hoc looking at the time and space of where and when the event occurs, they are not meant to be monitoring tools in normal conditions as these maps are only created in case of an emergency or humanitarian situation. For example, Figure 9 shows the geographical areas of India and Bangladesh with higher population displacement due to the Cyclone Fani.

**FIGURE 9.** Cities with the highest percentage of displacement are shown in red and in blue those cities with the lowest percentage of displacement due to Cyclone Fani who hit India and Bangladesh in May 2019.
**Source:** https://research.fb.com/blog/2020/01/facebook-releases-improved-displacement-maps-crisis-response/

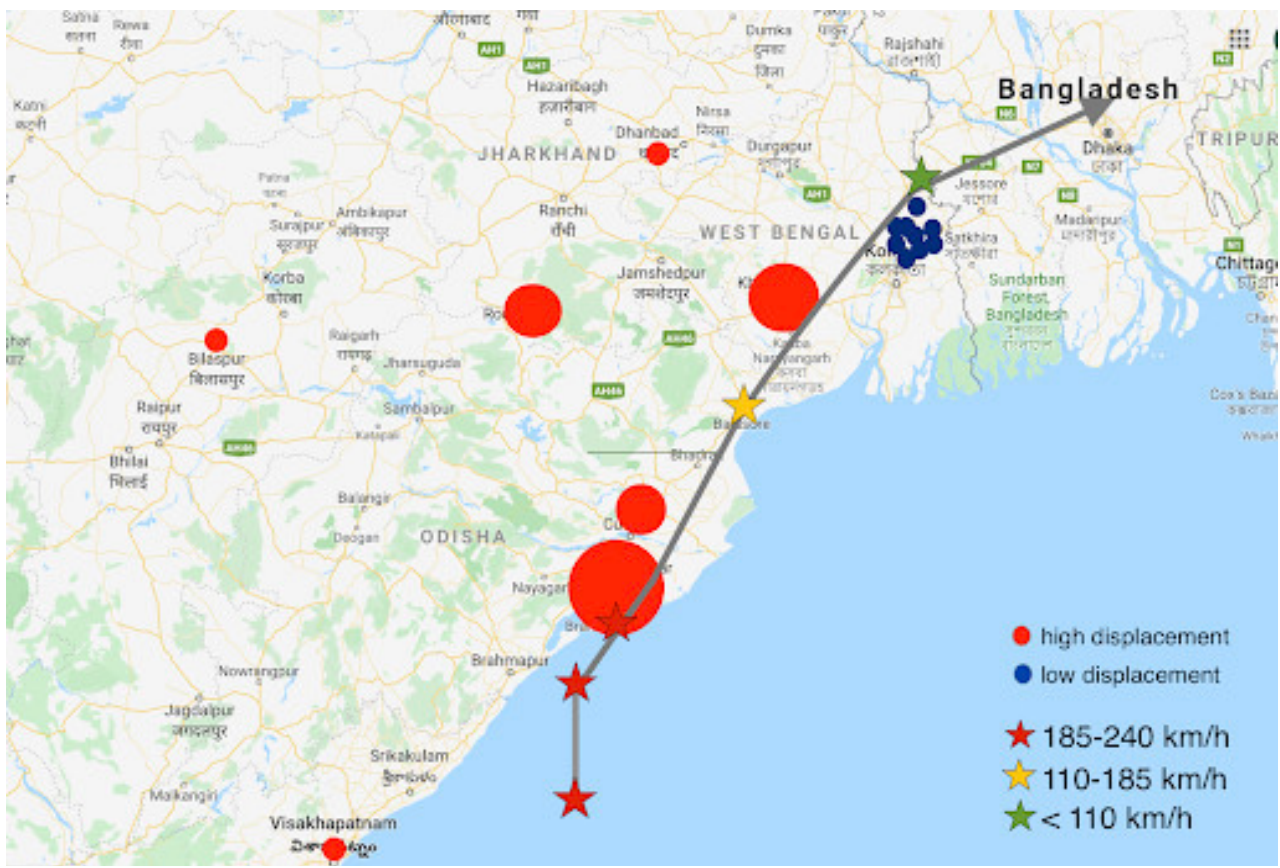

---

71    https://dataforgood.facebook.com/
72    https://dataforgood.facebook.com/dfg/tools/displacement-maps

## 7.2 SOCIAL CONNECTEDNESS INDEX

The Social Connectedness Index (SCI)[73] measures the strength of connectedness between two geographic areas as represented by Facebook friendship ties. SCI is very relevant for migration research since past migration patterns are highly correlated with the presence of increased Facebook friendship ties (Bailey et al., 2018). SCI can be used to measure the density of digital social connections around the world at national and sub-national level as well as up to zip code level for the US. The SCI between two locations i and j is estimated using the following formula:

$$\text{Social Connectedness}_{i,j} = \frac{\text{FB Connections}_{i,j}}{\text{FB Users}_i * \text{FB User}_j}$$

Where, FB Users$_i$ and FB Users$_j$ are the number of Facebook users in locations i and j, and FB Connections$_{i,j}$ is the number of Facebook friendship connections between the two.[74]

Bailey et al. (2020) used the SCI to understand how digital social connections in Europe are related to patterns of migration, past and present political borders, geographic distance, language, and other demographic characteristics. For example, as Figure 10 shows Facebook users in the Czech Republic are highly connected with Facebook users in Slovakia. Bailey et al. (2020) used the SCI to understand the digital social connections of Facebook users living in zip codes areas of the New York metro area to other zip codes of New York as well as to certain foreign countries. One of the conclusions of this study was that the degree of social connectedness of different New York zip codes is determined to a substantial degree by the presence of migrants from these countries in the respective zip codes. In a recent study (Tjaden et al., 2021) IOM combined the SCI with the United Nations Department of Economic and Social Affairs (UNDESA) estimates of international migrant stocks to explore whether these data have the potential to predict migration from one country to another. They tested the correlation between the SCI and bilateral stock of migrants from each country pair and they found that an increase in the strength of connections along the Facebook network, leads to an increase of migrants stock at destination, i.e., countries with a large migrant population like Mexico-United States or Morocco-Spain have a larger SCI value than other combination of countries with lower bilateral migration numbers like India-Argentina or Nigeria-Norway. More specifically, their model shows that an increment of 1% of the SCI corresponds to an increment of 0.7% of migrants stock (see also Figure 11).

**FIGURE 10.** Facebook Social Connectedness Index values for the Czech Republic. Source https://africapolis.org/sci/

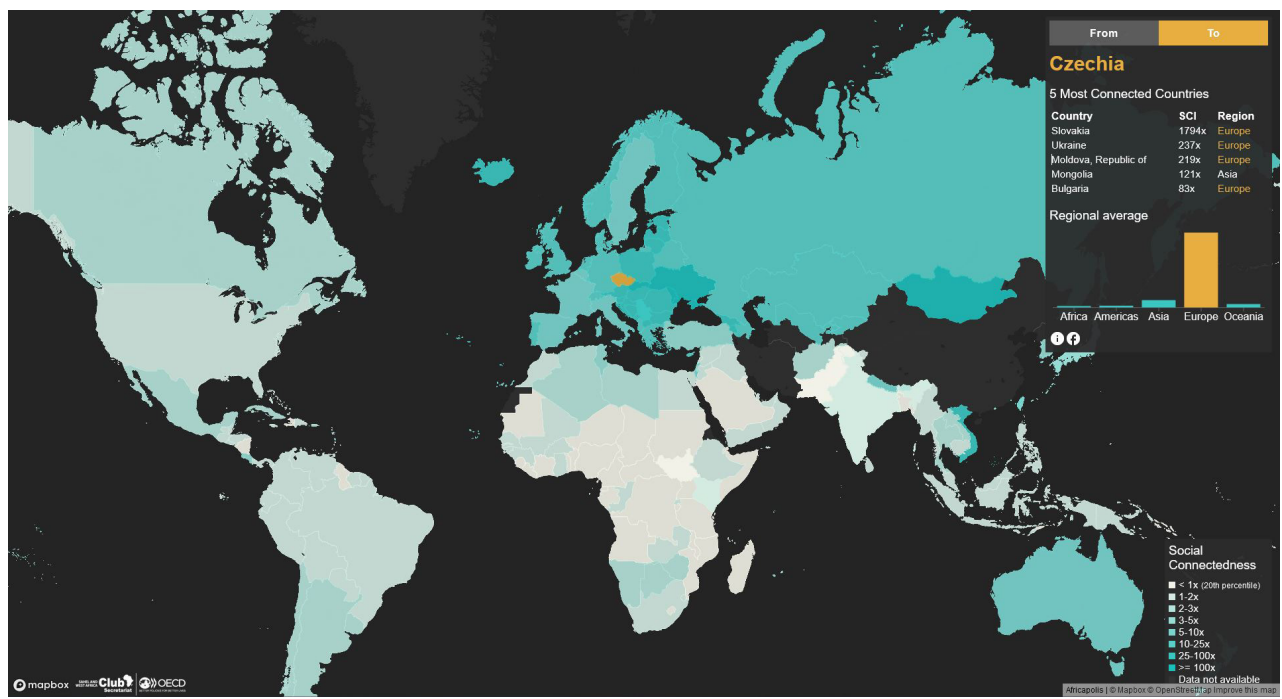

73    https://dataforgood.facebook.com/dfg/tools/social-connectedness-index
74    https://dataforgood.facebook.com/dfg/docs/methodology-social-connectedness-index

**FIGURE 11.** Higher levels of migrants from one country living in another (bilateral stocks) are associated with a higher probability of Facebook friendship links between users in both locations. A combination of countries with a large migrant population like Mexico-United States or Morocco-Spain have a larger Facebook Connectedness Index than other combination of countries with lower bilateral migration numbers like India-Argentina or Nigeria-Norway. Source: Figure 3 from Tjaden et al. (2021).

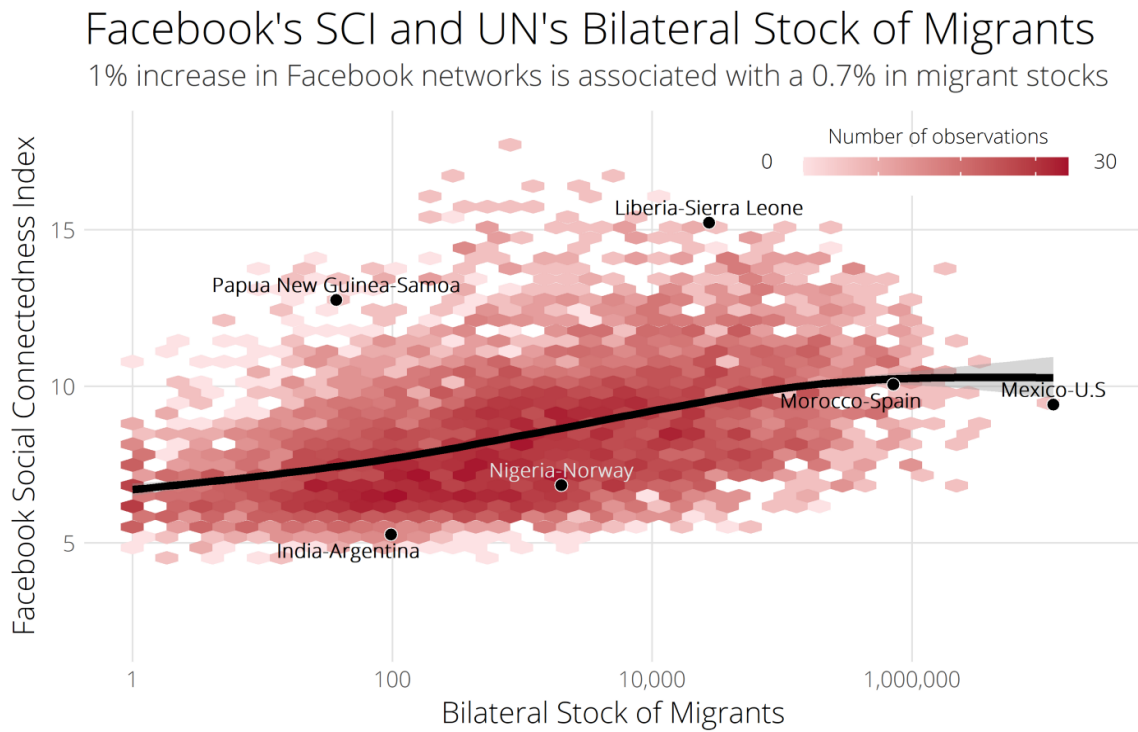

## Facebook's SCI and UN's Bilateral Stock of Migrants
1% increase in Facebook networks is associated with a 0.7% in migrant stocks

Data: United Nations Population Division, 2019 | Facebook SCI, November 2020 | Chart: GMDAC

## 7.3 COMMUTING ZONES

Facebook Commuting Zones[75] are geographic areas defined by the mobility patterns of Facebook users. These geographical areas represent areas where Facebook users spend most of their time and, for example, they can be used to understand where people live and work, local economies, and how disease might be transmitted. Facebook Commuting Zones, a concept similar to the Mobility Functional Areas (Iacus et al., 2021), can also be used for demographic and migration research. For example, Facebook Commuting zones can be used to estimate different populations of interest such as daytime and night-time populations as well as to identify disadvantaged and geographically isolated areas characterised by an aging population and depopulation. Apart from each Commuting Zone's shape, Facebook provides additional information about its economic and commuting characteristics. Facebook Commuting Zones can cross political boundaries and can be updated every few months in order to capture mobility changes in a timely manner and at almost global level. An example representation of Facebook commuting zones is shown in Figure 12.

## 7.4 TRAVEL PATTERNS

Facebook's Travel Patterns[76] provide a daily count of the total Facebook mobile app users who have opted to share location data and have moved from one country to another, see for example Figure 13. Du et al. (2021) used Facebook's Travel Patterns data to explain the initial spread of the SARS-CoV-2 Alpha variant to 15 countries by travellers from the United Kingdom. Travel patterns can be used to quantify tourist as well as migration flows.

## 7.5 MOVEMENT RANGE MAPS

Facebook's Movement Range Maps[77] are metrics that describe whether Facebook users are moving or they stay within a small area surrounding their home for an entire day compared to the pre-COVID 19 period. This data is estimated on a daily basis and cover 153 countries at national and sub-national level. These datasets aim to inform researchers and public health experts about how populations are responding to physical distancing measures. Cortés et al. (2021) used Movement Range

---

75    https://dataforgood.facebook.com/dfg/tools/commuting-zones
76    https://dataforgood.facebook.com/dfg/tools/travel-patterns
77    https://dataforgood.facebook.com/dfg/tools/movement-range-maps

Maps to explore the potential of using aggregated and anonymised mobility data to fight the COVID-19 pandemic. Various studies have used Facebook Movement Range Maps data to understand the impact of mobility restrictions on public health and the economy (Spelta et al., 2020; Pérez-Arnal et al., 2021; Mena et al., 2021).

**FIGURE 12.** Facebook Commuting Zones in North America. Source: Facebook, 2021.

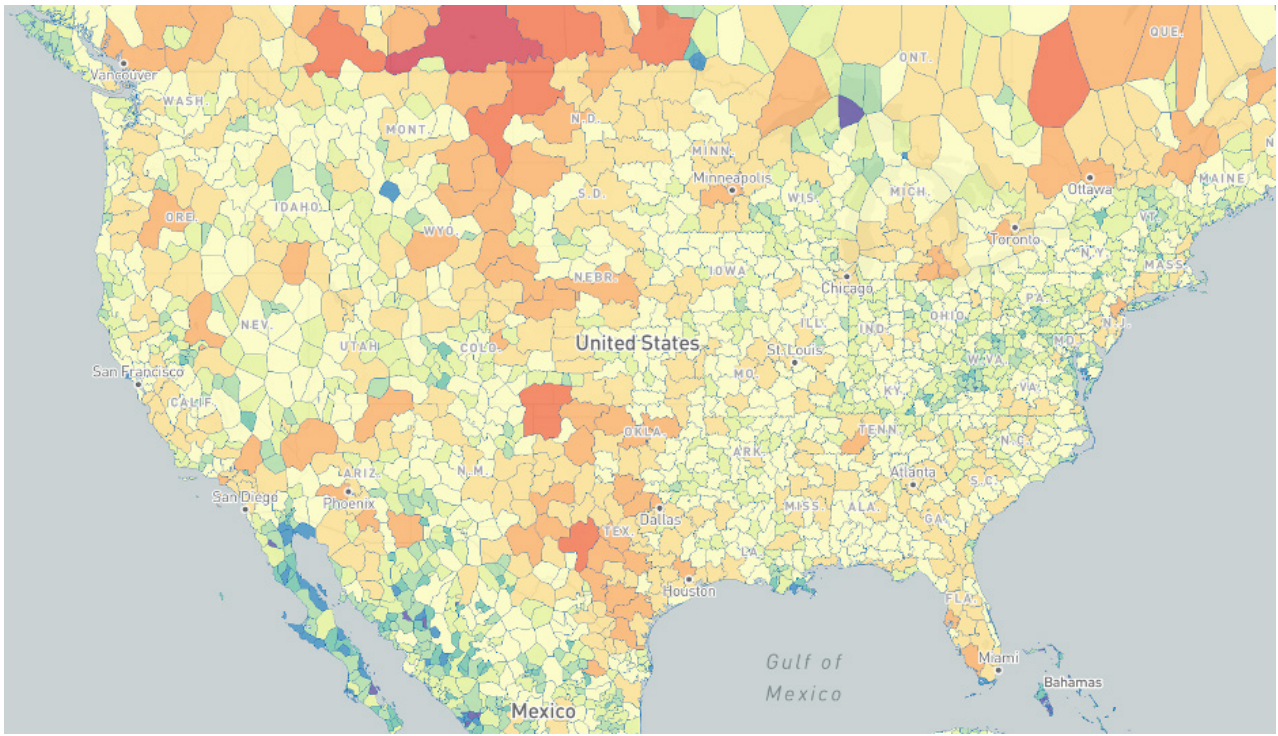

**FIGURE 13.** Travel across countries during the COVID-19 pandemic. Source: Facebook, 2021.

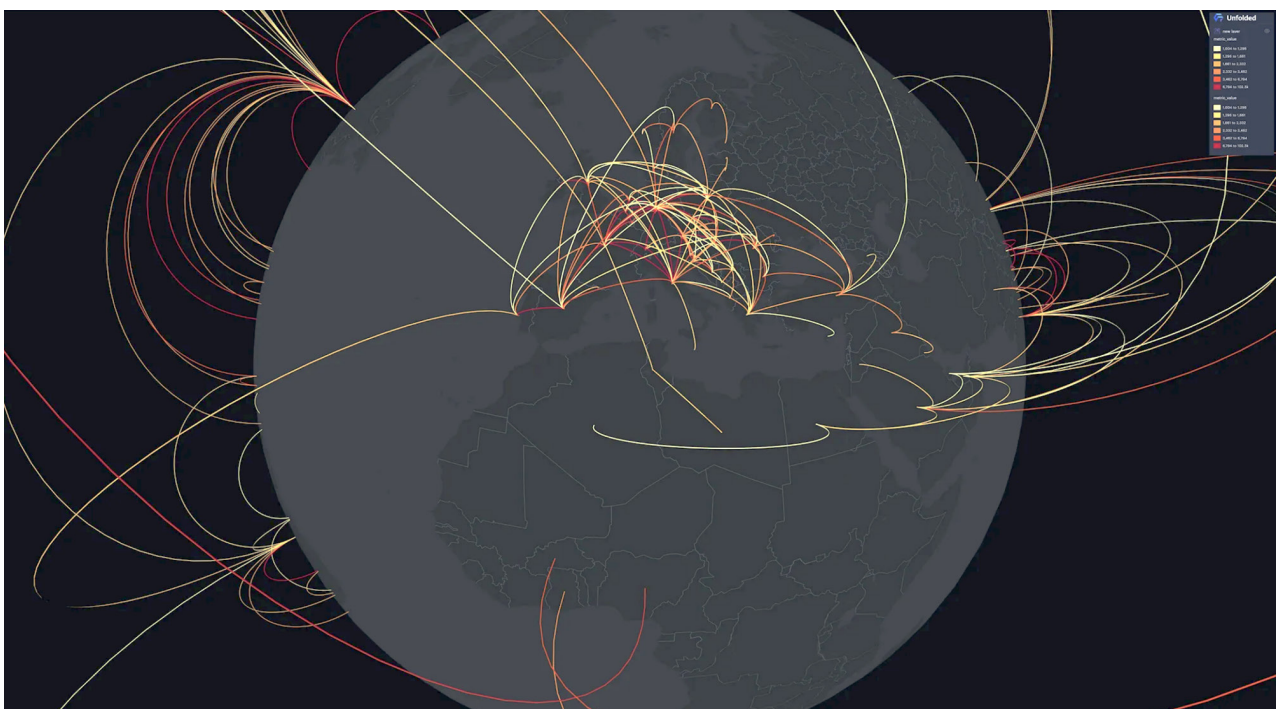

# 8. FACEBOOK SURVEY DATA

Facebook application is being used to deploy large scale online surveys of specific demographic groups. Grow et al. (2021) conducted an online survey in seven European countries (Belgium, France, Germany, Italy, the Netherlands, Spain, and the United Kingdom) and in the United States, with the goal of collecting information about people's attitudes and behaviours in response to the COVID-19 pandemic. A total of 137 224 questionnaires with complete information on respondents' sex, age, and region of residence were collected between March 13 and August 12, 2020. Across countries, approximately 86% – 93% of respondents' answers matched the Facebook Advertising Platform's categorisation on user sex, age, and region of residence. As a result, the Grow et al. (2021) study concluded that the use of Facebook is a valuable tool that under certain conditions can be used to recruit participants in online surveys. Pötzschke and Braun (2017) and Pötzschke and Wei (2021) used the Facebook Advertising Platform to respectively carry out a targeted on-line survey on Polish Migrants in Four European Countries and German emigrants in countries beyond Europe. Both studies concluded that Facebook can be considered to be both accurate and cost effective in recruiting people at international level that belong to geographically dispersed demographic groups and which are generally otherwise hard to reach.

Even though most researcher using the Facebook Advertising Platform were able to correctly target the desired audiences, selection bias is still present. The precision in most studies was at 90% level. However, the recall/sensitivity is not known. This means, for example, that groups of people who do not use or are prevented from using social media like women living in specific communities will never be reached by these surveys. As a result, the representativeness of Facebook surveys remains a challenge.

## 8.1 SURVEY ON GENDER EQUALITY AT HOME

Facebook's Survey on Gender Equality at Home[78] is a survey that reached over 460,000 Facebook users in 208 countries in July 2020. This survey generated a global snapshot of women and men's access to resources, their time spent on unpaid care work, and their attitudes about equality. As shown in Figure 14, women are less likely to fully cover their own expenses compared to men. It should be mentioned here that women who live in households and societies with high gender inequality are less likely to use Facebook (Fatehkia et al., 2018) and so are not captured by Facebook surveys.

**FIGURE 14.** Percentage of survey respondents by gender and geographical region who answer that they do not fully cover their own expenses.
**Source:** https://www.equalityathome.org/charts

Question B.3: During the last 12 months, which of the following -most closely- reflects your current financial situation?

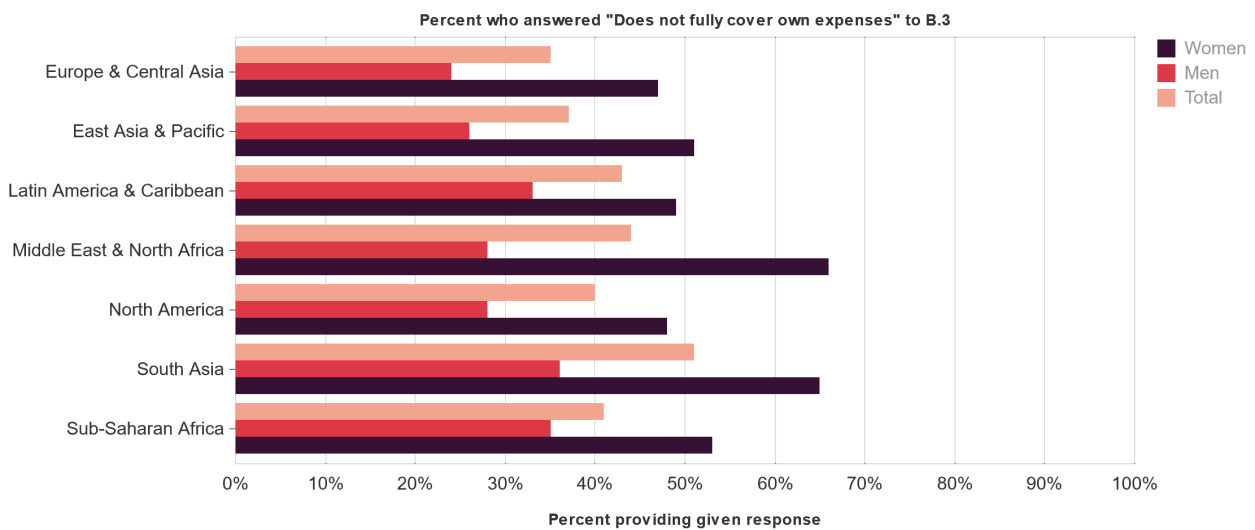

# 9. INSTAGRAM

Instagram is a platform of the Meta company which focuses on visual content. It is a platform which has become increasingly relevant for political discourse. Users share pictures or short videos - called stories - which other users can re-share, comment on, like, etc., in a similar way to other SNSs. However, compared to other platforms, Instagram content has not been extensively researched (Highfield and Leaver, 2015) and only a few studies have focused on migration and the refugee crisis (Chouliaraki, 2017; Guidry et al., 2018) using digital methods (Geboers et al., 2016; Sánchez-Querubín et al., 2018). Like Twitter, Instagram allows the user to record geographic information in the metadata of the posts.

## 9.1 THE VISUAL FRAMING REPRESENTATION OF THE SYRIAN REFUGEES

Radojevic et al. (2020) analysed 367 images posted from the refugee camp in Idomeni between 7 March 2016 and 15 March 2016. In April 2016 Instagram changed the policy and heavily restricted access to their Application Programming Interface (API). The authors classified the images in several categories based on the visual framing classification introduced in Chouliaraki (2017). They could identify the following frames: *victimisation* of refugees, *humanitarian aid*, *messages from Idomeni*,[79] *securitisation*[80] and *solidarity*. But in the end their analysis shows that refugees remained on the lens side of the camera, retaining their status as subjects rather than actors.

## 9.2 MONITORING THE SYRIAN REFUGEE CRISIS

A related study by Iacus and Teocharis (2017) analysed 5 000 public posts collected from the same Idomeni camp during the period 11-21 February 2016. By manually analysing about 3 000 user profiles[81] corresponding to the collected posts, they found that they mostly belong to individuals/volunteers, NGOs and press reporters as well as migrants. The authors tried to study both the present content and the future timelines of these accounts.

To follow these account in the future, it was necessary to explicitly ask for permission to look at users timelines and this was agreed in 28% of the cases, among them only a relatively small portion (1%) of the open accounts were identified as migrants and another 30% were only likely to be migrants. The post-crisis analysis involved analysing posts from these users from September to November 2017. Among the present stock of posts in 2016, a language analysis on the posts identified that most of the migrants were from Syria, then generic Arabic speaking accounts, followed by Iranians and Iraqi people and finally Kurdish and Afghans with about the same proportions. In terms of shared content, three main topics emerge: stories about personal life in the camps by migrants, migrant life documented by reporters, and protest or politically related posts. Finally, the ex-post analysis of migrants accounts in late 2017 showed that some of them have been relocated to several EU countries. Their distribution correlates with the cumulative flows of migrants over the period 2010-2016 quite well (correlation about 0.75).

Both studies are limited in their scope and one problem lies in the availability of the data source due to the restriction in the usage of the data though this can be addressed as other platforms[82] did through aggregation and anonimisation of the raw data and differential privacy approaches (see for example Dwork, 2008). In all events when it comes to the identification of individuals, the GDPR regulation is in force, and these analyses should be carried out in accordance with the GDPR regulation.

---

79  Posts by volunteers or activists commenting on current events and taking a stand in the discourse.

80  This frame includes images of anticipating refugee crowds, rioting and/or protesting groups, police force, and border fences/barbed wire fences which marked the border crossing.

81  This analysis ended in 2017 and was conducted in accordance with GDPR rules even though the European Data Protection Regulation was in force since 25 May 2018. Data have been collected through the official Instagram API's. Accounts have been pseudoanonymised, the text of the post was used only to identify the language and the image to classify the visual content, the source data have been deleted when these information were obtained and aggregated. Only aggregated results were retained.

82  See the SocialScienceOne initiative at https://socialscience.one/.

# 10. FLICKR

Flickr (now a Yahoo! company) is a platform like Instagram mainly used to share photos. Compared to Instagram, Flickr has mostly been used by professional photographers. However, its popularity has been declining with the passage of time. Today it has about 87 million users compared to the 800 million of Instagram.

## 10.1 MONITORING THE SYRIAN REFUGEE PATHWAY

A few attempts were made to study the refugee crisis especially in 2015-2016. For example, Curry et al. (2019) explored the discourses around refugees in several countries and, for example, found that most of the posts identified as images about refugees were posted from Greece, Turkey, Spain, Hungary, Italy, Germany, and the UK as expected, in line with what has been observed on Instagram in parallel. However, on Flickr, Algeria, Sudan, and Iraq had more substantial photograph density then the previous set of data which suggests other significant migration processes.

## 10.2 CHARACTERISING INTERNATIONAL MOBILITY

Yuan and Medel (2016) analysed a Flickr dataset of 100 million photos released by Yahoo! in 2014 (Thomee et al., 2016). This study mainly aimed at validating the magnitude of travel outflows from 12 countries[83] by using official statistics. Based on a gravity model approach, they found that apart from China, the Flicker based statistics and official statistics correlated well. Using the same data, Beiró et al. (2016) analysed domestic travel flows in the US, while Barchiesi et al. (2015) specifically analyzed the travel flows to the United Kingdom.

---

83    USA, UK, Spain, Germany, Italy, France, Canada, Australia, Brazil, Japan, Netherlands, and China.

# 11. TWITTER

Twitter is a micro-blogging platform that has been available since 2006 and has been increasingly used worldwide since 2010. One of the main advantages of large-scale datasets like those that can be collected from this platform is their continuous updating. On the other hand, like for any SNS these data also have some intrinsic limitations. First of all, users of these platforms are not a representative sample of the whole population though adjusting procedures can be applied as per Iacus et al. (2020) in order to make the results more general but, above all, and despite their limited representativeness, SNS can be considered to be a sort of opinion-making arena, where ideas expressed affect or anticipate collective sentiment and trends. As Salganik (2017) stated, there is always a risk of *drifting* in constructing indicators based on social media data due to the fact that, not only the reference Twitter population is not representative, but it might also change composition through time or the users may post different volumes of tweets at different times, etc.

The main advantage of using Twitter data is that it has public and free of charge APIs that allow the data to be collected without the need of doing unofficial scraping. As per the official documentation, the standard Twitter search API only provides a 10% sample of all tweets though the company does not disclose any information about the representativeness of the sample compared to the whole universe of tweets posted on the social network. Nevertheless, in large scale battery of experiments Hino and Fahey (2019) confirmed that the coverage of topics and keywords is quite accurate and appears to be randomly selected, therefore it is possible to consider Twitter data obtained through the standard API as a representative sample of what is posted on Twitter. On the other hand, a recent academic programme[84] allows for more flexibility in accessing large volumes of data and offers a refined search tool. A commercial version of the APIs is also available.

Another advantage of Twitter data is that it allows to study networks of relationships, which may help different types of migration studies such as migrants' integration, diversity, on-line discourse and mood on migration topics, as well as the so-called digital demography.

Indeed, most Twitter-based studies either apply textual analysis or network analysis or a combination of both. The following sections report a few examples of Twitter data being used in the context of migration studies.

## 11.1 CULTURAL INTEGRATION OF MIGRANTS

Kim et al. (2021) suggested that indicators of cultural integration should be defined from the linguistic perspective, starting from textual analysis of and geographic information in Twitter posts. These authors exploit the link between cultural traits and the country of origin or the country of destination and created two indexes call home attachment (HA) and destination attachment (DA). The two indicators HA and DA were defined by taking the proportions of country-specific hashtags that either belongs to the country of residence (DA) or the country of nationality (HA). The study finds that the proficiency of the language of the host country corresponds to a higher level of DA while for countries that share borders, both the DA and HA level increase. At the same time, the further the destination country, the higher the DA level. The study also characterize these indexes in terms of the Hofstede's cultural dimensions (Hofstede et al., 2010). It has been found that the higher the differences between the origin and destination countries in terms of individualism, masculinity and uncertainty, the higher the level of DA. The opposite occurs for the HA index. This study was not meant to apply causal models, so the proficiency of the language of the host country could facilitate higher DA levels, but this relationship could also be true the other

......................................................................................................

**FIGURE 15.** Left: Box-plots for the DA and HA index of immigrants in the United States. Right: Scatter plot of HA vs. DA indicating approximate integration types for immigrants in the US. Source: Figure 8 from Kim et al. (2021).

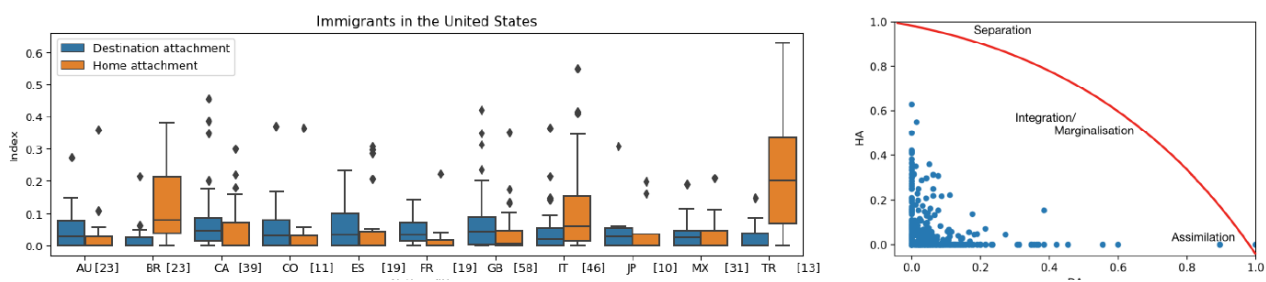

way round. The study has a somewhat limited validity in that it only analyses three thousand Twitter users out of 60 million tweets collected because of a lack of geographic information, but the methodology used to produce the HA and DA indicators seems to be widely applicable and interesting. Figure 15 shows an example of analysis of the HA and DA indexes for the United States.

## 11.2 MIGRANTS SEGREGATION

A strongly related topic but with a more geographic analysis approach, is found in the study conducted by Institute for Cross-Disciplinary Physics and Complex Systems, Palma de Mallorca (Spain), aimed at proposing a metric to assess spatial segregation of immigrant communities in a subset of the 53 most populated cities in the world[85] (Lamanna et al., 2018). The study analysed the geo-localised tweets sent from each city between October

2010 and December 2015. The metadata associated to each tweet include the user ID, the geographical coordinates (latitude and longitude), the date and time and the text of the tweet. Non-human tweeters, i.e. Twitter bots, were filtered out leaving a total of 350.9 million tweets posted by 14.5 millions of users in the 53 cities. Twitter users were assigned to a location according to their most frequented grid cell between 8pm and 8am, their regular movements around the city, a minimum number of consecutive months of activity, etc. Users are then identified as being in a given community of immigrants according to a machine learning model based on the language most frequently used in tweeting. For example[86], if a user regularly tweets in Spanish from an address in Boston, the model would label the user as a Spanish immigrant. The spatial segregation measure for each immigrant community and city is obtained starting from a bipartite network diagram that links each language to each city. The spatial integration measure is based

......................................................................................................................

**FIGURE 16.** Clusters of cities and Power of Integration. In (a), three groups of cities show similar behaviour in the number of communities detected and in their levels of integration. The height of the bars represents the number of languages (communities) detected in each city. The colour scale is representative of the decay of the entropy metric. The Power of Integration metric evaluate the potential of each city to uniformly integrate immigrant communities into its own urban area according to entropy values. In (b), decay of the normalised Shannon entropy ($h_{l,c}$) for the cities in each cluster.
**Source:** Figure 4 from Lamanna et al. (2018).

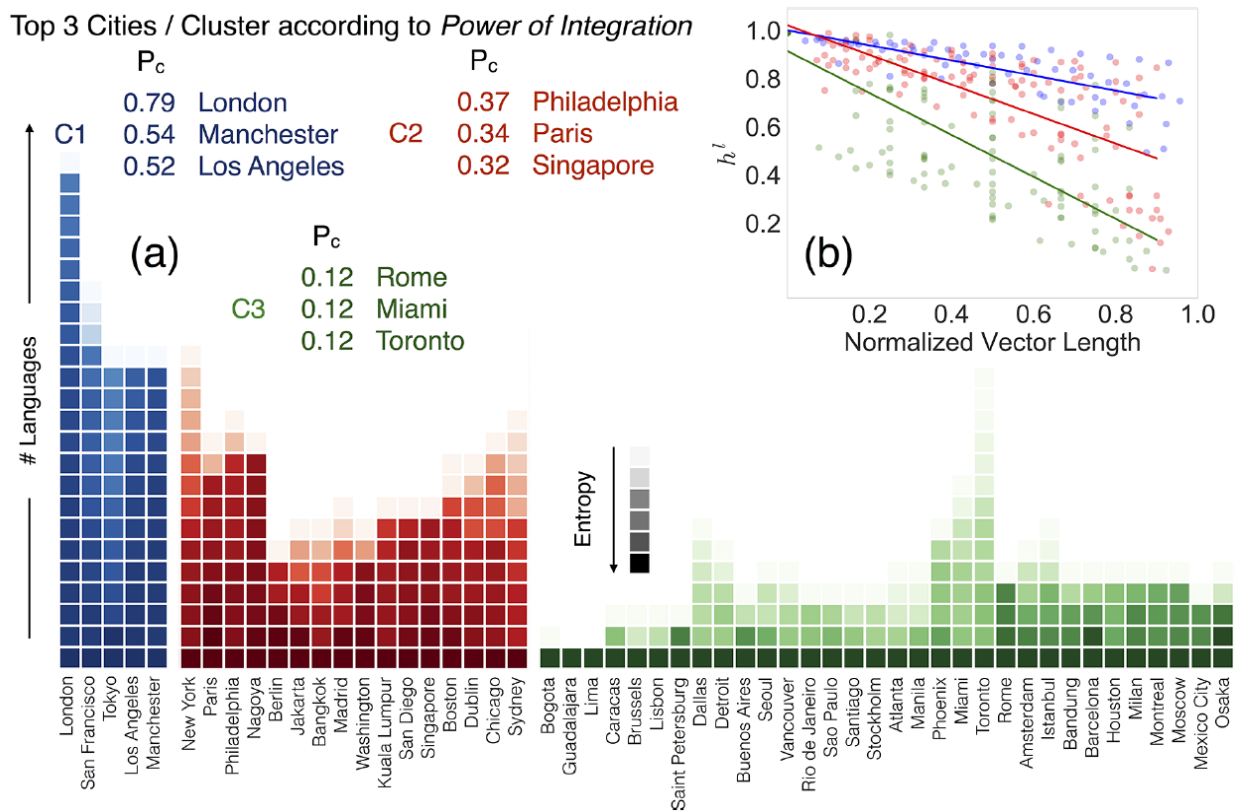

---

on the concept of Shannon entropy.[87] Researchers were able to identify 35 immigrant communities in 50 cities and assigned the neighbourhoods in those cities where each likely immigrant community lives. The immigrant community is considered to be well-integrated into the city if large portion of native residents also live in the same neighbourhoods. Using the normalised entropy index, the concept of *Power of Integration* is calculated for each city *c* as follows:

$$P_c = \frac{L_c}{L_{\max}} Q_2(1 - IQR)$$

where $L_c$ is the number of languages spoken in the city *c*, $L_{\max}$ is the maximum number of languages across the whole set of cities, $Q_2$ is the median value of the entropy measure and *IQR* is its interquartile range ($Q_3 - Q_1$) used as a measure of dispersion. The index $P_c$ is maximum when the median of the normalised entropy measure is one or larger, *IQR* = 0 and the number of languages hosted by city c is the maximum. On the other hand, it tends to zero when there are no hosted language, the languages are spatially isolated with $Q_2 = 0$, or when the *IQR* = 1 covering the full range of values. The $P_c$ index is designed to capture the contribution in the spatial integration process within each urban area. Figure 16 shows an example of outcome. The study also found patterns and distinctions among immigrant groups as a whole, regardless of the cities where they settled. For example, Korean, West-Slavic, and Dutch communities tend to receive the highest integration scores. Spanish-speaking immigrants make up the most common immigrant communities in world cities and generally have high integration scores, and so forth.

## 11.3 SUPERDIVERSITY INDEX

In a series of studies Pollacci (2019), Pollacci et al. (2021), and Sîrbu et al. (2021) introduced the notion of Superdiversity based on Twitter data. Superdiversity means a new level of cultural diversity due to immigration and cultural differences between the immigrants themselves (Vertovec, 2007). The authors create a Superdiversity Index (SI) based on the idea that "*different cultures assign different emotional valence to different words*". The SI is calculated as the distance between the "standard" emotional valence of a set of words and the "*used*" valence in the population of a region or a community.

**FIGURE 17.** Superdiversity index (left) and immigration levels (right) across UK regions at NUTS2 level.
**Source:** Figure 2 from Sîrbu et al. (2021).

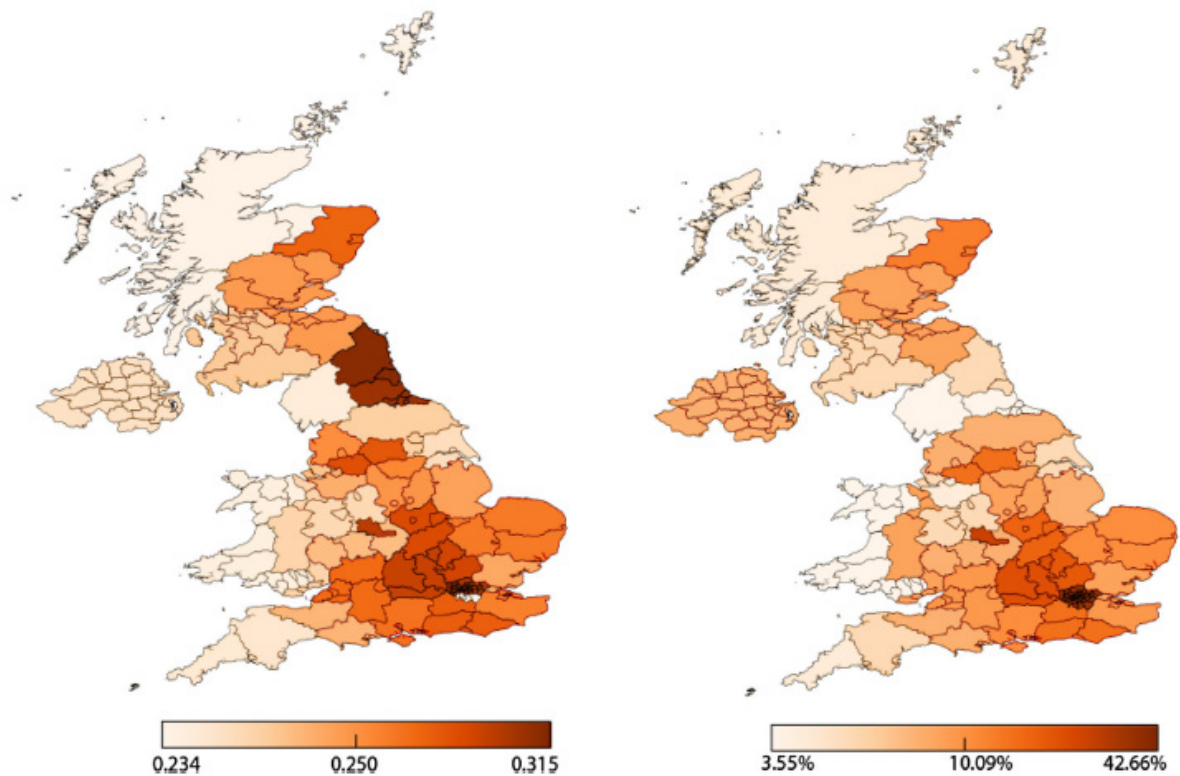

| | | |
|---|---|---|
| 0.234 | 0.250 | 0.315 |
| 3.55% | 10.09% | 42.66% |

---

87  Entropy is mathematically defined as a measure of information present in a communication exchange: information is low if no surprise is contained in the message, and high if the message brings in something unexpected. In statistical terms, if there is high dispersion of a statistical variable the entropy is high, and it is low in case of high concentration. Shannon entropy-like descriptors have been used before in the context of spatial segregation of ethnic minorities (see White, 1986).

Without going into too much details on how to extract valence and meaning from tweets,[88] its calculation is based on the following formula:

$$SI = \frac{1 - \bar{r}}{2},$$

where $\bar{r}$ is the average value of the similarity measurement between ''*standard*'' and ''*community*'' use of a language.

The index is such that:

$SI$ = 0, if there is no diversity, population uses the language in a standard way;
$SI$ = 0.5, very large diversity, the emotional content of words is not related to the standard one;
$SI$ = 1, emotional content of words is reverted compared to standard language.

The authors calculated the $SI$ using a sample of geolocalised Twitter data for the United Kingdom, Italy, the Netherlands, Germany, France, Spain, and Ireland. They analysed the pattern of SI at different granularities (NUTS0 to NUTS3) comparing it with high resolution data on immigration rates from the JRC D4I dataset (Alessandrini et al., 2017). Figure 17 shows the results for the United Kingdom. The analysis shows strong correlation between the SI and the corresponding index based on D4I data allowing the authors to propose their measure as a nowcasting tool for estimating migration stocks.

## 11.4 MOOD TOWARDS MIGRANTS AND MIGRATION

Coletto et al. (2017) analysed the sentiment toward the refugee crisis in 2015. They collected tweets written in English posted from mid-August to mid-September 2015. They collected about 1.2 million posts from 47 824 unique users. They classified each user as one with either positive or negative sentiment towards the refugee crisis and created the ratio index $\rho$ as follows:

$$\rho = \frac{\text{number of accounts expressing positive sentiment}}{\text{number of accounts expressing negative sentiment}}$$

and they classify several countries in Europe according to this metric. The result is shown in Figure 18. The study revealed that Europeans mostly express positive sentiments toward the refugees, but this attitude changes when a country is more exposed to migration flows. Righi et al. (2021) collected around 2 400 tweets per day in Italian language in the period January 2015 to October 2018 and classified their mood using unsupervised sentiment analysis to derive an index of migration mood (DIV) based on the ratio between the number of positive tweets over the sum of positive and negative tweets.

$$DIV = \frac{\text{number of tweets expressing positive sentiment}}{\text{number of tweets expressing negative or positive sentiment}}.$$

Their analysis shows that the mood toward migration, as measured through this system, seems to move from an initial positive area to a negative one during the summer 2016 crisis when the arrivals of migrants consistently increased, with negative sentiment deepening after March 2018 (see also Figure 19).
The above mentioned results are just a few examples of various studies (Scettri, 2019; Kim et al., 2020; Arcila-Calderón et al., 2021; Pitropakis et al., 2020; Rowe, 2021b; Bollen et al., 2021; Freire-Vidal et al., 2021; Rowe et al., 2021) investigating the use of Twitter data to measure the mood towards migration.

· · · · · · · · · · · · · · · · · · · · · · · · · · · · · · · · · · · · · · · · · · · · · · · · · · · · · · · · · · · · · · · · · · · ·

**FIGURE 18.** Index ρ across European countries: red corresponds to a higher predominance of positive sentiment, yellow indicates lower ρ. (a) Refers to the whole dataset. (b) Is limited to users when mentioning locations in their own country. (c) Is limited to the remaining users
**Source:** Figure 8 from Coletto et al. (2017).

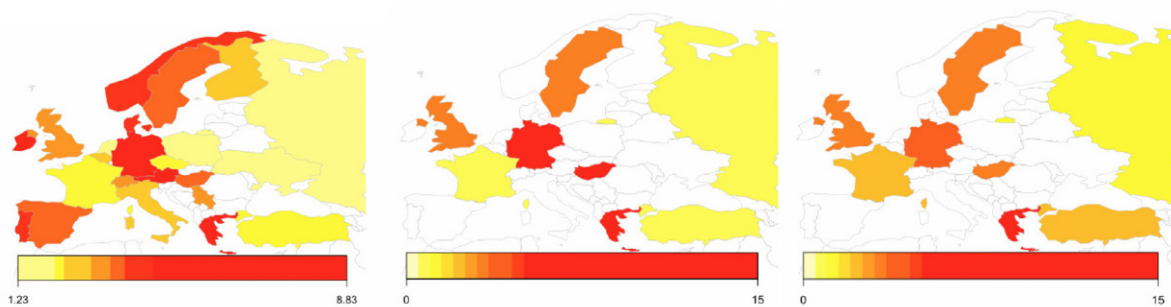

**FIGURE 19.** The Daily Value Index of sentiment towards migration. Source: Righi et al. (2021).

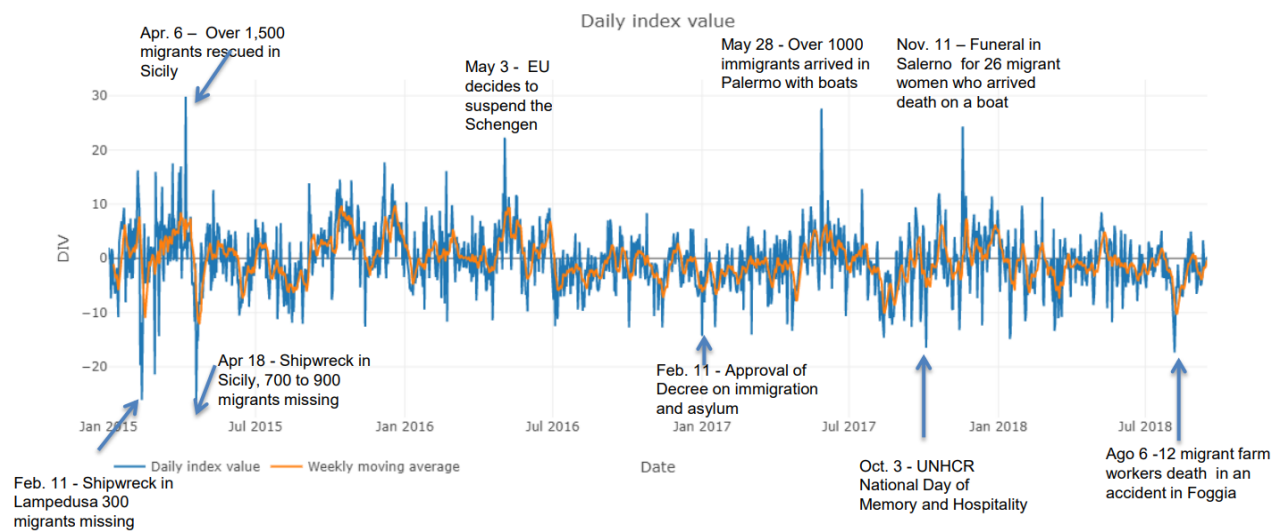

## 11.5 MONITORING HUMAN MOBILITY

Among other authors (Azmandian et al., 2013; Jurdak et al., 2015; Comito, 2018; Liao et al., 2019; Osorio Arjona and García Palomares, 2020; Huang et al., 2020; Bisanzio et al., 2020; Xu et al., 2020; Jiang et al., 2021), Hawelka et al. (2014), exploited the geo-location feature of Twitter data. Their analysis relies on one full year of geo-located tweets posted by users all over the world between January 1, 2012, and December 31, 2012. The database consists of 944M records generated by a total of 13M users. Their study shows that despite the unequal distribution over the different parts of the world and possible bias toward a certain part of the population, in many cases geo-located Twitter can and should be considered as a valuable proxy for human mobility, especially at the level of country-to-country flows. In their approach they proposed capturing mobility and the nationality of travellers based on the simple method of assigning each user to his or her country of residence. Results showed that increased mobility measured in terms of probability, diversity of destinations, and geographical spread of travel, is characteristic of more developed countries such as the Western European ones. The travelling distance was additionally affected by the geographic isolation of a country as in the case of Australia or New Zealand. The analysis of temporal patterns indicated the presence of a globally universal season of increased mobility at the end of the year, clearly visible regardless of the nationality of travellers. Figure 20 shows the radius of gyration of Twitter accounts.

The radius of gyration[89] indicates a tendency to travel locally, and higher values indicate more long-distance journeys.

Fiorio et al. (2021) recently proposed a method to convert digital traces left on Twitter into estimates of migration transitions and for systematically analysing their variation along a quasi-continuous time scale, analogous to a survival function.

## 11.6 SHORT-TERM MOBILITY VERSUS LONG-TERM MIGRATION

The definition of migrant is sensitive to the time and space, as highlighted by the EMN,[90] who built on the definition by the IOM. For example, Fiorio et al. (2017) put the question very straightforwardly: *"when does a tourist become a migrant?"* This is a subtle definition especially for internal movements in territories like the European Union or the United States that has implications on the way the statistics are built. In fact, surveys typically only ask a limited set questions about migration (e.g., where did you live a year ago? where did you live five years ago?). The authors leveraged a sample of geo-referenced tweets for about 62 000 users, collected in the period 2010-2016 and studied the US internal migration flows under varying time intervals and durations. They introduce the concept of *"migration curves"* to describe the relationships between short-term mobility and long-term migration. Figure 21 shows the relationship between migration rate, interval, and duration. Furthermore, their results can be used to produce probabilistic estimates of long-term migration from the short-term mobility (and vice versa) and to nowcast mobility rates at different levels of spatial and temporal granularity using a combination of previously published American Community Survey data and up-to-date data from a panel of Twitter users (for a detailed explanation see Fiorio et al., 2017). Once again this study raises the issue of harmonisation of migration statistics in terms of duration, especially true for innovative data sources. A possible solution has been addressed, e.g., in the probabilistic framework proposed by Nowok and Willekens (2011) based on continuous time data generating processes.

---

89   The radius of gyration is defined as: $r_g = \sqrt{\frac{1}{n}\sum_{i=1}^{n} |\bar{a}_i - \bar{a}_{cm}|}$ , where $n$ is the number of tweeting locations, $\bar{a}_i$ cm is the location of a particular tweet and $\bar{a}_{cm}$ is a user's centre of mass.

90   https://ec.europa.eu/home-affairs/pages/glossary/migrant_en

**FIGURE 20.** Average radius of gyration of users from different countries compared to (A) percentage of mobile Twitter users and (B) number of countries visited. As expected New Zealand and Australia are outliers.
**Source:** Figure 4 in Hawelka et al. (2014).

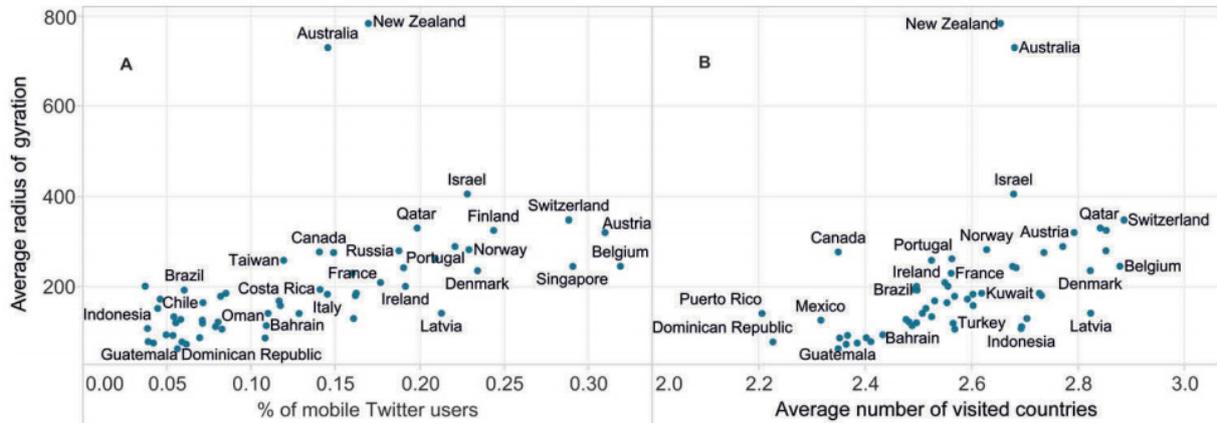

**FIGURE 21.** Plot of estimated migration rate as a function of interval and duration length. Rates were estimated fixing July 1st 2012 as the starting point.
**Source:** Figure 6 from Fiorio et al. (2017).

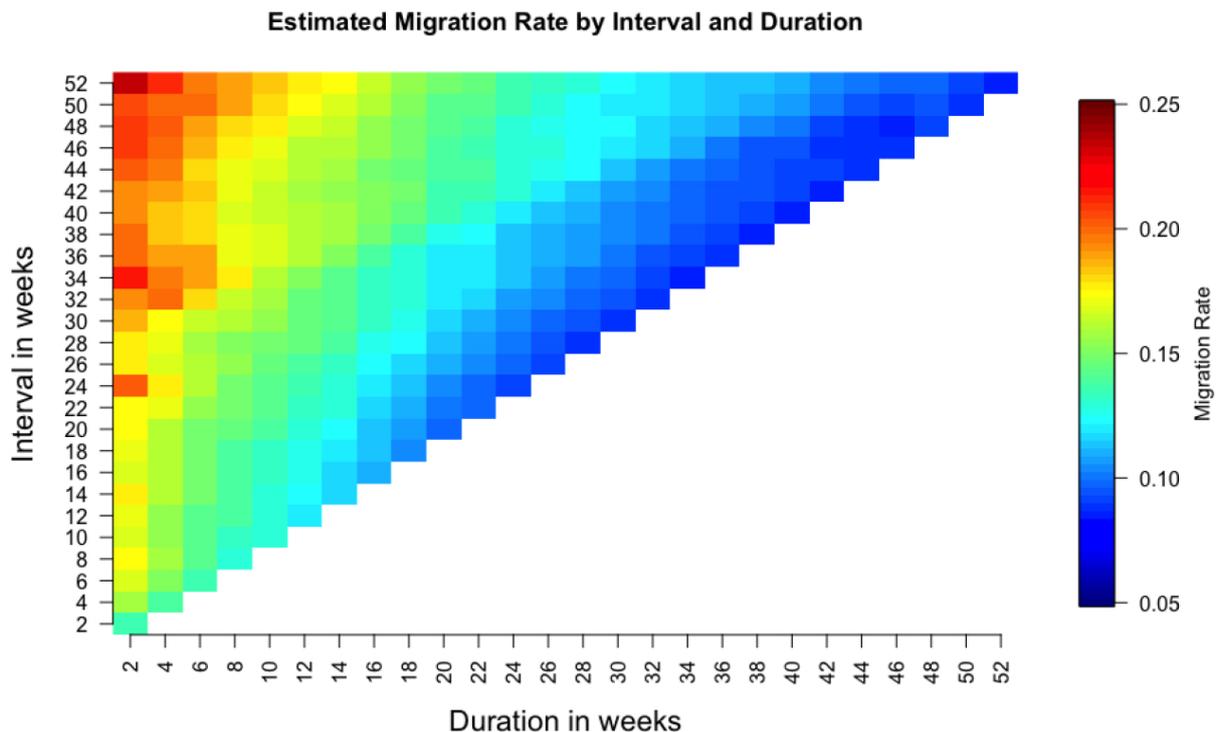

## 11.7 ATTITUDES TOWARDS FERTILITY AND PARENTHOOD

Although still far from being fully exploited to nowcast fertility as, for example, the Google search data in Chapter 14 in, Twitter may also offer good opportunities. A few studies have tried to capture the intention to have children and/or parenthood. For example, Ryan et al. (2021)

**FIGURE 22.** Top: distribution of topics related to tweets around pregnancy and parenting. Down: Distribution of emotions (BECOMEPA = become parent; TOBEPA = being parent; TOBESO = to be son; JUDGOTHERPA = judgment of other parents behaviour; FUTURE = children' future; DAILYLIFE = daily life experience parent/children).
**Source:** Figure 3 from Ryan et al. (2021) and Figure 4 from Sulis et al. (2016) respectively.

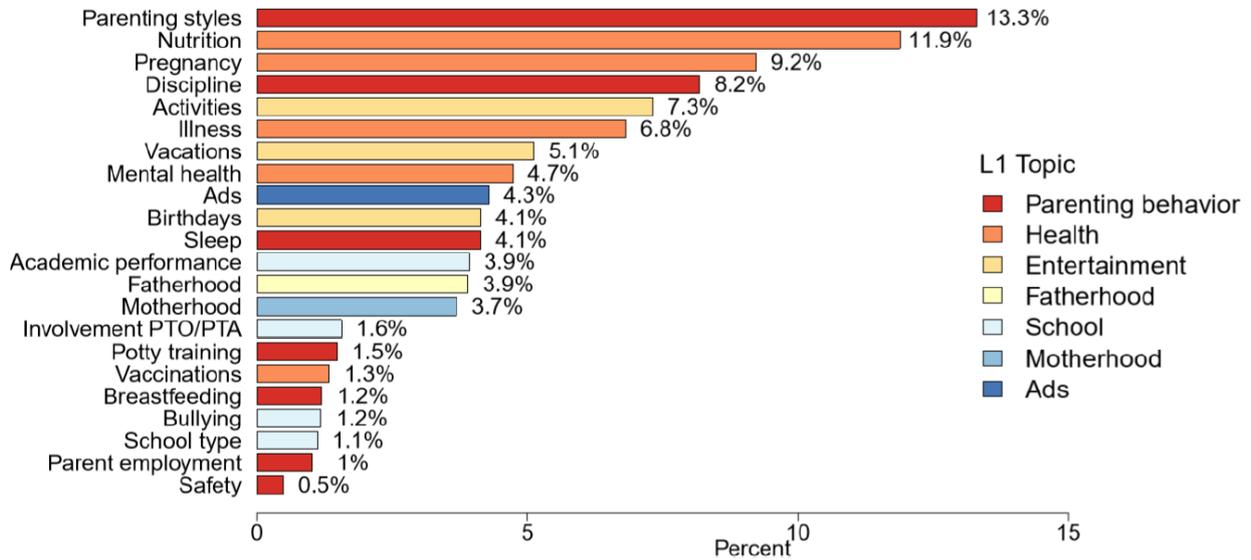

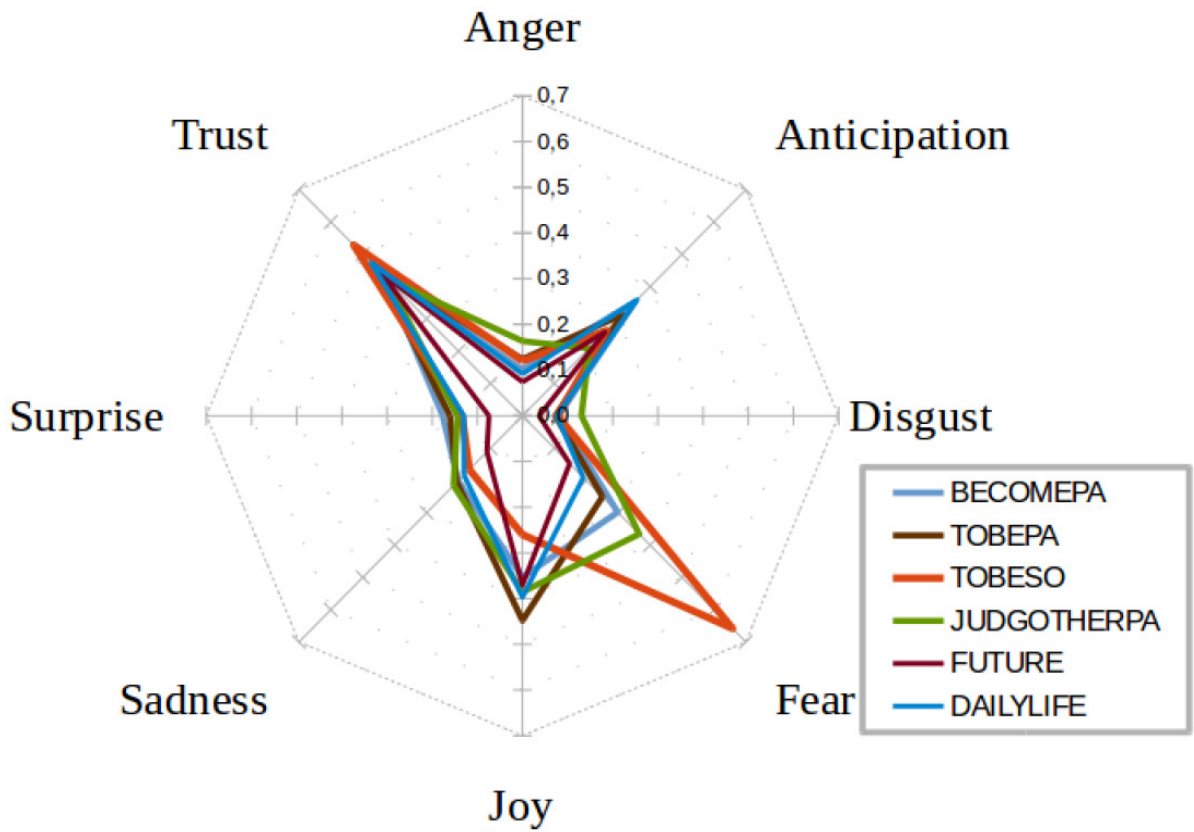

investigated the content of parenting information shared on Twitter by identifying the range and frequency of topics shared by parenting-focused accounts. They analysed 675 069 tweets collected from January 2016 to June 2018 and classified their content into several categories including parenthood and pregnancy (see top panel of Figure 22). Smith et al. (2021) used Social Network Analysis (SNA) to investigate the propensity to become a parent (as a proxy for fertility). They found that the overall Twitter sentiment regarding fertility was a little more positive during (64.3%) than before the pandemic (58.1%). Sulis et al. (2016) annotated 2 760 416 tweets around parenthood, written in Italian and posted in 2014, along several dimensions: becoming a parent, being a parent, etc. They ran the Linguistic Inquiry and Word Count (LIWC) algorithm (see Pennebaker et al., 2001) for each group of tweets to extract emotions related to joy, sadness, fear, disgust, anticipation, trust, anger, and surprise. The bottom panel of Figure 22 shows the results. The analysis shows a positive polarity for messages concerning TOBEPA (to be parents), and a more negative polarity for BECOMEPA (to become parents). Messages judging the behaviour of other parents (JUDGOTHERPA) contain a high frequency of anger and disgust terms. Finally, the category TOBESO (to be sons) is more controversial, having the highest frequency of negative terms as well as the lowest frequency of joy terms. In summary, the authors interpret these results as showing that children seem more critical towards parents whereas parents seem to express a more positive attitude towards children.

## 11.8 ADVERSE PREGNANCY OUTCOMES

Sarker et al. (2017) applied Natural Language Processing (NLP) techniques to identify tweets discussing pregnancy condition and hence cohorts of pregnant women starting from the idea that pregnancy exposure registries are developed for new medications only[91] and that these registries enroll women prospectively (eg, after exposure but before childbirth) in a voluntary fashion and follow them for the entire duration of the pregnancy or longer. The conclusion of Sarker et al. (2017) was that it is indeed possible to identify large scale cohorts of pregnant women who can be eventually enrolled in a clinical study.

Klein et al. (2020) built on the results above and found evidence that, e.g., in the United States, 17% of pregnancies end in foetal loss (miscarriage or stillbirth), and that preterm birth affects 10% of live births which are the leading cause of neonatal death globally. Despite their prevalence, the causes of miscarriage, stillbirth, and preterm birth are still largely unknown. Klein et al. (2020) tried to use Twitter data to shed some light on the phenomena. Indeed, the initial result is that women report not only pregnancy status but also miscarriage, stillbirth, and preterm birth, among others, on Twitter. This has allowed the researcher to identify possible participants in a large-scale observational study. Starting from a similar approach, it seems that these adverse pregnancy outcomes can be nowcast using Twitter data and appropriate selection bias removal techniques.

---

91 Also due to the fact that pregnant women are actively excluded from clinical trials during the development of new medications because of concerns over foetal safety.

# 12. REDDIT

Reddit, is a discussion-based platform that has only being explored very recently possibly because it is still difficult for the researchers to exploit its potential applications. Reddit posts lack georeferencing information and this is a further limitation. Here we review a few pioneering works.

## 12.1 REPRESENTATION OF REFUGEES

Coppin (2020) analysed the content of online articles shared on Reddit over a two-month period (from the 10th November, 2019 to the 10th January, 2020) regarding refugees, asylum seekers, immigrants and migrants (RASIM). The results showed that they were framed in the articles as either victims or as an *(economic) burden*. They are mostly on the receiving end of actions and represented as quite a homogeneous population. Several metaphors, different emphases, and mitigations were found, accompanied by arguments referring to their growing number and the need for management. Many discursive strategies dehumanised RASIM except for refugees who are more likely to be depicted as human beings. On the whole, the representations of RASIM in the articles agree with previous literature.

## 12.2 ATTITUDES TOWARDS MIGRANTS AND ECHO CHAMBERS

While studying the echo chamber phenomenon on Reddit for pro-Trump and anti-Trump supporters, Morini et al. (2021) found a strong association between attitudes towards migrants and echo chambers. In addition to these very few studies, Fox et al. (2021) recently explored the unexploited potential of this platform in social science studies which still remains to be fully explored.

# **13.** LINKEDIN

LinkedIn is a popular professional social network. It has about 774 million members as of 2021 in more than 200 countries and territories worldwide.[92] LinkedIn's Economic Graph team[93] is collaborating with various organisations around the world and using LinkedIn data studies the rapidly changing employment market. For example, LinkedIn collaborated with the World Bank[94] to provide public statistics about talent migration using data generated by LinkedIn's members in 100+ countries for the period between January 2017 and December 2019.

## **13.1** HIGHLY-SKILLED MIGRANTS

By analysing LinkedIn data and comparing with official statistics, State et al. (2014) concluded that LinkedIn can provide important insights about recent trends in migrations of highlyskilled migrants to the United States. An additional interesting outcome of the State et al. (2014) study is that the global coverage of the LinkedIn dataset revealed that Asia became more attractive as a major professional migration destination from 2000 to 2012. Barslund and Busse (2016) used LinkedIn data to study the labour mobility of IT professionals within and beyond the EU in 2013 and 2014. They revealed that IT professionals tend to migrate from eastern and southern EU member States to the western and norther EU member states as well from the EU to the United States. Finally, Johnson et al. (2021) used aggregate-level information on LinkedIn users open to work-related international relocation to assess the relative attractiveness of prospective relocation between countries within the European Union, European Free Trade Association, and the United Kingdom (see Figure 23).

**FIGURE 23.** Percentage error between predicted and observed values of LinkedIn users open to relocate between two countries. Each cell is coloured based on the quintile of percentage error, with observed values much lower than predicted shown in blue and observed values much higher than expected in red. Countries are sorted by the total LinkedIn users. White cells indicate a lack of data.
**Source:** Johnson et al. (2021).

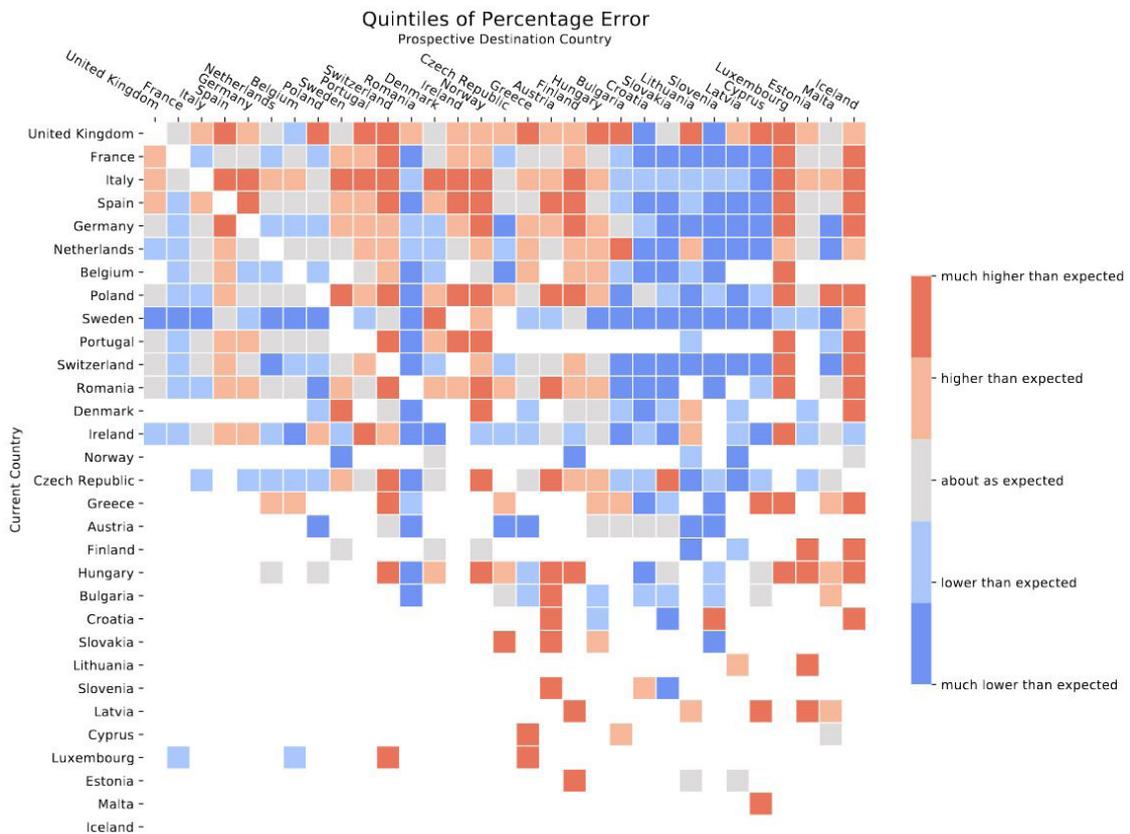

---

92   https://about.linkedin.com/
93   https://economicgraphchallenge.linkedin.com/
94   https://linkedindata.worldbank.org/about

**FIGURE 24.** Number of LinkedIn users as of February 2018 who hold a University degree from a Syrian University and live in Europe.
**Source:** KCMD elaboration.

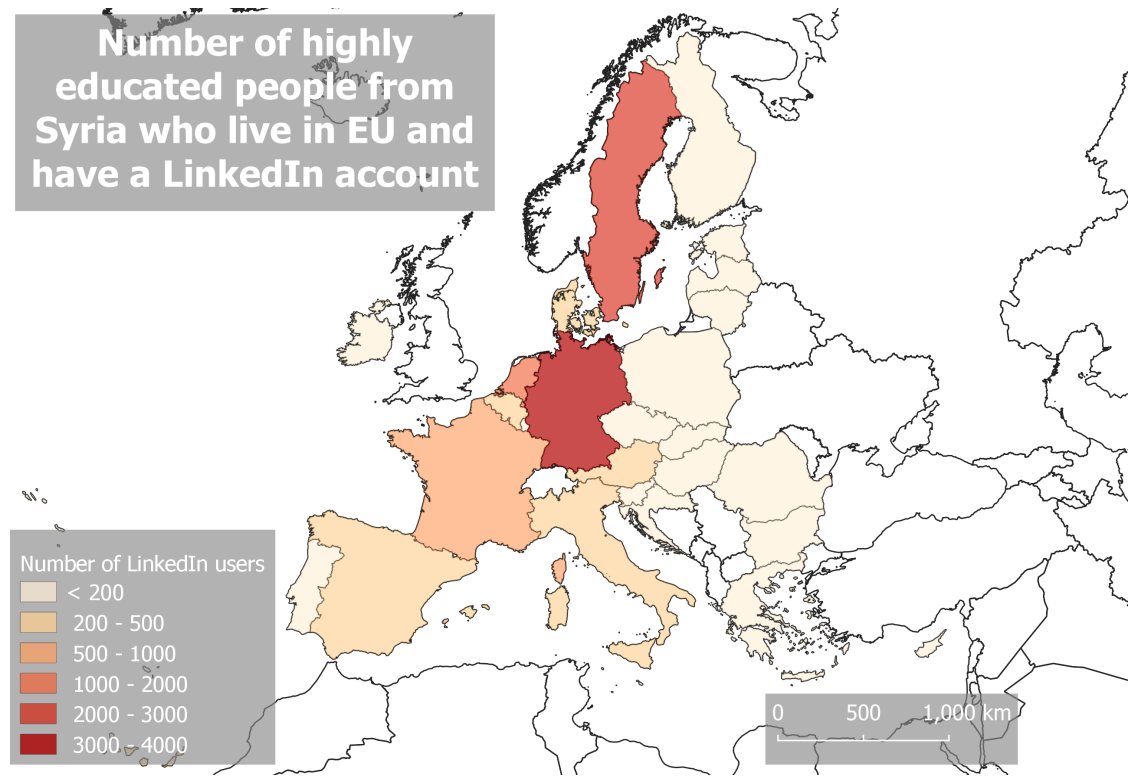

## 13.2 ESTIMATING BRAIN GAIN

The KCMD has explored the re-use of data from the LinkedIn Campaign manager[95] for estimating Brain gain/drain in Europe. The LinkedIn Campaign manager allows the user to design targeted advertisement campaigns to users based on their characteristics, for example their location, age, the sector of their employment, and the University they graduated from, among other. For example, Figure 24 presents the number of people on LinkedIn who hold a university degree from a University in Syria and who now live in Europe. These users are very likely to be Syrian migrants.

---

# 14 GOOGLE SEARCH DATA

Google Trends[96] tracks relative changes in Google search queries worldwide starting from January 2004. There are different types of search that can be tracked: web, news, image, product or YouTube searches. The system aggregates the volume of searches by different geographical areas and levels (national, state and metropolitan area level) based on the originating IP address of the users. Google Trends data represents the total number of web searches for a particular keyword or topic (i.e., group of keywords), relative to the total number of searches in particular geographical area over time, therefore the indicator only provides information about the likelihood of a random user searching for a particular keyword/topic on Google from a certain location at a certain time on a relative basis. This is a crucial feature of the instrument in that it can be used to monitor trends but unlikely to estimate actual figures. This index, called the Search Volume Index (SVI), changes on a daily basis as it is calculated using daily Google logs.

Another feature of the SVI is that the number of searches is only included in the calculation if it exceeds a certain threshold; multiple queries from a single user/IP over a short period of time are disregarded. The SVI is normalised on a 0-100 scale and is further aggregated at monthly, weekly, daily, and intra-daily frequency depending on the type of search made by the user. The SVI for region $r$ and week $\tau$ is constructed by aggregating the daily data for each day $t$. Given the search volume on a term $V_{t,r}$ in region $r$ on day $t$ and the total search volume in that region $T_{t,r}$ the following indicators are obtained:

$$ S_{t,r} = \frac{V_{t,r}}{T_{t,r}}, \quad \text{and} \quad S_{\tau,r} = \frac{1}{7} \sum_{t=\text{Sunday}}^{\text{Saturday}} S_{t,r} $$

and the SVI for week $\tau$, $SVI_{\tau,r}$ is given by

$$ \text{SVI}_{\tau,r} = 100 \cdot \frac{S_{\tau,r}}{\max_\tau S_{\tau,r}} $$

for $\tau$ the calendar week (starting from Sunday) that spans the indexes 1 to the number of weeks. So it is clear that the maximum value of the SVI may vary through time and only the trend is informative.

## 14.1 FORECASTING FERTILITY AND ABORTION

In their pioneering work, Billari et al. (2018) found a good relationship between birth rate and Google searches. In particular, they constructed three indexes based on web searches around the keywords ''maternity'' (GI1), ''pregnancy'' (GI2) and ''ovulation'' (GI3). Figure 25 shows the empirical evidence of this link between Google search and actual birth rates. The researchers also attempted a forecasting exercise to predict the growth of birth rate, i.e. the quantity $d(br_t) = br_t - br_{t-1}$, using a battery of models (AR, ARMA, ARMAX) including additional macroeconomic variables such as GDP growth, unemployment rates, etc, and found that including the Google indexes, the relative performance of the models improves at all time horizons of 6, 12, 18, and 24 months compared to the same models without the GIs. In a related work, Ojala et al. (2017) found similar patterns and moreover they found relationship between fertility-related Google searches and socio-economic and demographic characteristics of births. As noted by the authors, ''*the availability of information via the Web can affect demographic choices (…). At the same time, Web searches partially reflect social structures*''. Because of issues related to social desirability and conformity, attitudes towards fertility are difficult to measure using surveys alone whereas ''*Web searches are less likely to suffer from desirability biases*''. The authors also point out that the identification of the queries which are strongly related to fertility (yet still to come) would, for example, help to focus public health information campaigns. Moreover, ''*in the context of developing countries that are still undergoing the fertility transition, web searches may provide relevant proxies about attitudes and offer information useful to predict the pace of future fertility reductions*''. Based on a similar approach, Reis and Brownstein (2010) found that the volume of Internet searches for abortion worldwide is inversely proportional to local abortion rates and directly proportional to local restrictions on abortion. These findings are inline with the evidence collected from other data showing that ''*local restrictions on abortion lead individuals to seek abortion services outside of their area*''.

## 14.2 CAPTURING INTENTION TO MIGRATE

Based on previous scientific attempts to exploit Google trends data, the ''*Improving Migration Management*''

---

**FIGURE 25.** GI1 (top-left) is the monthly average of the google index for ''maternity'', GI2 (top-right) is the monthly average of the GI for ''*ovulation*'', GI3 (bottom-left) is the monthly average of the GI for ''*pregnancy*'', and GIPC1 (bottom-right) is the first principal component of the previous three Google indices. Time span is from January 2004 (due to Google limits) till December 2009.
The indexes are normalized with respect to the maximum value of GI1. Blue lines show actual birth rates, whereas red lines show the GIs. Shaded areas identify NBER recessions.
**Source:** Figure 2 from Billari et al. (2018).

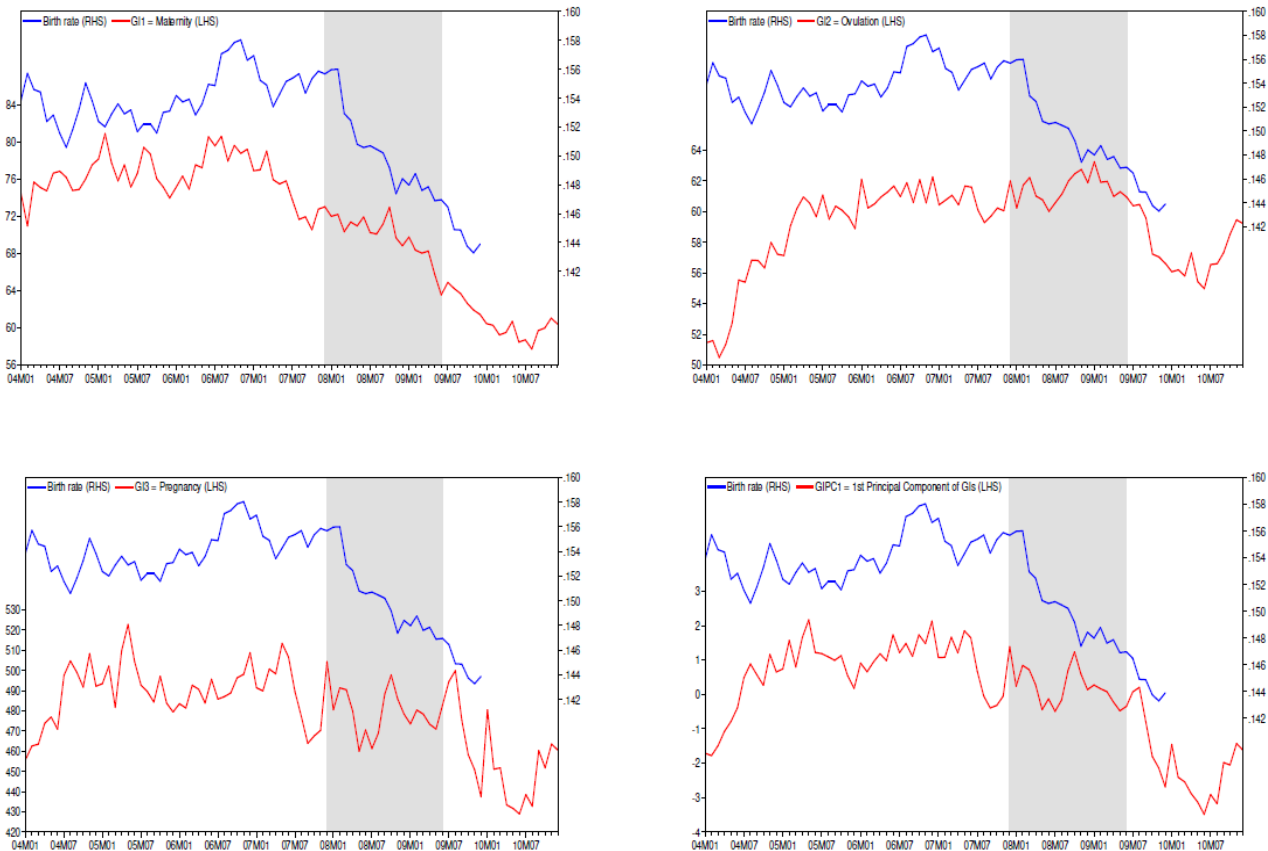

project[97] of the Bertelsmann Stiftung institute, tried to collect Google search data in the EU (plus United Kingdom and Switzerland) by using around 60 keywords/keyword groups (roughly 150 terms in total) that are immigration/mobility related and always used in conjunction with the term ''*Germany*'' in each national language, English and German.

Preliminary findings seem to show that while migration inflow to Germany has a clear seasonal pattern, Google search data do not exhibit this pattern. In general, countries with less than 5 million inhabitants do not produce reliable times series because of Google's policy of hiding low numbers of queries compared to population. Finally, there is a mixed evidence on how much of this data is able to predict inflows: there are clusters of countries for which intuition seems to hold and other countries for which does not. This was also seen in the model presented in Chapter 17 which take into account the ability to predict flows through a combined data science approach. Similarly Lin et al. (2019).

## 14.3 FORECASTING FLOWS

Böhme et al. (2020) studied the additional contribution of Google Search data to predictive models of migration flows to several OECD countries looking at an extensive set of keywords in English, French, and Spanish related to migration. Wladyka (2013) used this data to capture the intention to migrate from Latin America to Spain and found that it is a good predictor of these particular migration flows. Fantazzini et al. (2021) investigated the suitability of Google Trends data for the modelling and forecasting of interregional migration in Russia. Their empirical analysis did not provide evidence that the more people search online, the more likely they are to relocate to other regions, though the inclusion of Google Trends data in a model improves the forecasting of the migration flows. Indeed, this seems to be a common conclusion of the literature reviewed as Google trends is only useful as an ingredient in more sophisticated modelling as will be shown in detail in Chapter 17.

---

97   The web page of the project can be found here: https://www.bertelsmann-stiftung.de/en/our-projects/making-fair-migration-a-reality/project-topics/improving-migration-management. This research is still an ongoing project.

**FIGURE 26.** Relationship between abortion search volume and abortion rates across 37 countries. Data reveals an inverse correlation between abortion search volume and local abortion rates ($\rho = -0.484$, p-value = 0.004), with certain geographic regions showing general trends. Marker size indicates the number of restrictions (from 0 to 7) that exist on abortion in each country according to UNStats.
**Source:** Figure 3 from Reis and Brownstein (2010).

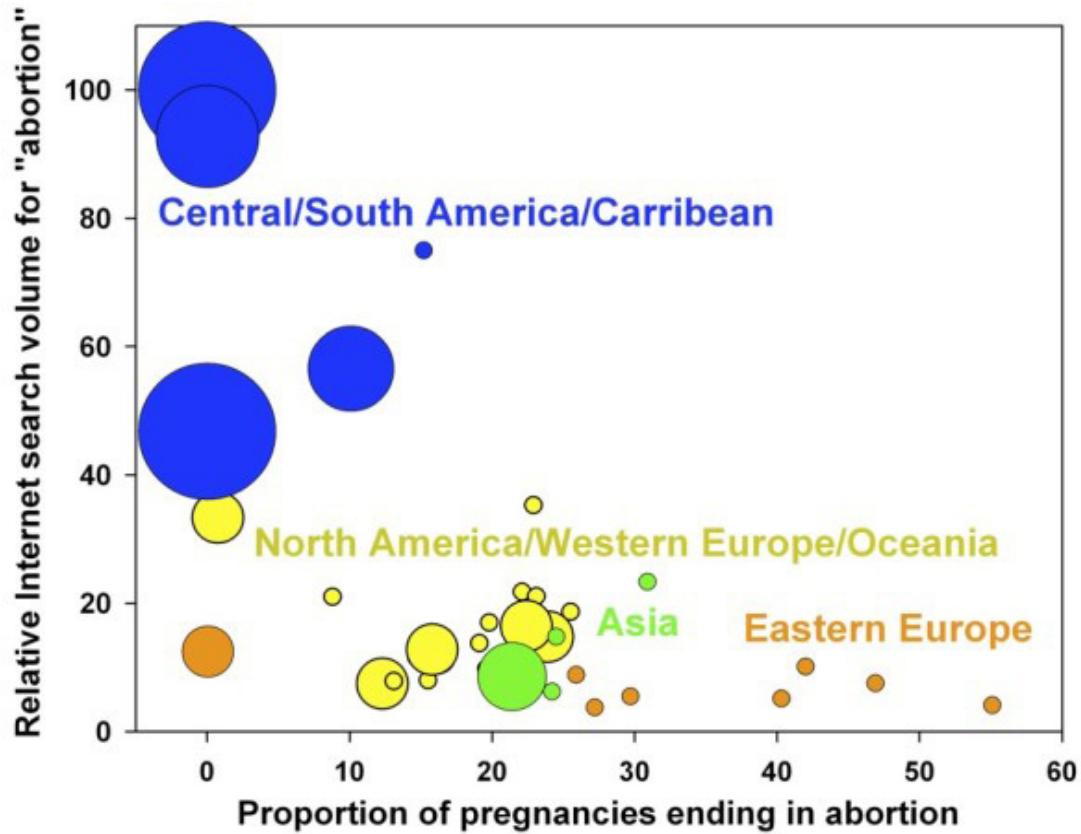

A general final comment on the use of Google search data is that Google trends data can be used to approximate the salience of a topic and the trend in a phenomenon but the absolute value of the index is not very informative per se unless it is combined with other sources of information. Moreover, as observed by in the case of the Google Flu study failure (Lazer et al., 2014), a recent study by Tjaden et al. (2021) found that there is no universal approach at the global scale and each migration flow needs to have ad hoc queries. This is true to the extent in which search data is used in models as is rather than using data science approaches like in Carammia et al. (2022) or when keywords are too strict.

As a remark, it can be added that Lin et al. (2019) also explored the potential of using Bing.com[98] search data to forecast U.S. domestic migration. Having access to disaggregated search data by age and other socio-economic characteristics, they were not only able to capture the intention to migrate but also the reasons. In summary, although Bing does not have a public tool like the Google Trends dashboard and therefore setting aside the issue of accessing this data, what has been said above for Google search data is likely to be the case for the Bing search data as well. Issues concerning data accessibility will be discussed in Chapter 26.

---

98    Bing is the web search engine developed by Microsoft.

# 15. GOOGLE LOCATION HISTORY DATA

With movements of people varying from short, repeated movements (for example to work or school), to migratory movements across country borders, it is difficult to collect detailed mobility data across different temporal and spatial scales. The typical sources of mobility data (such as travel history surveys and GPS tracker data) contain information on different typologies of movement but almost no source of data can address all types of movement at the same time (Ruktanonchai et al., 2018)

Google location history (GLH) data is a novel source of information capable of linking fine scale mobility with long distance mobility and international trips. It is a unique source of data, passively collected by android smartphones, that spans large temporal scales with high spatial granularity (Ruktanonchai et al., 2018). When enabled within android smartphones, ''*Location History*" passively and continuously collects location data using technologies that include GPS, Wi-Fi and cellular positioning. In some recent studies (Ruktanonchai et al., 2018; Yu et al., 2019) the possibility of using Google location data to characterise fine-scale human mobility was investigated. Results suggest that Google location history data could provide unmatched individualised human movement information and also address some key gaps in data that are currently available.

GLH data is functionally similar to GPS tracker data, but easier to collect with a survey and less problematic for users because it is collected passively and is easily retrieved. In many middle and low income settings, Android phones are becoming the first device purchased to access Internet. It makes GLH data increasingly more appropriate for answering to a wide range of scientific questions.

Kraemer et al. (2020) mapped global human mobility using data from 300 million smartphone users around the world who opted in for the GLH option on their smartphone. Figure 27 shows the probability of short range movements by country. Although not exploited by more extensive migration studies yet, this data looks very promising for studying internal migration, secondary movements, and return migration within Europe.

**FIGURE 27.** Aggregated human movement data from 70 010 784 unique location flows (~ 300 million participants) covering 65% of the Earth's populated surface reveal variability in human movements across the world. The map shows the variation in connectivity (relative frequency of staying at short distances within 5 × 5 km cells) across countries with at least 100 cells containing information about human mobility. Warm colours indicate high frequency of "movements longer than 5 km, whereas cool colours correspond to countries with low frequency of "movements longer than 5 km.
**Source:** Figure 2 in Kraemer et al. (2020).

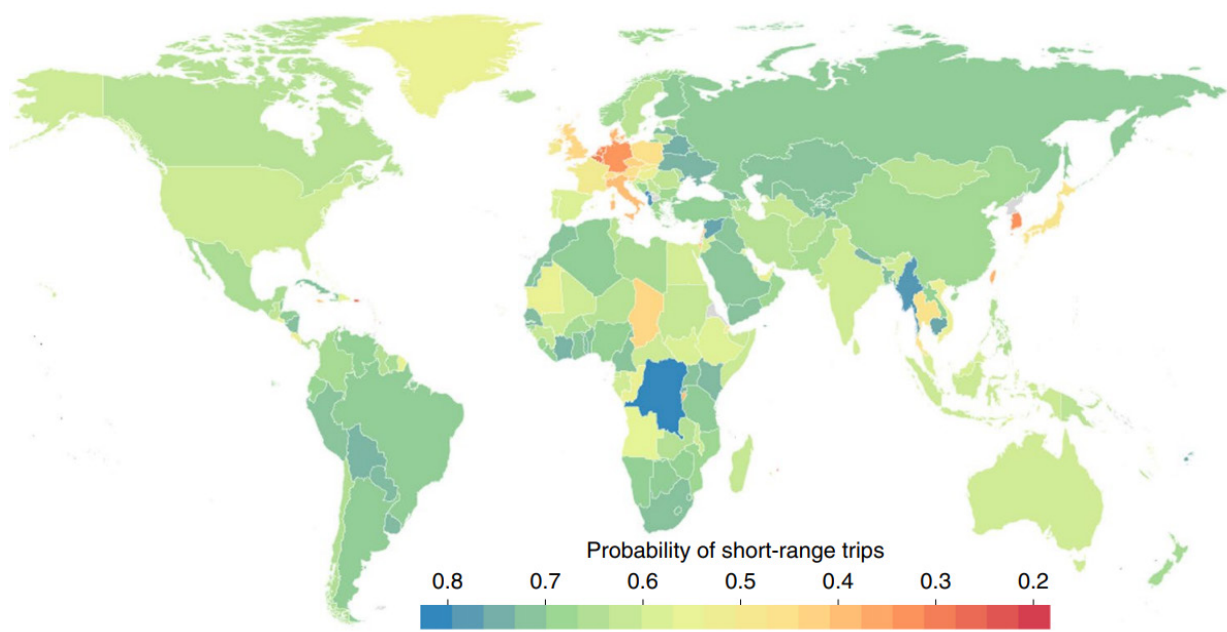

Unfortunately, access to Google location history is still very limited. For example, the global data base used in Kraemer et al. (2020) is not currently available for any research goal other than for COVID-19 related studies.

# 16. CONFLICT AND POLITICAL VIOLENCE DATA

Conflicts are certainly related to displacement of people and hence migration. In fact, conflict is one of the classical migration drivers (Migali et al., 2018). In recent years several databases of conflict data at very high frequency and at high geographical resolution have been collected. More generally, these are called *event data* which is collected using semiautomatic annotation of events as they appear in various types of news. They are annotated using code books widely approved by the scientific community. Indeed, most of the databases reviewed in the next sections use the same common framework to code event data (typically used for events that merit news coverage, and generally applied to the study of political news and violence) called Conflict and Mediation Event Observations (CAMEO).[99] Other alternative codebook exist, like the World Event/Interaction Survey (WEIS) coding system (Goldstein, 1992; McClelland, 2006) and the Conflict and Peace Data Bank (COPDAB) coding system (Azar, 2009). Here a brief summary of the available on-line resources is presented although how data is generated is not explained in full detail.

## 16.1 UCDP

The Uppsala Conflict Data Program Georeferenced Event Dataset Global (UCDP) provides a series of datasets on organised violence (incidents of lethal violence at a given time and place) according to three types of violence: state-based armed conflicts, non-state conflicts, and one-sided violence (Wallensteen, 2011). The global version of the UCDP Georeferenced Event Dataset (GED) is the most disaggregated dataset, covering events of organised violence geocoded at the village level and with duration disaggregated to individual days, for the period 1989+. Using UCDP data at the national level, Conte and Migali (2019) found that the intensity of the conflict and where the fighting is taking place explain an essential portion of the variation in flows of asylum applications and stocks of refugees (see also Figure 28).

## 16.2 GDELT

The Global Database of Events, Language, and Tone (GDELT) project[100] ''*monitors the world's broadcast, print, and web news from different countries in over 100 languages and identifies the people, locations, organizations, themes, sources, emotions, counts, quotes, images and events driving society daily*''.

Starting from this event data, EASO built synthetic indicators related to the intensity of political events, social unrest, conflicts, economic events, and governance-related events and then aggregated them at country level producing the Push Factor Index (PFI) (see Constantinos et al., 2020). Figure 29 shows the PFI, and the same index for Africa normalized by the resident population. The same data has also been used in the early warning and forecasting system of EASO as shown in Chapter 17.

## 16.3 ACLED

The Armed Conflict Location & Event Data Project (ACLED)[101] ''*is a disaggregated data collection, analysis, and crisis mapping project. ACLED collects the dates, actors, locations, fatalities, and types of all reported political violence and protest events across Africa, the Middle East, Latin America & the Caribbean, East Asia, South Asia, Southeast Asia, Central Asia & the Caucasus, Europe, and the United States of America*''. ACLED is the main source of data on political violence and protests around the world from 1997 to the present. Data is derived from a variety of sources, mainly focusing on reports from war zones. New data is available in real time.[102]

## 16.4 GTD

The Global Terrorism Database (GTD) is an open-source database including information on domestic and international terrorist attacks around the world from 1970 to 2019 and includes more than 200 000 cases. Information on each event is available including the date and location of the incident, the weapons used, the nature of the target, the number of casualties, and, when identifiable, the group or individual responsible. Foubert and Ruyssen (2021) investigated the impact of terrorist attacks on migration intentions around the World. They crossed survey data on migration intentions in and from 133 countries in the period 2007–2015. They found that terrorist attacks have an impact on both internal and international migration intentions, though the effect is stronger for the latter. They also found that migration intentions are not necessarily related to the frequency of terrorist attacks, but are related to their intensity measured in terms of the number of killed and of those wounded.

---

99   The full codebook is available at https://parusanalytics.com/eventdata/data.dir/cameo.html.
100  https://www.gdeltproject.org
101  https://acleddata.com/
102  With the ongoing COVID-19 Disorder Tracker (CDT) project, ACLED is looking at the impact of the pandemic on political violence, protests and unrest around the world using real-time data. For more information, see https://acleddata.com/analysis/covid-19-disorder-tracker/.

**FIGURE 28.** The evolution of the refugee population of the 2017 top six sending countries to the EU-28.
**Source:** Figure 5 in Conte and Migali (2019).

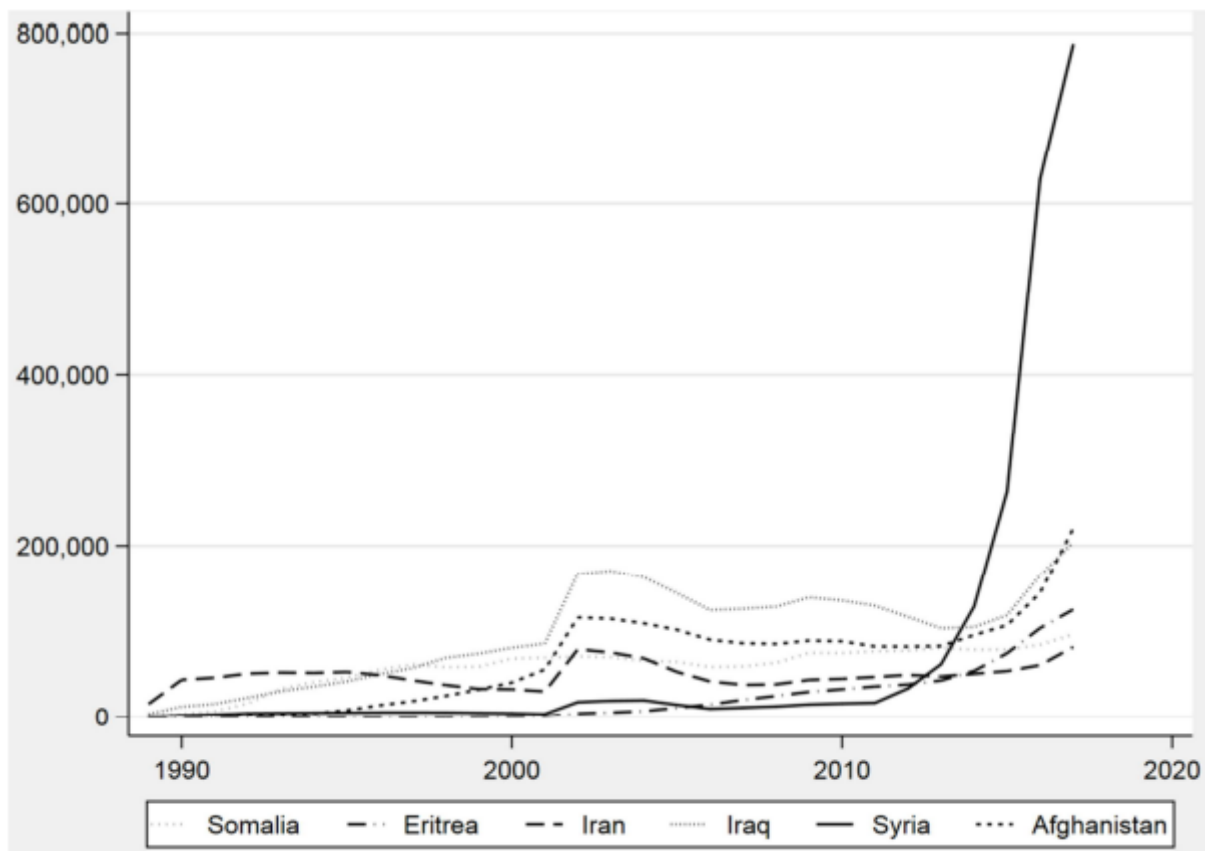

## 16.5 ICEWS

The Integrated Crisis Early Warning System (ICEWS)[103] produce event data for the Defense Advanced Research Projects Agency (DARPA) and Office of Naval Research (ONR). It consists of events coded as interactions between socio-political actors. Events are extracted from news articles from 1995 to the present. ICEWS uses a mixed methods approach to forecast instability and combines heterogeneous statistical and agent-based models to produce forecasts (O'Brien, 2010). Lastly, in a case study on Iraq, Singh et al. (2019) used this data together with Twitter data to forecast forced migration (see also Figure 30).

## 16.6 CLINE

Cline Center Historical Phoenix Event Data (CLINE) includes 8.2 million events extracted from 21.2 million news items for the period 1945–2019 and analyses content from the New York Times, BBC Monitoring's Summary of World Broadcasts, Wall Street Journal and the Foreign Broadcast Information Service of the Central Intelligence Agency. It provides information on actors and locations for a variety of conflict, cooperation and communication events in the Conflict and Mediation Event Observations ontology (Althaus et al., 2020).

## 16.7 SPEC

The Spark-based Political Event Coding (SPEC) is a system developed by the University of Texas in Dallas[104] "*that extracts and processes political event data from over 380 different English written news media on a daily basis, and has been doing so since October 2017*" (Solaimani et al., 2016). This database of event data shares the same codebook as GDELT but it is based on a different semantic approach to automatically extract event data from the news.

---

**FIGURE 29.** EASO Push Factor Index in Africa for 2019 (left panel) and population-adjusted Push Factor Index for 2019 (right panel), with darker shades indicating higher values of the Index.
**Source:** Figure 8.6 from Constantinos et al. (2020).

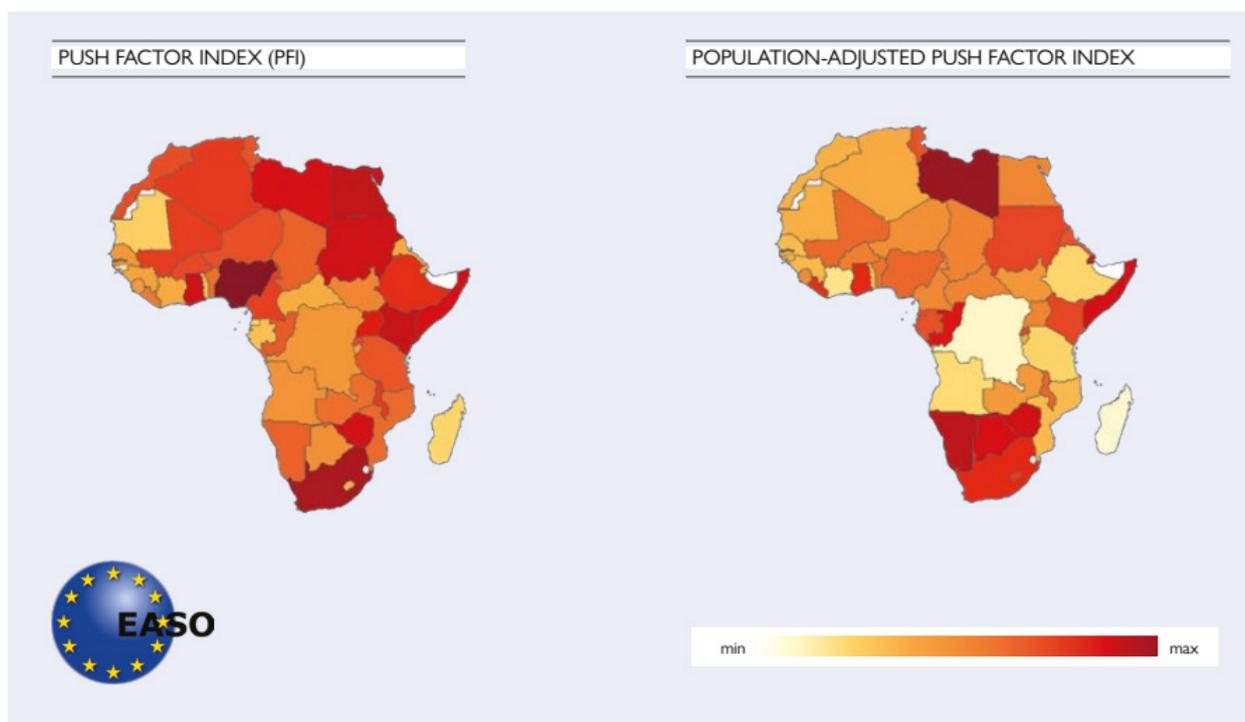

Although the ICEWS, CLINE and SPEC datasets have not been applied directly to migration studies yet, this data sources can be used as an alternative to, or in combination with, one the other event data bases. The interested reader can also refer to the Open Event Data Alliance (OEDA)[105] project that collects references to the different additional events data repositories.

## 16.8 CONFLICTS, LITERACY AND STUNTING

Bosco et al. (2017) used ACLED data to estimate the distance from conflicts and this information was exploited jointly with other data sources to produce maps of male and female literacy (Figure 31) and stunting in children (under age of five) in Nigeria.

Bayesian geostatistical and machine learning modelling methods (Artificial Neural Networks (ANNs)) were applied to produce the maps of literacy and stunting with a spatial resolution of 1 x 1 km. The modelling architecture applied is based on the Geospatial Semantic Array Programming (GeoSemAP) paradigm (Bosco and Sander, 2015) with focus on computational reproducibility and semantic modularization of the many data-transformation components.

## 16.9 CONFLICT AND CHILD DEATHS

Wagner et al. (2018) exploited the high resolution of the UCDP conflict data to test the assumption that a substantial proportion of child deaths in Africa take place in countries with recent histories of armed conflict and political instability. They took the time series of geo-referenced armed conflicts from 1995 to 2015 together with the location, timing, and survival of infants younger than 1 year (their primary outcome) in 35 African countries (see left-hand panel of Figure 32). The authors found that a child born within 50 km from an armed conflict had a risk of dying before reaching age 1 year of 5.2 per 1 000 births higher than being born in the same region during periods without conflict (95% CI: 3.7 – 6.7), a 7.7% increase above the baseline. This increased risk of dying ranged from a 3.0% increase for armed conflicts with one to four deaths to a 26.7% increase for armed conflicts with more than 1 000 deaths. The right-hand panel of Figure 32 shows the outcome of the estimated statistical models used to test the assumptions. The authors also found evidence of increased mortality risk from an armed conflict up to 100 km away, and for 8 years after conflicts, with cumulative increase in infant mortality two to four times higher than the contemporaneous increase. This study also has the merit of quantifying the impact of conflict on child deaths and highlight the importance of developing interventions to address child health in areas of conflict.

---

105   The web page of the project is available at http://openeventdata.org/.

**FIGURE 30.** Correlation matrix between Twitter extracted sentiment (buzz) in Arabic (left) and English (right) and other sources including ICEWS data on deaths in Iraq.
**Source:** Figure 4 in Singh et al. (2019).

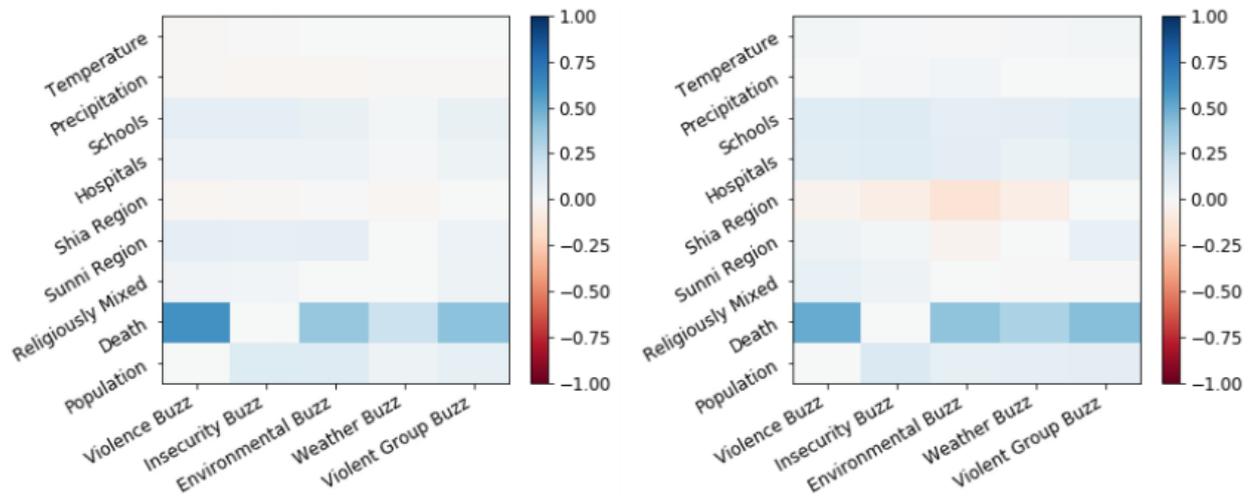

**FIGURE 31.** Map of the mean predicted proportion of literacy in Nigeria for women in the age 15 – 49 at 1 km² resolution.
**Source:** Bosco et al. (2017b).

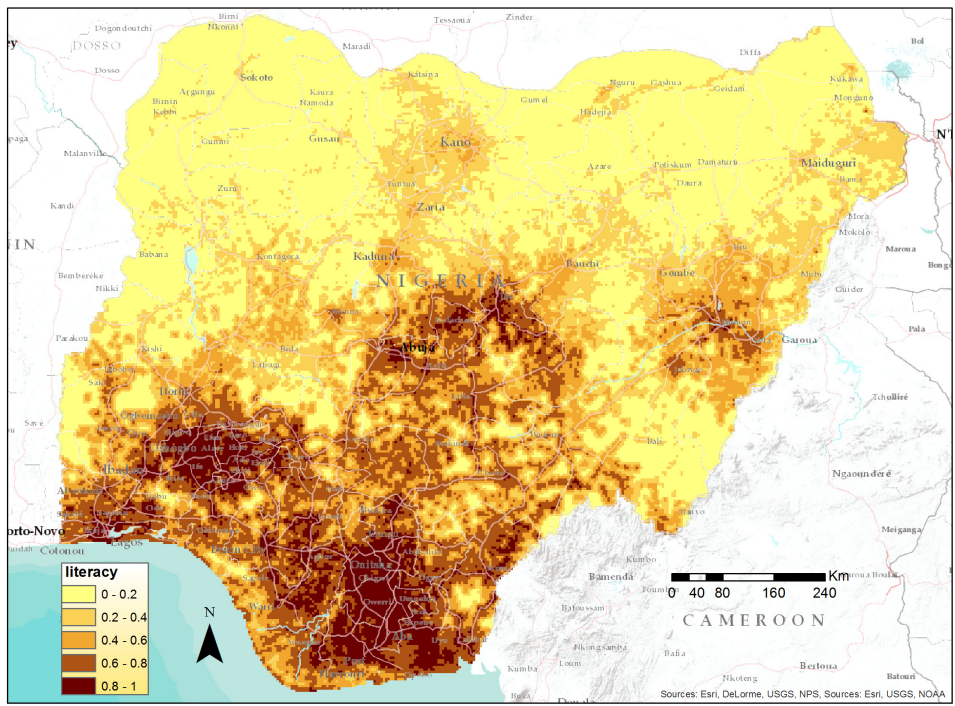

## 16.10 SIMULATING REFUGEE MOVEMENTS

Suleimenova et al. (2017) used ACLED data along with data from United Nations High Commissioner for Refugees (UNHCR) and Bing Maps to construct agent-based simulations of refugee movements by studying three major African conflicts (Burundi, Mali and the Central African Republic), estimating the distribution of incoming refugees across destination camps, given the expected total number of refugees in the conflict. Figure 33 shows the simulation scheme developed by the authors. The same institution is also exploiting artificial intelligence to predict internal displacement.[106] Extensive applications of agent based modelling to simulate migration scenarios can be found in Bijak et al. (2021).

---

106   See, e.g., https://www.unglobalpulse.org/project/using-artificial-intelligence-to-mod el-displacement-in-somalia/.

- - - - - - - - - - - - - - - - - - - - - - - - - - - - - - - - - - - - - - - - - - - - - - - - - - - - - -

**FIGURE 32.** Left: The distribution of armed conflict events in Africa, 1995-2015 as seen from UCDP. Right: The cumulative mortality risk over time for conflicts within 0-50 km, 51-100 km, and 101-250 km from a child's place of birth, estimated in one regression, showing an attenuating effect at greater distances. The blue shading represents 95% CIs.
**Source:** Figures 1 and 3, respectively left to right, from Wagner et al. (2018).

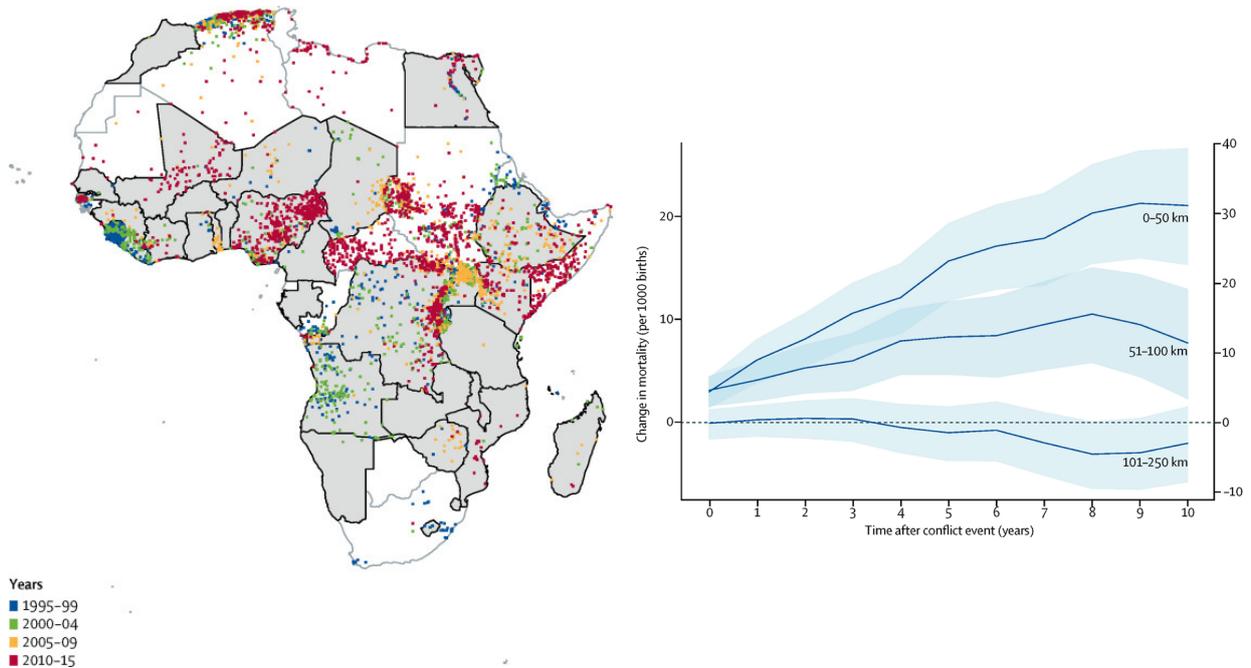

- - - - - - - - - - - - - - - - - - - - - - - - - - - - - - - - - - - - - - - - - - - - - - - - - - - - - -

**FIGURE 33.** Simulation development approach to predict the distribution of refugee arrivals across camps using ACLED data along with other operational data.
**Source:** Figure 1 in Suleimenova et al. (2017).

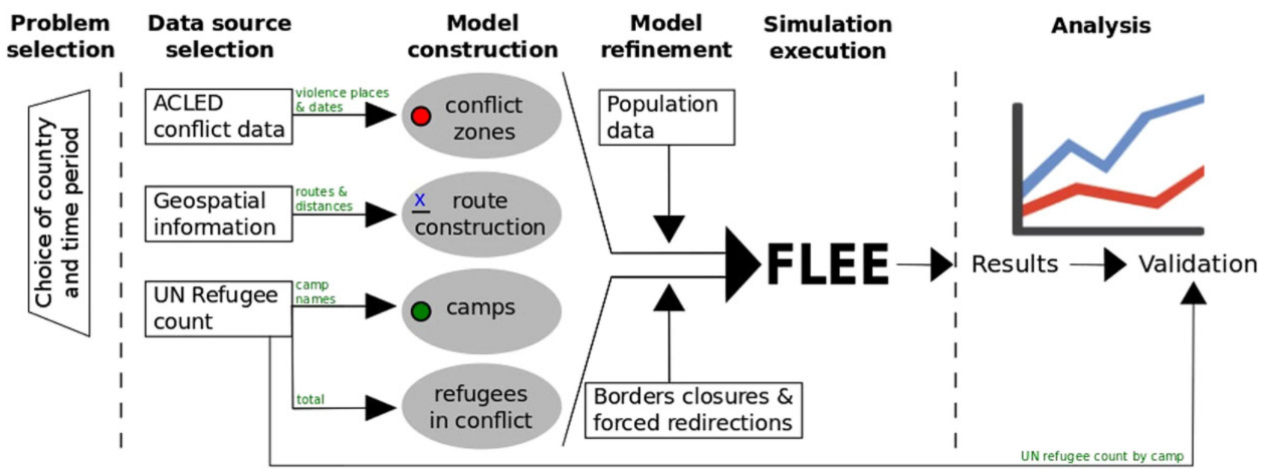

# 17. COMBINING EVENT DATA, GOOGLE SEARCH AND OPERATIONAL DATA

Carammia et al. (2022) presented an integrated approach to mix information coming from operational data, big data, and official statistics to create an early warning and forecasting system for asylum seekers in European countries, which was developed for EASO. The idea is to take the complexity of migration flows into account by looking at possible drivers and pull factors using a data science approach while relying on migration theory.

Other examples of early warning of asylum seeker flows based on traditional data recently appeared (e.g.Napierała et al., 2021) but these are not going to be reviewed here since our focus is on data innovation only. The EASO early warning and forecasting system is aimed at a) monitoring migration drivers in countries of origin and destination to detect the early onset of change; b) estimating the effects of individual drivers, including lagged effects; then on the bases of these two, c) producing forecasts of asylum applications in countries of destination up to four weeks ahead and d) assessing how patterns of drivers shift over time to describe the functioning and change of migration patterns. The system mixes the following sources of data:

- operational data on weekly applications from EASO as well as recognition rates;

- monthly irregular border crossings by the European Border and Coast Guard Agency (FRONTEX);

- event data from the GDELT project about political events, social unrest, conflicts, economic events, and governance-related events;

- Google trends data on a selection of 17 topics (clusters of keywords) related to international migration and travelling in general (visa or passport), asylum seeking (right of asylum or refugee), countries of transit (e.g., Jordan or Turkey for searches that take place in Syria) and countries of destination (e.g. Germany, France, or the EU), etc.

This data is harmonised in time and then a model selection procedure selects from the hundreds of potential drivers and generates early warning signals and forecasting for the four weeks ahead.

**FIGURE 34.** Overview of the Early Warning and Forecasting System workflow developed for EASO.
**Source:** Figure 1 from Carammia et al. (2022).

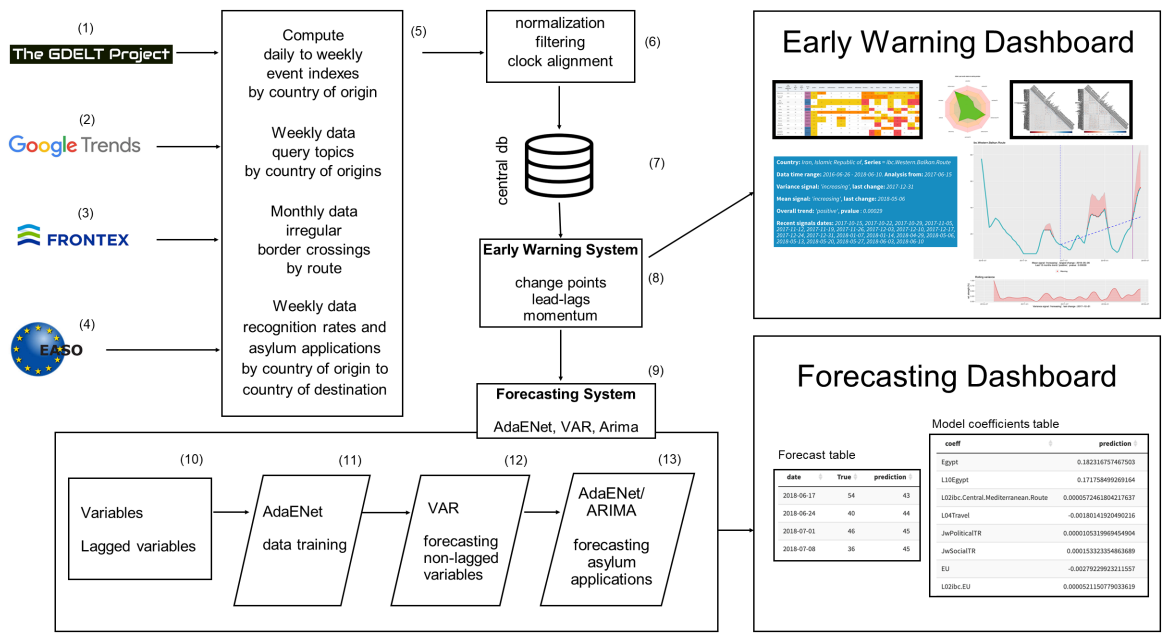

**FIGURE 35.** Back-testing performance of the system forecasting applications by Syrians in Germany (a). The black line shows the actual number of applications lodged by Syrian nationals in Germany. The dotted blue line is the moving average of the process. The red dashed line shows the 4-week ahead forecast at each time point. The pink shaded area represents a ±2 standard errors confidence bands around the moving average. Summary statistics for the relative error (b) and for the absolute error (b). Arima is a benchmark model which is only based on the auto-correlation of the application timeseries.
**Source:** Figure 4 in Carammia et al. (2022).

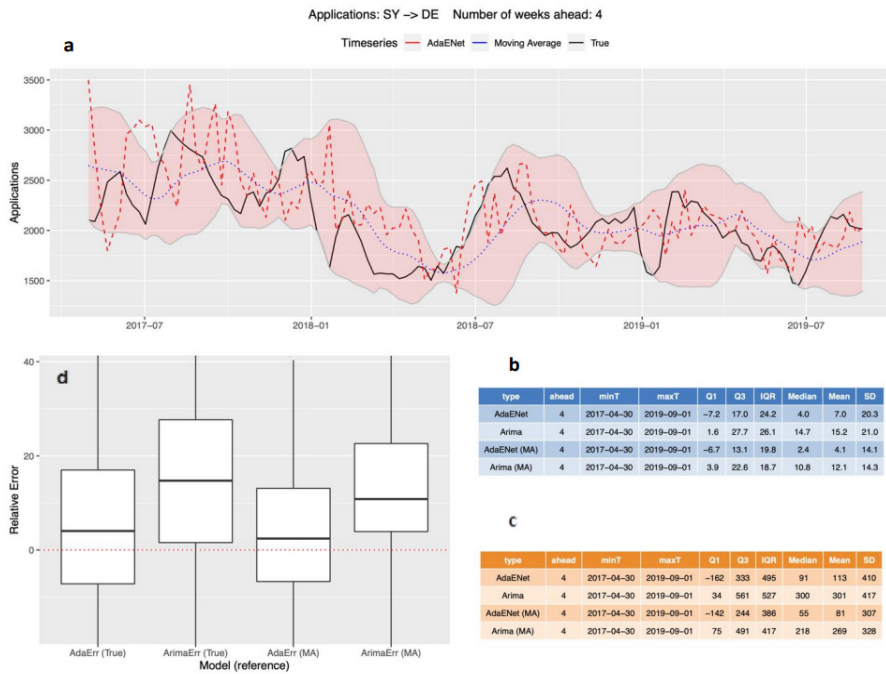

**FIGURE 36.** The factors that impact the EASO forecasting model for Syria towards Germany in the period January 2017 – September 2019. The model adapts over time, some effects are persistent in the first period, then other become more important. The colour scale only represents the relative importance of the variable. The coloured variables are all included in the model, the others have been dropped by model selection.
**Source:** Figure 5 in Carammia et al. (2022).

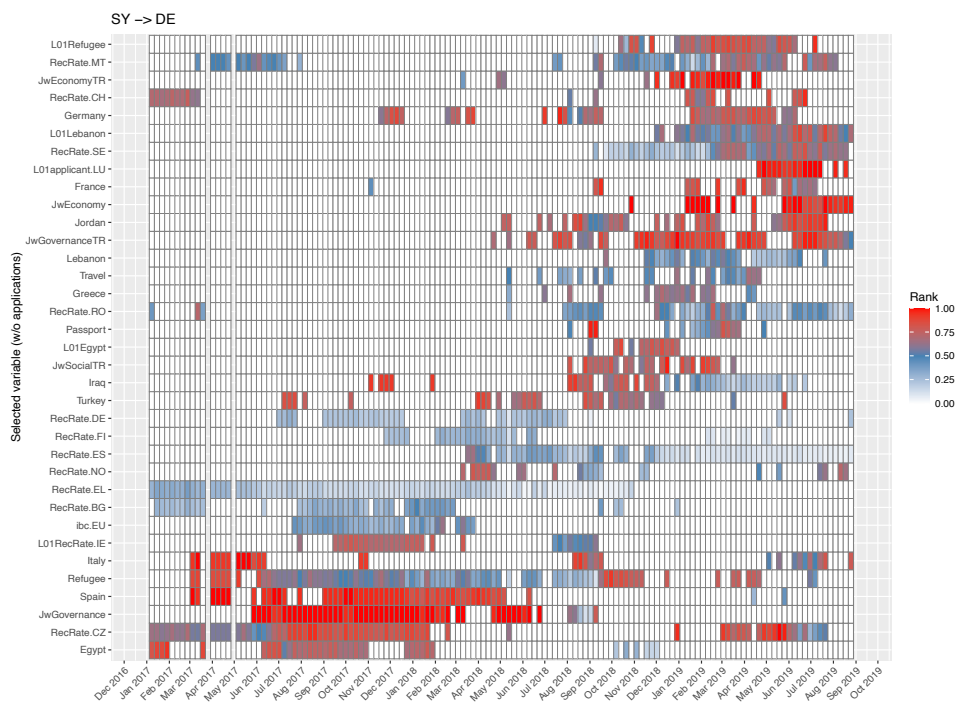

Figure 34 summarises the outcome and Figure 35 shows the typical performance of the system. It is also interesting to note that one of the outputs of the EASO system is that it selects the relevant migration determinants. Figure 36 shows the heatmap of selected determinants through time. Clusters of determinants are clearly visible in different periods of time, showing once again that migration is a complex phenomenon.

As mentioned in the previous section, other types of conflict data such as ACLED have been used to simulate short-term displacement (Suleimenova et al., 2017) (see also Figure 33).

# 18. AIR PASSENGER DATA

Even though migrants and refugees only constitute a tiny portion of the whole number of people crossing national borders daily (Recchi et al., 2019), air passenger data can be used to capture migration patterns such as seasonal migration. There are various sources of air passenger data suchas those by the Sabre Corporation (Sabre, 2021) as well as national agencies; for example the Bureau of Transportation Statistics (2021) which collects data from carriers operating between airports located within the boundaries of the United States and its territories, and the Office for National Statistics (2021) which conducts the International Passenger Survey and collects information about passengers entering and leaving UK.

## 18.1 MAPPING INTERNATIONAL MOBILITY

Gabrielli et al. (2019) used monthly air passenger traffic data between 239 countries worldwide from Sabre (2021) from 2010 to 2018 and successfully identified seasonal work migration (Figure 37).

## 18.2 MAPPING INDUCED MIGRATION

Rayer (2018) used Flight Passenger data from the Bureau of Transportation Statistics (2021) to estimate the Hurricane-induced migration from Puerto Rico to Florida in 2017. The study successfully captured the excess passenger flows between Puerto Rico and Florida but also highlighted the limitation of using flight passenger data to capture migration movements.

## 18.3 DISENTANGLING FACTORS BEHIND AIR TRAFFIC DATA

There are various factors that influence air passenger traffic between two areas such as touristic attractions, distance, cost of the flight as well as the stock of migrants. Choo (2018) found that a 10% increase in number of foreign-born Canadian residents leads to an increase of around 3% in inbound travel demand. In addition the same study found that recently immigrants have a greater impact in generating inbound air travel demand than in the past. Finally, Bilge et al. (2020), disentangled and quantified the impact of international migration to the air traffic volumes by analysing air passenger traffic data from 2004 to 2014 at province level in Turkey.

**FIGURE 37.** Air traffic routes predominately used for individual tourism and temporary work migration. Red ties denote intraregional mobility, black ties denote interregional mobility.
**Source:** Gabrielli et al. (2019).

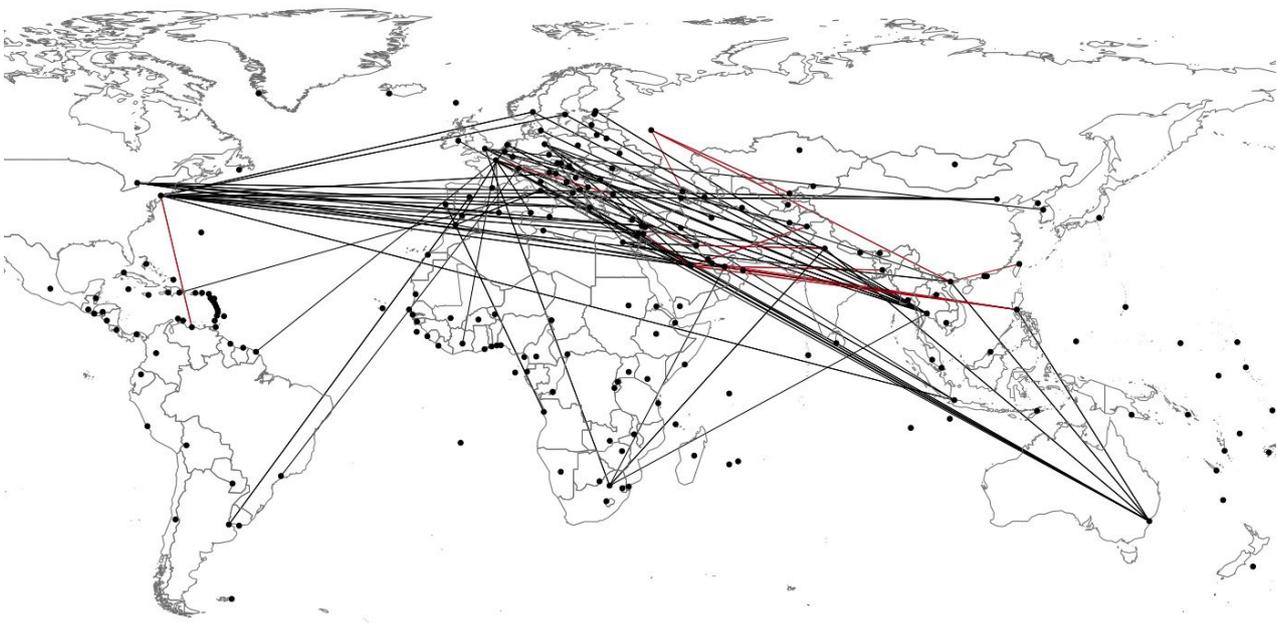

# 19. MOBILE PAYMENT SERVICES

In COM(2005) 390 final, remittances are defined as ''*financial transfers from migrants to beneficiaries in their countries of origin*''. Remittances represent an important tool for economic growth and poverty alleviation by ensuring a flow of financial resources from migrants and diasporas to households and communities in countries of origin (Kalantaryan and McMahon, 2021). In low and middle income countries, remittances often outpace Official Development Assistance (ODA)[107] as well as Foreign Direct Investment (FDI)[108] (Bank, 2006). Nonetheless, the conditions of sending remittances, and more generally the conditions under which they are sent, often remains unsatisfactory (COM(2005) 390 final). Moreover, many of the top remittance recipient countries have large unbanked populations, particularly in Least Developed Countries (LDCs) (Nurse, 2018).

One key recommendation towards financial inclusion is that of mobile money and mobile payments which could potentially reach the unbanked recipients and alleviate the costs of remittances. According to the 2018 State of the Industry Report on Mobile Money by GSMA,[109] the introduction of mobile payment services reaches millions of unbanked people and businesses in developing and emerging economies. First launched in Kenya in 2007, mobile money services are now active in about 90 countries and count more than 866 million registered accounts, processing 1.3 billion US dollars per day. The same document reports that the largest share of 45.6% of global registered mobile money users in 2018 come from Sub-Saharan Africa, 33.2% from South Asia, and 11% are located in East Asia and the Pacific. In 2018, the number of registered mobile money accounts grew by 20% compared to 2017 levels.

The use of mobile payments could potentially help meeting goal 10.c of the SDGs,[110] as digital channels are known to lower the cost of sending remittances. Furthermore, when data is available, the use of micro data on remittances from a mobile money transfer service may overcome the main measurement errors related to survey data (memory, misdating, etc.). This is because the data is made available at the transfer level rather than at the migrant or at the household level, also including a higher level of detail (e.g. low-value remittances) which is not possible with aggregated data from traditional sources. This level of detail also makes it possible to quantify how much short of reaching the SDG 10.c goal current remittances are. Last but not least, the COVID-19 pandemic has clearly indicated that having access to digital channels allows sending money (vital for the recipients), even if the physical mobility is restricted (Kalantaryan and McMahon, 2020).

Major limitations for the spread and use of this technology are financial illiteracy,[111] technological shortfalls, and lack of financial resources, in particular for people living in isolated rural contexts. However, the increase in access to mobile phones worldwide (and so to mobile payment services to a certain extent), may reduce some of these limitations.

Finally, another relevant topic related to the use of this kind of data is that concerning about data protection, privacy and confidentiality. In fact, the dataset might consist of personal data information on the remitter (first name, surname, age, profession, etc.), the recipient (first name, surname, country, etc.), and the transfer (amount, date, etc.) which would bring it within the scope of the GDPR in the EU and analogous regulations in other countries. As such, it should undergo an anonymization process before being used (see also Section 26).

## 19.1 MONITORING FINANCIAL INCLUSION

Recent studies (Evans, 2018; Emara and Zhang, 2021) have observed that Internet and mobile phones have a significant positive correlation with financial inclusion, meaning that rising levels of Internet access and mobile phones use are associated with increased financial inclusion, although the relation with remittances seems to be non-linear. In particular, the marginal effect of digital penetration seems to be larger as this increases from a low level, until it becomes negligible when it reaches a given threshold of digitisation.

## 19.2 INTERNATIONAL REMITTANCES

Bounie et al. (2013) analysed migrant's international remittances from a mobile money transfer service of

---

107  Official development assistance (ODA) is defined by the OECD Development Assistance Committee (DAC) as government aid that promotes and specifically targets the economic development and welfare of developing countries. https://www.oecd.org/dac/financing-sustainable-development/development-finance ~standards/official-development-assistance.htm

108  Foreign direct investment (FDI) is a category of cross-border investment in which an investor resident in one economy establishes a lasting interest in and a significant degree of influence over an enterprise resident in another economy. https://www.oecd-ilibrary.org/finance-and-investment/foreign-directinvestment-fdi/indicator-group/english_9a523b18-en

109  https://www.gsma.com/mobilefordevelopment/resources/2018-state-of-the-industry-rep ort-on-mobile-money/

110  SDG goal 10.c aims at reducing the transaction costs of migrant remittances to less than 3% and eliminating remittance corridors with costs greater than 5% by 2030 (https://sdgs.un.org/goals/goal10, last accessed 24 November 2021)

111  Financial literacy: ''the knowledge and skills needed to make important financial decisions'' (Source: https://ec.europa.eu/info/business-economy-euro/banking-and-finance/consumer-finance-a nd-payments/financial-literacy_en, last accessed 24 November 2021)

**TABLE 2.** Remittance characteristics by region over the period 2004-2009.
**Source:** Table 2 in Bounie et al. (2013).

| | Regions | | | | |
|---|---|---|---|---|---|
| | Sub-Sahara Africa | Eastern Europe | Madagascar | North Africa and Middle East | Overall |
| Number of migrants | 1 370 | 191 | 202 | 1 531 | 3 294 |
| Number of transfers | 8 660 | 1 662 | 1 283 | 7 718 | 19 323 |
| Transfer per migrant | 6.32 | 8.70 | 6.35 | 5.04 | 5.87 |
| Amount of transfers (Euro) | 7 050 878 | 1 662 967 | 1 368 805 | 8 975 391 | 19 058 041 |
| Average amount per transfer (Euro) | 814 | 1 001 | 10 667 | 1 163 | 986 |
| Average amount per migrant (Euro) | 5 147 | 8 707 | 6 776 | 5 862 | 5791 |

**TABLE 3.** Remittances and gender of remitters and recipients.
**Source:** Table 3 in Bounie et al. (2013).

| | Gender of the remitter | |
|---|---|---|
| | Man | Woman |
| **Gender of the recipient: man** | | |
| Number of transfers | 11 401 | 678 |
| Amount of transfers (Euro) | 12 215 738 | 641 544 |
| Amount per transfer (Euro) | | |
| **Gender of the recipient: woman** | 1 071 | 946 |
| Number of transfers | 2 742 | 4 499 |
| Amount of transfers (Euro) | 2 293 565 | 3 988 694 |
| Amount per transfer (Euro) | 836 | 867 |
| **Total** | 14 143 | 5 177 |
| % | 73.2 | 26.8 |

Société Générale, one of the major French banks. Their data consists of 19 323 transaction-level transfers recorded on the mobile phone made in France by 3 294 migrants from 2004 to 2009 to recipients located in four main regions: Sub-Saharan Africa, the Middle East, Eastern Europe, and Madagascar. They compared their sample with the National Migrant Workforce data the National Institute of Statistics and Economics (INSEE) in France to analyse the existence of a potential selection bias, as well as the representativeness of the sample analysed. The overall distribution of migrants by gender and age turned out to be very close to the official statistics. Using this type of dataset they were able to measure the volume of remittances and report descriptive statistics (see Tables 2 and 3).

## 19.3 AFTER-SHOCK REMITTANCES

Blumenstock et al. (2011) analysed billions of mobile phone-based transactions before and after the early 2008 destructive earthquake in the Western Lake Kivu region of Rwanda in order to understand why people remit in times of dire needs. Their dataset was provided by what at the time of writing was the monopoly mobile phone provider in Rwanda, which had launched a rudimentary mobile money system in 2006 that allowed mobile subscribers to transfer airtime from one person to another, free of charge. It consistsed of over 50 billion individual mobile phone transactions between early 2005 and late 2008, including roughly 10 million person-to-person transfers. Using this dataset, the authors were able to analyse both which types of individuals were most likely to receive shock-induced transfers, and the relationships that support interpersonal transfers between pair of users (dyads). Combining these two types of analysis is seldom possible because researchers typically only have either aggregated or survey data.

Because of the sensitivity of the information collected, this data needs to be handled according to appropriate data protection, privacy and ethical rules. Table 4 shows the level of detail and the summary statistics of the dataset.

**TABLE 4.** Summary statistics of mobile network data.
**Source:** Table 2 from Blumenstock et al. (2011).

| Dates covered | All dates 10/1/2006-7/1/2008 | Earthquake window 1/3/2008-3/3/2008 |
|---|---|---|
| **Panel A: Aggregate traffic** | | |
| Number of Calls | 46 000 000 000 | 868 786 684 |
| Number of interpersonal transfers | 9 202 954 | 362 053 |
| Number of unique users | 1 084 085 | 119 745 |
| Number of people who send airtime | 870 099 | 48 295 |
| Number of people who receive airtime | 946 855 | 101 351 |
| Number of people who both send and receive | 732 869 | 29 901 |
| Number of unique dyads | 646 713 | 159 204 |
| **Panel B: Basic statistics (12/1/2007–4/1/2008)** | Mean | S.D. |
| Transactions per user (send+receive) | 6.05 | 12.05 |
| Average distance per transaction (km) | 13.51 | 27.67 |
| Average transaction value (RWF) | 223.58 | 652.02 |

# 20. MOBILE PHONE DATA

Unique mobile phone service subscribers[112] represent about 65% of the population across Europe (GSMA, 2020) and 66.6% worldwide according to recent statistics and as shown in Figure 38.

This section contains evidence that, generally speaking, mobile phone data can be reliably used to capture the individual and aggregate mobility patterns. The use of mobile phone data is explored to estimate significant demographic features (such as total population and population density), and human migration. According to Newman and Matzke (1984), the concepts of migration and circulation are both included under the broader heading of population mobility.

Many studies have made use of mobile phone data from Mobile Network Operator (MNOs) either alone or combined with other data sources (e.g. satellite imagery, population censuses, etc.) in order to study population density and distribution (Krings et al., 2009; Dan and He, 2010; Kang et al., 2012; Douglass et al., 2015). In addition, recent interest in studying and predicting human mobility patterns using mobile phone has increased since the penetration rate of mobile phones across the globe has become high, with rapid increases in ownership in low-income countries (Blumenstock, 2012; Lai et al., 2019).

Cell phone traces have been successfully used to model aggregated human mobility. Mobile network data provides a novel tool to quantify directionality and seasonality of migration patterns on both local and national scales (Lu et al., 2016; Chi et al., 2020).

Mobile phone data offers a novel and effective "mean to measure migration at multiple temporal and spatial scales, where traditional migration flow statistics typically remain constrained by the logistics of infrequent censuses or surveys (Douglass et al., 2015). Enabled devices in a more sophisticated Global Positioning System (GPS) continuously record the exact geographic coordinates of the subscriber, but even the most rudimentary mobile phone can be roughly located in space according to the position of the cell towers which are used to route calls and data sent between the device and the network (Blumenstock, 2012). It is a compelling possibility that new sources of insight on human behaviour and development processes can be found in the data automatically generated through the everyday use of common technologies, which requires no special device nor software to be by the user. Consequently, we could cost-effectively provide accurate and detailed information of population distribution over national scales and any time period (Deville et al., 2014).

**FIGURE 38.** Digital connections around the world as of January 2021.
**Source:** WeAreSocial https://datareportal.com/reports/digital-2021-global-overview-report.

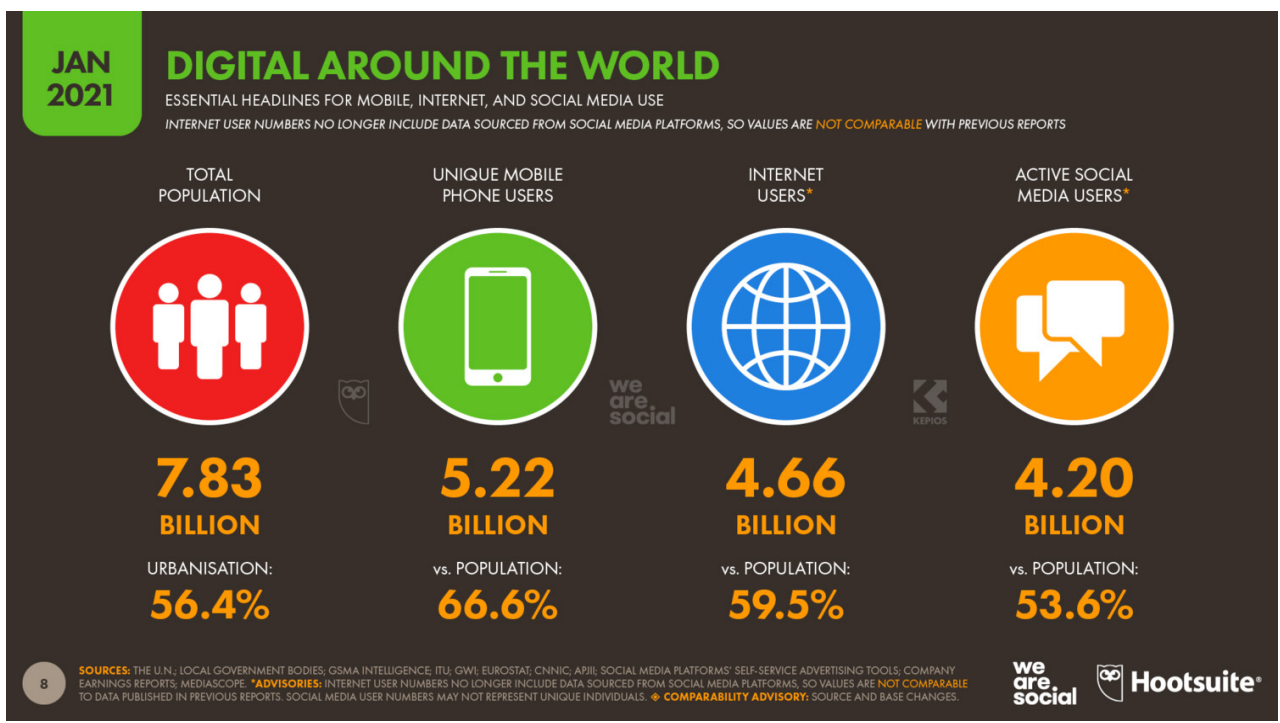

---

112  All mobile services subscribers, including IoT, are about 86% of the world population, 76% of which are smartphone users.

One of the major limitations of this technology is that it is rarely possible to use mobile phone data for cross-countries (or more precisely cross-MNOs) studies. In fact, each MNO generally collects data on a single country basis, using a methodology that differs from that of other MNOs, and only rarely information on cross-boundary users is available, making it difficult to compare data across different countries and different MNOs. Eurostat and other members of the European Statistical System are working towards the definition of a general Reference Methodological Framework for processing MNO data for official statistics (Ricciato et al., 2020), that could potentially overcome such limitation and move the use of mobile phone data from the stage of explorative research to a more operational one.

Three different types of mobility data can be derived from the use of mobile network operator data: the Call Detail Record (CDR), the Origin Destination Matrix (ODM), and the eXtended Detail Record (XDR). Their application to demography and migration will be respectively presented in Sections 21, 21.6, and 22. They are presented in this study as individual data sources although they are clearly related to each other.

# 21. CALL DETAIL RECORDS

CDRs capture information on a call made through a telephone system.[113]

CDRs come in an industry standard format. They are generated whenever a cell phone connected to the network makes or receives a phone call. A CDR includes a timestamp of the call, the mobile phone number, and the identifier of the mobile tower used to route the call. Note that no information about the exact position of a cell phone user is known. Nevertheless, this data can be used to analyse how phones move during towers between calls (González et al., 2008).

Although some information may be lost in the process of anonymisation to protect user privacy (Hughes et al., 2016), this data still contains useful information to study mobility trends. In urban areas, where towers are quite dense, the geographic precision of tower-based locational inference can be quite precise, though the resolution decreases in rural areas where towers are sparser (Blumenstock, 2012).

CDRs have a wide range of applications: they have been used to infer internal migration (Blumenstock (2012); Lai et al. (2019), see Section 21.1), to model aggregated human mobility in response to natural disasters, such as cyclones (Lu et al. 2016), droughts (Isaacman et al., 2018), or earthquakes (Wilson et al. 2016; Flowminder Foundation 2021; Lu et al. 2012, see Section 21.2), and to study the behavioural dynamics of rural and urban societies (Eagle et al. 2009, see Section 21.3).

Pioneering work using mobile phone data suggests that the movements of individuals are surprisingly predictable in stable social conditions (Song et al., 2010). Lu et al. (2013), analysed the movement of 500 000 mobile phone users from Côte d'Ivoire using CDR, founding that a family of mathematical models is able to predict human mobility with great accuracy, and also to predict daily population movements. Their results suggest that individuals' movements are highly influenced by their historical behaviour.

Predicting human mobility patterns through the use of mobile data (in particular, CDR) has also proven to be effective during unstable social conditions such as large-scale natural disasters.

CDRs do not come without limitations though. Firstly, they do not contain identifiers, nor demographic information on any of the subscribers, and so do not provide sufficient deducible information on the socio-demographic characteristics of the users. Indeed, they are generally complemented by other data sources to overcome this limitation (e.g. official statistics such as censuses, surveys, and data generated by various remote sensing techniques). Secondly, there are concerns regarding data protection, privacy and confidentiality. As such, they should undergo an anonymisation process, a difficult task that requires more than the removal of identifying information (Bayardo and Agrawal (2005); Zang and Bolot (2011), also see Section 26). There is also the issue of potential sampling bias due to the difference between the mobile phone subscribers and those that do not own mobile phones; this phenomenon used to be frequent in developing countries (Blumenstock and Eagle, 2012), although this has been changing in recent years (Lai et al., 2019). Another problem is linked to widespread SIM sharing. The potential for SIM sharing to destabilise results based on the use of CDR data is widely acknowledged but little investigated (Jahani et al., 2017; Bosco et al., 2019). Data heterogeneity is yet another intrinsic problem when dealing with CDRs; in particular, when collecting data from different MNOs or different countries, records are not entirely standardised and will vary depending on the operator, the software supplier and configuration (Ricciato et al., 2017). Finally, if the phone is not actively using (making or receiving) mobile network services, there will be no information in the CDRs.

Models based on CDR alone often tend to have high error rates due to these limitations. Therefore, they are typically accompanied by other forms of data. Steele et al. (2017) and Bosco et al. (2019) used CDR data along with satellite imagery and geo-located surveys to model and map spatial variations and gender-based inequalities for a number of development indicators (e.g. literacy, poverty, agriculture-based-occupations, and births in health facilities). These studies show evidence of correlation between the set of covariates derived from CDR data and the indicators modelled. It was also proven that CDR data produce rather accurate, high-resolution estimates in urban areas, not possible using data derived from remote sensing alone (Steele et al., 2017). This new source of information is capable of improving model performance at local level (Bosco et al., 2019). With surveys and covariates derived from CDR data available for multiple time intervals, there is the potential for multi-temporal mapping to measure progresses towards meeting the SDGs at fine spatial disaggregation. Recent work investigating whether indicators extracted from mobile phone usage can reveal information about mobile phone users, focused on deriving gender (Jahani et al., 2017; Bosco et al., 2019), employment status (Almaatouq et al., 2016; Sundsøy et al., 2016), education (Sundsøy, 2016), household wealth (Šćepanović et al., 2015), and individual income (Blumenstock et al., 2015; Sundsøy et al., 2016a).

---

113 https://ec.europa.eu/eurostat/cros/content/Glossary:Call_detail_record_(CDR)_en

It is worth noting that MNOs log additional data besides CDR (e.g. mobility management, network troubleshooting). Such data contain additional information about the mobility patterns of individual terminals that is not captured by CDR (Ricciato et al., 2017) (see Section 22). However, the vast majority of previous research has been limited to the analysis of CDR, notwithstanding the technical challenges involved in the acquisition and processing of other network-based data sources (Valerio et al., 2009; Janecek et al., 2015).

Despite the abundance of research on mobility analyses in the current literature, little has so far been done to analyse international migration phenomenon using CDR (Sîrbu et al., 2021). This is mostly because CDRs typically span only one nation, and they may or may not contain information on the nationality of the users. An exception to this is represented by initiatives like the Data for Refugees (D4R) challenge in Turkey (Salah et al., 2019), where an indication of the broad activity and mobility patterns of refugees was provided to improve the conditions of Syrian refugees in Turkey, the largest refugee population in the world.

Lastly, CDR can be used to estimate high resolution ODM (see Section 21.6).

## 21.1 INFERRING INTERNAL MIGRATION

In his work, Blumenstock (2012) developed a new quantitative measure of inferred mobility to compute rates of temporary and circular migration. He used the phone records of 1.5 million users from 2005 to 2008 from (at the time of writing) Rwanda's primary telecommunications operator to demonstrate the proposed empirical analysis. The anonymous CDR dataset was supplemented with demographic information from a phone survey conducted in Rwanda between 2009 and 2010. The work highlights a number of aspects of internal migration, particularly with respect to temporary and circular migration, and to the heterogeneity of migrants that are quite difficult to investigate using the data typically collected in government censuses and household surveys. For example, Table 5 shows the full set of migration and mobility metrics for a selected sample, and disaggregates the measures by different demographic groups.

Lai et al. (2019) used 72 billion anonymised CDRs in Namibia from October 2010 to April 2014 to explore how internal migration estimates can be derived and modelled at subnational and annual scales, and analysed how they compared with census-derived migration statistics. Their dataset came from the leading network operator in Namibia with a 76% market share and providing network spatial coverage for 95% population.[114] They supplemented the CDR dataset with potential migration-related demographic, socioeconomic, geographic, and environmental variables. First of all in their analysis, they identified migrants from the dataset as being those mobile users who changed residence between two reference years (corresponding to the time frame of official census data published by the Namibia Statistics Agency). Figure 39 shows the correlation between CDR- and census-derived migrants for the years 2011 and 2012.

After this, they fit three types of models to census data to explore whether CDR-derived migration data can accurately replicate traditional census-derived migration statistics. Moreover, the best fit was found using CDR-based linear models based on CDR-derived migrating user data alone (without other covariates) used in gravity models. Finally,

**TABLE 5.** Average mobility and migration metrics by demographic type.
**Source:** Table 3 from Blumenstock (2012).

| | Gender | | | Education | | Wealth | |
|---|---|---|---|---|---|---|---|
| | All | Men | Women | Did not finish primary school | Finished secondary school | Top 10% | Bottom 10% |
| ROG (km) | 15.32 | 15.7 | 14.5 | 11.1*** | 20.4*** | 17.7 | 14.8 |
| # towers | 14.02 | 14.1 | 13.7 | 9.01*** | 23.7*** | 24.6+++ | 14+++ |
| Max distance (km) | 71.62 | 72.4 | 69.7 | 51.6*** | 96.3*** | 83+ | 70.9+ |
| 3-month migration | 0.23 | 0.22 | 0.25 | 0.12*** | 0.48*** | 0.38+++ | 0.19+++ |
| 12-month migration | 0.016 | 0.016 | 0.016 | 0* | 0.032* | 0.044 | 0.022 |
| N | 901 | 645 | 256 | 139 | 95 | 90 | 90 |

114  Mobile Telecommunications www.mtc.com.na/coverage, last accessed 3 June 2018

they used this model, calibrated using 2011 census data, to predict internal migration flows in Namibia for 2012 (adjusting the increasing penetration rate for 2012 by region, see Figure 40).

## 21.2 HUMAN MOBILITY DURING NATURAL DISASTERS

Lu et al. (2016) studied the effect of climate induced migration in Bangladesh in the short- (hours-week), and long-term (years) using individual mobility trajectories of six million de-identified mobile phone users from the largest mobile network operator in the country. The analysis focused on both the environmentally stressed areas in the Southern delta region of Bangladesh before and after Cyclone Mahasen (May 2013) as well as on longterm national level migration patterns (almost 2 years between 2012 and 2013). They observed considerable mobility during cyclone passing, at a time when all people should have moved to cyclone shelters, which provided refuge and saved lives (see Figure 41). These findings exemplify how mobile network data can enable identification of areas where high risk behaviours are observed. In particular, there was a clear increase of users arriving in Chittagong City, the second largest city in Bangladesh, beginning approximately two days after the cyclone and continuing throughout the remaining one and a half months. Temporary migration to urban areas for short-term employment is a common adaptation strategy for rural households affected by disasters or other economic hardship (Hugo, 1996; Tacoli, 2009).

**FIGURE 39.** Logarithmic relationship between census-derived populations in 2011 and the number of CDR-derived mobile phone users.
**Source:** Figure 2 in Lai et al. (2019).

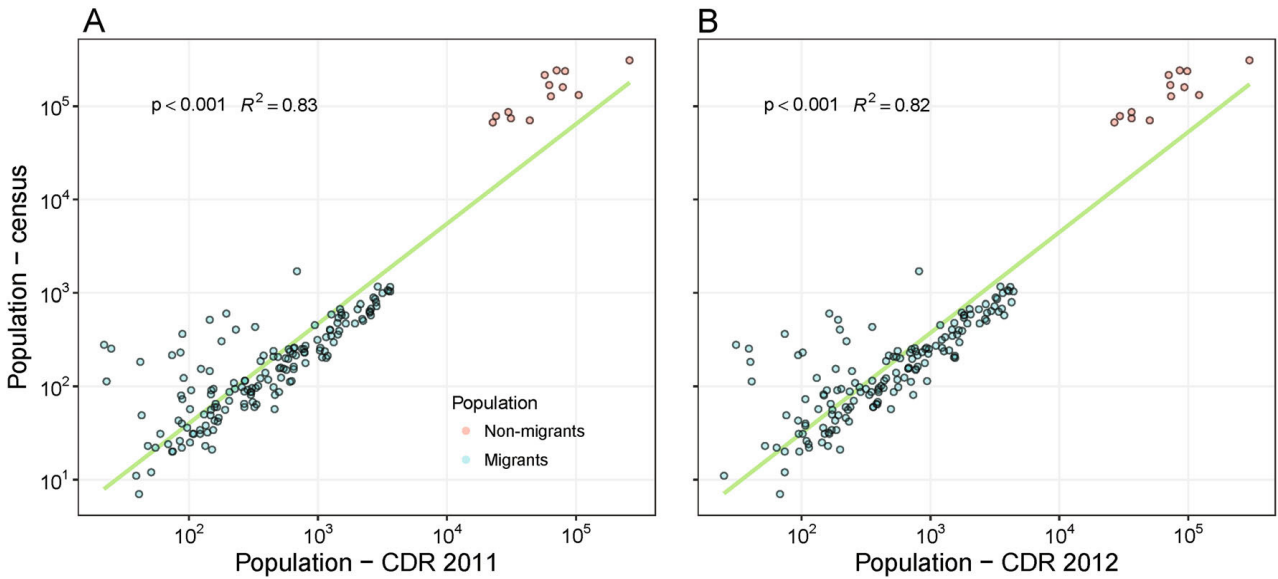

**FIGURE 40.** Relative difference of regional outflow (A) and inflow (B) between Period 2011 and Period 2012. The migrations were estimated by the CDRLM using only CDRs, and the adjusted CDR data of Period 2012 were used to offset the impact of the increasing mobile phone ownership across periods. The numbers of migrants by region in Period 2011 and Period 2012 are presented under the name of each region, respectively, and the Zambezi region as a significant outlier is excluded in the analysis and so coloured grey.
**Source:** courtesy of Shengjie Lai.

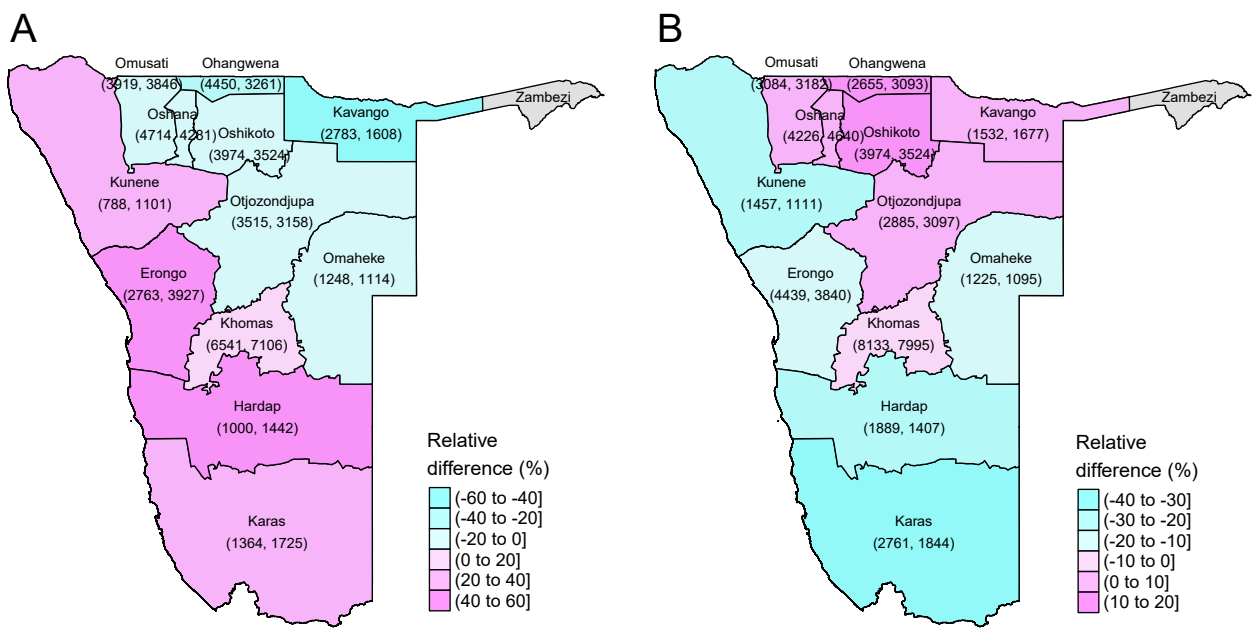

**FIGURE 41.** High-risk mobility of mobile phone users during cyclone landfall and passing (00:00 06:00 a.m., 16 May). The thickness of each link represents the number of moving SIM cards, and the size of each node is proportional to the number of incoming subscribers to the node.
**Source:** Figure 1 from Lu et al. (2016).

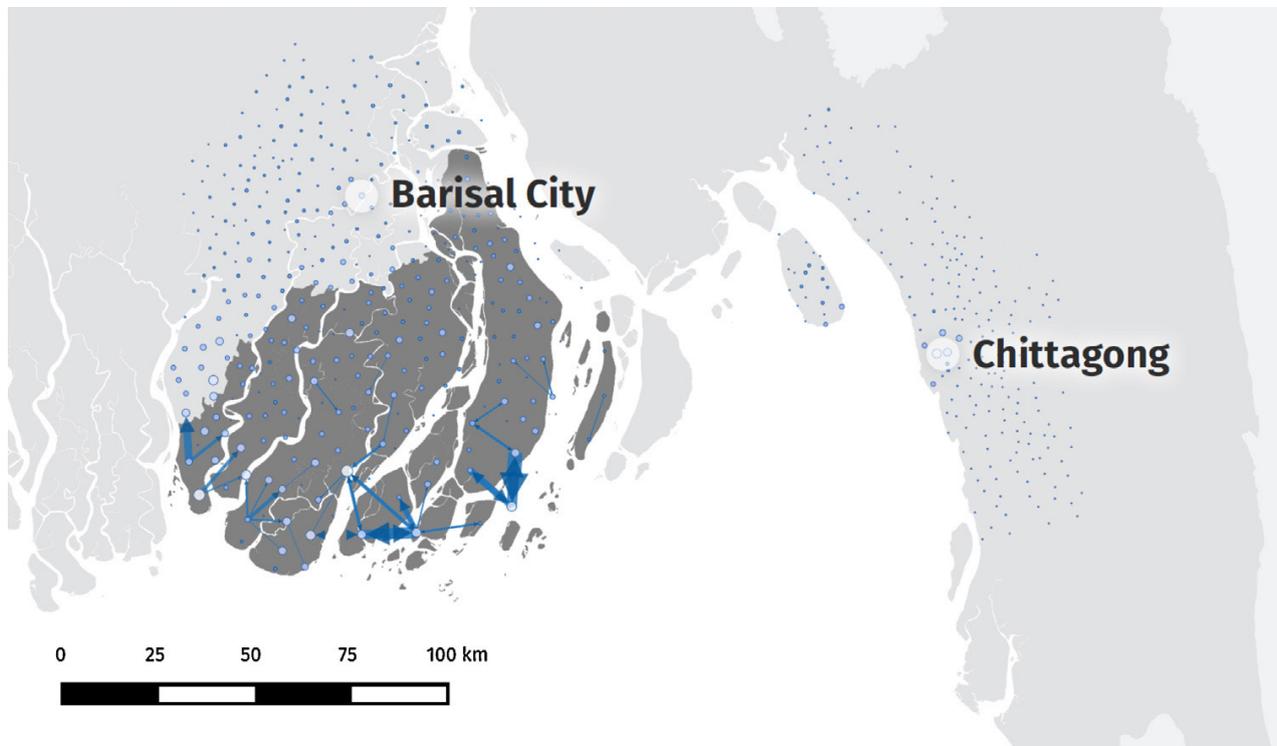

Overall, these results seem to highlight the importance of mobile phone data for studying human migration at different spatial scales. In particular, their work shows how they can be used to focus on small areas and short-time intervals that are more related to particular climate events (as also stated by Isaacman et al. 2018).

Furthermore, for the long-term and national scale analysis, they identified migration as a change in residence lasting between one and twenty-three months. Unexpectedly, they found that the major inflow to Chittagong City largely originated from outside the cyclone affected area, questioning the causal relation between the cyclone event and the migrant flow. In other words, the observed increase of subscribers in Chittagong City could be attributed to non climate-change related reasons such as regular seasonal migration patterns (Bryan et al., 2014).

The two analyses taken together firstly highlight the spatio-temporal detail and scale with which migration as an adaptive response to climate change can be studied across a country. Secondly, the findings illustrate the profound significance of seasonality in migration patterns in general and the importance of taking these into account in the planning, execution, and interpretation of migration surveys in particular.

Lu et al. (2012) analysed the movements of 1.9 million mobile phone users from Haiti before and after the tragic earthquake of 2010. They focused on the aggregate mobility of individuals of the severely hit capital Port-au-Prince, and discovered that, despite large changes in the population distribution across the country, most of their movements remained highly predictable (mainly influenced by people's social support structures). Figure 42 shows that the movement patterns primarily changed for directly affected people and returned back to normal after 4 to 5 months.

**FIGURE 42.** Proportion of individuals who travelled more than *d* km between day t − 1 and t. Distances are calculated by comparing the person's current location with his or her latest observed location.
**Source:** Figure 1B in Lu et al. (2012).

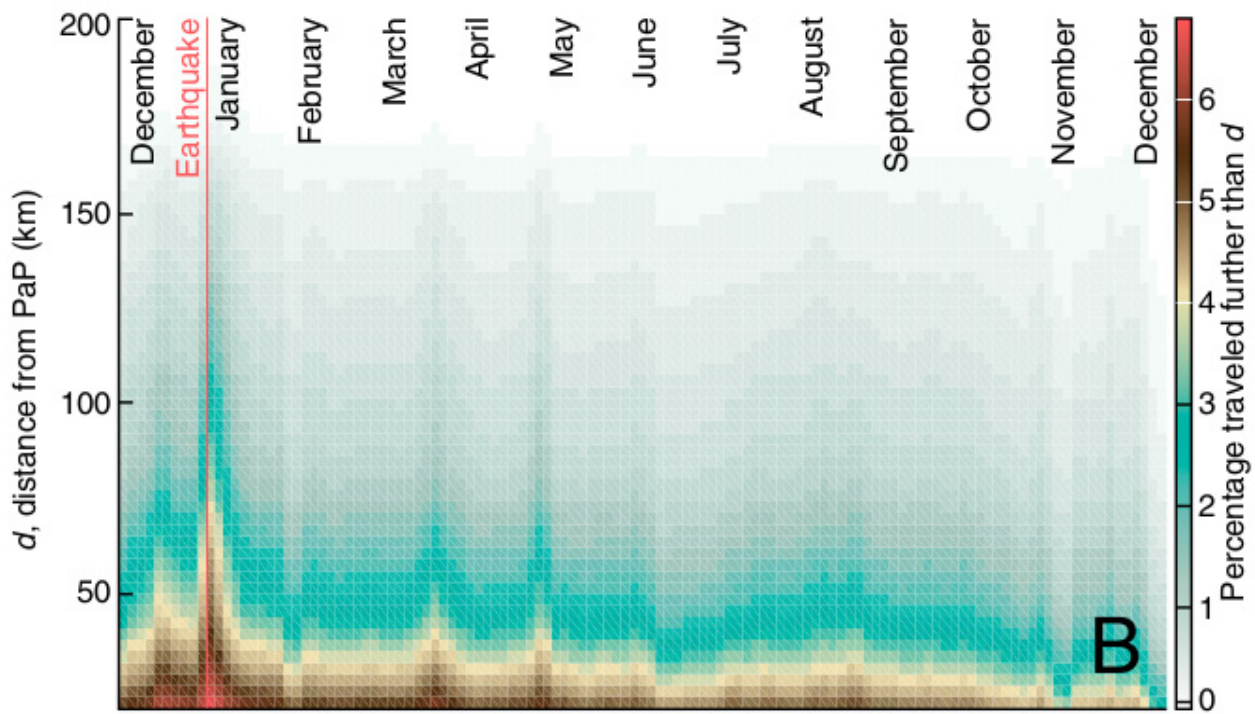

Isaacman et al. (2018) used anonymised and aggregated cell phone traces and weather data to model the migrations occurring during a severe drought in La Guajira, Colombia, in 2014. They used Home Detection algorithms to identify home locations in the CDR dataset on a weekly basis and measure the change in home location each week. They found a linear decrease of 10% of the population in Uribia (the most damaged municipality of La Guajira department) during the period of study. The majority of those climate migrants were found to stay in the same department, relocating to other municipalities where access to help, food, and water was probably easier. Only 10% of them went to other departments, suggesting that migrants tend to go to neighbouring states, with the exception of states with high population density, such as the capital city Bogotá (see Figure 43).

Nine days after the Gorkha earthquake in Nepal of 25th April 2015, a computational architecture and analytical capacity were rapidly deployed to provide spatiotemporally detailed estimates of population displacements from CDRs based on movements of 12 million de-identified mobile phones users. In the analysis made by Wilson et al. (2016), the authors estimated flows by recording a location for each user at two different times and then counting the number of users who moved from one location to another. So they produced a transition matrix containing the flow of users between each possible pair of locations. To remove noise from the dataset (e.g. short movements, commuting, etc.), they calculated a "home location" for each individual by calculating the modal daily location over a certain period. These home locations were then used to calculate transition matrices describing the countrywide mobility between two points in time. Using these matrices, they were able to observe the population displacement due to the earthquake, comparing the difference of flows before (normal flows), and after the natural disaster. The flows calculation produced a transition matrix giving the anomalous flow (number of users, above and below normal) moving between each pair of locations which could be used to observe above-normal inflows and outflows for each region. Another important point addressed in their analysis following the earthquake was to identify regions to which people had not returned, which could potentially indicate regions in which recovery had not reached sufficient levels at the time of the analysis. To identify users who did not come back to their place of origin, they took account of those users who had spent at least seven consecutive days away from their pre-earthquake home location in a two week period after the earthquake. Figure 44 illustrates the percentage of displaced population by region as of 19th August 2015.

A similar approach has been applied in a more recent report by Flowminder Foundation (2021). They used CDR data from Digicel MNO to estimate population movements following the Haiti earthquake on 14th August 2021 and Tropical Depression Grace. In their report dated 27 August 2021, they took into account data collected until 23rd August 2021 and estimated 90,000 people in the South,

**FIGURE 43.** Heatmap of the relocation of individuals changing home during the second week of January 14 in Colombia at a department level.
**Source:** Figure 5 in Isaacman et al. (2018).

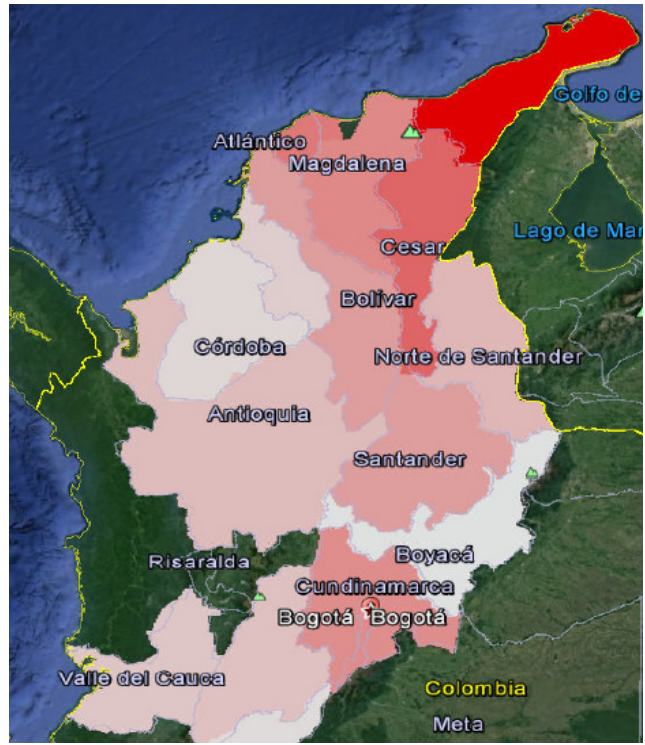

**FIGURE 44.** Percentage of people displaced by the earthquake who remained away as of the 19th August, shown spatially for the focus Districts, at Village Development Committee (VDC) level.
**Source:** Figure 7 in Wilson et al. (2016).

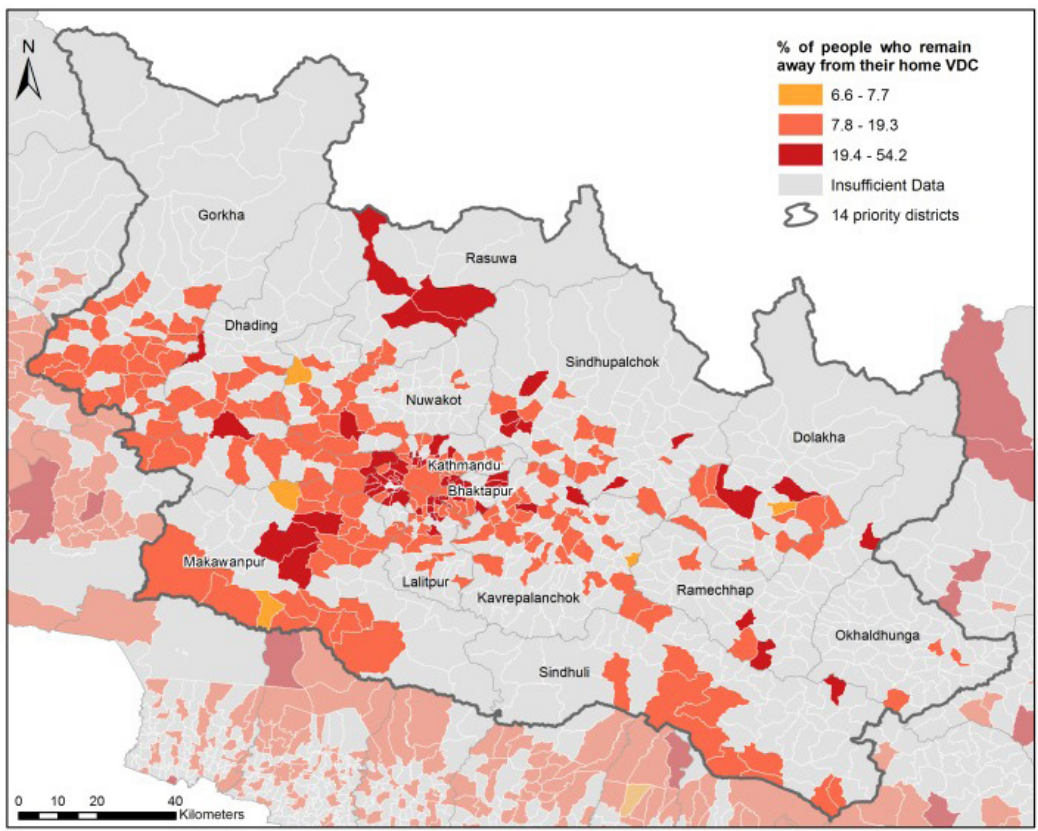

Nippes, and Grand'Anse departments had moved from their usual pre-earthquake locations. They mainly reported short distance movements, mostly within commune boundaries, with no major changes in population numbers at communal level.

## 21.3 BEHAVIOURAL DYNAMICS OF RURAL AND URBAN SOCIETIES

In their study, Eagle et al. (2009) used CDR data from 1.4 million subscribers within a small country between January 2005 and January 2008 in order to compare the behavioural patterns that differentiate urban and rural communities, and the individuals who move between them. The authors coupled the mobile phone records with data from regional census and airtime sharing data. Their analysis shows that individuals in the rural areas travel significantly more per month than individuals living in the cities. One reason for this could simply be the small potential distances that can be travelled within the capital and the much larger distances within rural areas (see Figure 45).

The analysis also empirically confirmed the theory that people in rural areas have relatively strong ties to fewer people, whereas individuals in urban areas tend to have more, but weaker ties (Tonnies and Loomis, 2002; Simmel, 2012; Wirth, 1938; Milgram, 1970; Fischer, 1982). In general, the work by Eagle et al. (2009) showed how using mobile phone data can be useful to test, develop, and quantify classical hypotheses in sociology, social psychology, and economics about behavioural changes and human and social adaptations as a result of life in large cities versus smaller urban areas and rural settings.

## 21.4 POPULATION DISTRIBUTION

In their work, Kang et al. (2012) discussed the issues in estimating population distribution based on mobile call data. They analysed the call activity of nearly two million mobile subscribers in Harbin, a large city in north-eastern China, obtained from a dominant mobile phone carrier, and over one hundred million communications over seven consecutive days. They used the LandScan 2008 Global Population Dataset,[115] the community standard for global population distributions, using an innovative approach with geographic information systems and remote sensing, taking into account census data, road proximity, slope, land cover, and night-time lights to evaluate the process of estimating the population from mobile call data. Among other analyses presented in their work, they investigated the correlation between the caller volume and the underlying population in different time intervals, and found that the correlation is higher in two periods of the day (referred to as "call-active", and shown in red in Figure 46). This period is preferred for understanding population distribution (i.e. static population). Another interesting period of the day is the one they refer to as "motion-active" (shown in green in Figure 46), which is less suitable for estimating static population. However,

**FIGURE 45.** Average amount of monthly travel by subscribers from each region.
**Source:** Figure 3 from Eagle et al. (2009).

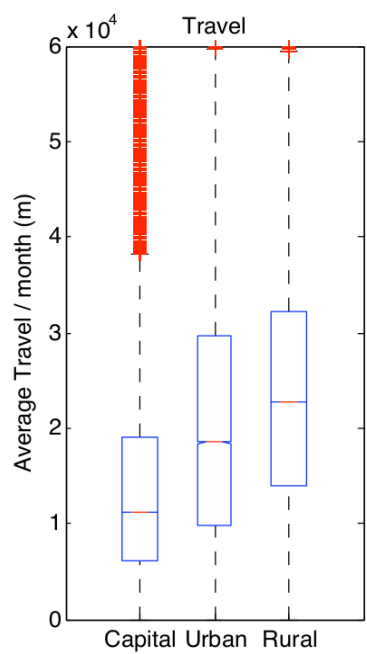

. . . . . . . . . . . . . . . . . . . . . . . . . . . . . . . . . . . . . . . . . . . . . . . . . . . . . . . . . . . . . . . . . . . . . . . . . .

**FIGURE 46.** Correlations between the caller volume and the underlying population in different time intervals.
**Source:** Figure 8 in Kang et al. (2012).

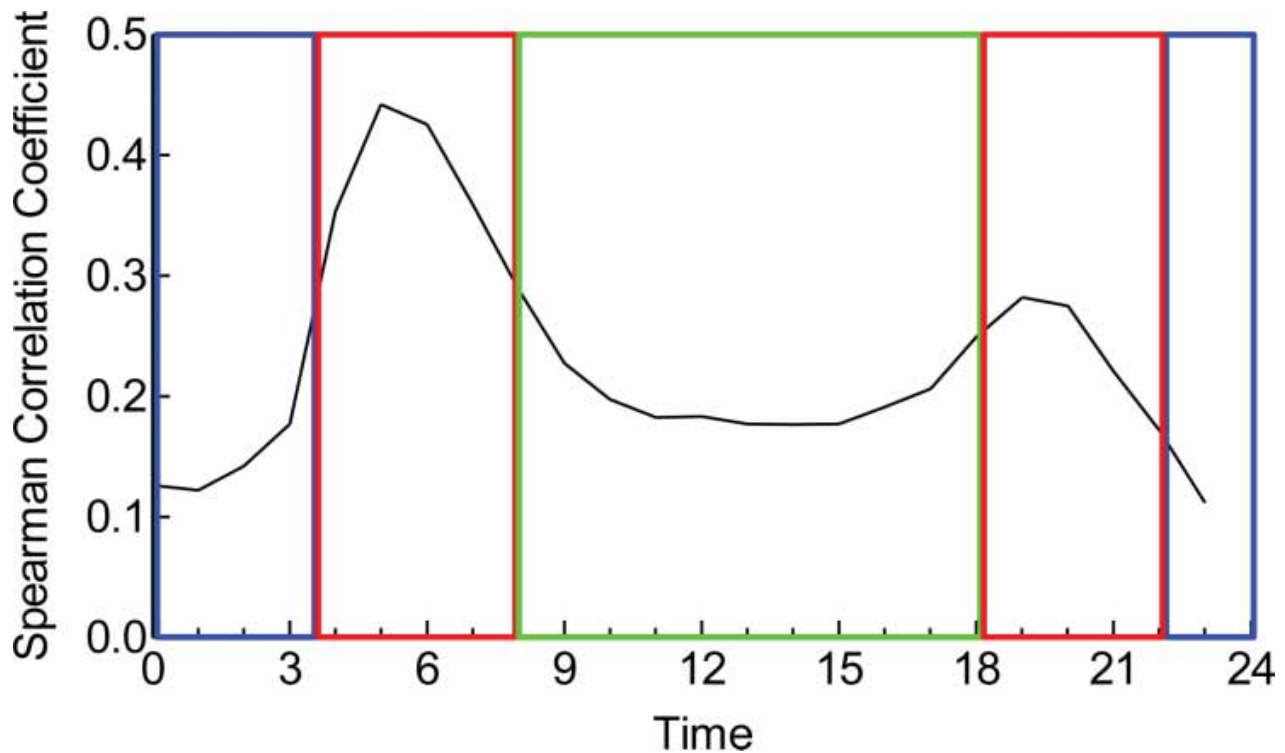

it reveals the dynamics of the entire population. Finally, they stress the importance of considering the influence of the spatial and temporal characteristics of the raw data in explaining the information that can be obtained from the call activity.

## 21.5 POPULATION DENSITY

The use of an alternative source of data to estimate population is particularly important in the context of Haiti, where the most recent census dates back to 2003. In a recent study, Zagatti et al. (2018) used CDR from the main MNO as well as antennae location data in Haiti from 1st March to 30th May 2016 to investigate night and daytime population densities and commuting patterns. The authors used an approach similar to that used by Isaacman et al. (2011) for Los Angeles and New York to estimate the home and work locations of the population. The study focused on urban areas, in particular the metropolitan regions of Port-au-Prince and Cap-Haïtien where this methodology performs much better than the countryside of Haiti, as mobile phone ownership and usage are higher. Using a dataset containing approximately 5.2 million users, they identified home and work clusters, and run a linear model to estimate stationary population distributions for two different categories. The first category is weekday daytime, which for most adults is likely to be work location. The second category is weekend and weekday evening, which for most people is likely to be home location. Figure 47 depicts the Port-au-Prince population distribution during day and evening time.

The final result is limited by the lack of actual observations with the exception of the 2016 population predictions produced by the Institut Haitien de Statistique et D'Informatique (Institut Haitien de Statistique et D'Informatique, 2009). These predictions are based on an outdated population census from 2003 which has mainly been updated using administrative records. However, they do not take into account the major events that have happened since 2003, such as the 2010 earthquake, a cholera outbreak in 2011, and Hurricane Matthew in 2016, which are likely to have caused a major redistribution of the population. Nevertheless, the work demonstrates the usefulness of using CDRs in data-poor environments as they provide enough information for an initial assessment of the conditions on the ground and provides useful information to help inform policy and investment decisions.

Ricciato et al. (2017) addressed the problem of estimating the population density from mobile phone data. They proposed a novel framework for interoperable MNO data analyses that goes beyond the traditional "CDR-only, single MNO" paradigm. Their method integrates other data such as Visitor Location Register (VLR) and service coverage maps. They proposed an interoperable framework to enable data from different MNOs to overlay a common reference grid and to address the problem of estimating population densities. Their simulation results show that the approach increases the accuracy of traditional models based on CDR-only data.

**FIGURE 47.** Population distribution day versus evening and all versus commuters Port-au-Prince. Concentric rings at 1, 5, and 25 km from city centre.
**Source:** Figure 14 in Zagatti et al. (2018).

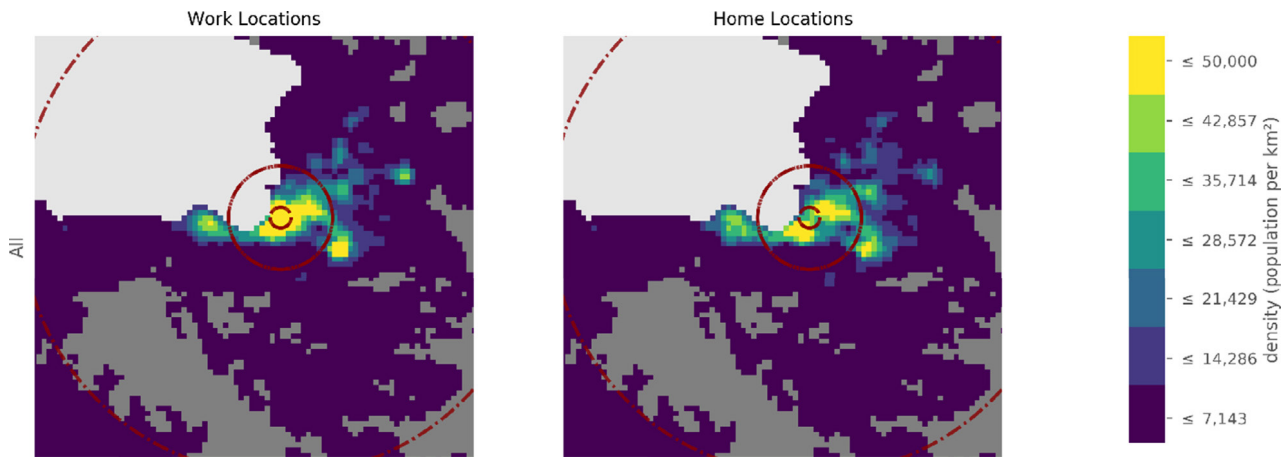

## 21.6 MOBILE NETWORK ORIGIN-DESTINATION MATRICES

ODM describe people movement in a certain area. In ODM, the rows represent the origins and the columns are the destinations. As reported by Wheeler (2005), they may measure work trips between census tracts, planning areas within a city, migration flows, and population distribution.

Traditionally, the estimation of ODM is done using household questionnaires or a census and road surveys. However, these traditional sources are generally limited in terms of time and space resolution, are expensive, hard to update, and the calculation of ODM using these types of data can take years. More recent studies have demonstrated that it is also possible to estimate ODMs using mobile phone location data (Calabrese et al., 2011; Bonnel et al., 2018; Mamei et al., 2019; Fekih et al., 2021). Mamei et al. (2019) analysed algorithms to extract ODM from CDR data, while Fekih et al. (2021) used passively-collected cellular network signalling data for millions of anonymous mobile phone users in the Rhône-Alpes region of France. Unlike CDR data, signalling data include all network-based records, providing higher spatio-temporal granularity than CDR, which only rely on phone usage (see Section 22). Huang et al. (2021) used a large-scale signal data (200 million cell phone users and a total of almost 50 million signal records) in Changchun (China) from July 3, 2017 to July 7, 2017 to predict human Origin-Destination flows in an urban area.

Although the concept of migration can be included under the broader heading of human mobility, it is nevertheless non-trivial to infer migratory flows from ODM. As a migration trip takes place much less frequently compared to other types of movements (e.g. commuting), it requires a longitudinal set of data for the analysis. Typically, long distance trips, which are largely associated with temporary or permanent migration, account for a much smaller portion, are not so regular, and are not well represented in most regional as well as national travel demand models (Hankaew et al., 2019).

Robinson and Dilkina (2018) derived ODM from the USA Migration dataset (IRS Tax-Stats data, Pierce (2015)), and the Global Migration dataset (World Bank Global Bilateral Migration Database, Özden, et al. (2011)). To complement the original datasets, they also incorporated exogenous features that influence the decision to migrate. They used the extracted Origin-Destination data to develop an innovative general method to predict migrations with Machine Learning (ML). They compared the predictions of migration flows between US counties as well as international migrations using both ML and traditional models (i.e. gravity models or the more recent radiation model), and their results seem to indicate that ML performed better (see Figure 48). They discovered that the intervening number of counties and the population of origin are the features that contribute to migration the most in the case of migrations between US counties, whereas population growth of origin and intervening population[116] were the most important features in international migrations. Figure 48 shows the modelling errors of the actual versus the predicted migrants between US counties for both traditional and ML models as reported by Robinson and Dilkina (2018). Moreover, the artificial neural network model (ANN) with extended features was best at capturing the incoming migrant distributions per county.

---

116  The intervening features are calculated on the basis of the idea of intervening opportunities presented in the radiation model. For any given county-level variable, $x$, e.g. population, the intervening amount of that variable between counties $i$ and $j$ is defined as $S^x_{ij}$, the sum of all $x$ in the intervening counties that fall within the circle centred at county $i$ with a radius to county $j$ (excluding $x_i$ and $x_j$).

**FIGURE 48.** Difference between the actual and predicted numbers of incoming migrants by county for traditional models (top row), and ML models (middle and bottom rows), trained with (left), and without (right) extended feature set. Blue corresponds to overestimation by the model, red to underestimation by the model, and white if the model accurately predicts the correct number of incoming migrants.
**Source:** Figure 1 in Robinson and Dilkina (2018).

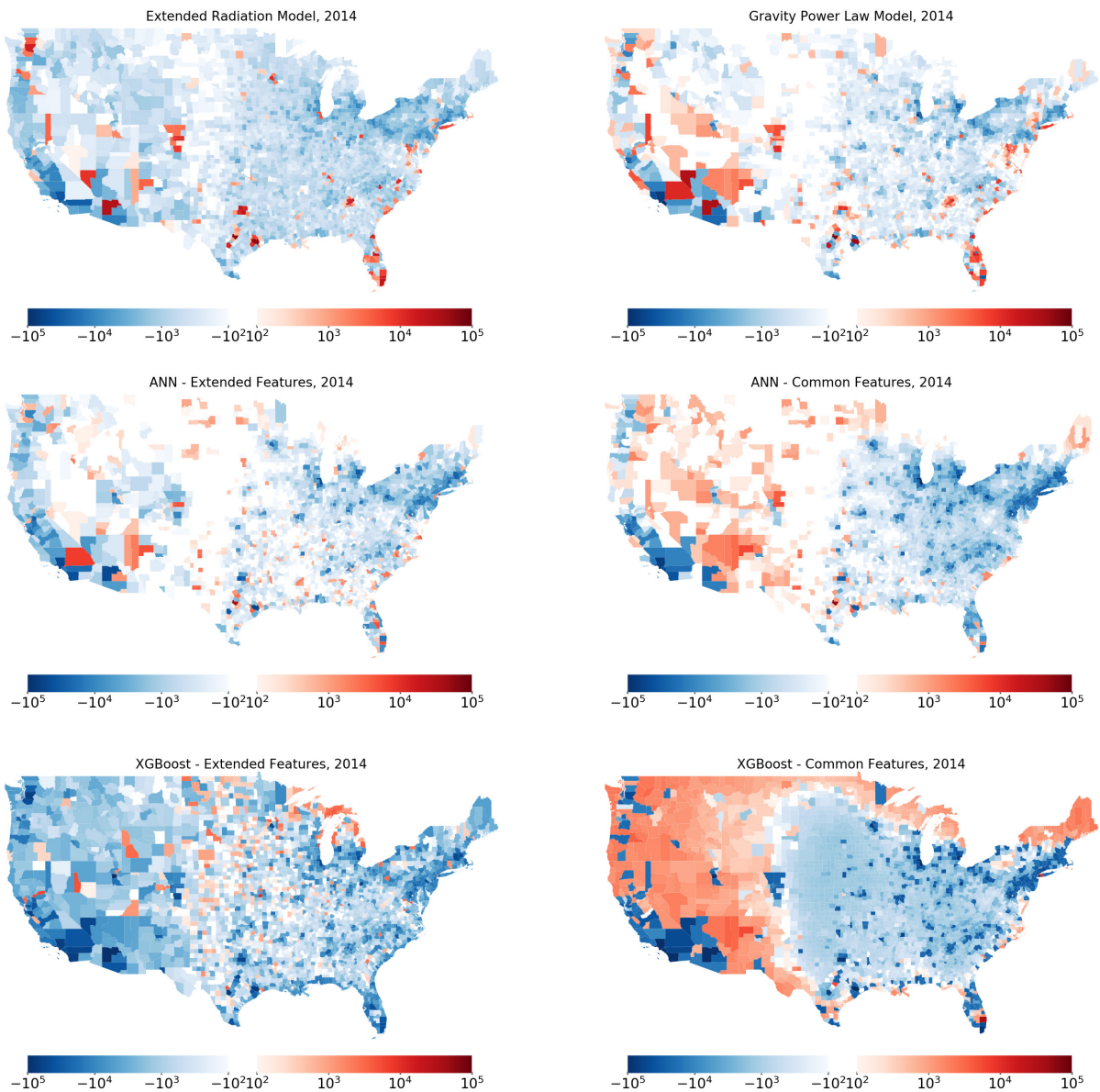

Guo and Zhu (2014) presented a new generalised approach for the computational analysis and flow mapping of large spatial mobility data. It is based on a flow-based density estimation method and a flow selection method and is used to normalise and smooth flows with controlled neighbourhood size, and it should faithfully represent the major flow patterns. Their proposed approach could be used to analyse flow patterns at different scales and disaggregation levels from both point-based flow data (such as taxi trips with GPS locations), and area-based flow data (such as county-to-county migration). They used the U.S. internal (county-to-county) migration dataset from the 2000 Census as a case study to evaluate their methodology. Their new approach was able to confirm and explain the overall net migration patterns. It was also able to provide insights at different scales (multi-resolution migration mapping), and it succeeded in mapping the migration flows of different age groups separately and in comparing their flow patterns. For example, using their approach, they were able to produce a map showing the smoothed flow maps for age group 25-29 (see Figure 49).

## 21.7 HUMAN MOBILITY AND EXCESS DEATHS DURING COVID-19

In order to fight the spread of COVID-19, the European Commission and the European GSMA made an unprecedented Business to Government (B2G) agreement to share ODM data for 22 member states of the Union plus Norway (Vespe et al., 2021). Looking at ODM data from France and Italy, Iacus et al. (2020) found that mobility alone can explain up to 92% of the excess deaths in the initial phase of the pandemic, meaning that mobility can be used to nowcast excess deaths in pandemics by using epidemiological models until official statistics become available (see Figure 50).

**FIGURE 49.** Smoothed net migration flows for age 25-29, with population threshold = 1 000 000. The background map shows the net migration rate for age group 25-29.
**Source:** Figure 7 in Guo and Zhu (2014).

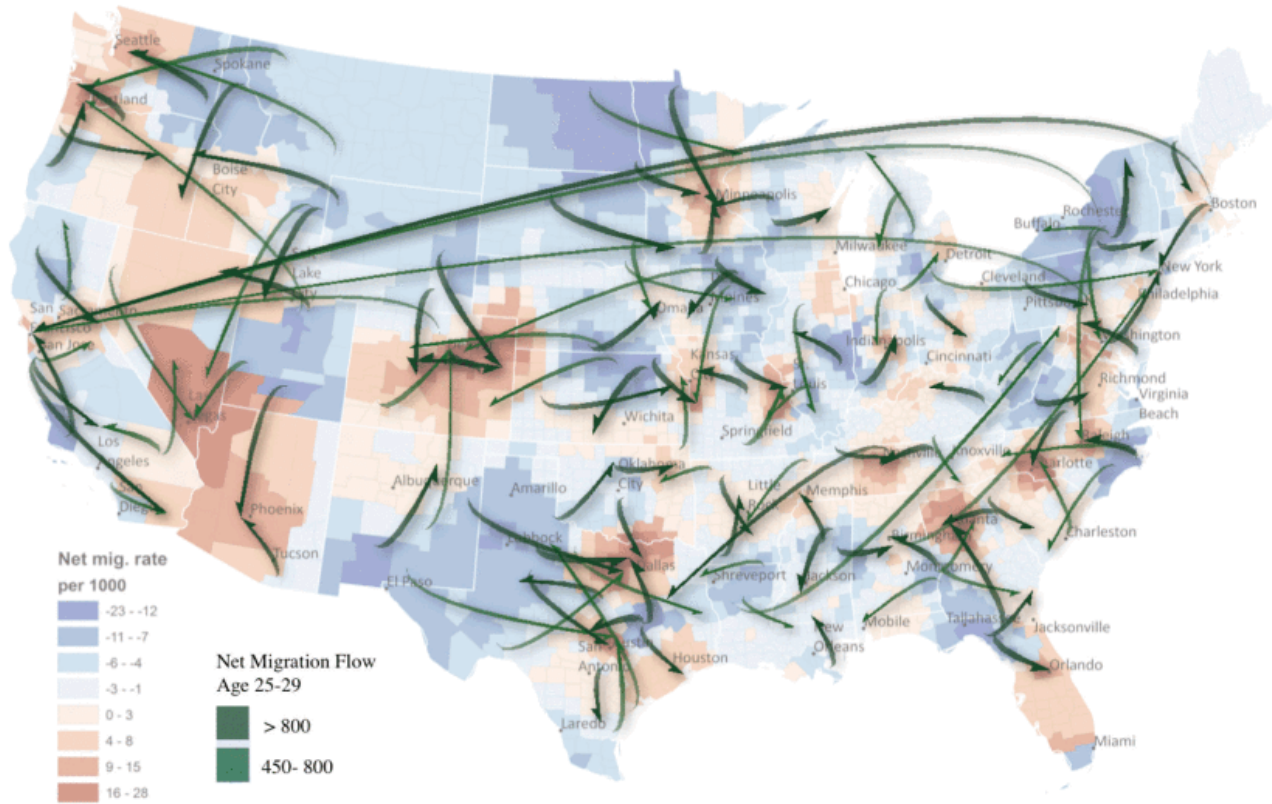

**FIGURE 50. Left**: the data for log-mobility and cumulative excess deaths on 16 March 2020. In red the selected subset of positive excess deaths and mobility. The other points are the residual data. Red and light blue together represent the whole data. **Right**: $R^2$ evolution for each mobility week using the model based on mobility alone on the selected data set of positive excess deaths and mobility index. The peak of correlation was on 16 March 2020 and the mobility week was between 15 and 21 March 2020. A similar pattern is shown for all weeks, explaining that mobility has no effect before the outbreak of COVID-19 and has a fast decay after the lockdown measures were put in place (left and right tails of the figure). On the contrary, during the initial emergency phase mobility is a key element of the spread of the virus and therefore in fatalities.
**Source:** Figures 10 and 11 in Iacus et al. (2020).

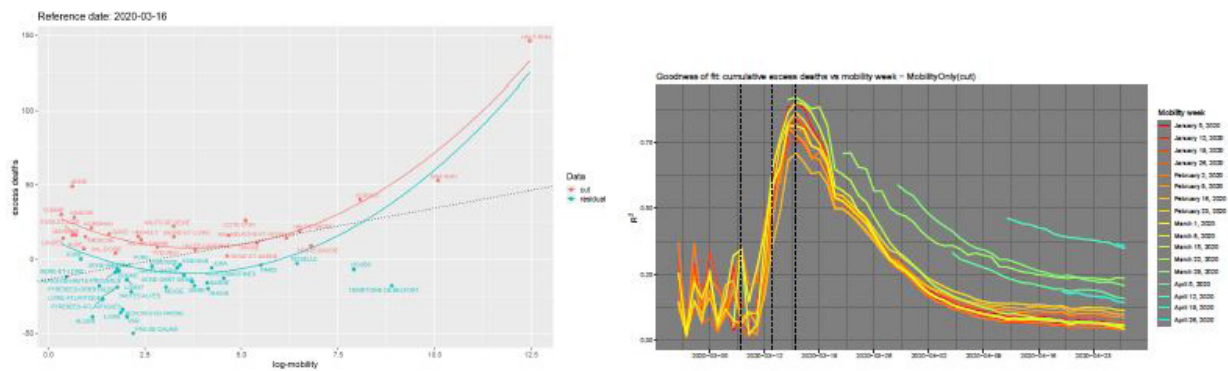

# 22. EXTENDED DETAIL RECORDS

One of the limits of CDRs is that the number of observable devices represents a small fraction of the whole population as they are purely based on SMS and call logs, only generated when people use these services (see Section 21). In other words, data from CDR derive from an active interaction between the users and their mobile devices. An extended version of these records is called XDR, which also includes mobile data records. XDR include all of the signals generated by the transmission of data packages by smartphones, generated by both idle and active devices. For example, when a user uploads or downloads data from the Internet using their phone's connection, they generate an XDR.

XDRs are not as frequent in the literature as CDRs mainly because the advent of the mobile data connection is relatively recent (Luca et al., 2020). They contain much more information than CDRs alone, but one of their major downsides is that they contain some noise than CDRs, making it harder to find signals of interest. Nevertheless, they become useful for applications that need more frequent location updates (every 15 to 30 minutes), such as real-time road traffic monitoring (Janecek et al., 2015). With regards to CDRs, the spatial granularity of XDRs is approximated on the basis of the location of the Radio Base Station used for the telecommunication activity.

## 22.1 HOME DETECTION

Pappalardo et al. (2020) analysed the accuracy of home detection algorithms based on ground truth observations, and discovered that the choice of XDR stream performed the best with an hour-of-day algorithm, compared to CDR and Control Plane Record (CPR) (a network type of stream which consists of purely machine-triggered events, as opposed to CDR which is human-triggered). Overall, their work underlines that CDRs, the stream most used in analysing human mobility using mobile phone data in the literature, could lead to low accuracy and stability.

Home detection is particularly important for measuring population changes in response to natural disasters. Isaacman et al. (2018) used home detection algorithms to identify changes in population during drought conditions in La Guajira (Colombia). They found that 90% of the climate migrants stayed in La Guajira relocating to other municipalities where access to help, food and water was probably easier, while the other 10% moved to other departments. Their results seem to indicate that only short-distance moves appear to be affected by climatic factors, which agrees with other studies (Henry et al., 2004).

## 22.2 URBAN POPULATION ESTIMATION

Chen et al. (2018) used a large dataset of mobile phone locations and a neural networkbased model to predict fine-grained urban population at a large spatiotemporal scale. The spatial resolution of the estimates is a 500x500 m grid, and the temporal resolution is 30 minutes. Their dataset was collected by a dominant cell phone operator that accounts for approximately 65% of the entire mobile phone market in Shanghai, covering 17 million users. Each recorded event (e.g. phone communication, regular update, periodic update, cellular handover, power on, and power off) was recorded along with the location of the nearest mobile phone tower and the time it occurred. Their results suggest that their method and the dataset is suitable for capturing the future distributions of a population in cities, and it could improve understanding of the complex interplay between the urban environment and human activities compared to traditional datasets (such as social life data, health data, business and commercial data, transportation and traffic data, scientific research data, etc.). Figure 51 shows the predicted and observed grid population values and their absolute differences over time for the central city of Shanghai, which is the historic and commercial centre of Shanghai.

## 22.3 TEMPORARY POPULATION AND GENDER GAPS IN URBAN SPACES

Jo et al. (2020) used temporary population to analyse gender gaps in the use of urban space in Seoul, Korea. They used the term ''temporary population'' to refer to a population that exists in a certain area during a specific period of time. In particular, they used daily temporary population released publicly by Geovision, a division of SK Telecom, which is Korea's largest telecommunication company. The data to estimate the daily population was collected from a pCell measuring 50x50 m, which is a spatial unit that aggregates mobile phone signals such as phone use, text messaging, and data communication, excluding Wi-Fi use. Figure 52 shows the estimated total daily population averaged using the month of September in Seoul

**FIGURE 51.** Prediction results of population by grid in the central city of Shanghai.
**Source:** Figure 5 from Chen et al. (2018).

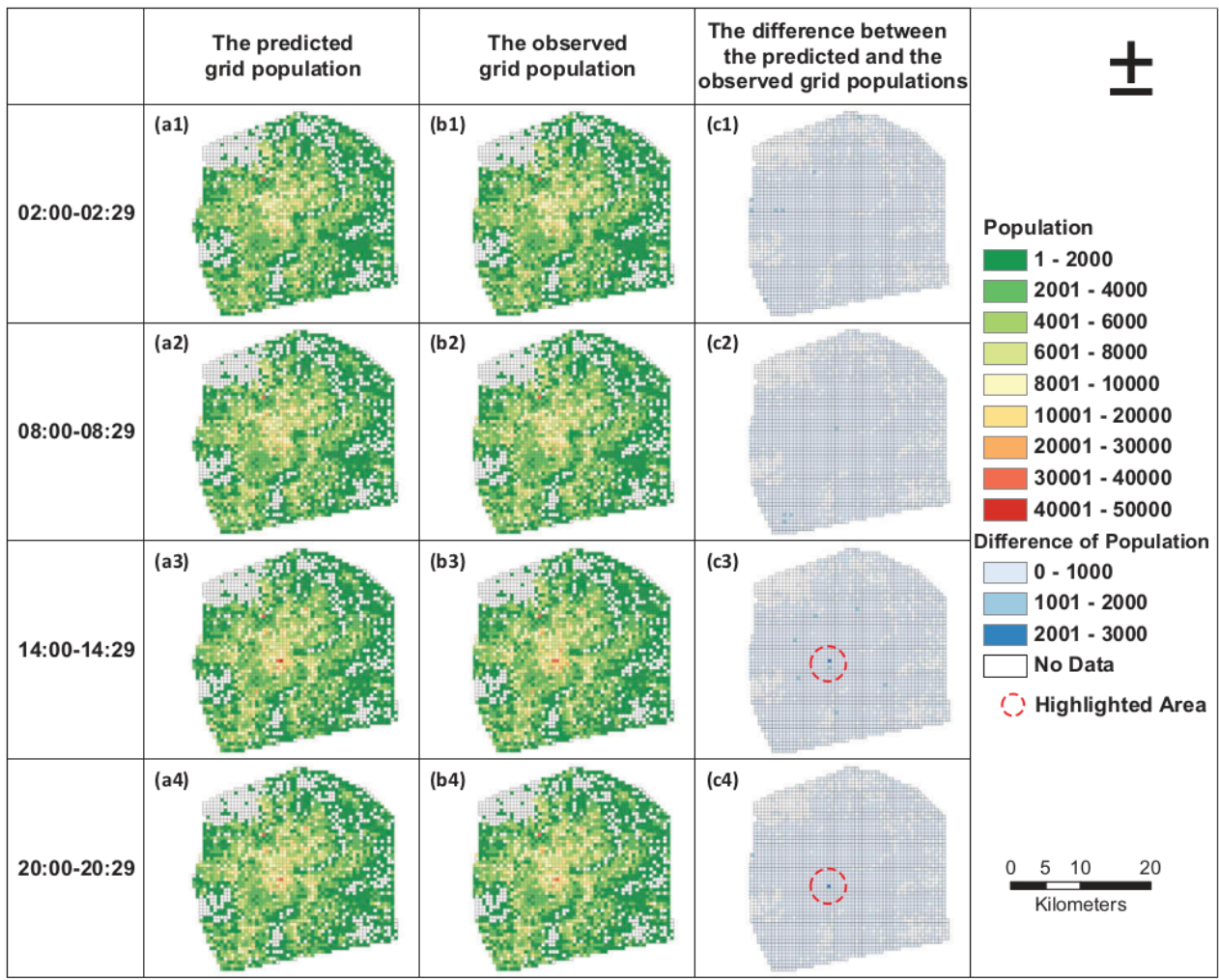

**FIGURE 52.** Estimated mean total daily population for September in the three main centres of Seoul.
**Source:** Figure 2 in Jo et al. (2020).

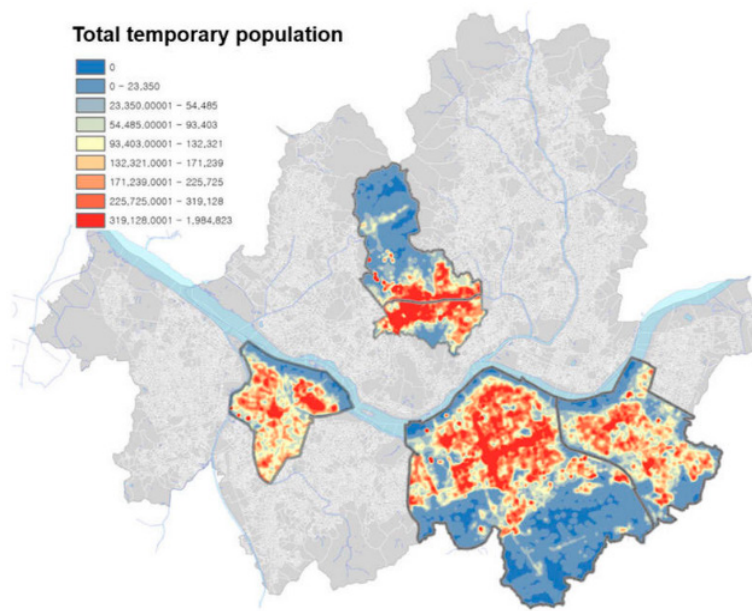

# 23. OTHER MOBILITY DATA

The COVID-19 pandemic has stimulated a lot of data for good initiatives in order to fight the SARS-CoV2 global crisis. Many private companies have released different types of mobility data to help the policy makers to study the impact of mobility on the spread of the virus and to assess compliance with mobility restriction measures. Among these initiatives are the following:

- *Apple Mobility Trends Reports*. This data is based on the use of Apple devices. It reports mobility changes relative to 13 January, 2020. It distinguishes between three types of mobility: transit, walking and driving. The geographical resolution is either country or regional level. Data is available at: https://covid19.apple.com/mobility. Several studies have exploited these data (see, e.g., Cot et al., 2021; Kurita et al., 2021; Snoeijer et al., 2021).

- *Google Community Mobility Reports*. This data provides insights which are essentially based on Google maps and reports movement trends over time by geography, across different categories of places such as retail and recreation, groceries and pharmacies, parks, transit stations, workplaces, and residential locations. The data shows these different types of mobility at city, region, or country level. The data is available at: https://www.google.com/covid19/mobility/. This data has been used in several studies (see, e.g., Yilmazkuday, 2021; Ilin et al., 2021; Kishore et al., 2021).

- *EnelX*. For a few countries in the world (for example, see the Italian case: https://www.enelx.com/it/it/istituzioni/servizi-citta-digitale/dashboard-covid-19) EnelX produces interactive maps of mobility at city, provincial, and regional level. In contrast to most of the other initiatives, this data also provides inflow and outflow data from a given location, i.e. ODM information. The movement statistics are obtained by analysing digital traces produced by vehicle data, navigation systems, maps, and mobile apps and then statistically reweighted to be representative of the reference population data. This data has been used in only a few studies (Borri et al., 2021) because of their specific geographical target.

- *Baidu mobility data* is derived from their location-based service (LBS) which is in turn based on information from the global positioning system (GPS), IP addresses, locations of signaling towers, Wi-Fi, online searching and mapping, and a large variety of apps and software on mobile devices. This data has been described and used in Hu et al. (2020) and Lai et al. (2020) especially to study the early spread of the SARS-CoV2 virus but not only for this purpose (Kraemer et al., 2019). Although this data can be scraped from the Baidu Migration website, a live collection of this data is also available at: https://dataverse.harvard.edu/dataset.xhtml?persistentId=doi:10.7910/DVN/FAEZIO.

# 24. SATELLITE IMAGERY

Many indices of population health and well-being show a strong correlation with variables describing the surrounding environmental, geographical, and socio-economic conditions (Bosco et al., 2017a). Earth observation data, including satellite images, are being increasingly used for monitoring population characteristics and are a clear example of a big data source which can be obtained for long time series at limited cost (Figure 53). Satellite images can be used to identify features of interest such as agricultural land, forests, urban areas, roads, and water bodies based on how they appear in the images.

The existence, strength, and nature of spatial relationships between observations can be identified and quantified using a number of measurements. Along with the spatial autocorrelation present, spatial interpolation techniques were developed to exploit this correlative relationship to predict the indicators of interest at locations where data from surveys are not available (Alegana et al., 2015; Sedda et al., 2015; Bosco et al., 2019). Spatial autocorrelation occurs when a variable is correlated with itself in space. This spatial correlation continues to exist after the effect of other variables has been considered (Hefley et al., 2017).

The availability of high-resolution geographical data is fundamental for this approach. This data can be used to describe conditions at survey locations and to make predictions across the area of interest where survey data is not available. For example, satellite images are normally used to create maps of land cover types, and land cover has been found to be a good proxy for estimating female literacy (Watmough et al., 2013). Another example is the study of Kinyoki et al. (2016) that found the Enhanced Vegetation Index (EVI) to be a significant common factor in describing rates of stunting in children. Furthermore, since early 2000 when nighttime light data became available, nighttime lights were found to be correlated with total population or population density (Lo, 2001; Sutton et al., 2003) and a few years later nighttime lights were also found to be a proxy for poverty (Noor et al., 2008; Elvidge et al., 2009).

Nevertheless, some studies show that the same type of geospatial data can have a different correlation with demographic aspects of a population in different countries and, occasionally, also within the same country (Bosco et al., 2017a, 2018). For example, in Bosco et al. (2017a), female literacy rates in Nigeria have a relatively high correlation (0.53) with urbanisation (normally derived from satellite images) whereas the female literacy rate in Kenya is poorly correlated with urbanisation (0.17).

**FIGURE 53.** Example of remote sensing derived layers showing the mean Gross Primary Productivity and mean Normalised Difference Vegetation Index in Nepal.
**Source:** Bosco et al. (2019).

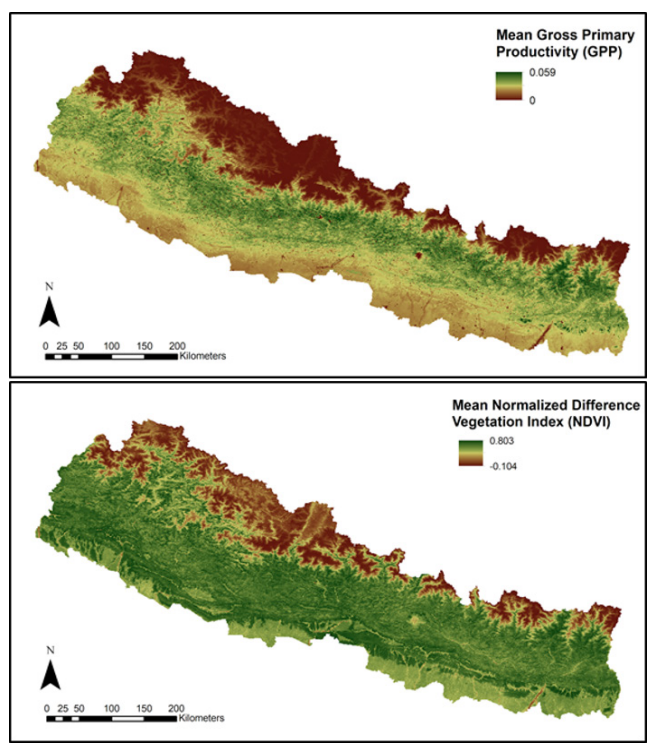

**FIGURE 54.** Map of the mean predicted proportion of male stunting at 1 km² resolution in Nigeria (the map in the NE of the country presents an higher uncertainty rate).
**Source:** Bosco et al. (2017).

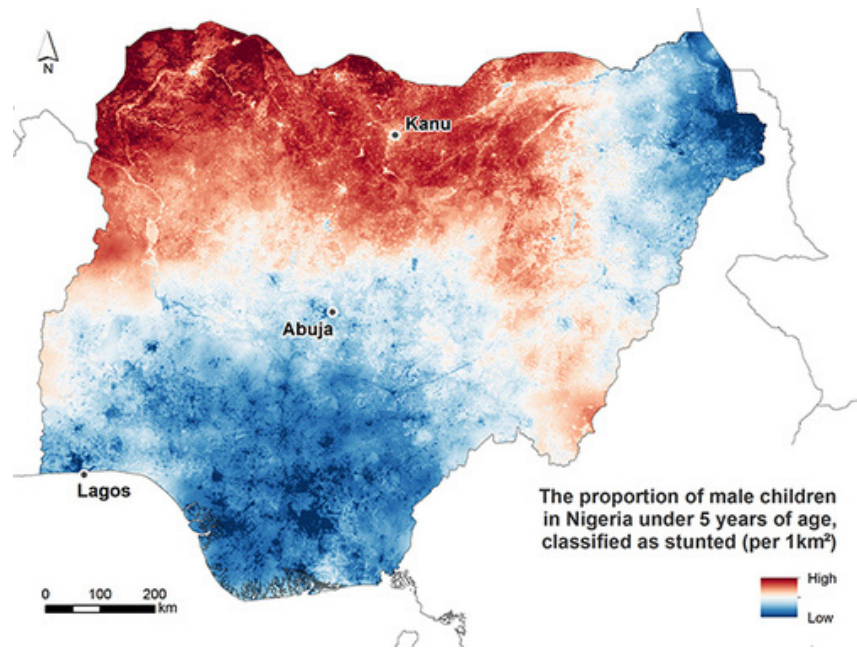

There are many reasons for this and further studies are necessary to understand these differences better. All the same, the value added by using remote sensing data in demography was widely recognised by both the scientific community and policy makers.

In recent years, spatial interpolation approaches exploiting spatial relationships between cluster-located survey data and geospatial covariates were tested to estimate many different demographic characteristics. Among other things, these techniques were largely used to map population count and age structures (Alegana et al., 2015), poverty (Tatem et al., 2014; Steele et al., 2017), child mortality (Carioli et al., 2017), stunting (Figure 54) and literacy (Figure 31) (Bosco et al., 2017a, 2017b, 2018, 2019). Satellite imagery, jointly with de-identified Facebook connectivity data and other nontraditional data sources, were also used to estimate the relative wealth index[117] that is an estimate of relative wealth of the people living in different areas within the same country.
In the next sections the use of satellite images for estimating the main demographic aspects of a population is explored more in detail.

## 24.1 POPULATION MAPPING AT HIGH SPATIAL RESOLUTION

Population counts at local levels are fundamental for many different reasons including the delivery and planning of services, responses to natural disasters or epidemics, and preparation of elections. The absence of updated and accurate national population and housing census data in many developing countries jointly with growing requests for spatially disaggregated population data led to an increase in exploring different data sources in efforts to produce spatially disaggregated population estimates at different time periods and geographical scales.

Census data is typically available aggregated at the level of large administrative areas (typically districts or subdistricts). Dasymetric mapping is a well-known cartographic approach to disaggregate large areal units to more granular representations (Eicher and Brewer, 2001; Mennis, 2003, 2009). This technique is based on the use of ancillary data (also called "covariates") to redistribute data at finer scales by exploiting a relationship between population counts/density and the available ancillary data (Mennis and Hultgren, 2006). This approach to disaggregate population is normally called top-down population mapping. From the mid-90s (Balk et al., 2006) researchers have used Geographic Information System (GIS) and satellite data for the dasymetric mapping of population, producing high spatial-resolution maps of population distributions.
One of the first examples of top-down census disaggregation includes the Gridded Population of the World (GPW) V1, a proportional allocation of population counts to grid cells within each administrative area, and then a smoothing across boundaries (GPW V1b) (Balk et al., 2006; Tobler et al., 1997) or allocation of population counts based on covariates that correlate with population density (e.g., LandScan 1998) (Balk et al., 2006; Bhaduri et al., 2002; Dobson et al., 2000).

---

117  https://dataforgood.facebook.com/dfg/tools/relative-wealth-index

**FIGURE 55.** Disaggregated census data for population mapping using random forests with remotely sensed and ancillary data.
**Source:** Stevens et al. (2015).

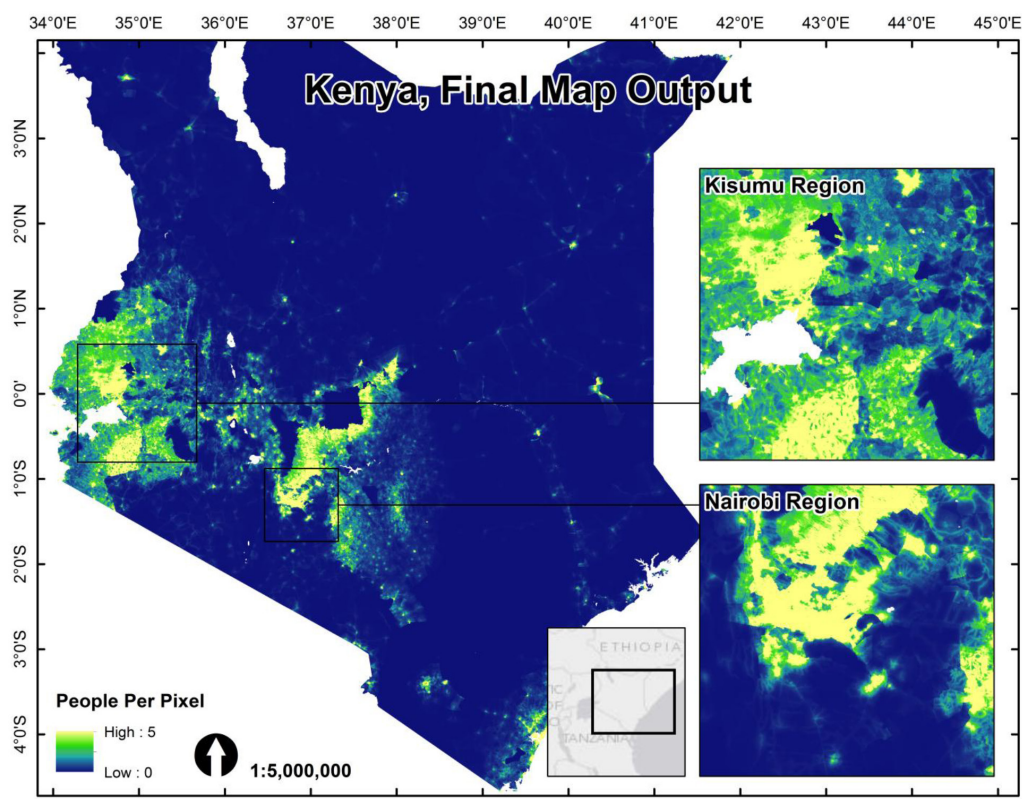

In the last twenty years, increased computational power, advances in methodological approaches and increased data availability have enabled the production of gridded population maps at finer spatial resolutions and global coverage (Freire et al., 2016; Stevens et al., 2015; Sorichetta et al., 2015) (Figure 55). Another recent advance includes the use of very high-resolution satellite imagery and the introduction of machine-learning techniques to identify human-built structures used to allocate population to these potentially settled sites (Gros and Tiecke, 2016; Corbane et al., 2021).

The lack of national population census data in many developing countries and the growing request for spatially disaggregated population estimates mean that other sources of data are increasingly being explored to produce population maps at different geographical scales and for different time periods. Here the concept of a bottom-up approach to population estimation is introduced (Wardrop et al., 2018). This modelling approach shares the common objective with top-down approaches of producing population estimates for small areas or uniform, high spatial-resolution grids (Wardrop et al., 2018).

Bottom-up approaches for population mapping is a significant area of active research. In Nigeria, settlement mapping obtained using very high-resolution satellite images, jointly with small-area microcensus surveys, geolocated national household surveys, and a range of different geospatial layers, were used to estimate population characteristics (counts, age and sex) at 90 m spatial resolution (Weber et al., 2018). The results of this study were used to improve the efficiency and effectiveness of vaccination planning and were also adopted for humanitarian needs assessments (Lee, 2013). Bottom-up population mapping approaches were also used in Afghanistan to derive population estimates across all parts of Afghanistan. The purpose of this work was to update existing estimates based on projections using the 1979 census and 2003 ‒‒ 2005 household listing (Wardrop et al., 2018; Office of Chief of Staff for the President, 2017). With the use of geospatial statistical techniques for quantifying the relationships between microcensus-derived population data and a number of spatial covariates (also including detailed settlement information derived from satellite imagery), it has been possible to predict population in areas where recent enumeration has not been possible.

Recently, Meta's HRSL[118] have been developed by Meta and the Center for International Earth Science Information Network (CIESIN) at Columbia University. By combining satellite images to identify buildings and census data, Meta and Columbia University have estimated populations counts at 30-meter resolution for 160+ countries around the world. Fries et al. (2021) compared the accuracy of population density estimates from various sources in the presence of a gold-standard dataset in Bioko island, Equatorial Guinea, and concluded that Meta's HRSL is very similar to the gold-standard. Smith et al. (2019) used Meta's HRSL population data to estimate flood risk in 18 developing countries spread across Africa, Asia, and Latin America. World Vision is using three high resolution population data-sets, one of which is Meta's HRSL, to bring clean water to the most vulnerable in Rwanda and Zambia (World Vision Water Team, 2021). A Malawian social enterprise ''BASEflow'' has been using Facebook HRSL population data to determine the location and usage of water points in rural parts of Malawi (Baseflow, 2021).

---

118   https://dataforgood.facebook.com/dfg/tools/high-resolution-population-density-maps

## 24.2 ESTIMATING MORTALITY RATES

Recent studies have shown the potential of using satellite imagery (often in combination with geospatial covariate layers from different sources) to assess mortality rates. Evans et al. (2013) carried out a study to estimate the mortality rate attributable to air pollution (premature mortality) at global level. They demonstrated the feasibility of using pollution concentrations derived from remote sensing to assess the impact of air pollution at global scale. As a result of this approach, global estimates of mortality attributable to Particulate Matter (PM)(2.5) are greater than those based on ground-level measures of PM(2.5) in urban areas.

Another recent study (Carioli et al., 2017) exploited geospatial covariates derived from satellite data for testing the possibility of mapping child mortality rates in Nigeria, Nepal and Bangladesh. They applied a Bayesian hierarchical spatial model (Press, 2002) (applying the Integrated Nested Laplace Approximations (INLA) approach (Rue et al., 2009)), combined with a stochastic partial differential equations (SPDE) approach (Lindgren et al., 2011), to produce continuous maps of mortality in children under five years of age. The results of this study show how satellite images, jointly with modern geostatistical techniques and survey data, have the potential to estimate the child mortality rate in developing countries. The availability of updated and spatially detailed data-sets that accurately depict the distribution of child mortality is especially important for family planning and decision-making purposes.

Furthermore, satellite images were recently used to estimate excess mortality during the COVID-19 epidemic period in Yemen (Besson et al., 2020). Activity across all identifiable cemeteries within the Aden governorate (population approximately 1 million) were identified by analysing very high-resolution satellite imagery and then compared with estimates from the Civil Registry office records. Satellite imagery burial analysis appears a promising approach for monitoring effects of epidemics and the impact of other crises, especially where real data is difficult to collect.

## 24.3 MAPPING FEMALE FERTILITY

Satellite imagery are also at the basis of some studies linked with estimation of the fertility rate. In their work in Nepal, Nigeria, and Bangladesh, jointly with child mortality, Carioli et al. (2017) analysed the possibility of estimating the Total Fertiltiy Rate (TFR). Results from this study show the potential of using satellite images to estimate fertility rates in developing countries. For example, in Nigeria satellite derived data such as night lights, gross cell production, elevation, and land cover were used to map TFR across the country. The results in two out of the three countries used as a case study were promising, though this work is still a preliminary study to test the strength and limits of using spatial interpolation approaches in this field.

Predicting TFR is important to understand where to concentrate maternal care, vaccination, or family planning campaigns and, as is often the case in developing countries, where there is an unmet need for contraception (Bongaarts, 2002).

Satellite derived maps of settlements and land cover were also used by Tatem et al. (2014) to redistribute areal census counts to produce detailed maps of the distributions of women of childbearing age, then converted, by using additional data, into datasets of gridded estimates of births and pregnancies. These detailed maps of settlement extents (with a spatial resolution of 30 metres) were derived from Landsat satellite imagery through semi-automated classification approaches (Tatem et al., 2004, 2005) or analyses based on expert judgement.

These estimates, based on standard demographic projections, can be produced for current, past, or future years, can provide the basis for strategic planning services, and can be used to track progress (Tatem et al., 2014).

## 24.4 INITIAL ATTEMPTS TO PREDICT HUMAN MIGRATION

In many different studies population size and GDP variables are used in models to estimate migration population size (White and Lindstrom, 2005; Brown and Bean, 2005; Karemera et al., 2000). Traditional demographic data-gathering methods (censuses, surveys, or official statistical reports) form the basis of the collection of these classic predictors. Accurate population or GDP statistics are often impossible to be collected in small areas using traditional data-gathering methods, it makes difficult to estimate migrations occurring in small areas, such as counties or towns.

**FIGURE 56.** The maps show the difference between estimated arriving and departing migrants in Bangladesh. **Source:** Davis et al. (2018).

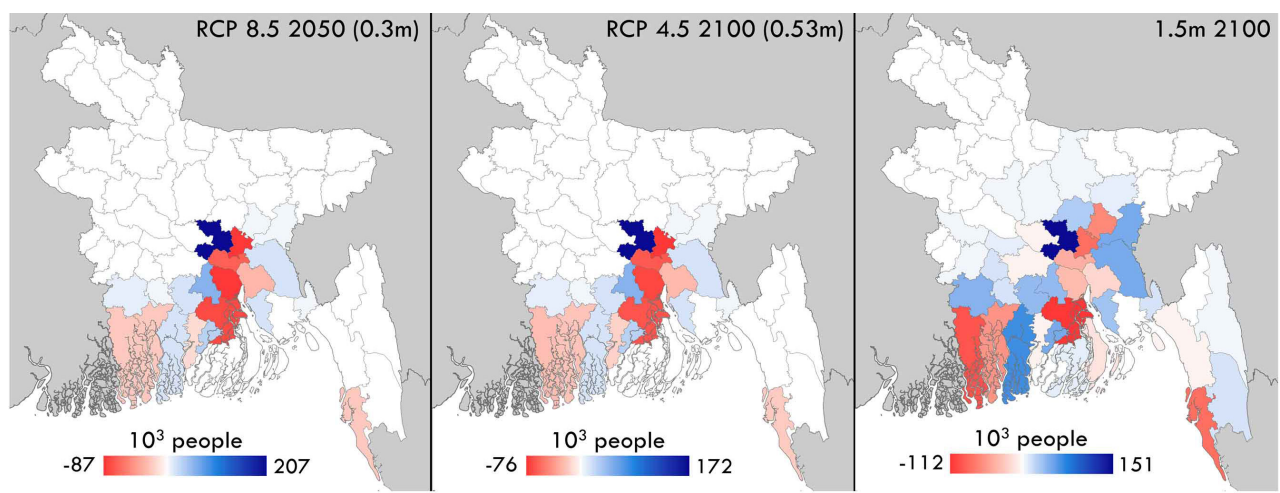

In a recent study by Chen (2020) focusing on EU countries, Visible Infrared Imaging Radiometer Suite (VIIRS) nighttime lights were found to be useful for estimating migrant population. The most extensive application of nighttime lights appears in the field of economics, where rigorous statistical models have been developed during the last decade using lights to estimate national and regional GDP and income (Nordhaus and Chen, 2015; Ghosh et al., 2010), and for population counts (Elvidge et al., 2009; Sutton et al., 2001). As previously mentioned, population density and general economic measurements of destinations (mainly derived from GDP) have been used as predictors of migration in many empirical studies on migration (Chen, 2020).
The analyses performed by the author used ordinary least squares regression and spatial lag models. Compared to population size and GDP, nighttime lights showed to be better predictors of migration in small regions in many European countries. Large variations in how these variables can be used to estimate migration flows exist across countries. This is probably due to the characteristics of migration. It is such a dynamic phenomenon that no set of common variables can provide equal explanation for all countries (Chen, 2020).

Night-time satellite data, integrated with population data derived from the JRC's Global Human Settlement Layer (GHSL), were used by Corbane et al. (2016) to assess the humanitarian impact of the Syrian conflict. A total of 11.9 million affected people was obtained using this data. These estimates overlap with the estimated displaced persons reported by the Syrian Observatory for Human Rights, UN-OCHA and Worldvision (Corbane et al., 2016). GHSL data were also used in the CLICIM project (McMahon et al., 2021) to estimate net migration at high spatial resolution. Population data from GHSL were combined with demographic data from UN DESA to estimate five-year net migration from 1975 to 2015 at a spatial resolution of about 25 km and with a global coverage (Alessandrini et al., 2020). The analysis adopted a residual indirect method to provide estimates of migration component as in De Sherbinin et al. (2011, 2012).

Satellite data can also help in estimating migration related to climate (Sîrbu et al., 2021). For example, Davis et al. (2018) using only maps of population and elevation for Bangladesh (the elevation was taken from the Shuttle Radar Topography Mission (SRTM)), estimated that inundations could displace 0.9 million people (by year 2050) to 2.1 million people (by year 2100) depending on the sea level rise scenario, with the large majority of this movements occurring within the southern half of the country (Figure 56). However, despite a number of existing estimates of the 'numbers of climate migrants', migration is a multi-causal phenomenon and it is always problematic to assign a proportion of migrants as moving as a direct result of environmental changes (Black et al., 2011).

# 25. THE CROWDSOURCING APPROACH

During mass migration events like the recent Syrian crisis,[119] information is an essential commodity. Access to on-line information about pathways, refugee camps and other social information (e.g. refugee activities and networking) is crucial to migrants. Access to data and generated content about refugees is another potential source in understanding migration routes and migration determinants according to Mahabir et al. (2018) and Curry et al. (2019). This approach has mainly been developed in disaster and risk management studies (see, e.g., Schimak et al., 2015) and it still remains in its infancy due to the burden of collecting this data, often qualitative, in a systematic way. In their review paper, Curry et al. (2019) show the value of crowd-generated data (especially open data and volunteered geographic information) to study migration events. Using a set of specific case studies, they show how migration-related information can be dynamically mined and analysed to study the migrant pathways from their home countries to their destination sites, as well as the conditions and activities that evolve during the migration process. Of course this approach, which exploits a combination of the data

**FIGURE 57.** Migrant deaths documented and crowdsourced by journalists and international relief organizations since 2000, based on data from: Files (2014) and IOM (2016).
**Source:** Figure 5 in Curry et al. (2019).

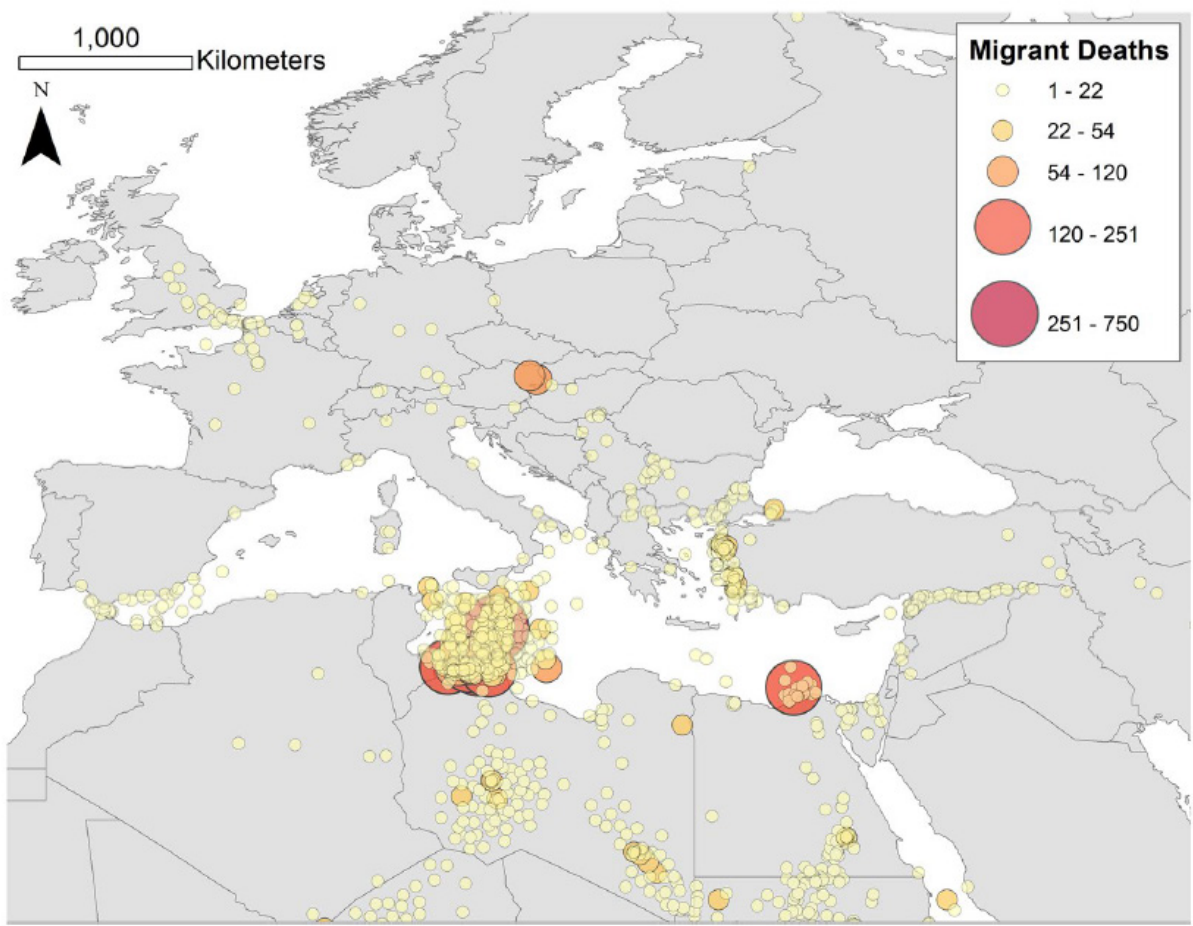

---

sources presented in the previous sections with qualitative data, also raises additional methodological and technical challenges. Among these data sources, the authors list the OpenStreetMap (OSM) collection of data about refugee camps or crowdsourced data, mainly collected through social media (Twitter, Flickr, Instagram, etc) about living conditions, on how to engage in activities or things to avoid during the trip, as well as blog posts dedicated to refugees conditions and reports from NGOs, war reporters, etc in which geographical information is disclosed. Figure 57 is an example of how sparse crowdsourced data can fit a quantitative representation that can possibly be used for further modelling.

# 26. CHALLENGES AND LIMITATIONS

Some of the challenges and limitations in the usage of innovative data sources in the context of demography, migration, and human mobility that have emerged in this review are summarised in this section. While Figure 58 graphically summarises these highlights by data source, here the common shortfalls and limitations are reported.

- *Data sharing* is of paramount importance in data innovation. Current data sharing practices vary considerably across data providers and even across products of the same data provider. In some cases, non-traditional data can support policymaking in a highly cost-effective manner: examples are publicly available and free of charge Facebook insights and Twitter APIs. In other cases, access to data is very limited due to a series of regulatory and economic barriers (e.g. Google location history data, MNO data), or even prohibited (e.g. Instagram). Data access should therefore be regulated to some extent and not limited to *ad hoc* philanthropic initiatives only.

- *Representativeness*. For some data sources (e.g., MNO data, social media data, etc) it is difficult to know the extent of the demographic representativeness of these data sources either because the data owner does not disclose this information for privacy or commercial reasons, or inevitably because a part of the population does not use the tools or devices that generate them. However, in some cases it has been shown that, for example, *selection bias* can be taken into account by using some instrumental variables and using reweighting techniques (see, e.g., Zagheni et al., 2017; Spyratos et al.,

**FIGURE 58.** The potential of non-traditional data in the fields of demography and migration: main strengths and challenges for different non-traditional data sources. The picture is inspired by IOM's GMDAC 2017 report available at https://www.migrationdataportal.org/themes/big-data-migration-and-human-mobility.

2019; Iacus et al., 2020). In other cases, as in satellite imaging data applications, a *geographical bias* exists because, for example, on field demographic data is not available or, in the case of MNO data, because cities are better covered than rural areas. In other cases it is a matter of *technological bias*, i.e., when the switch to the 5G network has been completed, historical MNO data based on previous technologies will not be comparable.

- *Transparency*. For reasons related to proprietary algorithms and for industrial reasons, the way the data is collected and processed before being distributed is not made known to the researchers who are going to use the data, and this may lead to invalid inferences.

- *Time inhomogeneity*. It might happen that some definitions or even the way the data is generated or collected at the source change due to technological advances (e.g. the new 5G network), or for commercial reasons (e.g. the target audience of an advertising platform changes according to market evolution or segmentation) or because users may change behaviour or platform through time (see, e.g., Salganik, 2017; Ricciato et al., 2020). This may lead to abrupt changes in the statistics which are difficult to adjust.

- *Heterogeneity*. Data heterogeneity arises when collecting the same type of data from different providers (e.g. different MNOs operating in the same country or in different countries).

- *Comparability*. As shown in Figure 1, the same target can be observed or measured using different data sources, but to date, no systematic studies have been carried out on the comparability of the outcomes. Differences in the results may arise because of different forms of bias, different definitions (e.g. migrants), and different analytical tools being used in the analysis.

- *Privacy*. Although there are privacy-preserving methods to achieve the goal of producing aggregated statistics or even individual records that make re-identification of individuals virtually impossible (see, e.g., Ram Mohan Rao et al., 2018; Dwork, 2008; De Montjoye et al., 2018), there are also limits imposed by the legislation in each country.

- *Ethics*. Data analysis of web scraped or API accessed individual data made explicitly public by users (e.g., via social media) may be potentially harmful in many ways[120] (Freelon, 2018). De Montjoye et al. (2013, 2018) show that four spatio-temporal points are enough to uniquely identify at least 95% of the individuals in their samples sharing personal mobile phone locations. Furthermore, selection bias in some data (namely social media) may lead to specific groups of users being tracked and then migration policies being steered in directions that unwillingly perpetuate discrimination or neglect the needs of invisible groups (Sîrbu et al., 2021). This is quite concerning, particularly if the target is minority groups which are more easily identified. We remind here that the GDPR requires that all data processing is lawful, fair and transparent and this has to be demonstrated by the analyst. Data revealing racial or ethnic origin, political opinions, religious or philosophical beliefs, or trade union membership data, must be processed in agreement with Article 9 of the GDPR. This and other types of bias in input data will become quite critical with the advent of artificial intelligence that will increasingly make use of innovative data sources. In general, ethics should not be considered ex-post with data at hand, but should be part of the design starting from the very initial data collection step (for example, see FAIR, 2021). Indeed, when processing personal data there is a requirement to carry out data protection impact assessment (Article 35 GDPR) where it is likely to result in a high risk to the rights and freedoms of persons. More specifically, Article 25 of the GDPR prescribes also that the controller should implement appropriate technical and organizational measures in order to integrate safeguards into the processing from the beginning.

- *Data linkage*. As emerged in the COVID-19 pandemic, the need to cross different data sources (e.g. individual level mobility and epidemiological data) immediately raises the problem of licensing, privacy, and data protection. New frameworks and platforms that enable trustworthiness to all stakeholders (citizens, private sector, government, researchers, etc) are needed. One possible example of datatrust services is the concept of Trusted Research Environment (TRE) (Boniface et al., 2021).

- *Technological development*. Due to lack of digital skills or limited accessibility to technology in a given country or region, not all data sources are available at the same granularity in terms of geographical and demographic dimensions.

- *Technical infrastructure*. Big data is often stored in high performance and scalable infrastructure at the source with a logic that is purely designed for administrative purposes. Extracting relatively complex statistics (like creating ODM from CDR records) has costs at the origin that cannot be ignored. Furthermore, not all of the receivers of the data are necessarily capable of receiving and handling large volumes of data. Investment should be considered.

- *Data literacy*. Even when data is released by the private sector for research or policy purposes, the analysis and quality assessment of this data is not an easy task and requires special skills in the fields of computer science, statistics, and computational social science.

---

120 This means that the data isn't truly anonymized and that it constitutes personal data bringing it, within an EU context, within the scope of the GDPR.

# 27. THE EUROPEAN UNION DATA POLICY

Harnessing the potential of data-driven innovation for social and economic development, while guaranteeing individual's fundamental rights has been high on the European Commission's agenda for almost a decade now.

A series of important policy initiatives have already been taken under the Juncker Commission (2014-2019) to improve the framework conditions for the EU's data-agile economy. The GDPR (2016/679) was certainly one of the most relevant legislative initiatives which has even set a global benchmark in terms of data privacy and security law (EU 2016/679, 2016). The GDPR has strengthened the individuals' fundamental rights in the digital age, consequently laying a framework for digital trust. Moreover, in 2017, for the first time the Commission raised the issue of barriers limiting the access and re-use of private sector data in the Communication "*Building a Europe data Economy*" (COM(2017) 9 final). A stakeholder dialogue, however, concluded that grounds for a horizontal legislative intervention on this issue are lacking and, in contrast, considered it more appropriate to provide a set of guidelines for data sharing in business-to-business (B2B) and business-to-government (B2G) situations. Following this the Communication "*Towards a common European data space*" (COM(2018) 232 final) and the accompanying Staff Working Document "*Guidance on sharing private sector data in the European data economy*" (European Commission, final) defined key principles in support of Business to Business (B2B) and B2G data-sharing initiatives. These key principles include:

- transparency in contractual agreements as to who has access to data, to what type of data, and for what purpose;

- recognition of shared value creation in contractual agreements;

- respect of commercial interests of both data holders and data users;

- guaranteeing undistorted competition when exchanging commercially sensitive data and;

- minimising data-lock in by enabling data portability.

Under the current von der Leyen Commission (2019-2024), building "*A Europe fit for the digital age*" able to grasp the opportunities of the digital revolution has become one of six policy priorities, with data at the core of this process. The EU's Digital Strategy is developed around three main aims, as identified by the Commission in the Communication "*Shaping Europe's Digital Future*" (COM(2020) 67 final):

- technology that works for the people;

- a fair and competitive digital economy;

- an open, democratic and sustainable society.

Specifically, under the pillar of "*A fair and competitive digital economy*", the Commission sets out a series of actions for the purpose of strengthening the EU's position as global leader in the new data-agile economy, including by adopting a European Data Strategy.

## 27.1 A EUROPEAN STRATEGY FOR DATA

The over-arching vision of the European Strategy for Data (COM(2020) 66 final) is that of a European society empowered by data to make better decisions which can support Europe's economic growth and global competitiveness. To realise that vision, the European Strategy for Data sets the ground for creating a single European market for data where data can flow across sectors and Member States, the European rules and values are fully respected, and the rules for access and re-use of data are fair, practical and clear with trustworthy data governance mechanisms in place. The Strategy acknowledges that there are currently numerous regulatory and economic barriers limiting the innovative re-use of data and the development of artificial intelligence, which is fueled by data. For example, these issues include market imbalances affecting the access to and use of data, interoperability issues that hinder the possibility of combining data from different sources, lack of data governance structures, and technological dependency on data infrastructures, etc.

In order to address these challenges and realise the vision of a society empowered by data, by creating a single European market for data, the Strategy for Data is developed on the basis of four pillars/streams of action. The first pillar brings together policy initiatives that aim to develop "*A cross-sectoral governance framework for data access and use*". Specifically, this pillar brings together actions that horizontally aim to enable and stimulate data sharing within and across sectors while ensuring a level playing field for all actors involved and avoiding the fragmentation of the internal market. The second pillar concerns ''Enablers'' and focuses on actions in the area of investments, interoperability, standardisation, and next-generation infrastructure as prerequisites for making rapid progress on data-driven innovation and ensuring Europe's technological sovereignty. The third pillar focuses on "Competencies" and comprises initiatives for investment in general data literacy, addressing lack of skilled labour, up/reskilling of the European workforce as well as dedicated capacity building for SMEs. In addition to horizontal policy initiatives, the Strategy's fourth pillar proposes the rollout of common European data spaces in crucial economic sectors supporting their implementation through the Digital Europe Programme. The Strategy identifies the following domain-specific common European data spaces:

- A Common European industrial (manufacturing) data space;

- A Common European Green Deal data space;

- A Common European mobility data space;

- A Common European health data space;

- A Common European financial data space;

- A Common European energy data space;

- A Common European agriculture data space;

- Common European data spaces for public administration;

- A Common European skills data space.

Among these domain-specific data spaces, the development of the *Common European Health Data Space* is also relevant for advancing demographic research in areas of mortality and population ageing not only by promoting access and use of health data not only for supporting delivery of healthcare services (i.e. primary use of data) but also for research and evidence-based policymaking (i.e. secondary use of data). Likewise, other domain-specific data spaces, and especially both the Common European mobility data space and the Common European skills data space, represent an opportunity to advance the research and evidence-based policymaking in the area of migration management and migrant integration. The Strategy also states that the Commission could launch additional data spaces in other sectors, this being the case of the recently announced data space on cultural heritage.

## 27.2 THE DATA GOVERNANCE ACT

The Proposal for a Regulation on European Data Governance, the so-called *Data Governance Act* (COM(2020) 66 final), presented on 25 November 2020, was the first set of measures adopted under the Strategy for data. To be specific, the Data Governance Act is proposed as an instrument to strengthen the re-use of public sector data that cannot be made available as public data. The Data Governance Act also aims to increase the trust in data-sharing by, among others things, introducing the category of data intermediaries. The Regulation will create the basis for a new European way of data governance that is in line with EU values and principles such as personal data protection (GDPR), consumer protection, and competition rules. Thanks to the Regulation, more data will be available and exchanged in the EU across sectors and Member States. It will boost data sharing and the development of common European data spaces, as announced in the European strategy for data.

## 27.3 THE DATA ACT

The European Strategy for Data announced a second major legislative initiative called the Data Act. It will complement the proposal for a Regulation on data governance in order to maximise the value of data for the economy and society by addressing issues in accessing and using data in business-to-business (B2B), business-to-consumer (B2C) and business-to-government (B2G) situations. In addition, the Data Act will propose measures to allow cloud users in the EU to switch more easily between different providers of data-processing services as well as a framework for efficient data interoperability. These are pre-conditions for efficient data sharing within and across sectors. The future Data Act will clarify as to who can create value from data. It aims to ensure fairness in the allocation of data value among the actors in the data economy while respecting the legitimate interests of companies and individuals that invest in the generation of data. It builds on and makes the personal data protection (GDPR) rules more effective.

# 28. CONCLUSIONS

This report aims to contribute to the discussion on the potential of harnessing data innovation in order to advance knowledge and enhance evidence-based policymaking in the areas of demography, migration, and human mobility. Consequently, the report provides examples of the current state of data innovation applications for measuring vital population events, population change, population structure, as well as migration, and human mobility, and related economic, and social aspects. The examples are drawn from a review of over 300 articles and scientific reports as well as numerous tools related to one of the following non-traditional data sources: social media (Flickr, Twitter, LinkedIn, Reddit, and the Meta family of apps such as Facebook and Instagram), Internet activity (Google trends, online news), mobile payment apps, mobile phone data (CDR, ODM, and XDR), air passenger data, and satellite imagery.[121] As emphasised in the introduction, this is a non-exhaustive state-of-the-art review and the authors also intend to expand this study in future editions to sub-topics of demography and migration currently not covered such as marriage and divorce, ageing and health, etc.

The report shows that the growing demand for updated and spatially detailed datasets that accurately depict the demographic and migration-related phenomena has led to an increase in the number of studies exploring the use of non-traditional data. Data innovation applications are still not part of the mainstream demographic, migration and mobility-related literature. Nevertheless, this review of current scientific efforts to explore non-traditional data highlights the latter's great potential in supporting social sciences and policymaking.

The current landscape of non-traditional data sources offers insights into a variety of demographic, human mobility, and migration-related topics (see Figure 1). However, the scope of applicability of each innovative data source largely varies in relation to the specific target market of each data provider. Therefore, some innovative data sources have a wider scope of application: for example, satellite images can contribute to measuring fertility and mortality, but also various migration- and mobility-related aspects. Other innovative data sources, especially those operating in a niche market, offer information limited to specific sub-topics. For example, LinkedIn as an employment-oriented online service platform which is particularly suitable for analysing the migration of highly skilled.

In general, compared to traditional data, the current competitive advantage of innovative data sources lies in their greater geographic and temporal granularity, (near-)real time availability, and their extensive coverage which facilitates more immediate international comparisons. According to the review presented in this paper, innovative data is most extensively used to *fill the gaps in traditional statistics*, with migration-related estimates at multiple temporal and spatial scales being the most frequent example. In some cases non-traditional data has also the potential to *overcome measurement errors* such as reconstruction errors and memory distortion related to survey data . This is the case for transfer-level microdata on remittances derived from the use of mobile money transfer services which overcomes the issues of misdating, incorrect reporting of the transferred money, etc.

At the same time, this review found that there is an increase in the literature adopting mixed methodologies, based on the *integration of traditional data with innovative ones*, to study demographic and migration phenomena. Among the literature surveyed, examples of integration of mobile phone data and/or satellite imagery with (micro)census-derived population data were found most frequently. However, data linkage still remains an important challenge as the application of these methodologies brings into question issues such as licensing, privacy and data protection.

Moreover, this report found that the definitions of migration and human mobility in studies based on innovative data do not always comply with conventional definitions adopted by official statistics. One of the main reasons for such fluidity in definitions is that private sector data is based on algorithms that are often proprietary and do not necessarily reflect conventional statistical standards.

In order to grasp the *potential of harnessing data innovation for agile and evidence-based policymaking* the extent to which the current data innovation literature contributes to the following four domains was analysed (Verhulst, 2021): 1) situational awareness, nowcasting and response; 2) knowledge creation through better cause and effect analysis; 3) prediction and forecasting; and 4) impact assessment, evaluation and experimentation.

The review identifies the greatest potential of data innovation lies in the domain of *situational awareness, nowcasting and response*. Indeed, it was observed that all innovative data sources analysed have the potential to provide (almost) real-time, accurate, and detailed information of demographic trends and/or public opinion at different geographical scales.

---

121    The authors have also reviewed the studies using car traffic data (TomTom and Waze), but found that its use is still rather limited in the demographic and migration literature. Therefore, a specific section on car traffic data was not included in this report.

Another domain with extensive employment of data innovation is that of *prediction and forecasting*. It was found that studies mainly rely on combining traditional and nontraditional data sources for the purposes of prediction and forecasting. EASO's early warning and forecasting system for the number or asylum seekers, which is based on the combination of event data, Google search and operational data, is just one of the examples.

The domain in which we observe the most limited use of non-traditional data is that of *knowledge creation through better cause and effect analysis*. The existing examples largely relate to use of event-level data to study the effect of "external shocks" such as the causal effect of conflicts on infant mortality or the effect of natural disasters on population displacement. Several explanations could be advanced on this point. The small number of cause–effect analyses using innovative data could suggest that the exploration, understanding, and use of non-traditional data sources are in their early stages. Another concurrent explanation could be that the application of non-traditional data sources has been dominated by non–social science disciplines with little tradition in the application of statistical inference approaches.

In addition, data innovation has enormous, although not yet fully exploited potential for *impact assessment, evaluation and experimentation*. Examples of this potential are numerous such as satellite images which jointly with modern geostatistical techniques and survey data have the potential to estimate mortality and fertility rates in developing countries. These estimates can serve as a basis for strategic family planning and health services, as well as for monitoring and evaluation. Likewise, mobile phone data can support impact assessment and evaluation of rural development programmes and urban planning by providing insights into population dynamics in specific areas. Last but not least, data from social media could provide significant inputs for monitoring and evaluation of integration measures and programmes.

Moreover, this report shows that not all innovative data sources have been explored to the same extent in the demographic and migration-related literature. Some types of non-traditional data (e.g. MNO data and satellite imagery) have been more extensively explored in the literature than others (e.g. social media or google search data). Therefore, it can be argued that only some types of non-traditional data is potentially mature enough to exit the exploratory stage.[122] Moreover, in parallel to the scientific advances in the use of non-traditional data, other important factors (e.g. ethics, privacy, data governance, data quality, bias, data access, etc.) influence the advent of an actual data innovation transition.

Indeed, the bottom line of harnessing the potential of innovative data sources is that of enabling a *transition from exploratory to regular use of non-traditional data for official statistics and policymaking*. However, a series of conditions have to exist in order for this transition to take place.

Legislation is undoubtedly a *sine qua non* condition of data innovation transition. Among other things, legislation must regulate access to data held by the private sector in a way that guarantees the fundamental rights of the individual. Legislation has to also mandate national statistical offices to collect, analyse, and publish data from non-traditional data sources. However, there is currently no universal stance toward the data innovation transition. In fact, the regulatory framework and experience with using non-traditional data for policymaking and official statistics is highly heterogeneous across the World. There are countries where the legislation does not support the use of innovative data sources, and in contrast there are countries where statistical offices are pressured into using non-traditional data due to resource constraints (UNESCAP, 2021b).

The EU's data policy can certainly be considered to be one of the most advanced frameworks for setting the ground for data-driven innovation. In fact, the EU through its data policy has been implementing a series of measures to horizontally support the consolidation of an ecosystem that enables more data sharing from both public and private sectors. The most widely known example of these measures is the GDPR which has set a global benchmark in terms of laying the framework for digital trust by strengthening the individuals' fundamental rights in the digital age.

Yet, a favourable regulatory framework for data innovation transition alone is not sufficient. In order to enable safe and trusted data sharing and statistics, investments must be made in developing operational models and secure technical systems too (European Commission, 2021). In this context, investments aimed at fostering collaborations between data owners and the private and public research sectors become equally important. Indeed, collaborations can help trigger improvements in the ethics of data handling and privacy preservation that could in turn feed into the development of principles of trusted and smart official statistics for decision making. As previously mentioned, selection bias in some data (namely social media) may lead to the tracking of specific groups of users and then policies being steered in directions that unwillingly perpetuate discriminations or neglect the needs of invisible groups (Sîrbu et al., 2021).

---

122  It should be stressed that, at the current stage, the authors believe that an actual ranking of the abovementioned innovative data sources for investment purposes is not feasible for several reasons. Firstly, the scope of applicability of each data source is different, reflecting the specific target market of each data provider. Secondly, digital technologies are advancing rapidly and may at any time yield new, more optimum, sources of innovative data. Finally, companies may choose to stop providing access to data at any time for business-related reasons.

In addition to the above, an investment needs to be made in data managing and analysis skills to empower the institutions of professionals (with a mix of knowledge in computer science, statistics, and social science) able to assess the quality of the data, develop methodologies and extract insights and meaningful statistics from innovative data sources.

To conclude, in order to enable the data innovation transition and harness its full potential for policymaking, the European Commission's KCMD will continue to monitor and report on the state of data innovation applications in the fields of demography, human mobility, and migration as well as foster collaborations between the policymakers, statistical offices, expert communities, and data owners.

# INDEX

# REFERENCES

[1]     Acosta, R. J., N. Kishore, R. A. Irizarry, and C. O. Buckee (2020). Quantifying the dynamics of migration after hurricane maria in puerto rico. *Proceedings of the National Academy of Sciences 117(51), 32772--32778*. https://www.pnas.org/content/117/51/32772.

[2]     Alegana, V., P. Atkinson, C. Pezzulo, S. A., D. Weiss, T. Bird, E. Erbach-Schoenberg, and A. Tatem (2015). Fine resolution mapping of population age-structures for health and development applications. *J. R. Soc. Interface 12(20150073),* 1--11.

[3]     Alessandrini, A., D. Ghio, and S. Migali (2020). *Estimating net migration at high spatial resolution.* Number KJ-NA-30261-EN-N (online),KJ-NA-30261-EN-C (print). Luxembourg (Luxembourg): Publications Office of the European Union.

[4]     Alessandrini, A., F. Natale, F. Sermi, and M. Vespe (2017). *High resolution map of migrants in the EU.* Number KJ-NA-28770-EN-C (print),KJ-NA-28770-EN-N (online). Luxembourg (Luxembourg): Publications Office of the European Union.

[5]     Alexander, M., K. Polimis, and E. Zagheni (2020, Aug). Combining social media and survey data to nowcast migrant stocks in the united states. *Population Research and Policy Review.*

[6]     Almaatouq, A., F. Prieto-Castrillo, and A. Pentland (2016). Mobile communication signatures of unemployment. *In Social Informatics.SocInfo 2016. Lecture Notes in Computer Science,* Volume 10046.

[7]     Althaus, S., J. Bajjalieh, J. Carter, B. Peyton, and D. Shalmon (2020). Cline center historical phoenix event data. cline center for advanced social research. v1.3.0. may 4.

[8]     Antoniou, V., J. Morley, and M. Haklay (2010). Web 2.0 geotagged photos: Assessing the spatial dimension of the phenomenon. *GEOMATICA 64(1),* 99--110.

[9]     Araujo, M., Y. Mejova, I. Weber, and F. Benevenuto (2017). Using facebook ads audiences for global lifestyle disease surveillance: Promises and limitations. *In Proceedings of the 2017 ACM on Web Science Conference,* WebSci '17, New York, NY, USA, pp. 253 257. Association for Computing Machinery.

[10]    Arcila-Calderón, C., D. Blanco-Herrero, M. Fr as-VÆzquez, and F. Seoane-PØrez (2021). Refugees welcome? online hate speech and sentiments in twitter in spain during the reception of the boat aquarius. *Sustainability 13(5).*

[11]    Aurambout, J., F. Batista E Silva, C. Bosco, A. Conte, D. Ghio, S. Kalantaryan, M. Kompil, C. Perpiæa Castillo, P. Proietti, M. Scipioni, P. Sulis, and G. Tintori (2021). *The Demographic Landscape of EU Territories.* Number KJ-NA-30498-EN-N (online),KJ-NA-30498-EN-C (print). Luxembourg (Luxembourg): Publications Office of the European Union.

[12]    Azar, E. E. (2009). Conflict and peace data bank (copdab), 1948-1978.

[13]    Azmandian, M., K. Singh, B. Gelsey, Y.-H. Chang, and R. Maheswaran (2013). Following human mobility using tweets. In L. Cao, Y. Zeng, A. L. Symeonidis, V. I. Gorodetsky, P. S. Yu, and M. P. Singh (Eds.), *Agents and Data Mining Interaction,* Berlin, Heidelberg, pp. 139--149. Springer Berlin Heidelberg.

[14]    Bailey, M., R. Cao, T. Kuchler, J. Stroebel, and A. Wong (2018, Summer). Social Connectedness: Measurement, Determinants, and Effects. *Journal of Economic Perspectives 32(3),* 259--280.

[15]    Bailey, M., P. Farrell, T. Kuchler, and J. Stroebel (2020). Social connectedness in urban areas. *Journal of Urban Economics 118,* 103264.

[16]    Bailey, M., D. Johnston, T. Kuchler, D. Russel, B. State, and J. Stroebel (2020). The determinants of social connectedness in europe. In S. Aref, K. Bontcheva, M. Braghieri, F. Dignum, F. Giannotti, F. Grisolia, and D. Pedreschi (Eds.), *Social Informatics,* Cham, pp. 1--14. Springer International Publishing.

[17] Balk, D. L., U. Deichmann, G. Yetman, F. Pozzi, S. I. Hay, and A. Nelson (2006). Determining global population distribution: methods, applications and data. *Advances in parasitology 62*, 119--156.

[18] Bank, W. (2006). Global economic prospects: Economic implications of remittances and migration. *World Bank, Washington DC*.

[19] Barchiesi, D., H. Moat, C. Alis, S. Bishop, and T. Preis (2015, 07). Quantifying international travel flows using flickr. *PloS one 10*, e0128470.

[20] Barslund, M. and M. Busse (2016). How mobile is tech talent? a case study of it professionals based on data from linkedin. *CEPS Special Report No 140*.

[21] Baseflow (2021, Jun). Cracking a tough nut: Using high resolution population estimates from facebook to improve water infrastructure planning in Malawi.

[22] Bayardo, R. and R. Agrawal (2005). Data privacy through optimal k-anonymization. In *21st International Conference on Data Engineering (ICDE'05)*, pp. 217--228.

[23] Beiró, M. G., A. Panisson, M. Tizzoni, and C. Cattuto (2016, Oct). Predicting human mobility through the assimilation of social media traces into mobility models. *EPJ Data Science 5(1), 30*.

[24] Besson, E. S. K., A. Norris, T. Ghouth, A. S. B. nad Freemantle, M. Alhaffar, Y. Vazquez, and F. Checchi (2020). Excess mortality during the covid-19 pandemic in Aden Governorate, Yemen: a geospatial and statistical analysis. *BMJ Global health*.

[25] Bhaduri, B., E. Bright, P. Coleman, and J. Dobson (2002). Landscan. *Geoinformatics 5(2), 34--37*.

[26] Bijak, J., J. J. Forster, and J. Hilton (2017). Quantitative assessment of asylum-related migration: a survey of methodology. Technical report, EASO: Publication Office of the European Union.

[27] Bijak, J., P. Higham, J. Hilton, M. Hinsch, S. Nurse, T. Prike, O. Reinhardt, P. Smith, A. Uhrmacher, and T. Warnke (2021). *Towards Bayesian Model-Based Demography: Agency, Complexity and Uncertainty in Migration Studies. Methodos Series*. Springer International Publishing.

[28] Billari, F., F. D'Amuri, and J. Marcucci (2018). Forecasting births using google. In *Proceedings of the PAA Annual Meeting*. https://paa2013.princeton.edu/papers/1 31393.

[29] Bircan, T., A. Wali, A. Yar, D. Purkayastha, and S. Yilmaz (2021, January). Addressing the international migration data gaps. https://hummingbird-h2020.eu/news/news-items /D2.3policy.

[30] Bisanzio, D., M. Kraemer, I. Bogoch, T. Brewer, J. Brownstein, and R. Reithinger (2020, Jun.). Use of twitter social media activity as a proxy for human mobility to predict the spatiotemporal spread of covid-19 at global scale. *Geospatial Health 15(1)*. https://geospatialhealth.net/index.php/gh/article/view/882.

[31] Black, R., N. Adger, N. Arnell, S. Dercon, A. Geddes, and D. Thomas (2011). Migration and global environmental change: future challenges and opportunities.

[32] Blumenstock, J. E., G. Cadamuro, and R. On (2015). Predicting poverty and wealth from mobile phone metadata. *Science 350(6264), 1073--1076*.

[33] Blumenstock, J. E., N. Eagle, and M. Fafchamps (2011). Charity and reciprocity in mobile phone-based giving: Evidence from rwanda.

[34] Blumenstock, J. E. (2012). Inferring patterns of internal migration from mobile phone call records: evidence from rwanda. *Information Technology for Development 18(2), 107--125*.

[35] Blumenstock, J. E. and N. Eagle (2012). Divided we call: disparities in access and use of mobile phones in Rwanda. *Information Technologies & International Development 8(2), pp--1*.

[36]   Bohon, S. A. (2018, Jun). Demography in the big data revolution: Changing the culture to forge new frontiers. *Population Research and Policy Review 37(3)*, 323--341. https://doi.org/10.1007/s11113-018-9464-6.

[37]   Bollen, J., H. Mao, and A. Pepe (2021, Aug.). Modeling public mood and emotion: Twitter sentiment and socio-economic phenomena. *Proceedings of the International AAAI Conference on Web and Social Media 5(1)*, 450--453. https://ojs.aaai.org/index.p hp/ICWSM/article/view/14171.

[38]   Bongaarts, J. (2002). The end of the fertility transition in the developed world. *Population and development review*, 419--443.

[39]   Boniface, M., L. Carmichael, W. Hall, B. Pickering, S. Stalla-Bourdillon, and S. Taylor (2021, June). The social data foundation model: facilitating health and social care transformation through datatrust services. *Author's Original.*

[40]   Bonnel, P., M. Fekih, and Z. Smoreda (2018). Origin-destination estimation using mobile network probe data. Transportation Research Procedia 32, 69--81. T*ransport Survey Methods in the era of big data:facing the challenges.*

[41]   Borri, N., F. Drago, C. Santantonio, and F. Sobbrio (2021). The "great lockdown": Inactive workers and mortality by covid-19. *Health Economics 30(10)*, 2367--2382.

[42]   Bosco, C., V. Alegana, T. Bird, C. Pezzulo, L. Bengtsson, A. Sorichetta, J. Steele, G. Hornby, C. Ruktanonchay, N. Ruktanonchai, E. Wetter, and A. J. Tatem (2017a). Exploring the high-resolution mapping of gender-disaggregated development indicators. *J.R.Soc.Interface 14.*

[43]   Bosco, C., V. Aleganaa, T. Bird, C. Pezzulo, G. Hornby, A. Sorichetta, J. Steele, C. Ruktanonchai, N. Ruktanonchai, E. Wetter, L. Bengtsson, and A.J. Tatem (2017b). Mapping indicators of female welfare at high spatial resolution. Technical report, WorldPop project, Flowminder Foundation, Stockholm, Sweden.

[44]   Bosco, C., D. de Rigo, A. Tatem, C. Pezzulo, R. Wood, H. Chamberlain, and T. Bird (2018). Geostatistical tools to map the interaction between development aid and indices of need. *AidData: Washington, DC, USA.*

[45]   Bosco, C. and G. Sander (2015). Estimating the effects of water-induced shallow landslides on soil erosion. *ArXiv preprint arXiv:1501.05739.*

[46]   Bosco, C., S. Watson, A. Game, C. Brooks, D. de Rigo, S. Qader, and Bengtsson (2019). Towards high-resolution sex-disaggregated dynamic mapping. *The Flowminder Foundation 1(1),* 1--85. https://data2x.org/resource-center/towards-high-reso lution-sex-disaggregated-dynamic-mapping/.

[47]   Boswell, C. and A. Geddes (2010). *Migration and Mobility in the European Union.* The European Union Series. Palgrave Macmillan.

[48]   Bounie, D., D. Diminescu, and A. Fran ois (2013). On the effect of mobile phone on migrant remittances: A closer look at international transfers. *Electronic Commerce Research and Applications 12(4),* 280--288. Social Commerce- Part 2.

[49]   Brown, S. and F. Bean (2005). International migration. In Springer (Ed.), *Handbook of Population,* pp. 347--382.

[50]   Bryan, G., S. Chowdhury, and A. M. Mobarak (2014). Underinvestment in a profitable technology: The case of seasonal migration in bangladesh. *Econometrica 82(5),* 1671-1748.

[51]   Bureau of Transportation Statistics (2021). Data bank 28dm - t-100 domestic market data (world area code). https://www.bts.gov/browse-statistical-productsand-data/bts-publications/data-bank-28dm-t-100-domestic-market-data. Accessed: 2021-11-15.

[52]   Böhme, M. H., A. Gröger, and T. Stöhr (2020). Searching for a better life: Predicting international migration with online search keywords. *Journal of Development Economics 142*, 102347. Special Issue on papers from "10th AFD-World Bank Development Conference held at CERDI, Clermont-Ferrand, on June 30 - July 1, 2017".

[53]   Calabrese, F., G. Di Lorenzo, L. Liu, and C. Ratti (2011). Estimating origin-destination flows using mobile phone location data. *IEEE Pervasive Computing 10(4),* 36--44.

[54]    Carammia, M., S. M. Iacus, and T. Wilkin (2022). Forecasting asylum-related migration flows with machine learning and data at scale. Nature Scientific Reports (SREP-21-01036)  https://www.nature.com/articles/s41598-022-05241-8

[55]    Carioli, A., C. Bosco, C. Pezzulo, and A. J. Tatem (2017). High resolution mapping of fertility and mortality from national household survey data in low income settings. In PAA 2017 *Annual Meeting, pp.* 1--18. Population Association of America.

[56]    Cesare, N., H. Lee, T. McCormick, E. Spiro, and E. Zagheni (2018, Oct). Promises and pitfalls of using digital traces for demographic research. *Demography 55(5)*, 1979--1999.

[57]    Chen, J., T. Pei, S.-L. Shaw, F. Lu, M. Li, S. Cheng, X. Liu, and H. Zhang (2018). Fine-grained prediction of urban population using mobile phone location data. *International Journal of Geographical Information Science 32(9)*, 1770--1786.

[58]    Chen, X. (2020). Nighttime lights and population migration: Revisiting classic demographic perspectives with an analysis of recent european data. *Remote Sensing 12(1)*.

[59]    Chi, G., F. Lin, G. Chi, and J. E. Blumenstock (2020, Oct). A general approach to detecting migration events in digital trace data. *PloS one 15(10)*, e0239408--e0239408. 33007015[pmid].

[60]    Choo, Y. Y. (2018). Immigration and inbound air travel demand in canada. *Journal of Air Transport Management 71,* 153--159.

[61]    Chouliaraki, L. (2017). Symbolic bordering: The self-representation of migrants and refugees in digital news. *Popular Communication 15(2)*, 78--94. https://doi.org/10.1 080/15405702.2017.1281415.

[62]    Coletto, M., A. Esuli, C. Lucchese, C. I. Muntean, F. M. Nardini, R. Perego, and C. Renso (2017). Perception of social phenomena through the multidimensional analysis of online social networks. *Online Social Networks and Media 1,* 14--32.

[63]    COM(2005) 390 final (2005). Communication from the commission to the council, the european parliament, the european economic and social committee and the committee of the regions, ''migration and development: some concrete orientations''. https://eur-le x.europa.eu/LexUriServ/LexUriServ.do?uri=COM:2005:0390:FIN:EN:PDF.

[64]    COM(2017) 9 final (2017). Communication from the commission to the european parliament, the council, the european economic and social committee and the committee of the regions, ''building a european data economy''. https://eur-lex.europa.eu/le gal-content/EN/TXT/PDF/?uri=CELEX:52017DC0009&from=EN.

[65]    COM(2018) 232 final (2018). Communication from the commission to the european parliament, the council, the european economic and social committee and the committee of the regions, ''towards a common european data space''. https://eur-lex.europa.e u/legal-content/EN/TXT/PDF/?uri=CELEX:52018DC0232&from=en.

[66]    COM(2020) 66 final (2020). Communication from the commission to the european parliament, the council, the european economic and social committee and the committee of the regions, ''a european strategy for data''. https://eur-lex.europa.eu/legal-co ntent/EN/TXT/PDF/?uri=CELEX:52020DC0066&from=EN.

[67]    COM(2020) 67 final (2020). Communication from the commission to the european parliament, the council, the european economic and social committee and the committee of the regions, ''shaping europe's digital future''. https://eur-lex.europa.eu/legalcontent/EN/TXT/PDF/?uri=CELEX:52020DC0067&from=IT.

[68]    Comito, C. (2018). Human mobility prediction through twitter. *Procedia Computer Science 134,* 129--136. The 15th International Conference on Mobile Systems and Pervasive Computing (MobiSPC 2018) / The 13th International Conference on Future Networks and Communications (FNC-2018) / Affiliated Workshops.

[69]    Constantinos, M., M. Carammia, and T. Wilkin (2020). Using big data to estimate migration push factors from africa. In *Migration in West and North Africa and across the Mediterranean. Trends, Risks, Development and Governance, Geneva,* pp. 98--116. International Organisation on Migration.

[70]    Conte, A. and S. Migali (2019). The role of conflict and organized violence in international forced migration. *Demographic Research 41(14),* 393–424. https://www.demographi c-research.org/volumes/vol41/14/.

[71]    Coppin, J. (2020). The representations of migrants in articles shared on reddit and their related comments. Master thesis, University of Oslo.

[72]    Corbane, C., T. Kemper, S. Freire, C. Louvrier, and M. Pesaresi (2016). Monitoring the Syrian Humanitarian Crisis with the JRC's Global Human Settlement Layer and Night-Time Satellite Data. Number LB-NA-27933-EN-C (print),LB-NA-27933-EN-N (online). Luxembourg (Luxembourg): Publications Office of the European Union.

[73]    Corbane, C., V. Syrris, F. Sabo, P. Politis, M. Melchiorri, M. Pesaresi, P. Soille, and T. Kemper (2021). Convolutional neural networks for global human settlements mapping from sentinel-2 satellite imagery. *Neural Computing and Applications 33(12),* 6697–6720.

[74]    Cortés, U., A. Cortés, D. Garcia-Gasulla, R. Pérez-Arnal, S. Alvarez-Napagao, and E. Alvarez (2021, May). The ethical use of high-performance computing and artificial intelligence: fighting covid-19 at barcelona supercomputing center. *AI and Ethics.*

[75]    Cot, C., G. Cacciapaglia, and F. Sannino (2021, Feb). Mining google and apple mobility data: temporal anatomy for covid-19 social distancing. *Scientific Reports 11(1),* 4150.

[76]    Culora, A., E. Thomas, E. Dufresne, M. Cefalu, C. Fays, and S. Hoorens (2021). *Using social media data to 'nowcast' international migration around the globe.* Santa Monica, CA: RAND Corporation.

[77]    Curry, T., A. Croitoru, A. Crooks, and A. Stefanidis (2019, Mar). Exodus 2.0: crowdsourcing geographical and social trails of mass migration. *Journal of Geographical Systems 21(1),* 161–187.

[78]    Dan, Y. and Z. He (2010). A dynamic model for urban population density estimation using mobile phone location data. In *2010 5th IEEE Conference on Industrial Electronics and Applications*, pp. 1429–1433. IEEE.

[79]    Davis, K. F., A. Bhattachan, P. D'Odorico, and S. Suweis (2018). A universal model for predicting human migration under climate change: examining future sea level rise in Bangladesh. *Environmental Research Letters 13(6),* 064030.

[80]    De Montjoye, Y.-A., S. Gambs, V. Blondel, G. Canright, N. de Cordes, S. Deletaille, K. Engø-Monsen, M. Garcia-Herranz, J. Kendall, C. Kerry, G. Krings, E. Letouzè, M. Luengo-Oroz, N. Oliver, L. Rocher, A. Rutherford, Z. Smoreda, J. Steele, E. Wetter, A. S. Pentland, and L. Bengtsson (2018, Dec). On the privacy-conscientious use of mobile phone data. *Scientific Data 5(1),* 180286.

[81]    De Montjoye, Y.-A., C. A. Hidalgo, M. Verleysen, and V. D. Blondel (2013, Mar). Unique in the crowd: The privacy bounds of human mobility. *Scientific Reports 3(1),* 1376.

[82]    De Sherbinin, A., M. Levy, S. Adamo, C. Aichele, L. Pistolesi, B. Goodrich, T. Srebotnjak, G. Yetman, K. MacManus, L. Razafindrazay, C. Aichele, and V. Mara (2011). *MR4: Estimating net migration by ecosystem and by decade: 1970 2010.* UK Government Office for Science.

[83]    De Sherbinin, A., M. Levy, S. Adamo, K. MacManus, G. Yetman, V. Mara, L. Razafindrazay, B. Goodrich, T. Srebotnjak, C. Aichele, et al. (2012). Migration and risk: net migration in marginal ecosystems and hazardous areas. *Environmental Research Letters 7(4),* 045602.

[84]    Deville, P., C. Linard, S. Martin, M. Gilbert, F. R. Stevens, A. E. Gaughan, V. D. Blondel, and A. J. Tatem (2014). Dynamic population mapping using mobile phone data. *Proceedings of the National Academy of Sciences 111(45),* 15888–15893.

[85]    Dobson, J. E., E. A. Bright, P. R. Coleman, R. C. Durfee, and B. A. Worley (2000). Landscan: a global population database for estimating populations at risk. *Photogrammetric engineering and remote sensing 66(7),* 849–857.

[86]    Douglass, R. W., D. A. Meyer, M. Ram, D. Rideout, and D. Song (2015). High resolution population estimates from telecommunications data. *EPJ Data Science 4,* 1–13.

[87] Du, Z., L. Wang, B. Yang, S. T. Ali, T. Tsang, S. Shan, P. Wu, E. H. Lau, B. Cowling, and L. A. Meyers (2021). Risk for international importations of variant sars-cov-2 originating in the united kingdom. *Emerging Infectious Disease journal 27(5),* 1527.

[88] Dubois, A., E. Zagheni, K. Garimella, and I. Weber (2018). Studying migrant assimilation through facebook interests. In S. Staab, O. Koltsova, and D. I. Ignatov (Eds.), *Social Informatics,* Cham, pp. 51--60. Springer International Publishing.

[89] Dwork, C. (2008). Differential privacy: A survey of results. In *In Theory and Applications of Models of Computation,* pp. 1--19. Springer.

[90] Eagle, N., Y.-A. De Montjoye, and L. M. Bettencourt (2009). Community computing: Comparisons between rural and urban societies using mobile phone data. In *2009 International Conference on Computational Science and Engineering,* Volume 4, pp. 144-150.

[91] Eicher, C. L. and C. A. Brewer (2001). Dasymetric mapping and areal interpolation: Implementation and evaluation. *Cartography and Geographic Information Science 28(2),* 125--138.

[92] Elvidge, C., P. Sutton, T. Ghosh, B. Tuttle, K. Baugh, B. Bhaduri, and E. Bright (2009). Global poverty map derived from satellite data. *Comput. Geosci. 35,* 1652--1660.

[93] Elvidge, C. D., P. C. Sutton, B. T. Tuttle, T. Ghosh, and K. E. Baugh (2009). Global urban mapping based on nighttime lights. In *Global mapping of human settlement, pp. 157--172.* CRC Press.

[94] Emara, N. and Y. Zhang (2021). The non-linear impact of digitization on remittances inflow: Evidence from the brics. *Telecommunications Policy 45(4),* 102112.

[95] EU 2007/826 (2007). Regulation (ec) no 862/2007 of the european parliament and of the council of 11 july 2007 on community statistics on migration and international protection and repealing council regulation (eec) no 311/76 on the compilation of statistics on foreign workers (text with eea relevance).

[96] EU 2016/679 (2016). Regulation (eu) 2016/679 of the european parliament and of the council of 27 april 2016 on the protection of natural persons with regard to the processing of personal data and on the free movement of such data, and repealing directive 95/46/ec (general data protection regulation).

[97] EU 2019/1700 (2019). Regulation (eu) 2019/1700 of the european parliament and of the council of 10 october 2019 establishing a common framework for european statistics relating to persons and households, based on data at individual level collected from samples, amending regulations (ec) no 808/2004, (ec) no 452/2008 and (ec) no 1338/2008 of the european parliament and of the council, and repealing regulation (ec) no 1177/2003 of the european parliament and of the council and council regulation (ec) no 577/98 (text with eea relevance).

[98] European Commission (2018 125 final). Commission staff working document guidance on sharing private sector data in the european data economy accompanying the document communication from the commission to the european parliament, the council, the european economic and social committee and the committee of the regions "towards a common european data space" swd(2018) 125 final.

[99] European Commission (2021, 02). Towards a european strategy on business-togovernment data sharing for the public interest. Technical report, High-Level Expert Group on Business-to-Government Data Sharing.

[100] Evans, J., A. van Donkelaar, R. V. Martin, R. Burnett, D. G. Rainham, N. J. Birkett, and D. Krewski (2013). Estimates of global mortality attributable to particulate air pollution using satellite imagery. *Environmental research 120,* 33--42.

[101] Evans, O. (2018). Connecting the poor: the internet, mobile phones and financial inclusion in Africa. Digital Policy, *Regulation and Governance.*

[102] FAIR, G. (2021). Go fair initiative. http://www.go-fair.org/fair-principles/.

[103] Fantazzini, D., J. Pushchelenko, A. Mironenkov, and A. Kurbatskii (2021). Forecasting internal migration in russia using google trends: Evidence from moscov and saint petersburg. *Forecasting 3(4),* 774--803.

[104] Fatehkia, M., R. Kashyap, and I. Weber (2018). Using facebook ad data to track the global digital gender gap. *World Development 107,* 189--209.

[105] Fatehkia, M., I. Tingzon, A. Orden, S. Sy, V. Sekara, M. Garcia-Herranz, and I. Weber (2020, Jul). Mapping socioeconomic indicators using social media advertising data. *EPJ Data Science 9(1), 22.*

[106] Fekih, M., T. Bellemans, Z. Smoreda, P. Bonnel, A. Furno, and S. Galland (2021). A data-driven approach for origin--destination matrix construction from cellular network signalling data: a case study of lyon region (france). *Transportation 48(4),* 1671--1702.

[107] Files, T. M. (2014). Counting the dead. http://www.themigrantsfiles.com/.

[108] Fiorio, L., G. Abel, J. Cai, E. Zagheni, I. Weber, and G. VinuØ (2017). Using twitter data to estimate the relationship between short-term mobility and long-term migration. In *Proceedings of the 2017 ACM on Web Science Conference, WebSci'17,* New York, NY, USA, pp. 103 110. Association for Computing Machinery.

[109] Fiorio, L., E. Zagheni, G. Abel, J. Hill, G. Pestre, E. LetouzØ, and J. Cai (2021, 02). Analyzing the Effect of Time in Migration Measurement Using Georeferenced Digital Trace Data. *Demography 58(1),* 51--74.

[110] Fischer, C. S. (1982). *To dwell among friends: Personal networks in town and city.* University of chicago Press.

[111] Flowminder Foundation (2021). Population movements following the haiti earthquake on 14 august 2021 and the tropical depression grace, estimated with mobile operator data from digicel haiti: report from 27 august. Technical report, Flowminder Foundation.

[112] Foubert, K. and I. Ruyssen (2021). Leaving terrorism behind? the impact of terrorist attacks on migration intentions around the world. Technical report, UNU-CRIS - WORKING PAPER SERIES. https://cris.unu.edu/sites/cris.unu.edu/files/WP21 .07%20-%20Foubert%20and%20Ruyssen.pdf.

[113] Fox, N., L. J. Graham, F. Eigenbrod, J. M. Bullock, and K. E. Parks (2021). Reddit: A novel data source for cultural ecosystem service studies. *Ecosystem Services 50,* 101331.

[114] Freelon, D. (2018). Computational research in the post-API age. *Political Communication 35(4),* 665--668.

[115] Freire, S., K. MacManus, M. Pesaresi, E. Doxsey-Whitfield, and J. Mills (2016). Development of new open and free multi-temporal global population grids at 250 m resolution. In *Geospatial Data in a Changing World.* Association of Geographic Information Laboratories in Europe (AGILE).

[116] Freire-Vidal, Y., E. Graells-Garrido, and F. Rowe (2021). A framework to understand attitudes towards immigration through twitter. *Applied Sciences 11(20).*

[117] Fries, B., C. A. Guerra, G. A. Garc a, S. L. Wu, J. M. Smith, J. N. M. Oyono, O. T. Donfack, J. O. O. Nfumu, S. I. Hay, D. L. Smith, and A. J. Dolgert (2021, 09). Measuring the accuracy of gridded human population density surfaces: A case study in Bioko Island, Equatorial Guinea. *PLOS ONE 16(9),* 1--20.

[118] Gabrielli, L., E. Deutschmann, F. Natale, E. Recchi, and M. Vespe (2019, Aug). Dissecting global air traffic data to discern different types and trends of transnational human mobility. *EPJ Data Science 8(1), 26.*

[119] Garcia, D., Y. Mitike Kassa, A. Cuevas, M. Cebrian, E. Moro, I. Rahwan, and R. Cuevas (2018). Analyzing gender inequality through large-scale facebook advertising data. *Proceedings of the National Academy of Sciences 115(27),* 6958--6963.

[120] Geboers, M., J.-J. Heine, N. Hidding, J. Wissel, and D. van Zoggel, M.and Simons (2016). Engagement with tragedy in social media, digital methods winter school '16, amsterdam. https://wiki.digitalmethods.net/Dmi/WinterSchool2016EngagementWithTrag edySocialMedia.

[121] Geddes, A., L. Hadj-Abdou, and L. Brumat (2020). *Migration and Mobility in the European Union - 2nd edition.* Bloomsbury Academic.

[122] Ghosh, T., R. L Powell, C. D Elvidge, K. E Baugh, P. C Sutton, and S. Anderson (2010). Shedding light on the global distribution of economic activity. *The Open Geography Journal 3(1)*.

[123] Goldstein, J. S. (1992). A conflict-cooperation scale for weis events data. *The Journal of Conflict Resolution 36(2),* 369--385.

[124] González, M. C., C. A. Hidalgo, and A.-L. Barabási (2008, 06). Understanding individual human mobility patterns. *Nature 453,* 779--782.

[125] Gros, A. and T. Tiecke (2016). Connecting the world with better maps. Technical report, Facebook. https://code.facebook.com/posts/1676452492623525/connectingthe -world-with-better-maps/.

[126] Grow, A., D. Perrotta, E. D. Fava, J. Cimentada, F. Rampazzo, B. S. Gil-Clavel, E. Zagheni, R. D. Flores, I. Ventura, and I. G. Weber (2021). How reliable is Facebook's advertising data for use in social science research? Insights from a cross-national online survey. MPIDR Working Papers WP-2021-006, Max Planck Institute for Demographic Research, Rostock, Germany.

[127] GSMA (2020). The mobile economy 2020 report. Available at https://www.gsma.c om/mobileeconomy/.

[128] Guidry, J., L. Austin, K. Carlyle, K. Freberg, M. Cacciatore, S. Meganck, Y. Jin, and M. Messner (2018). Welcome or not: Comparing #refugee posts on instagram and pinterest. *American Behavioral Scientist 62(4),* 512--531. https://doi.org/10.1177/ 0002764218760369.

[129] Guo, D. and X. Zhu (2014). Origin-destination flow data smoothing and mapping. *IEEE Transactions on Visualization and Computer Graphics 20(12),* 2043--2052.

[130] Hankaew, S., S. Phithakkitnukoon, M. G. Demissie, L. Kattan, Z. Smoreda, and C. Ratti (2019). Inferring and modeling migration flows using mobile phone network data. IEEE Access 7, 164746--164758.

[131] Hawelka, B., I. Sitko, E. Beinat, S. Sobolevsky, P. Kazakopoulos, and C. Ratti (2014). Geo-located twitter as proxy for global mobility patterns. *Cartography and Geographic Information Science 41(3),* 260--271. https://doi.org/10.1080/15230406.2014.89 0072.

[132] Hefley, T. J., K. M. Broms, B. M. Brost, F. E. Buderman, S. L. Kay, H. R. Scharf, and M. B. Hooten (2017). The basis function approach for modeling autocorrelation in ecological data. *Ecology 98(3),* 632---646. https://doi.org/10.1002/ecy.1674/suppinfo.

[133] Henry, S., B. Schoumaker, and C. Beauchemin (2004). The impact of rainfall on the first out-migration: A multi-level event-history analysis in burkina faso. *Population and environment 25(5),* 423--460.

[134] Highfield, T. and T. Leaver (2015). A methodology for mapping instagram hashtags. *First Monday 20(1--5).* http://journals.uic.edu/ojs/index.php/fm/article/view /5563/4195.

[135] Hino, A. and R. A. Fahey (2019). Representing the twittersphere: Archiving a representative sample of twitter data under resource constraints. *International Journal of Information Management 48,* 175 -- 184. http://www.sciencedirect.com/science/article/pii/S0268401218306005.

[136] Hofstede, G., G. J. Hofstede, and M. Minkov (2010). *Cultures and Organizations Software of the Mind: Intercultural Cooperation and its Importance for Survival (3. ed.)* McGraw-Hill.

[137] Hu, T., W. W. Guan, X. Zhu, Y. Shao, L. Liu, J. Du, H. Liu, H. Zhou, J. Wang, B. She, L. Zhang, Z. Li, P. Wang, Y. Tang, R. Hou, Y. Li, D. Sha, Y. Yang, B. Lewis, D. Kakkar, and S. Bao (2020). Building an open resources repository for covid-19 research. *Data and Information Management 4(3),* 130--147.

[138] Huang, Q., Y. Yang, Y. Xu, E. Wang, and K. Zhu (2021). Human origin-destination flow prediction based on large scale mobile signal data. *Wireless Communications and Mobile Computing 2021.*

[139] Huang, X., Z. Li, Y. Jiang, X. Li, and D. Porter (2020, 11). Twitter reveals human mobility dynamics during the covid-19 pandemic. *PLOS ONE 15(11),* 1--21. https://doi.org/10.1371/journal.pone.0241957.

[140] Hughes, C., E. Zagheni, G. J. Abel, A. Sorichetta, A. Wi'sniowski, I. Weber, and A. J. Tatem (2016). Inferring migrations: traditional methods and new approaches based on mobile phone, social media, and other big data: feasibility study on inferring (labour) mobility and migration in the european union from big data and social media data. Technical report, European Union.

[141] Hugo, G. (1996). Environmental concerns and international migration. *International migration review 30(1)*, 105--131.

[142] Iacus, S. and Y. Teocharis (2017). Big data and migration: Limits and perspectives.

[143] Iacus, S. M., G. Porro, S. Salini, and E. Siletti (2020). Controlling for selection bias in social media indicators through official statistics: a proposal. *Journal of Official Statistics 36(2),* 315--338.

[144] Iacus, S. M., C. Santamaria, F. Sermi, S. Spyratos, D. Tarchi, and M. Vespe (2020, Aug). Human mobility and covid-19 initial dynamics. *Nonlinear Dynamics 101(3),* 1901--1919. https://doi.org/10.1007/s11071-020-05854-6.

[145] Iacus, S. M., C. Santamaria, F. Sermi, S. Spyratos, D. Tarchi, and M. Vespe (2021). Mobility functional areas and covid-19 spread. *Transportation.* https://doi.org/10.1007/s11116-021-10234-z

[146] Ilin, C., S. Annan-Phan, X. H. Tai, S. Mehra, S. Hsiang, and J. E. Blumenstock (2021, Jun). Public mobility data enables covid-19 forecasting and management at local and global scales. *Scientific Reports 11(1),* 13531.

[147] Institut Haitien de Statistique et D'Informatique (2009). Population totale, population de 18 ans et plus menages et densites estimes en 2009.

[148] IOM (2016). Mediterranean migrant arrivals reach 358 403; official deaths at sea: 4 913. https://www.iom.int/news/mediterranean-migrant-arrivals-reach-358403-official-deaths-sea-4913.

[149] Isaacman, S., R. Becker, R. CÆceres, S. Kobourov, M. Martonosi, J. Rowland, and A. Varshavsky (2011). Identifying important places in people's lives from cellular network data. In *International conference on pervasive computing,* pp. 133--151. Springer.

[150] Isaacman, S., V. Frias-Martinez, and E. Frias-Martinez (2018). Modeling human migration patterns during drought conditions in La guajira, Colombia. In *Proceedings of the 1st ACM SIGCAS Conference on Computing and Sustainable Societies,* COMPASS '18, New York, NY, USA. Association for Computing Machinery.

[151] Jahani, E., P. Sundsøy, J. Bjelland, L. Bengtsson, and Y. De Montjoye (2017). Improving official statistics in emerging markets using machine learning and mobile phone data. *EPJ Data Science 6.*

[152] Janecek, A., D. Valerio, K. A. Hummel, F. Ricciato, and H. Hlavacs (2015). The cellular network as a sensor: From mobile phone data to real-time road traffic monitoring. *IEEE transactions on intelligent transportation systems 16(5),* 2551--2572.

[153] Jia, S., S. H. Kim, S. V. Nghiem, P. Doherty, and M. C. Kafatos (2020, jul). Patterns of population displacement during mega-fires in california detected using facebook disaster maps. *Environmental Research Letters 15(7),* 074029.

[154] Jiang, Y., X. Huang, and Z. Li (2021). Spatiotemporal patterns of human mobility and its association with land use types during covid-19 in new york city. *ISPRS International Journal of Geo-Information 10(5).*

[155] Jo, A., S.-K. Lee, and J. Kim (2020). Gender gaps in the use of urban space in seoul: Analyzing spatial patterns of temporary populations using mobile phone data. *Sustainability 12(16),* 6481.

[156] Johnson, S., A. Grow, D. Perrotta, T. Theile, H. A. de Valk, and E. Zagheni (2021). Openness to migrate internationally for a job: Evidence from linkedin data. In *International Population Conference,* Hyderabad, India. International Union for the Scientific Study of Population.

[157] Juech, C. (2021). *Building the Field of Data for Good,* pp. 41--53. Cham: Springer International Publishing.

[158] Jurdak, R., K. Zhao, J. Liu, M. AbouJaoude, M. Cameron, and D. Newth (2015, 07). Understanding human mobility from twitter. *PLOS ONE 10(7),* 1--16.

[159] Kalantaryan, S. and S. McMahon (2020). *Covid-19 and Remittances in Africa.* Number KJ-NA-30262-EN-N (online),KJ-NA-30262-EN-C (print). Luxembourg (Luxembourg): Publications Office of the European Union.

[160] Kalantaryan, S. and S. McMahon (2021). *Remittances in North Africa: sources, scale and significance.* Number KJ-NA-30582-EN-N (online). Luxembourg (Luxembourg): Publications Office of the European Union.

[161] Kang, C., Y. Liu, X. Ma, and L. Wu (2012). Towards estimating urban population distributions from mobile call data. *Journal of Urban Technology 19(4),* 3--21.

[162] Karemera, D., V. Oguledo, and B. Davis (2000). A gravity model analysis of international migration to North America. Appl. Econ. 32, 1745--1755.

[163] Kağan Albayrak, M.B., Çağri Özcan, R. Can, and F. Dobruszkes (2020). The determinants of air passenger traffic at turkish airports. *Journal of Air Transport Management 86,* 101818.

[164] Kim, J., A. Sîrbu, F. Giannotti, and L. Gabrielli (2020). Digital footprints of international migration on twitter. In M. R. Berthold, A. Feelders, and G. Krempl (Eds.), *Advances in Intelligent Data Analysis XVIII,* Cham, pp. 274--286. Springer International Publishing.

[165] Kim, J., A. Srbu, G. Rossetti, F. Giannotti, and H. Rapoport (2021). Home and destination attachment: study of cultural integration on twitter. https://arxiv.org/abs/2102.113 98.

[166] Kinyoki, D. K., N.-B. Kandala, S. O. Manda, E. T. Krainski, G.-A. Fuglstad, G. M. Moloney, J. A. Berkley, and A. M. Noor (2016). Assessing comorbidity and correlates of wasting and stunting among children in Somalia using cross-sectional household surveys: 2007 to 2010. *BMJ Open 6(3).*

[167] Kishore, K., V. Jaswal, M. Verma, and V. Koushal (2021, Aug). Exploring the utility of google mobility data during the covid-19 pandemic in india: Digital epidemiological analysis. *JMIR Public Health Surveill 7(8),* e29957.

[168] Klein, A. Z., H. Cai, D. Weissenbacher, L. D. Levine, and G. Gonzalez-Hernandez (2020). A natural language processing pipeline to advance the use of twitter data for digital epidemiology of adverse pregnancy outcomes. *Journal of Biomedical Informatics 112,* 100076. Articles initially published in Journal of Biomedical Informatics: X 5-8, 2020.

[169] Kraemer, M. U. G., R. C. Reiner, O. J. Brady, J. P. Messina, M. Gilbert, D. M. Pigott, D. Yi, K. Johnson, L. Earl, L. B. Marczak, S. Shirude, N. Davis Weaver, D. Bisanzio, T. A. Perkins, S. Lai, X. Lu, P. Jones, G. E. Coelho, R. G. Carvalho, W. Van Bortel, C. Marsboom, G. Hendrickx, F. Schaffner, C. G. Moore, H. H. Nax, L. Bengtsson, E. Wetter, A. J. Tatem, J. S. Brownstein, D. L. Smith, L. Lambrechts, S. Cauchemez, C. Linard, N. R. Faria, O. G. Pybus, T. W. Scott, Q. Liu, H. Yu, G. R. W. Wint, S. I. Hay, and N. Golding (2019, May). Past and future spread of the arbovirus vectors aedes aegypti and aedes albopictus. *Nature Microbiology 4(5),* 854--863.

[170] Kraemer, M. U. G., A. Sadilek, Q. Zhang, N. A. Marchal, G. Tuli, E. L. Cohn, Y. Hswen, T. A. Perkins, D. L. Smith, R. C. Reiner, and J. S. Brownstein (2020, Aug). Mapping global variation in human mobility. *Nature Human Behaviour 4(8),* 800--810.

[171] Krings, G., F. Calabrese, C. Ratti, and V. D. Blondel (2009). Urban gravity: a model for inter-city telecommunication flows. *Journal of Statistical Mechanics: Theory and Experiment 2009(07),* L07003.

[172] Kurita, J., Y. Sugishita, T. Sugawara, and Y. Ohkusa (2021, Feb). Evaluating apple inc mobility trend data related to the covid-19 outbreak in Japan: Statistical analysis. *JMIR Public Health Surveill 7(2),* e20335.

[173] Lai, S., I. I. Bogoch, N. W. Ruktanonchai, A. Watts, X. Lu, W. Yang, H. Yu, K. Khan, and A. J. Tatem (2020). Assessing spread risk of wuhan novel coronavirus within and beyond China, january-april 2020: a travel network-based modelling study. *MedRxiv.*

[174] Lai, S., E. zu Erbach-Schoenberg, C. Pezzulo, N. W. Ruktanonchai, A. Sorichetta, J. Steele, T. Li, C. A. Dooley,

and A. J. Tatem (2019). Exploring the use of mobile phone data for national migration statistics. *Palgrave communications 5(1),* 1--10.

[175] Lamanna, F., M. Lenormand, M. Salas-Olmedo, G. Romanillos, B. Gon alves, and J. Ramasco (2018, 3). Immigrant community integration in world cities. *PLoS ONE 13(3),* 1--19. https://doi.org/10.1371/journal.pone.0191612.

[176] Lanzieri, G. (2019a). An alternative view on the statistical definition of migration. Technical report, United Nations Economic Commission for Europe. https://unece.org/ fileadmin/DAM/stats/documents/ece/ces/ge.10/2019/ mtg2/2.5_Definitions_E urostat.pdf.

[177] Lanzieri, G. (2019b). Towards a single population base in the eu. Technical report, United Nations Economic Commission for Europe. https://unece.org/fileadmin/DAM/ stats/documents/ece/ces/ge.41/2019/mtg1/WP2_Eurostat_Lanzieri.pdf.

[178] Lanzieri, G. (2021). Letter to the editors. *Journal of Official Statistics 37(3),* 543--545.

[179] Latour, B. (2007). Beware, your imagination leaves digital traces. *Times Higher Literary Supplement 6(4).*

[180] Lazer, D., R. Kennedy, G. King, and A. Vespignani (2014). The parable of google flu: Traps in big data analysis. *Science 343(14 March),* 1203--1205.

[181] Lee, D. (2013). CARBayes: An R Package for Bayesian Spatial Modeling with Conditional Autoregressive Priors. *Journal of Statistical Software, 55(13),* 1-24."

[182] Liao, Y., S. Yeh, and G. S. Jeuken (2019, Nov). From individual to collective behaviours: exploring population heterogeneity of human mobility based on social media data. *EPJ Data Science 8(1),* 34.

[183] Lin, A. Y., J. Cranshaw, and S. Counts (2019). Forecasting u.s. domestic migration using internet search queries. In *The World Wide Web Conference*, WWW'19, New York, NY, USA, pp. 1061 1072. Association for Computing Machinery.

[184] Lindgren, F., H. Rue, and J. Lindström (2011). An explicit link between gaussian fields and gaussian markov random fields: the stochastic partial differential equation approach. *Journal of the Royal Statistical Society: Series B (Statistical Methodology) 73(4),* 423--498.

[185] Lo, C. (2001). Modeling the population of China using dmsp operational linescan system nighttime data. *Photogramm. Eng. Remote Sens. 67,* 1037--1047.

[186] Lu, X., L. Bengtsson, and P. Holme (2012). Predictability of population displacement after the 2010 Haiti earthquake. *Proceedings of the National Academy of Sciences 109(29),* 11576--11581.

[187] Lu, X., E. Wetter, N. Bharti, A. J. Tatem, and L. Bengtsson (2013). Approaching the limit of predictability in human mobility. *Scientific reports 3(1),* 1--9.

[188] Lu, X., D. J. Wrathall, P. R. Sundsøy, M. Nadiruzzaman, E. Wetter, A. Iqbal, T. Qureshi, A. Tatem, G. Canright, K. Engł-Monsen, and L. Bengtsson (2016). Unveiling hidden migration and mobility patterns in climate stressed regions: A longitudinal study of six million anonymous mobile phone users in bangladesh. *Global Environmental Change 38,* 1--7.

[189] Luca, M., G. Barlacchi, B. Lepri, and L. Pappalardo (2020). Deep learning for human mobility: a survey on data and models. *ArXiv preprint arXiv:2012.02825.*

[190] Mahabir, R., A. Croitoru, A. Crooks, P. Agouris, and A. Stefanidis (2018, 11). News coverage, digital activism, and geographical saliency: A case study of refugee camps and volunteered geographical information. *PLOS ONE 13(11),* 1--22. https://doi.org/10.1 371/journal.pone.0206825.

[191] Mamei, M., N. Bicocchi, M. Lippi, S. Mariani, and F. Zambonelli (2019). Evaluating origin--destination matrices obtained from cdr data. *Sensors 19(20),* 4470.

[192] McClelland, C. (2006). World event/interaction survey (weis) project, 1966-1978.

[193] McMahon, S., G. Tintori, M. Perez Fernandez, A. Alessandrini, A. Goujon, D. Ghio, T. Petroliagkis, A. Conte, U. Minora, and S. Kalantaryan (2021). *Population exposure and migrations linked to climate change in Africa.* Number KJ-NA-30881-EN-N (online),KJNA-30881-EN-C (print). Luxembourg (Luxembourg): Publications Office of the European Union.

[194] Mejova, Y., I. Weber, and L. Fernandez-Luque (2018, Mar). Online health monitoring using facebook advertisement audience estimates in the united states: Evaluation study. *JMIR Public Health Surveill 4(1),* e30.

[195] Mena, G. E., P. P. Martinez, A. S. Mahmud, P. A. Marquet, C. O. Buckee, and M. Santillana (2021). Socioeconomic status determines covid-19 incidence and related mortality in santiago, chile. Science 372(6545), eabg5298.

[196] Mennis, J. (2003). Generating surface models of population using dasymetric mapping. *The Professional Geographer 55(1),* 31--42.

[167] Mennis, J. (2009). Dasymetric mapping for estimating population in small areas. *Geography Compass 3(2),* 727--745.

[198] Mennis, J. and T. Hultgren (2006). Intelligent dasymetric mapping and its application to areal interpolation. *Cartography and Geographic Information Science 33(3),* 179--194.

[199] Metzler, K., D. A. Kim, N. Allum, and A. Denman (2016). Who is doing computational social science? trends in big data research (white paper).

[200] Migali, S., F. Natale, G. Tintori, S. Kalantaryan, S. Grubanov-Boskovic, M. Scipioni, F. Farinosi, C. Cattaneo, B. Bendandi, M. Follador, S. McMahon, and T. Barbas (2018, 09). International migration drivers. a quantitative assessment of the structural factors shaping migration. Technical report, European Commission.

[201] Milgram, S. (1970). The experience of living in cities: A psychological analysis. In *Annual Meeting of the American Psychological Association., Sep, 1969, Washington, DC, US; This paper is based on an Invited Address presented to the Division of General Psychology at the aforementioned meeting.* American Psychological Association.

[202] Morini, V., L. Pollacci, and G. Rossetti (2021). Toward a standard approach for echo chamber detection: Reddit case study. *Applied Sciences 11(12).*

[203] Mrkic, S. (2021, September). Policy implications of the disruption of the implementation of the 2020 World Population and Housing Census Programme due to the COVID-19 pandemic. https://www.un.org/development/desa/dpad/publication/un-desa-p olicy-brief-118-policy-implications-of-the-disruption-of-the-implement ation-of-the-2020-world-population-and-housing-census-programme-due-to -the-covid-19-pandemic/.

[204] Napierała, J., J. Hilton, J. J. Forster, M. Carammia, and J. Bijak (2021). Toward an early warning system for monitoring asylum-related migration flows in europe. *International Migration Review 0(0),* 01979183211035736.

[205] Newman, J. and G. Matzke (1984). *Population--patterns, Dynamics, and Prospects.* Prentice-Hall.

[206] Noor, A. M., V. A. Alegana, P. W. Gething, A. J. Tatem, and R. W. Snow (2008). Using remotely sensed night-time light as a proxy for poverty in Africa. *Population Health Metrics 6(1),* 1--13.

[207] Nordhaus, W. and X. Chen (2015). A sharper image? Estimates of the precision of nighttime lights as a proxy for economic statistics. *Journal of Economic Geography 15(1),* 217--246.

[208] Nowok, B. and F. Willekens (2011). A probabilistic framework for harmonisation of migration statistics. *Population, Space and Place 17(5),* 521--533.

[209] Nurse, K. (2018). Migration, diasporas, remittances and the sustainable development goals in least developed countries. *Journal of Globalization and Development 9(2).*

[210] Nurse, S. and J. Bijak (2021). Syrian migration to Europe, 2011-21: Data inventory. https://www.baps-project.eu/inventory/data_inventory/.

[211] O'Brien, S. P. (2010). Crisis early warning and decision support: Contemporary approaches and thoughts on future research. *International Studies Review 12(1),* 87--104.

[212] Office for National Statistics (2021). International passenger survey. https://www. ons.gov.uk/surveys/ informationforhouseholdsandindividuals/householdand individualsurveys/internationalpassengersurvey. Accessed: 2021-11-15.

[213] Office of Chief of Staff for the President (2017). Afghanistan population estimation will be renewed. Technical report, Islamic Republic of Afghanistan. ocs.gov.af/english /4505.

[214] Ojala, J., E. Zagheni, F. C. Billari, and I. Weber (2017). Fertility and its meaning: Evidence from search behavior. *CoRR abs/1703.03935*.

[215] Osorio Arjona, J. and J. C. García Palomares (2020). Spatio-temporal mobility and twitter: 3d visualisation of mobility flows. *Journal of Maps 16(1),* 153--160.

[216] Özden, Ç., C. R. Parsons, M. Schiff, and T. L. Walmsley (2011). Where on earth is everybody? the evolution of global bilateral migration 1960--2000. *The World Bank Economic Review 25(1),* 12--56.

[217] Palotti, J., N. Adler, A. Morales-Guzman, J. Villaveces, V. Sekara, M. Garcia Herranz, M. Al-Asad, and I. Weber (2020, 02). Monitoring of the venezuelan exodus through facebook's advertising platform. *PLOS ONE 15(2),* 1--15.

[218] Panczak, R., E. Charles-Edwards, and J. Corcoran (2020). Estimating temporary populations: a systematic review of the empirical literature. *Humanit Soc Sci Commun 6(87).*

[219] Pappalardo, L., L. Ferres, M. Sacasa, C. Cattuto, and L. Bravo (2020). An individual-level ground truth dataset for home location detection. *ArXiv preprint arXiv:2010.08814.*

[220] Pennebaker, J. W., M. E. Francis, and R. J. Booth (2001). *Linguistic Inquiry and Word Count.* Mahwah, NJ: Lawerence Erlbaum Associates.

[221] Pierce, K. (2015). Soi migration data: A new approach: Methodological improvements for soi's united states population migration data, calendar years 2011-2012. *Statistics of Income. SOI Bulletin 35(1).*

[222] Pison, G. (2005). Population observatories as sources of information on mortality in developing countries. *Demographic Research 13,* 301--334.

[223] Pitropakis, N., K. Kokot, D. Gkatzia, R. Ludwiniak, A. Mylonas, and M. Kandias (2020). Monitoring users' behavior: Anti-immigration speech detection on twitter. *Machine Learning and Knowledge Extraction 2(3),* 192--215.

[224] Pollacci, L. (2019). Superdiversity: (big) data analytics at the crossroads of geography, language, and emotions.

[225] Pollacci, L., A. Sîrbu, F. Giannotti, and D. Pedreschi (2021). Measuring the salad bowl: Superdiversity on twitter. See also https://hummingbird-h2020.eu/images/projecto utput/d5-1-eind.pdf.

[226] Pollacci, L., A. Sîrbu, F. Giannotti, D. Pedreschi, C. Lucchese, and C. Muntean (2017). Sentiment spreading: An epidemic model for lexicon-based sentiment analysis on twitter. In F. Esposito, R. Basili, S. Ferilli, and F. A. Lisi (Eds.), AI*IA 2017 *Advances in Artificial Intelligence,* Cham, pp. 114--127. Springer International Publishing.

[227] Porway, J. (2021). Charting the 'data for good' landscape. https://data.org/news/ charting-the-data-for-good-landscape/.

[228] Press, S. (2002). Subjective and Objective Bayesian Statistics: Principles, Models, and Applications, 2nd Edition. *John Wiley & Sons.*

[229] Pérez-Arnal, R., D. Conesa, S. Alvarez-Napagao, T. Suzumura, M. Catal , E. AlvarezLacalle, and D. Garcia-Gasulla (2021). Comparative analysis of geolocation information through mobile-devices under different covid-19 mobility restriction patterns in spain. *ISPRS International Journal of Geo-Information 10(2).*

[230] Pötzschke, S. and M. Braun (2017). Migrant sampling using facebook advertisements: A case study of polish migrants in four european countries. *Social Science Computer Review 35(5),* 633--653.

[231] Pötzschke, S. and B. Wei (2021, Oct). Realizing a global survey of emigrants through facebook and instagram.

[232] Radojevic, R., D. Nguyen, J. Bajec, and F. I. (2020). Visual framing and migrant discourses in social media: The story of idomeni on instagram. In *Understanding Media and Society in the Age of Digitalisation*. Palgrave Macmillan.

[233] Ram Mohan Rao, P., S. Murali Krishna, and A. P. Siva Kumar (2018, Sep). Privacy preservation techniques in big data analytics: a survey. *Journal of Big Data 5(1),* 33.

[234] Rama, D., Y. Mejova, M. Tizzoni, K. Kalimeri, and I. Weber (2020). Facebook ads as a demographic tool to measure the urban-rural divide. In *Proceedings of The Web Conference 2020*, WWW '20, New York, NY, USA, pp. 327--338. Association for Computing Machinery.

[235] Rampazzo, F., J. Bijak, A. Vitali, I. Weber, and E. Zagheni (2021, 11). A Framework for Estimating Migrant Stocks Using Digital Traces and Survey Data: An Application in the United Kingdom. *Demography.* 9578562.

[236] Rampazzo, F., E. Zagheni, I. Weber, M. R. Testa, and F. Billari (2018, Jun.). Mater certa est, pater numquam: What can facebook advertising data tell us about male fertility rates? *Proceedings of the International AAAI Conference on Web and Social Media 12(1).*

[237] Rayer, S. (2018). Estimating the migration of puerto ricans to florida using flight passenger data. *Bureau of Economic and Business Research.* University of Florida.

[238] Recchi, E., E. Deutschmann, and M. Vespe (2019). Estimating transnational human mobility on a global scale. *Robert Schuman Centre for Advanced Studies Research Paper No. RSCAS 30.*

[239] Reis, B. Y. and J. S. Brownstein (2010, 2010). Measuring the impact of health policies using internet search patterns: the case of abortion. *BMC Public Health 10,* 514.

[240] Ricciato, F., G. Lanzieri, A. Wirthmann, and G. Seynaeve (2020). Towards a methodological framework for estimating present population density from mobile network operator data. *Pervasive and Mobile Computing 68,* 101263.

[241] Ricciato, F., P. Widhalm, F. Pantisano, and M. Craglia (2017). Beyond the "singleoperator, cdr-only" paradigm: An interoperable framework for mobile phone network data analyses and population density estimation. *Pervasive and Mobile Computing 35,* 65--82.

[242] Ricciato, F., A. Wirthmann, and M. Hahn (2020). Trusted smart statistics: How new data will change official statistics. *Data & Policy 2, e7.*

[243] Rigall-I-Torrent, R. (2010). Estimating overnight de facto population by forecasting symptomatic variables: an integrated framework. *Journal of Forecasting 29(7),* 635--654.

[244] Righi, A., D. Bianco, and M. Gentile (2021). Using twitter data to study the mood on migration. https://data4migration.org/articles/iom-un-migration-using-big-data-to-forecast-migration/index.html.

[245] Robinson, C. and B. Dilkina (2018). A machine learning approach to modeling human migration. In *Proceedings of the 1st ACM SIGCAS Conference on Computing and Sustainable Societies, pp. 1--8.*

[246] Rowe, F. (2021a, Oct). Big data and human geography.

[247] Rowe, F. (2021b, 04). Using twitter data to monitor immigration sentiment. Technical report, OSF Preprints. 10.31219/osf.io/sf7u4.

[248] Rowe, F., M. Mahony, E. Graells-Garrido, M. Rango, and N. Sievers (2021, Jul). Using twitter to track immigration sentiment during early stages of the covid-19 pandemic. SocArXiv pc3za, Center for Open Science.

[249] Rue, H., S. Martino, and N. Chopin (2009). Approximate bayesian inference for latent gaussian models by using integrated nested laplace approximations. *Journal of the royal statistical society: Series b (statistical methodology) 71(2),* 319--392.

[250] Ruktanonchai, N. W., C. W. Ruktanonchai, J. R. Floyd, and A. J. Tatem (2018). Using google location history data to quantify fine-scale human mobility. *International journal of health geographics 17(1),* 1--13.

[251] Ryan, R., P. Davis-Kean, L. Bode, J. Kr ger, Z. Mneimneh, and L. Singh (2021). Parenting online: Analyzing information provided by parenting-focused twitter accounts. *PsyArXiv.* https://psyarxiv.com/n3teh/.

[252] Sabre (2021). Market intelligence, global demand data. https://www.sabre.com/pr oducts/market-intelligence/. Accessed: 2021-11-15.

[253] Salah, A. A., A. Pentland, B. Lepri, E. LetouzØ, Y.-A. De Montjoye, X. Dong, . Da§delen, and P. Vinck (2019). Introduction to the data for refugees challenge on mobility of syrian refugees in turkey. In *Guide to Mobile Data Analytics in Refugee Scenarios, pp. 3--27.* Springer.

[254] Salganik, M. J. (2017). *Bit by Bit: Social Research in the Digital Age (Open Review Edition ed.).* Princeton, NJ: Princeton University Press.

[255] Sánchez-Querubín, N., , and R. Rogers (2018). Connected routes: Migration studies with digital devices and platforms. *Social Media + Society 4(1),* 2056305118764427. https://doi.org/10.1177/2056305118764427.

[256] Sarker, A., P. Chandrashekar, A. Magge, H. Cai, A. Klein, and G. Gonzalez (2017, Oct). Discovering cohorts of pregnant women from social media for safety surveillance and analysis. *J Med Internet Res 19(10),* e361. https://doi.org/10.2196/jmir.8164.

[257] Scettri, G. (2019). Studying national and international migration flows with twitter data. Technical report, SIC-UIB - Instituto de F sica Interdisciplinar y Sistemas Complejos (IFISC). http://hdl.handle.net/10261/189587.

[258] Schimak, G., D. Havlik, and J. Pielorz (2015). Crowdsourcing in crisis and disaster management challenges and considerations. In *Environmental Software Systems. Infrastructures, Services and Applications. ISESS 2015. IFIP Advances in Information and Communication Technology,* pp. 139--144.

[259] Sedda, L., A. J. Tatem, D. W. Morley, P. M. Atkinson, N. A. Wardrop, C. Pezzulo, A. Sorichetta, J. Kuleszo, and D. J. Rogers (2015, 02). Poverty, health and satellite-derived vegetation indices: their inter-spatial relationship in West Africa. *International Health 7(2),* 99--106.

[260] Simmel, G. (2012). *The metropolis and mental life.* Routledge.

[261] Singh, L., L. Wahedi, Y. Wang, Y. Wei, C. Kirov, S. Martin, K. Donato, Y. Liu, and K. Kawintiranon (2019). Blending noisy social media signals with traditional movement variables to predict forced migration. In *Proceedings of the 25th ACM SIGKDD International Conference on Knowledge Discovery & Data Mining, New York, NY, USA,* pp. 1975 1983. Association for Computing Machinery.

[262] Sîrbu, A., G. Andrienko, N. Andrienko, C. Boldrini, M. Conti, F. Giannotti, R. Guidotti, S. Bertoli, J. Kim, C. I. Muntean, et al. (2021). Human migration: the big data perspective. *International Journal of Data Science and Analytics 11(4),* 341--360.

[263] Smith, A., P. D. Bates, O. Wing, C. Sampson, N. Quinn, and J. Neal (2019, Apr). New estimates of flood exposure in developing countries using high-resolution population data. *Nature Communications 10(1),* 1814.

[264] Smith, M. B., J. K. Blakemore, J. R. Ho, and J. A. Grifo (2021). Making it (net)work: a social network analysis of fertility in twitter before and during the covid-19 pandemic. *F&S Reports.*

[265] Snoeijer, B. T., M. Burger, S. Sun, R. J. B. Dobson, and A. A. Folarin (2021, May). Measuring the effect of non-pharmaceutical interventions (npis) on mobility during the covid-19 pandemic using global mobility data. *Npj Digital Medicine 4(1),* 81.

[266] Sobotka, T., M. Winkler-Dworak, M. R. Testa, W. Lutz, D. Philipov, H. Engelhardt, and R. Gisser (2005, January). Monthly Estimates of the Quantum of Fertility: Towards a Fertility Monitoring System in Austria. VID Working Papers 0501, Vienna Institute of Demography (VID) of the Austrian Academy of Sciences in Vienna.

[267] Solaimani, M., S. Salam, A. M. Mustafa, L. Khan, P. T. Brandt, and B. Thuraisingham (2016). Near real-time atrocity event coding. In *2016 IEEE Conference on Intelligence and Security Informatics (ISI),* pp. 139--144.

[268] Song, C., Z. Qu, N. Blumm, and A.-L. Barabási (2010). Limits of predictability in human mobility. Science 327(5968), 1018--1021.

[269] Sorichetta, A., G. M. Hornby, F. R. Stevens, A. E. Gaughan, C. Linard, and A. J. Tatem (2015). High-resolution gridded population datasets for Latin America and the Caribbean in 2010, 2015, and 2020. *Scientific data 2(1),* 1--12.

[270] Spelta, A., A. Flori, F. Pierri, G. Bonaccorsi, and F. Pammolli (2020, Oct). After the lockdown: simulating mobility, public health and economic recovery scenarios. *Scientific Reports 10(1),* 16950.

[271] Spyratos, S., M. Lutz, and F. Pantisano (2014). Characteristics of citizen-contributed geographic information. In J. Huerta, S. Schade, and C. Granell (Eds.), *Connecting a Digital Europe through Location and Place.* Proceedings of the AGILE'2014 International Conference on Geographic Information Science, Castell n, June 3-6, 2014.

[272] Spyratos, S. and D. Stathakis (2018). Evaluating the services and facilities of european cities using crowdsourced place data. *Environment and Planning B: Urban Analytics and City Science 45(4),* 733--750.

[273] Spyratos, S., D. Stathakis, M. Lutz, and C. Tsinaraki (2017). Using foursquare place data for estimating building block use. *Environment and Planning B: Urban Analytics and City Science 44(4),* 693--717.

[274] Spyratos, S., M. Vespe, F. Natale, S. M. Iacus, and C. Santamaria (2020, 09). Explaining the travelling behaviour of migrants using facebook audience estimates. *PLOS ONE 15(9),* 1--16.

[275] Spyratos, S., M. Vespe, F. Natale, W. Ingmar, E. Zagheni, and M. Rango (2018). *Migration Data using Social Media: a European Perspective.* Number KJ-NA-29273-EN-N. Luxembourg (Luxembourg): Publications Office of the European Union.

[276] Spyratos, S., M. Vespe, F. Natale, I. Weber, E. Zagheni, and M. Rango (2019, 10). Quantifying international human mobility patterns using facebook network data. *PLOS ONE 14(10),* 1--22.

[277] State, B., M. Rodriguez, D. Helbing, and E. Zagheni (2014). Migration of professionals to the u.s. In L. Aiello and D. McFarland (Eds.), *Social Informatics. SocInfo 2014. Lecture Notes in Computer Science,* Volume 8851. Cham: Springer.

[278] Stathakis, D. and P. Baltas (2018). Seasonal population estimates based on night-time lights. *Computers, Environment and Urban Systems 68,* 133--141.

[279] Steele, J. E., P. R. Sundsøy, C. Pezzulo, V. A. Alegana, T. J. Bird, J. Blumenstock, J. Bjelland, K. Engø-Monsen, Y. De Montjoye, A. Iqbal, K. Hadiuzzaman, X. Lu, E. Wetter, A. Tatem, and L. Bengtsson (2017). Mapping poverty using mobile phone and satellite data. *Journal of The Royal Society Interface 14.* https://doi.org/10.1098/rsif.2016.0690.

[280] Stevens, F. R., A. E. Gaughan, C. Linard, and A. J. Tatem (2015, 02). Disaggregating census data for population mapping using random forests with remotely-sensed and ancillary data. *PLOS ONE 10(2),* 1--22.

[281] Suleimenova, D., D. Bell, and D. Groen (2017, Oct). A generalized simulation development approach for predicting refugee destinations. *Scientific Reports 7(1),* 13377. https://doi.org/10.1038/s41598-017-13828-9.

[282] Sulis, E., C. Bosco, V. Patti, M. Lai, D. I. H. Far as, L. Mencarini, M. Mozzachiodi, and D. Vignoli (2016). Subjective well-being and social media. a semantically annotated twitter corpus on fertility and parenthood. In *CLiC-it/ EVALITA.*

[283] Sundsøy, P. (2016). Can mobile usage predict illiteracy in a developing country? Technical report, *ArXiV.* https://arxiv.org/abs/1607.01337.

[284] Sundsøy, P., J. Bjelland, B. Reme, A. Iqbal, and E. Jahani (2016a). Deep learning applied to mobile phone data for individual income classification. In *Proceedings of the 2016 International Conference on Artificial Intelligence: Technologies and Applications,* pp. 96--99. https://dx.doi.org/10.2991/icaita-16.2016.24.

[285] Sundsøy, P., J. Bjelland, B.-A. Reme, E. Jahani, E. Wetter, and L. Bengtsson (2016). Estimating individual employment status using mobile phone network data. Technical report, *ArXiV*. https://arxiv.org/abs/1612.03870.

[286] Sutton, P., C. Elvidge, and T. Obremski (2003). Building and evaluating models to estimate ambient population density. *Photogramm. Eng. Remote Sens 69,* 545--553.

[287] Sutton, P., D. Roberts, C. Elvidge, and K. Baugh (2001). Census from heaven: An estimate of the global human population using night-time satellite imagery. *International Journal of Remote Sensing 22(16),* 3061--3076.

[288] Tacoli, C. (2009). Crisis or adaptation? Migration and climate change in a context of high mobility. *Environment and urbanization 21(2),* 513--525.

[289] Tank, E. P. T. (2017, October). Secondary movements of asylum-seekers in the eu asylum system. Technical report, European Parlament. https://www.europarl.europa. eu/thinktank/en/document.html?reference=EPRS_ BRI(2017)608728.

[290] Tatem, A., P. Gething, C. Pezzulo, D. Weiss, and S. Bhatt (2014). Development of high-resolution gridded poverty surfaces.

[291] Tatem, A. J., J. Campbell, M. Guerra-Arias, L. De Bernis, A. Moran, and Z. Matthews (2014). Mapping for maternal and newborn health: the distributions of women of childbearing age, pregnancies and births. *International journal of health geographics 13(1),* 1--11.

[292] Tatem, A. J., A. M. Noor, and S. I. Hay (2004). Defining approaches to settlement mapping for public health management in kenya using medium spatial resolution satellite imagery. *Remote Sensing of Environment 93(1-2),* 42--52.

[293] Tatem, A. J., A. M. Noor, and S. I. Hay (2005). Assessing the accuracy of satellite derived global and national urban maps in Kenya. *Remote sensing of environment 96(1),* 87--97.

[294] Thomee, B., D. A. Shamma, G. Friedland, B. Elizalde, K. Ni, D. Poland, D. Borth, and L.-J. Li (2016, January). Yfcc100m: The new data in multimedia research. *Commun. ACM 59(2),* 64--73.

[295] Tjaden, J. (2021, Dec). Measuring migration 2.0: a review of digital data sources. Comparative Migration Studies 9(1), 59.

[296] Tjaden, J., A. Arau, M. Nuermaimaiti, I. Cetin, E. Acostamadiedo, and M. Rango (2021). Using"big data" to forecast migration. https://data4migration.org/articles/iomun-migration-using-big-data-to-forecast-migration/index.html.

[297] Tobler, W., U. Deichmann, J. Gottsegen, and K. Maloy (1997). World population in a grid of spherical quadrilaterals. *International Journal of Population Geography 3(3),* 203--225.

[298] Tonnies, F. and C. P. Loomis (2002). *Community and society*. Courier Corporation.

[299] UNECE Task Force on Migration Statistics (2021). Use of new data sources for measuring international migration. Technical report, United Nations Economic Commission for Europe. https://statswiki.unece.org/ download/attachments/320996092/Use %20of%20new%20data%20sources%20for%20measuring%20 international%20migra tion%20-%20UNECE%20report%2021%20Oct.docx?version=2&modificationDate=16 34809274334&api=v2.

[300] UNESCAP (2021a). Big data for population and social statistics. Technical report, United Nations - Economic and Social Commission for Asia and the Pacific.

[301] UNESCAP (2021b). Report of the expert group meeting on the uses of big data for official statistics: Data governance and partnership models. Technical report, United Nations - Economic and Social Commission for Asia and the Pacific.

[302] United Nations Economic Commission for Europe (2019). Guidance on data integration for measuring migration.

[303]  Valerio, D., A. D'Alconzo, F. Ricciato, and W. Wiedermann (2009). Exploiting cellular networks for road traffic estimation: A survey and a research roadmap. In *VTC Spring 2009-IEEE 69th Vehicular Technology Conference, pp. 1--5. IEEE*.

[304]  Verhulst, S. G. (2021). The Value of Data and Data Collaboratives for Good: A Roadmap for Philanthropies to Facilitate Systems Change Through Data, pp. 9--27. Cham: Springer International Publishing.

[305]  Vertovec, S. (2007). Super-diversity and its implications. *Ethnic and Racial Studies 30(6),* 1024--1054. https://doi.org/10.1080/01419870701599465.

[306]  Vespe, M. and Santamaria Serna, C., A. Zampieri, D. Ivan, P. Juchno, G. Lanzieri, A. Wronski, G. Senchea-Badea, F. Roche, A. Gemma, A. Sola, P. Busiakiewicz, L. Aujean, and M. Rango (2018). *Towards an EU Policy on Migration Data.* Number KJ-NA-29351-EN-N (online). Luxembourg (Luxembourg): Publications Office of the European Union.

[307]  Vespe, M., S. M. Iacus, C. Santamaria, F. Sermi, and S. Spyratos (2021). On the use of data from multiple mobile network operators in europe to fight covid-19. *Data & Policy 3,* e8.

[308]  Vespe, M. and M. Rango (2017, 11). Big Data and alternative data sources on migration: From case-studies to policy support. Technical report, European Commission. https://www.researchgate.net/publication/323129786_Big_Data_and_alterna tive_data_sources_on_migration_from_case-studies_to_policy_support_-_S ummary_report.

[309]  Šćepanović, S., I. Mishkovski, P. Hui, J. Nurminen, and A. Ylä-Jääski (2015). Mobile phone call data as a regional socio-economic proxy indicator. *PLoS ONE 10(4).*

[310]  Wagner, Z., S. Heft-Neal, Z. A. Bhutta, R. E. Black, M. Burke, and E. Bendavid (2018). Armed conflict and child mortality in Africa: a geospatial analysis. *The Lancet 392(10150),* 857--865.

[311]  Wallensteen, P. (2011). *Peace research: Theory and practice.* London: Routledge.

[312]  Wardrop, N., W. Jochem, T. Bird, H. Chamberlain, D. Clarke, D. Kerr, L. Bengtsson, S. Juran, V. Seaman, and A. Tatem (2018). Spatially disaggregated population estimates in the absence of national population and housing census data. *Proceedings of the National Academy of Sciences 115(14),* 3529--3537.

[313]  Watmough, G. R., P. M. Atkinson, and C. W. Hutton (2013). Exploring the links between census andenvironment using remotely sensed satellite sensor imagery. *Journal of Land Use Science 8(3),* 284--303.

[314]  Weber, E. M., V. Y. Seaman, R. N. Stewart, T. J. Bird, A. J. Tatem, J. J. McKee, B. L. Bhaduri, J. J. Moehl, and A. E. Reith (2018). Census-independent population mapping in northern Nigeria. *Remote sensing of environment 204,* 786--798.

[315]  Wheeler, J. O. (2005). Geography. In K. Kempf-Leonard (Ed.), *Encyclopedia of Social Measurement,* pp. 115--123. New York: Elsevier.

[316]  White, M. (1986). Segregation and diversity measures in population distribution. *Population Index 52(2),* 198-221. https://doi.org/10.2307/3644339.

[317]  White, M. J. and D. P. Lindstrom (2005). Internal migration. In D. L. Poston and M. Micklin (Eds.), *Handbook of Population,* pp. 311--346. Boston, MA: Springer US. https://doi.org/10.1007/0-387-23106-4_12.

[318]  Wilson, R., E. Zu Erbach-Schoenberg, M. Albert, D. Power, S. Tudge, M. Gonzalez, S. Guthrie, H. Chamberlain, C. Brooks, C. Hughes, l. Pitonakova, C. Buckee, X. Lu, E. Wetter, A. Tatem, and L. Bengtsson (2016). Rapid and near real-time assessments of population displacement using mobile phone data following disasters: The 2015 nepal earthquake. *Plos currents 8.*

[319]  Wirth, L. (1938). Urbanism as a way of life. *American journal of sociology 44(1),* 1--24.

[320]  Wladyka, D. (2013, October). The queries to google search as predictors of migration flows from latin america to spain. Technical report, University of Texas Rio Grande Valley. https://core.ac.uk/download/pdf/294832951.pdf.

[321]  World Vision Water Team (2021, Jun). The geography of clean water for all.

[322]  Xu, P., M. Dredze, and D. A. Broniatowski (2020, Dec). The twitter social mobility index: Measuring social distancing practices with geolocated tweets. *J Med Internet Res 22(12),* e21499. https://doi.org/10.2196/21499.

[323]  Yilmazkuday, H. (2021). Stay-at-home works to fight against covid-19: International evidence from google mobility data. *Journal of Human Behavior in the Social Environment 31(1-4),* 210--220.

[324]  Yu, X., A. L. Stuart, Y. Liu, C. E. Ivey, A. G. Russell, H. Kan, L. R. Henneman, S. E. Sarnat, S. Hasan, A. Sadmani, et al. (2019). On the accuracy and potential of google maps location history data to characterize individual mobility for air pollution health studies. *Environmental pollution 252,* 924--930.

[325]  Yuan, Y. and M. Medel (2016, 5). Characterizing international travel behavior from geotagged photos: A case study of flickr. *PLOS ONE 13(5),* e0154885. https://doi.org/10.1371/journal.pone.0154885.

[326]  Zagatti, G. A., M. Gonzalez, P. Avner, N. Lozano-Gracia, C. J. Brooks, M. Albert, and et al. (2018). A trip to work: Estimation of origin and destination of commuting patterns in the main metropolitan regions of Haiti using CDR. *Development Engineering 3,* 133--165.

[327]  Zagheni, E. and I. Weber (2015, Jan). Demographic research with non-representative internet data. *International Journal of Manpower 36(1),* 13--25.

[328]  Zagheni, E., I. Weber, and K. Gummadi (2017). Leveraging facebook's advertising platform to monitor stocks of migrants. *Population and Development Review 43(4),* 721--734.

[329]  Zang, H. and J. Bolot (2011). Anonymization of location data does not work: a largescale measurement study. In *Proceedings of the 17th Annual International Conference on Mobile Computing and Networking, MobiCom '11, New York, NY, USA,* pp. 145--156. Association for Computing Machinery.

# LIST OF ABBREVIATIONS AND DEFINITIONS

**ACLED** Armed Conflict Location & Event Data Project ........ 57
**API** Application Programming Interface ........ 35
**ANN** Artificial Neural Network ........ 59
**AWS** Amazon Web Services ........ 13
**BD4M** Big Data for Migration Alliance ........ 14
**B2B** Business to Business ........ 101
**B2C** Business to Consumer ........ 102
**B2G** Business to Government ........ 85
**CAMEO** Conflict and Mediation Event Observations ........ 53
**CDC** Centers for Disease Control and Prevention ........ 8
**CDR** Call Detail Record ........ 75
**CIESIN** Center for International Earth Science Information Network ........ 94
**CLINE** Cline Center Historical Phoenix Event Data ........ 58
**COPDAB** Conflict and Peace Data Bank ........ 57
**CPR** Control Plane Record ........ 87
**data4sdgs** The Global Partnership for Sustainable Development Data ........ 14
**DGA** Data Governance Act ........ 102
**D4R** Data for Refugees ........ 76
**EASO** European Asylum Support Office ........ 10
**EC** European Commission ........ 7
**EFTA** European Free Trade Association ........ 8
**EMN** European Migration Network ........ 8
**ESOP** European statistics on population ........ 5
**EU** European Union ........ 5
**EVI** Enhanced Vegetation Index ........ 91
**FRONTEX** European Border and Coast Guard Agency ........ 63
**FDI** Foreign direct investment ........ 69
**GBD** Global Burden of Disease ........ 8
**GDELT** Global Database of Events, Language, and Tone ........ 57
**GDP** Gross Domestic Product ........ 5
**GeoSemAP** Geospatial Semantic Array Programming ........ 59
**GHSL** Global Human Settlement Layer ........ 95
**GIS** Geographic Information System ........ 14
**GMDAC** Global Migration Data Analysis Centre ........ 11
**GLH** Google location history ........ 55
**GovLab** Governance Lab ........ 15
**GPW** Gridded Population of the World ........ 92
**GSMA** Global System for Mobile Communications Association ........ 15
**GPS** Global Positioning System ........ 73
**GTD** The Global Terrorism Database ........ 57
**GDPR** General Data Protection Regulation ........ 3
**H2O2O** Horizon 2020 ........ 15
**HMD** Human Mortality Database ........ 8
**HRSL** High Resolution Settlement Layers ........ 13
**ICEWS** Integrated Crisis Early Warning System ........ 58
**IDP** Internally Displaced Person ........ 10
**ILO** International Labour Organisation ........ 10
**INED** French Institute for Demographic Studies ........ 8
**INLA** Integrated Nested Laplace Approximations ........ 94

# LIST OF FIGURES

# LIST OF TABLES

## The European Commission's science and knowledge service
Joint Research Centre

### JRC Mission
As the science and knowledge service of the European Commission, the Joint Research Centre's mission is to support EU policies with independent evidence throughout the whole policy cycle.

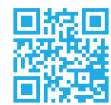

### EU Science Hub
ec.europa.eu/jrc

@EU_ScienceHub

EU Science Hub – Joint Research Centre

EU Science, Research and Innovation

EU Science Hub

EU Science

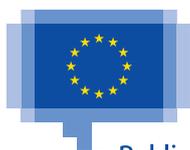

Publications Office
of the European Union